\documentclass[10pt, oneside]{article}
\usepackage{subscript} 
\usepackage{hyperref}
\usepackage{enumitem} 
\usepackage{easylist}  	
\usepackage{graphicx}			
\usepackage{amssymb}
\usepackage[usenames]{color}
\usepackage{pst-grad} 
\usepackage{url}
\usepackage{wrapfig}
\usepackage{float} 
\usepackage[abs]{overpic} 
\newenvironment{s-enumerate}{
\begin{enumerate}
  \setlength{\itemsep}{1pt}
  \setlength{\parskip}{0pt}
  \setlength{\parsep}{0pt}
}{\end{enumerate}}

\newenvironment{s-itemize}{
\begin{itemize}
  \setlength{\itemsep}{1pt}
  \setlength{\parskip}{0pt}
  \setlength{\parsep}{0pt}
}{\end{itemize}}

\let\OLDthebibliography\thebibliography
\renewcommand\thebibliography[1]{
  \OLDthebibliography{#1}
  \setlength{\parskip}{.5ex}
  \setlength{\itemsep}{0pt plus 0.3ex}
}
\title{Pulsing dynamics in randomly wired\\ glider cellular automata}
\author{Andrew Wuensche%
\thanks{andy@ddlab.org,  \url{http://www.ddlab.org}}%
\hspace{2ex}{\it \small  Discrete Dynamics Lab.}\\
Edward Coxon%
\thanks{edward.coxon@act.gov.au}%
\hspace{2ex}{\it \small Dept. of Anaethesia and Pain Medicine,}\\
{\it \small The Canberra Hospital, ACT, Australia.}
}

\date{}	

\begin{document}

\maketitle

\vspace{-3ex}
\begin{abstract}

\noindent Sustained rhythmic oscillations, pulsing dynamics, emerge
spontaneously when the local connection scheme is randomised in
3-value cellular automata that feature``glider'' dynamics.  Time-plots
of pulsing measures maintain a distinct waveform for each glider rule,
and scatter plots of entropy/density and the density return-map show
unique signatures, which have the characteristics of chaotic strange
attractors.  We present case studies, possible mechanisms, and
implications for oscillatory networks in biology.

\end{abstract}

\begin{center}
  {\it keywords:  cellular automata, glider dynamics, random wiring, pulsing, bio-oscillations,
    emergence, chaos, complexity, strange attractor, heartbeat, sympathetic centre, central pattern generator}
\end{center}

\section{Introduction}
\label{Introduction}

\noindent Arguably the most interesting manifestation of cellular
automata dynamics is the emergence of mobile (and stable)
configurations known as particles or gliders which interact by
collisions, possibly making compound emergent structures such as
glider-guns in an open ended hierarchy --- components which can
sometimes be rearranged to achieve universal
computation\cite{Adamatzky2002,Gardner1970}. Glider dynamics
arise within rare ``complex'' rules, which also include dynamics
apart from gliders, for example, dynamic patches, blinkers,
or mobile boundaries between domains. 
Otherwise the dynamics and rule
types, broadly speaking, are either ``ordered'' or ``disordered'' 
judged by subjective impressions of space-time patterns, but
also by objective measures such as input-entropy and its
variability\cite{Wuensche99}, basin of attraction
topology\cite{Wuensche92}, and Derrida
plots\cite{Derrida,Wuensche2016}.  By any evaluation, 
disorder comprises the vast majority of large rule-spaces.  

\enlargethispage{5ex}
We pose the question: while preserving a homogeneous rule, what kind
of dynamics would result if the regular local neighborhood connections
(the wiring) of classical CA are randomised?  --- an experiment
readily implemented in DDLab\cite{Wuensche2016,Wuensche-DDLab}, with
its functionality for toggling between regular and random wiring
on-the-fly, and where random wiring can be fully random or confined in
a local zone, or even re-randomised at each time-step.

The results of these ``random wiring'' experiments reveal a novel and
remarkable phenomenon --- for 3-value, k-totalistic, ``glider'' CA
rules, sustained rhythmic oscillations, pulsing behaviour, is the
inevitable outcome. The pulsing is obvious to the subjective eye when
observing space-time patterns, but is also characterised by objective
measures: the density of each value across the network, and the
collective input-entropy.  Time-plots of pulsing measures for each
glider rule maintain a particular
wavelength ($wl$), wave-height ($wh$, twice amplitude), 
and waveform (its shape or phase),
and scatter plots of entropy/density\cite{Wuensche99} and 
the density return-map\cite{Wuensche2016} show distinct signatures, which have 
the characteristics of chaotic strange attractors. 
We will use the term ``waveform'' to sum up these pulsing measures, and
the ``CA pulsing model'' for the system itself.
We demonstrate pulsing when the wiring is
fully (and sometimes partly) randomised. Pulsing is robust to
re-randomised wiring at each time-step, to noise, to boundary
conditions, and to asynchronous or sequential updating.  When random
wiring is confined in a relatively small local zone, spiral density
waves, reminiscent of reaction-diffusion, can emerge in a large enough
system, so local pulsing is still present (figure~\ref{spiral waves})
as waves sweep over local areas of the lattice.
 
Experiment shows that pulsing does not occur for ordered or disordered rules, 
or for complex rules that do not feature well defined gliders. Pulsing
is not discernible for glider rules in {\it binary} CA, such as the 1D rule
110, the 2D Game-of-Life\cite{Gardner1970}, or other
binary rules that support gliders and
glider-guns\cite{Gomez2015,Gomez2017,Gomez2018,Sapin2010}.
We can find large samples of complex rules by classifying
rule-space automatically according to the variability of
input-entropy\cite{Wuensche99,Wuensche05,Wuensche-DDLab}. Within these samples
a significant proportion are glider rules.


We focus on 3-value k-totalistic\footnote{Non-totalistic 3-value
  glider rules and 4-value k-totalistic glider rules, which are harder
  to find, will be examined in due course.} glider CA on a 2D
hexagonal lattice (and some extension to 3D) with
neighbourhoods\footnote{The CA pulsing model has also been
  demonstrated for neighbourhoods of 4 and 5.}  of 6 or 7, including
two well known CA rules that have been studied in depth, the Beehive
rule\cite{Adamatzky&Wuensche&Cosello2006,Wuensche05,beehivewebpage},
and the Spiral
rule\cite{Adamatzky&Wuensche2006,Wuensche&Adamatzky2006,spiralwebpage}.
The CA pulsing model is interesting in its own right, and may also
help to understanding and model oscillatory networks in biology.  We
address the questions that arise about possible mechanisms, thought
this paper is primarily concerned with presenting and documenting the
phenomena.

\enlargethispage{3ex}
The paper is organised as follows:
Section~\ref{CA and random wiring} describes CA and random wiring.
Section~\ref{k-totalistic rules} defines 3-value k-totalistic rules.
Section~\ref{The input-frequency and input-entropy} defines input-frequency
and input-entropy.
Section~\ref{Pulsing case studies} presents detailed pulsing case
studies, including different aspects of the waveform.
Sections~\mbox{\ref{Localised random wiring} --- \ref{3D systems}} examine the consequences
of freeing one wire from localised neighborhoods, including 3D systems.
In sections~\mbox{\ref{k-totalistic rules as reaction-diffusion systems} --- 
\ref{Questions on the pulsing mechanism}}
we discuss reaction-diffusion, asynchronous and noisy updating,
and possible pulsing mechanisms.
In sections~\ref{Relevance to bio-oscillations} and
\ref{Modeling to bio-oscillations}
we discuss the implications for bio-oscillations, ubiquitous
at many time/size scales in biology, and for modeling
oscillatory behaviour in mammalian tissue such as the heart and central nervous system.
\clearpage

\section{CA and random wiring}
\label{CA and random wiring}

\vspace{-2ex}
\begin{figure}[htb]
\begin{center}
\begin{minipage}[c]{.45\linewidth}
\begin{minipage}[c]{.25\linewidth} 
\includegraphics[width=1\linewidth,bb=94 136 168 206, clip=]{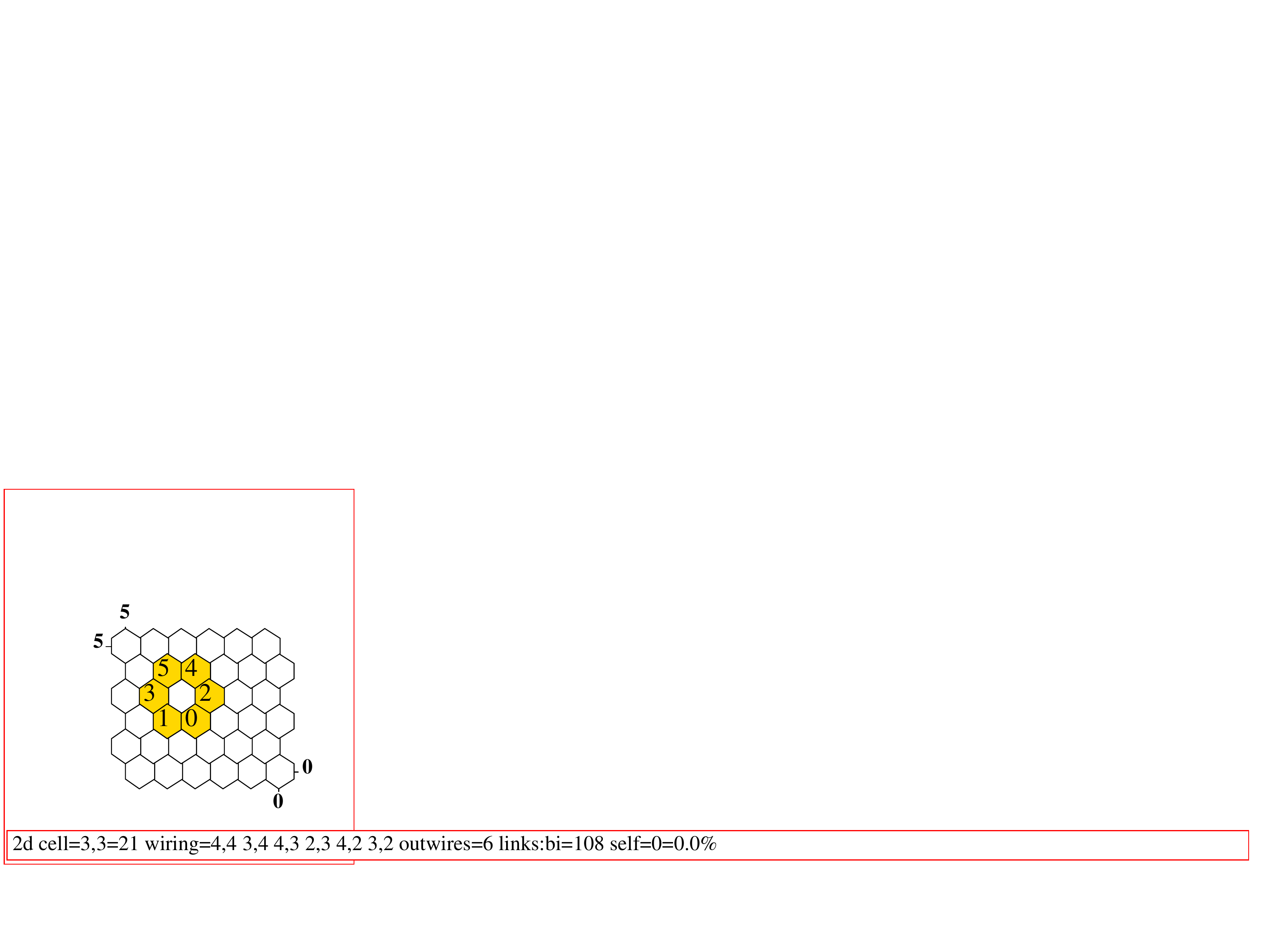}\\[-5ex]
\begin{center}$k$=6\end{center}
\end{minipage}
\hfill
\begin{minipage}[c]{.25\linewidth}
\includegraphics[width=1\linewidth,bb=94 136 168 206, clip=]{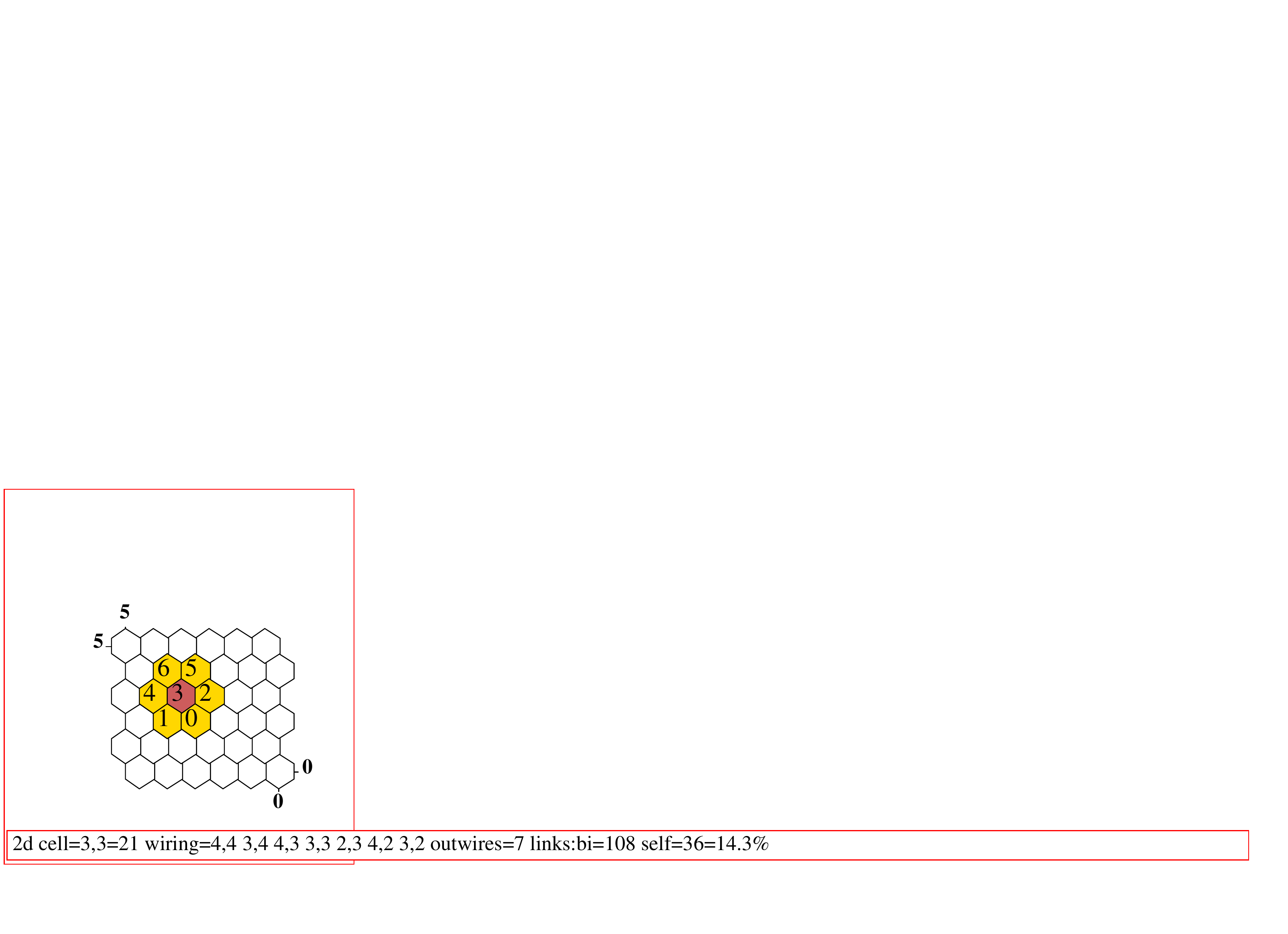}\\[-5ex]
\begin{center}$k$=7\end{center}
\end{minipage}
\end{minipage}
\end{center}
\vspace{-3ex}
\caption[The (pseudo)-neighborhood template 2D]
{\textsf{The (pseudo)-neighborhood template for hexagonal 2D CA, $k$=6 and $k$=7,
with template cells numbered as in DDLab
(for 3D see figure~\ref{3D wiring})}}
\label{The (pseudo)-neighborhood template 2D}
\vspace{-3ex}
\end{figure}
\begin{figure}[h]
\begin{center}
\begin{minipage}[c]{.95\linewidth}
\begin{minipage}[c]{.3\linewidth} 
\includegraphics[width=1\linewidth,bb=58 88 261 274, clip=]{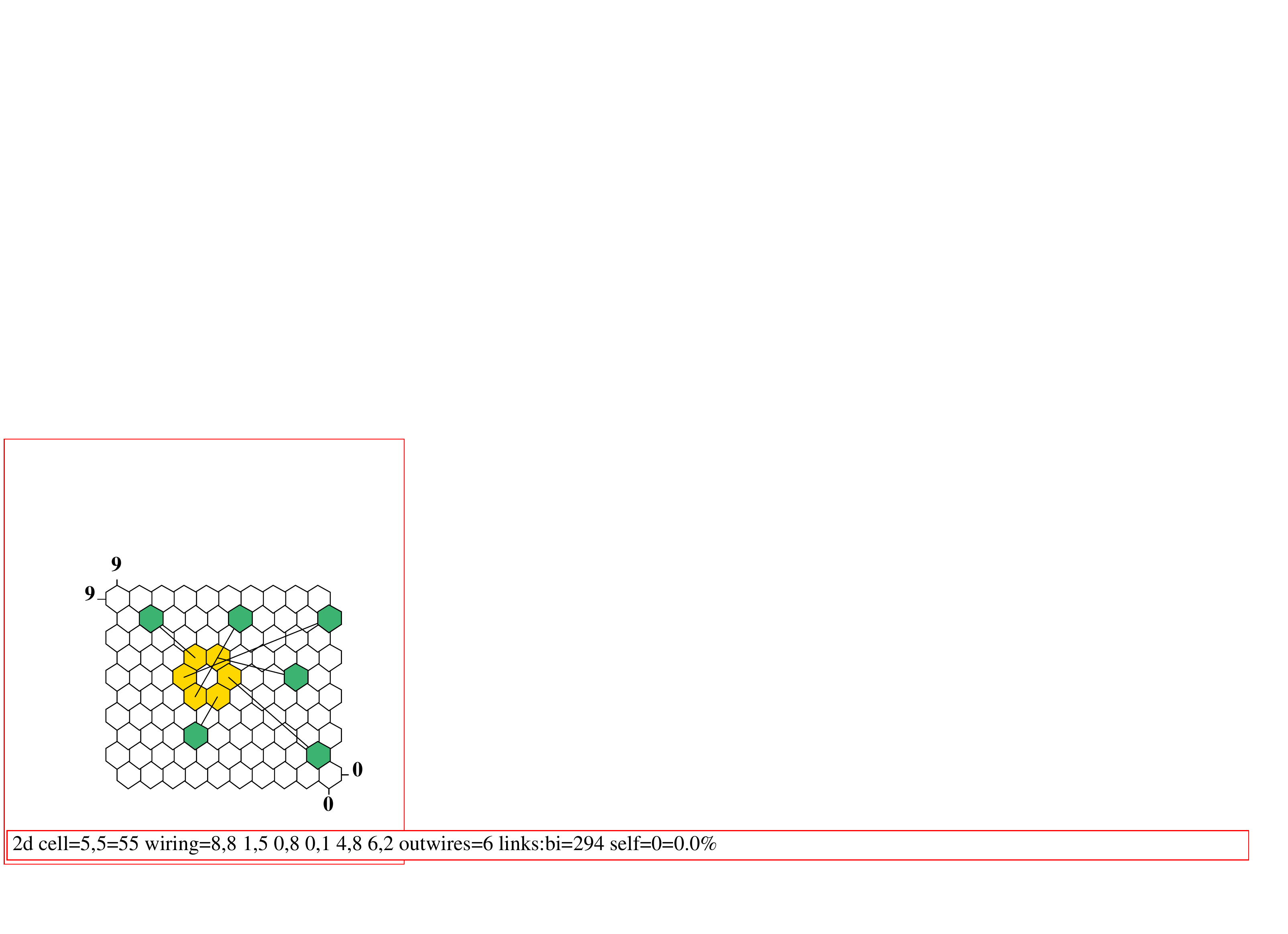}\\[-6ex]
\begin{center}(a)\end{center}
\end{minipage}
\hfill
\begin{minipage}[c]{.3\linewidth}
\includegraphics[width=1\linewidth,bb=58 88 261 274, clip=]{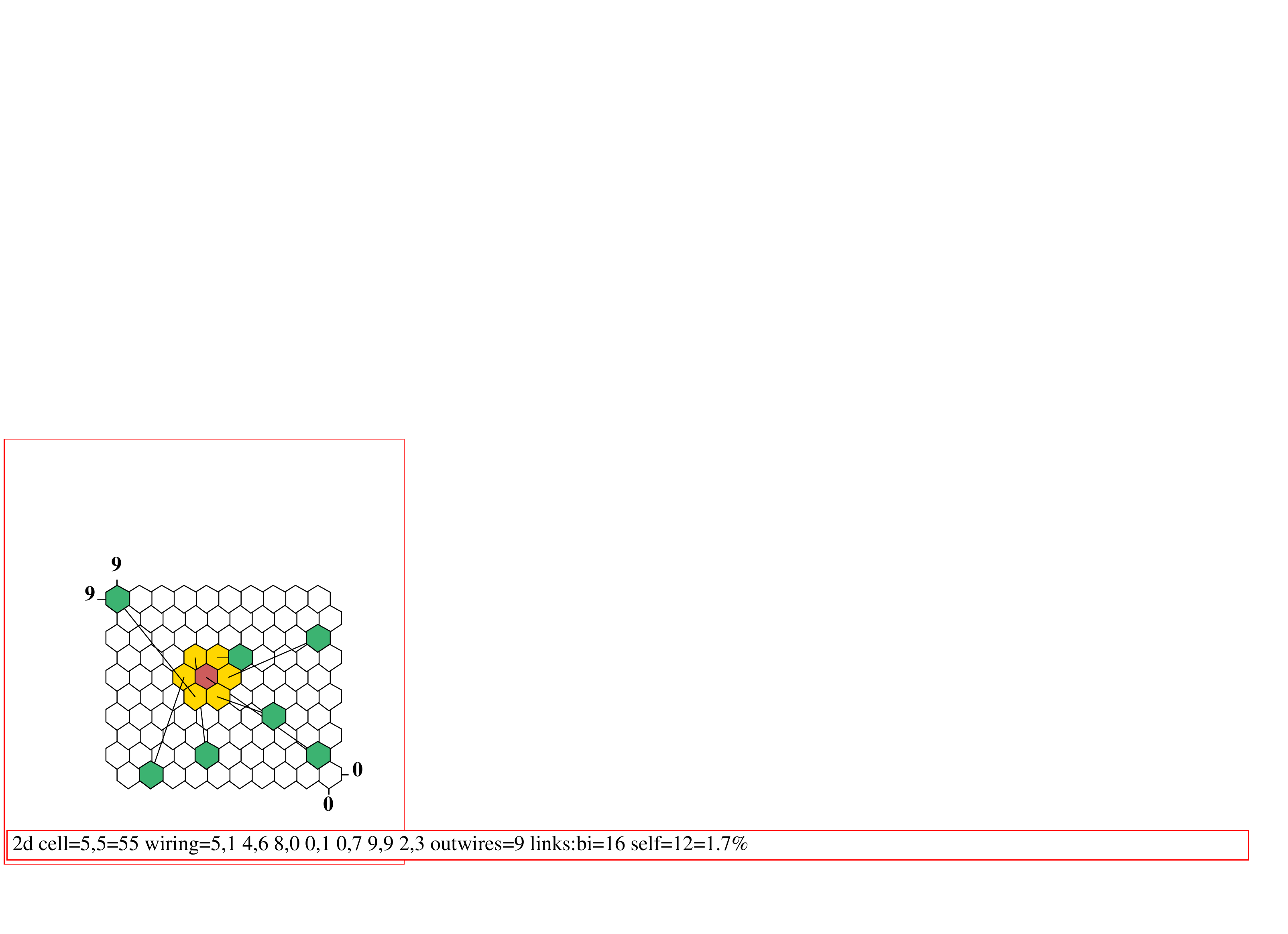}\\[-6ex]
\begin{center}(b)\end{center}
\end{minipage}
\hfill
\begin{minipage}[c]{.3\linewidth}
\includegraphics[width=1\linewidth,bb=58 88 261 274, clip=]{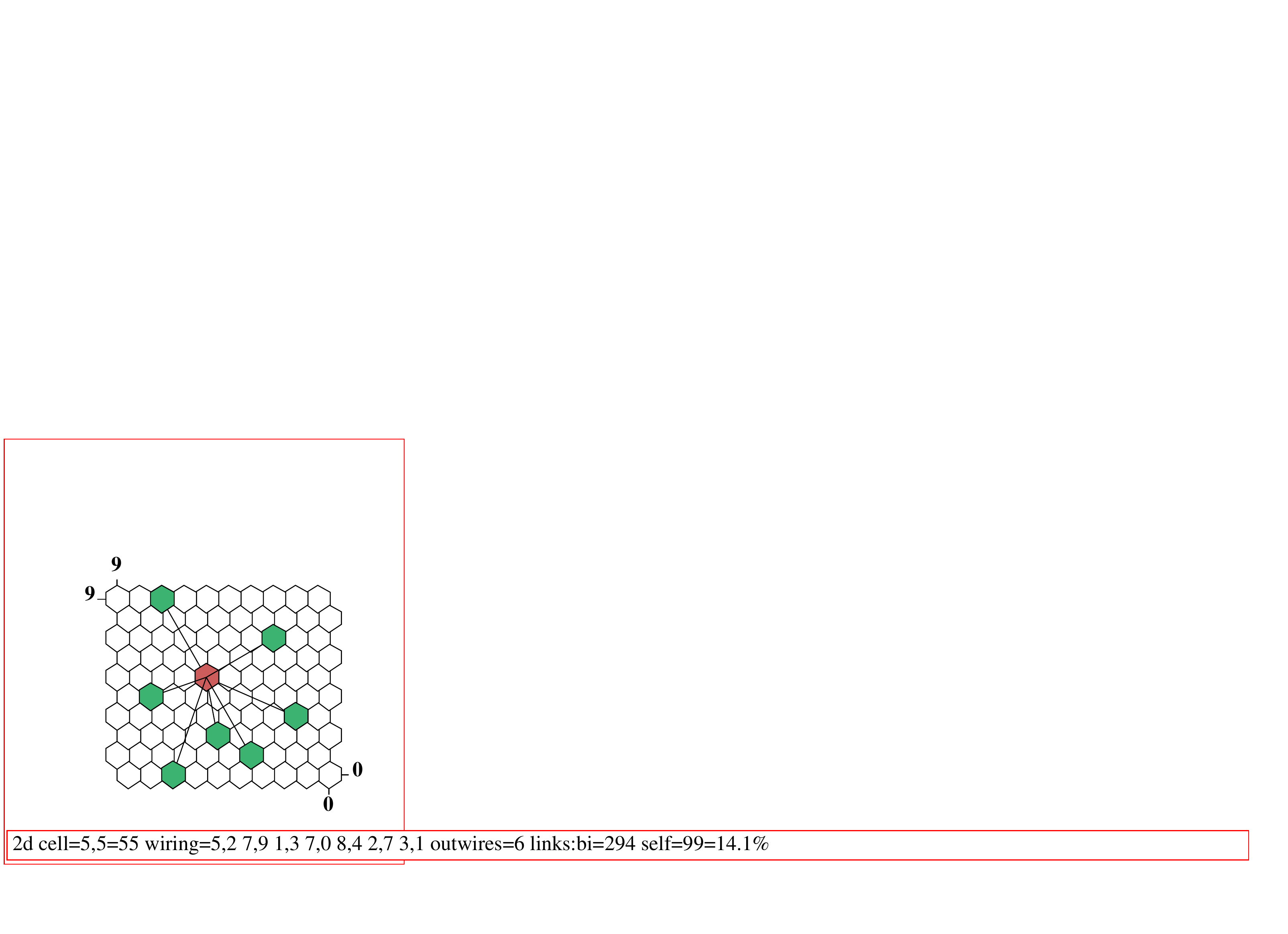}\\[-6ex]
\begin{center}(c)\end{center}
\end{minipage}
\end{minipage}
\end{center}
\vspace{-4ex}
\caption[Hexagonal lattice]
{\textsf{A hexagonal lattice 10$\times$10 showing $k$ random cells
(green) wired to the target cell, (a) and (b) via
a pseudo-neighborhood template (yellow and red), and (c) directly.
(a) For $k$=6 the target cell is not
included in its neighborhood. (b) For $k$=7 the target cell (red) is included.
For CA the actual neighborhood and pseudo-neighborhood are identical.
(c) For k-totalistic rules and random wiring, strictly speaking,
a pseudo-neighborhood template is not required.}}
\label{The hexagonal lattice}
\end{figure}

\noindent In classical CA, the pattern on the lattice updates (in discrete
time-steps) as each (target) cell 
synchronously\footnote{Asynchronous and noisy updating is discussed in
section~\ref{Asynchronous and noisy updating}.} updates its value
according to the values in its local neighborhood template.  In the
general case the updating function is a lookup-table (rule-table) of all $v^k$
possible neighborhood patterns, were $v$ is the value-range and $k$ is the
neighborhood size.

In this paper we focus mainly on 2D CA on a hexagonal lattice, with a local
neighborhood ``template'' of $k$=6 and $k$=7  
(figures~\ref{The (pseudo)-neighborhood template 2D} and \ref{The hexagonal lattice}),
and also 3D in section~\ref{3D systems}.
Boundary conditions are periodic (toroidal for 2D) --- effectively no boundaries,
but this is not significant in the CA pulsing model.
The $k$=6 template is shown in yellow surrounding the target cell,
whereas the $k$=7 template
includes the (red) target cell. Template cell numbers permit
a complete non-totalistic rule-table
to be assigned according to DDLab's convention\cite{Wuensche2016}.

To implement ``random wiring'', as in Kauffman's
``Random Boolean Networks'' \cite{kauffman69}, for each target cell, we take $k$ cells
at random in the lattice and ``wire'' them to distinct
cells in the pseudo-neighborhood template --- ``pseudo'' because the
actual template values are replaced by the values of the random cells.
Each target cell is assigned its own random wiring. The random wiring
can be biased in many ways in DDLab\cite{Wuensche2016}, one of which
is to confine random wiring within a local zone of arbitrary size
(section~\ref{Localised random wiring}). One or more wires can be ``freed'' 
from the zone, or from a CA neighborhood. Using DDLab, a single key
press enables switching between CA and any type of preset random wiring, or
between stable random wiring and re-randomising the wiring at each
time-step (within preset parameters) as in Derrida's
annealed model\cite{Derrida}.

The (pseudo)-neighborhood template allows a full rule-table, including isotropic rules, 
in various geometries and dimensions. For k-totalistic rules, however, 
though each incoming wire must connect to a distinct template cell, which one
is irrelevant.

\section{k-totalistic rules}
\label{k-totalistic rules}

\begin{table}[b]
\begin{center}
\begin{minipage}[b]{.9\linewidth}
\begin{minipage}[t]{.4\textwidth}{\scriptsize \baselineskip2ex
\begin{verbatim}
v3k6 kcodeSize=28
(hex) 0a0282816a0264
(kcode-table:2-0)
0022000220022001122200021210
vfreq=11+4+13=28
  27: 6 0 0  -> 0
  26: 5 1 0  -> 0
  25: 5 0 1  -> 2
  24: 4 2 0  -> 2
  23: 4 1 1  -> 0
  22: 4 0 2  -> 0
  21: 3 3 0  -> 0
  20: 3 2 1  -> 2
  19: 3 1 2  -> 2
  18: 3 0 3  -> 0
  17: 2 4 0  -> 0
  16: 2 3 1  -> 2
  15: 2 2 2  -> 2
  14: 2 1 3  -> 0
  13: 2 0 4  -> 0
  12: 1 5 0  -> 1
  11: 1 4 1  -> 1
  10: 1 3 2  -> 2
   9: 1 2 3  -> 2
   8: 1 1 4  -> 2
   7: 1 0 5  -> 0
   6: 0 6 0  -> 0
   5: 0 5 1  -> 0
   4: 0 4 2  -> 2
   3: 0 3 3  -> 1
   2: 0 2 4  -> 2
   1: 0 1 5  -> 1
   0: 0 0 6  -> 0
   \  - - -     \
    \ 2 1 0    kcode (outputs)
     \  \ 
      \  totals of 2s, 1s, 0s
       \  in the neighborhood
        \
       kcode index
\end{verbatim}}
\vspace{-2ex}
\textsf{\small(a) $v3k6$ kcode\\Beehive rule\cite{Wuensche05}}
\end{minipage}
\hfill 
\begin{minipage}[t]{.4\textwidth}
{\scriptsize \baselineskip2ex
\begin{verbatim}
v3k7 kcodeSize=36
(hex) 020609a2982a68aa64
(kcode-table:2-0)
000200120021220221200222122022221210
vfreq=18+6+12=36
  35: 7 0 0  -> 0
  34: 6 1 0  -> 0
  33: 6 0 1  -> 0
  32: 5 2 0  -> 2
  31: 5 1 1  -> 0
  30: 5 0 2  -> 0
  29: 4 3 0  -> 1
  28: 4 2 1  -> 2
  27: 4 1 2  -> 0
  26: 4 0 3  -> 0
  25: 3 4 0  -> 2
  24: 3 3 1  -> 1
  23: 3 2 2  -> 2
  22: 3 1 3  -> 2
  21: 3 0 4  -> 0
  20: 2 5 0  -> 2
  19: 2 4 1  -> 2
  18: 2 3 2  -> 1
  17: 2 2 3  -> 2
  16: 2 1 4  -> 0
  15: 2 0 5  -> 0
  14: 1 6 0  -> 2
  13: 1 5 1  -> 2
  12: 1 4 2  -> 2
  11: 1 3 3  -> 1
  10: 1 2 4  -> 2
   9: 1 1 5  -> 2
   8: 1 0 6  -> 0
   7: 0 7 0  -> 2
   6: 0 6 1  -> 2
   5: 0 5 2  -> 2
   4: 0 4 3  -> 2
   3: 0 3 4  -> 1
   2: 0 2 5  -> 2
   1: 0 1 6  -> 1
   0: 0 0 7  -> 0
\end{verbatim}}
\vspace{-2ex}
\textsf{(b) $v3k7$ kcode\\Spiral rule\cite{Wuensche&Adamatzky2006}}
\end{minipage}
\end{minipage}
\end{center}
\vspace{-3ex}
\caption[kcode] 
{\textsf{The kcode is a rule-table listing the output for every combination
of value totals in the neighborhood. For a system with 3 values (colors)
the list is ordered by the number of
2s, 1s, 0s, taken as a decimal number. The kcode is then a string listing each
output in descending order, from left to right,
which can a be converted to hexadecimal for compactness.
In DDLab these methods are implemented automatically,
for $v\leq8$ and $k\leq27$.
These examples show the kcode for the Beehive rule
and Spiral rule --- their pulsing dynamics are examined below.
}}
\label{kcode in vertical layout}
\end{table}

\enlargethispage{3ex}
\noindent We focus on 3-value k-totalistic rules for the following
reasons: their rule-tables are relatively short and thus tractable for
displaying the input-frequency histogram and its entropy
(input-entropy); the dynamics are isotropic so closer to nature; 
and the availability of samples of glider rules. At present gilder rules are found by 
looking at complex rules --- examining their space-time dynamics by eye, 
though pulsing itself could provide the basis for
an automatic search (section~\ref{Questions on the pulsing mechanism}).
Complex rules themselves are found automatically by classifying 
rule-space by the variability of input-entropy method\cite{Wuensche99,Wuensche05,Wuensche2016}.

In k-totalistic rules the output depends on
just the combination of totals, or frequencies, of the values in the neighborhood,
making k-totalistic rules a special case of isotropic rules --- the same output for
neighborhood rotation or reflection.
Each combination of totals make up the rule-table (kcode), which has
$S = (v + k - 1)! / (k! \times (v-1)!)$ entries.
Figure~\ref{kcode in vertical layout} explains the rule system for 
$v3k6$ where $S$=28, and for $v3k7$ where $S$=36, taking as examples the
Beehive rule\cite{Wuensche05} and the Spiral rule\cite{Wuensche&Adamatzky2006}.
The size of k-totalistic rule-space is $v^S$.

\begin{figure}[htb]
\textsf{\small
\begin{center}
\begin{minipage}[c]{1\linewidth} 
\begin{minipage}[c]{.11\linewidth}
\includegraphics[width=1\linewidth]{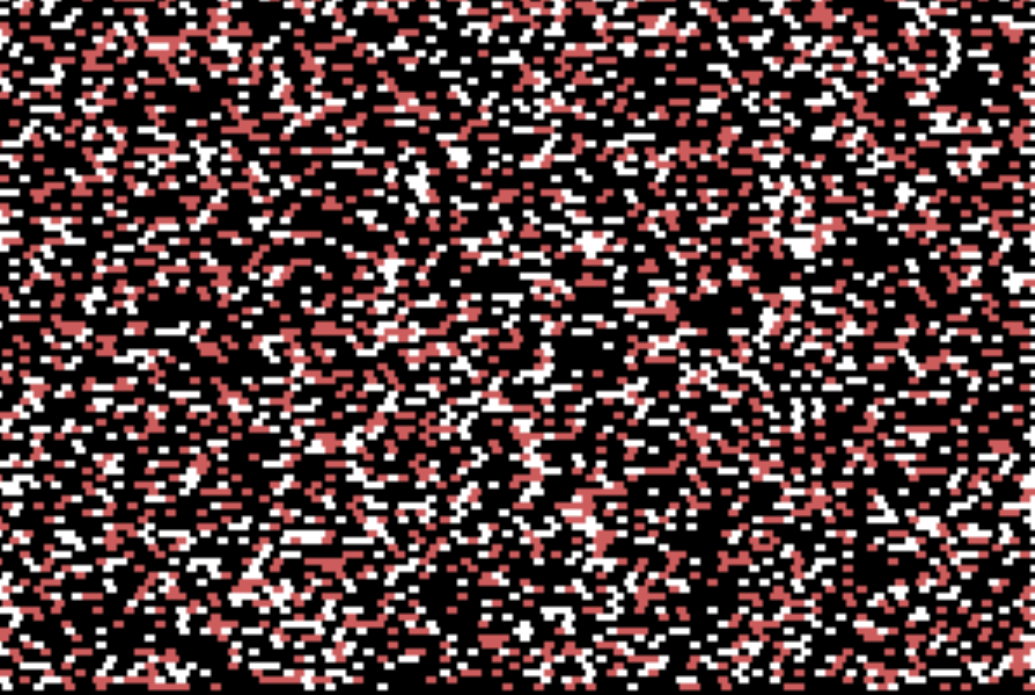}\\[2ex]
\includegraphics[width=1\linewidth]{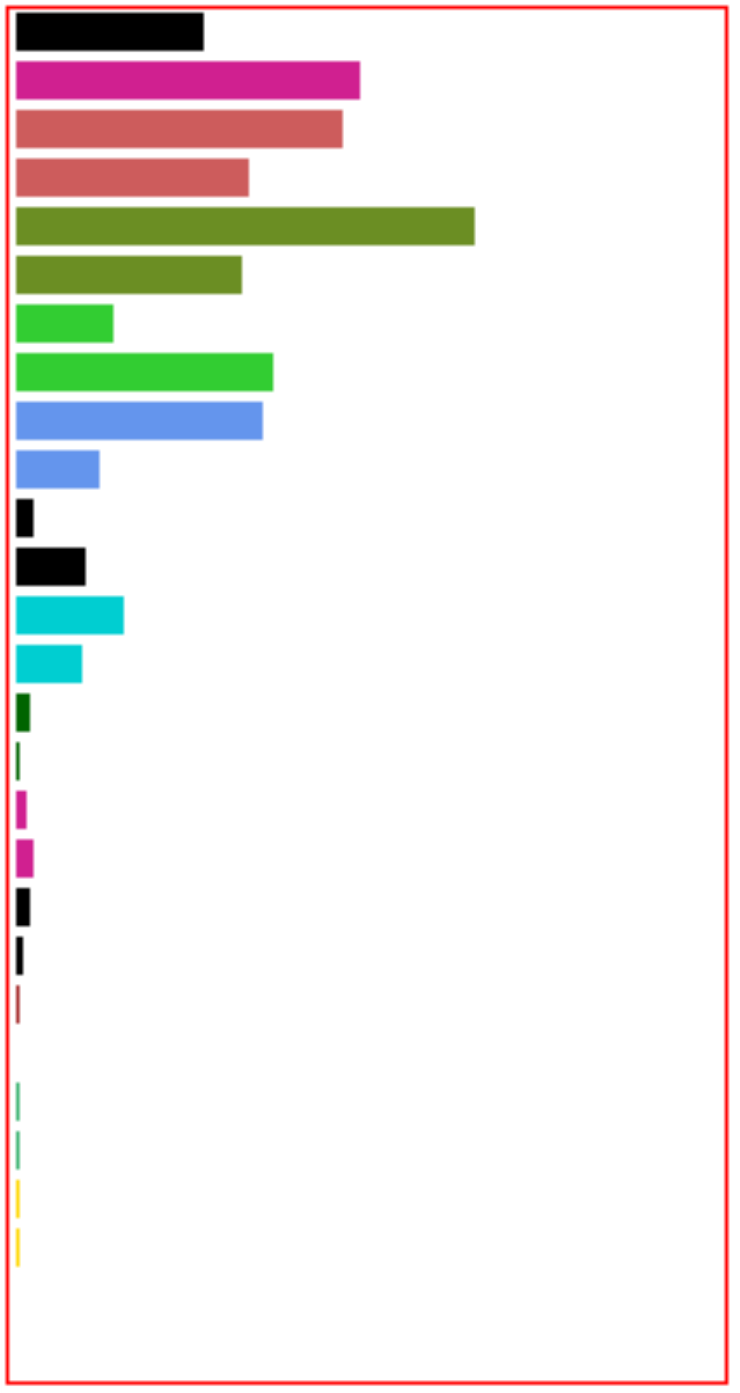}\\
1
\end{minipage}
\hfill
\begin{minipage}[c]{.11\linewidth}
\includegraphics[width=1\linewidth]{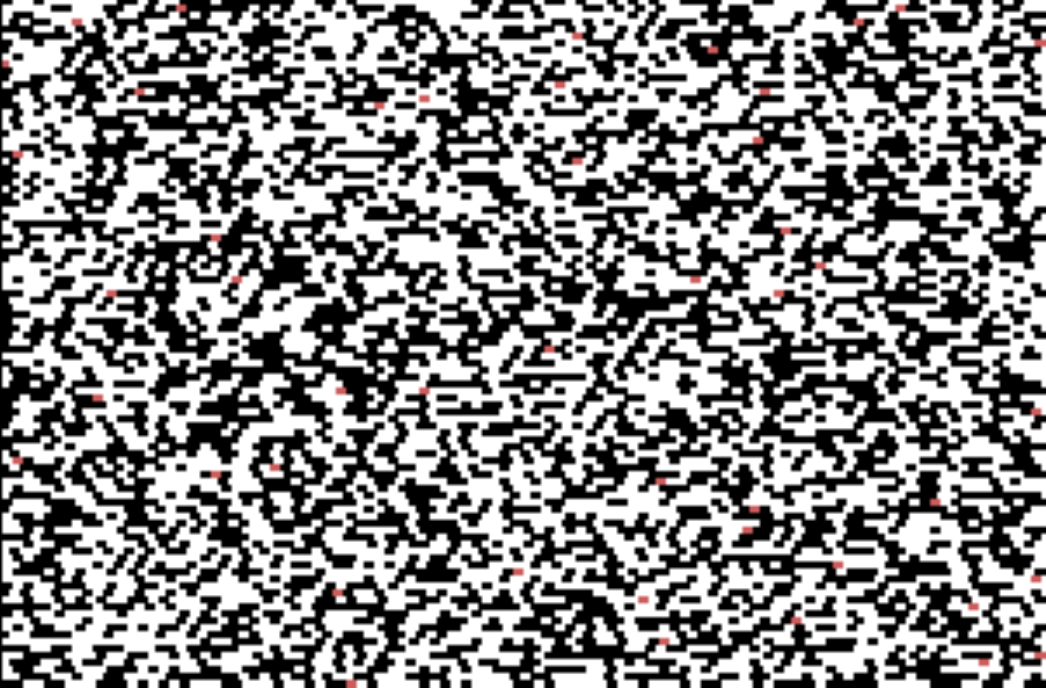}\\[2ex]
\includegraphics[width=1\linewidth]{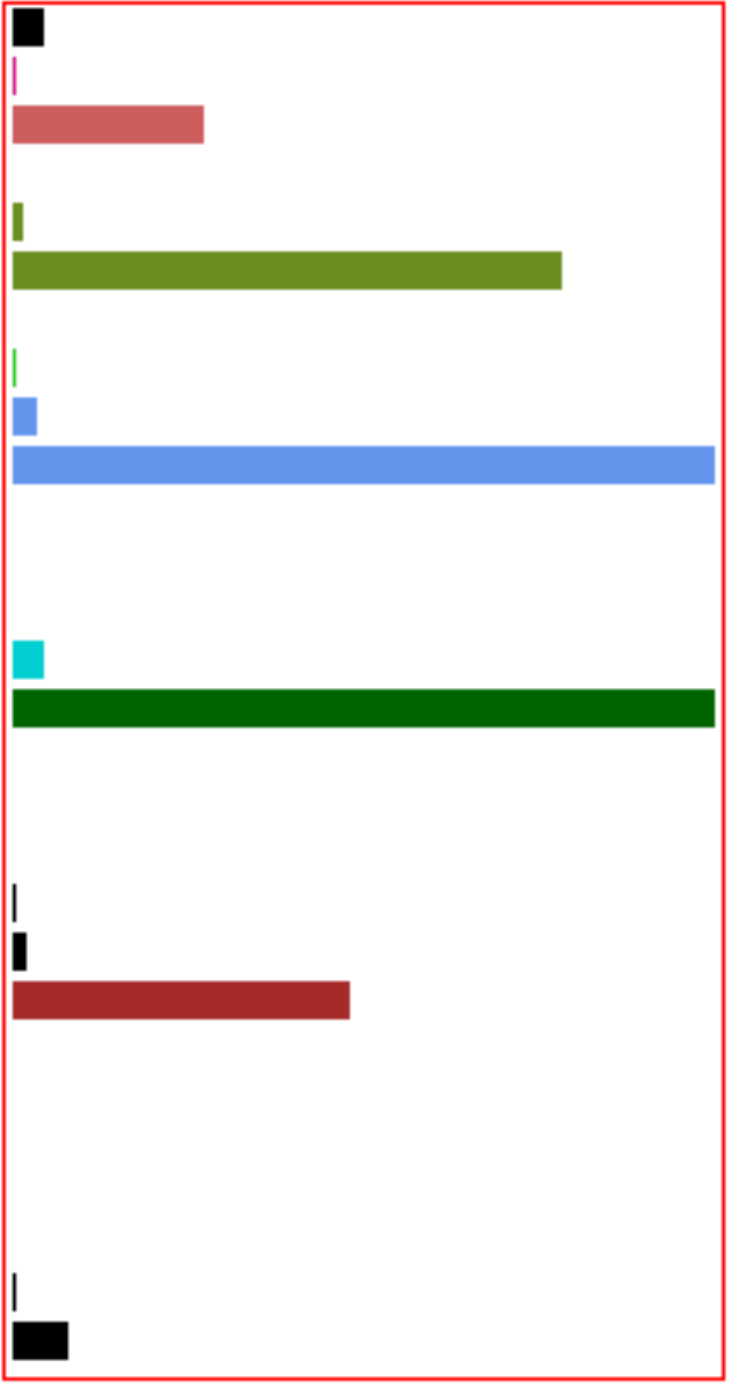}\\
2
\end{minipage}
\hfill
\begin{minipage}[c]{.11\linewidth}
\includegraphics[width=1\linewidth]{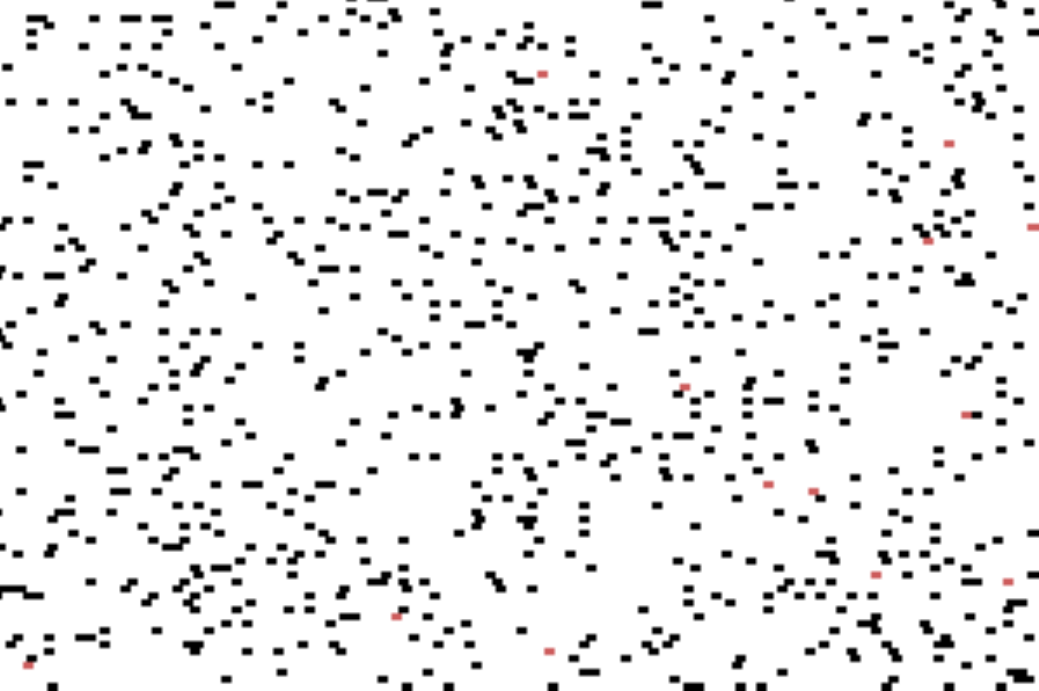}\\[2ex]
\includegraphics[width=1\linewidth]{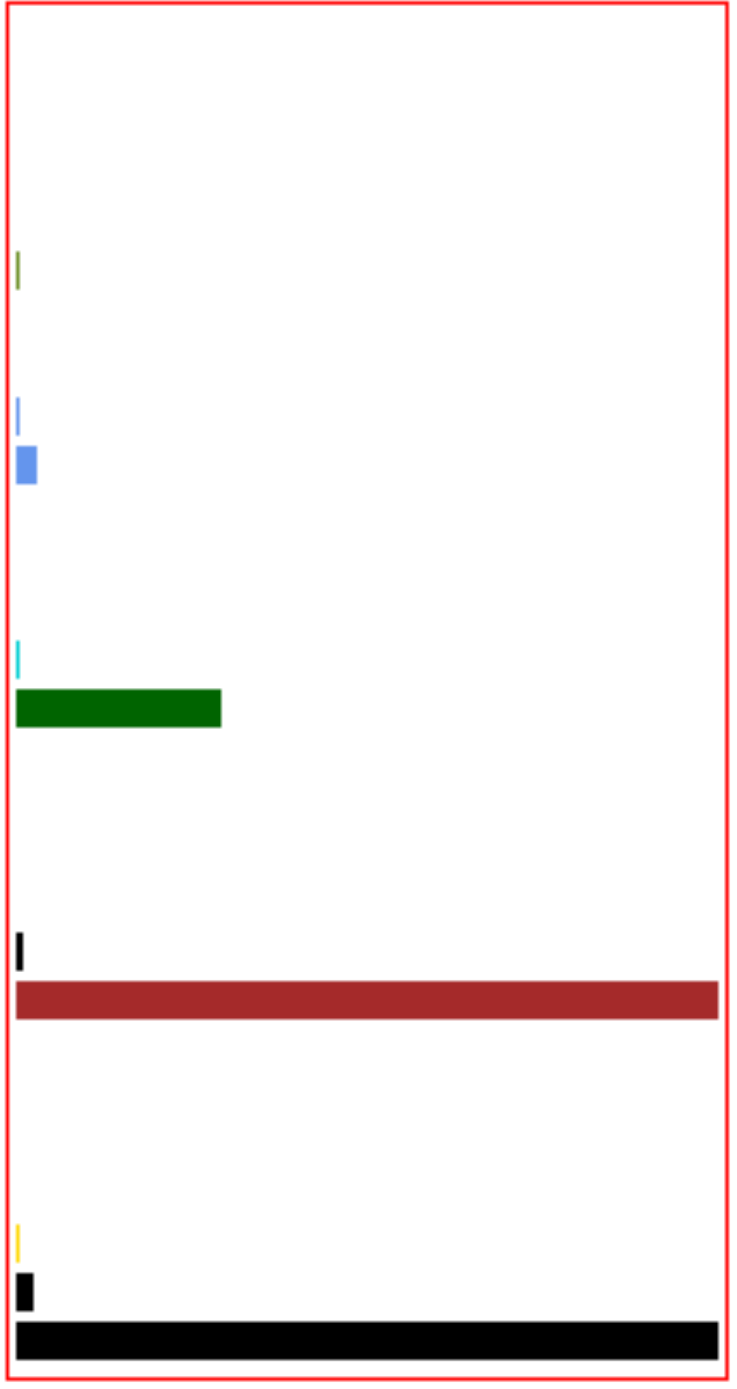}\\
3
\end{minipage}
\hfill
\begin{minipage}[c]{.11\linewidth}
\includegraphics[width=1\linewidth]{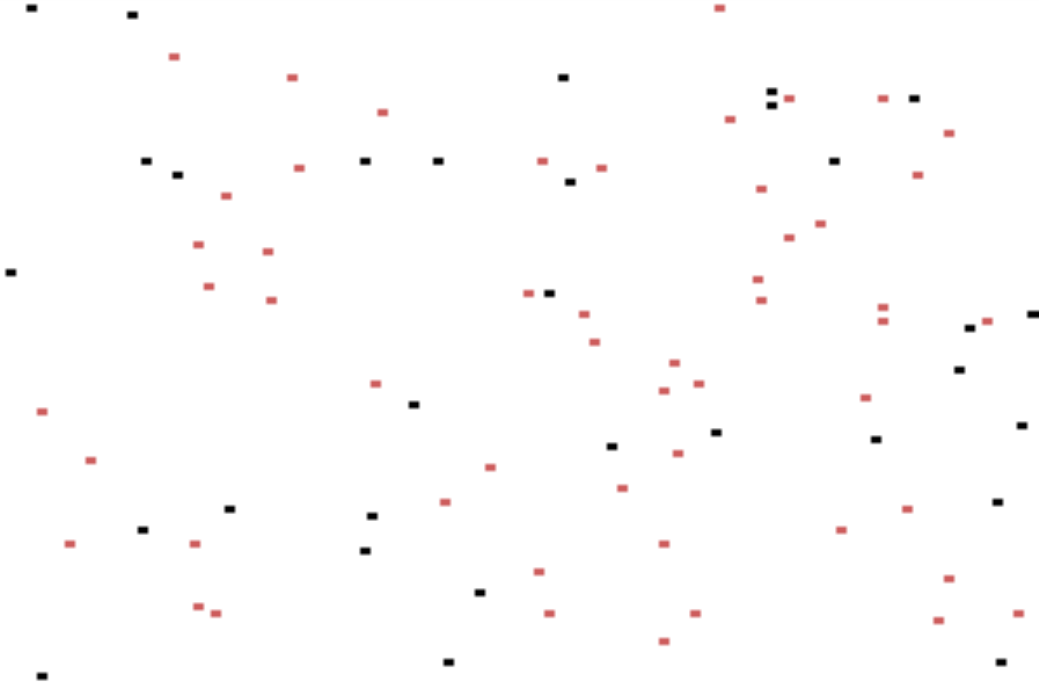}\\[2ex]
\includegraphics[width=1\linewidth]{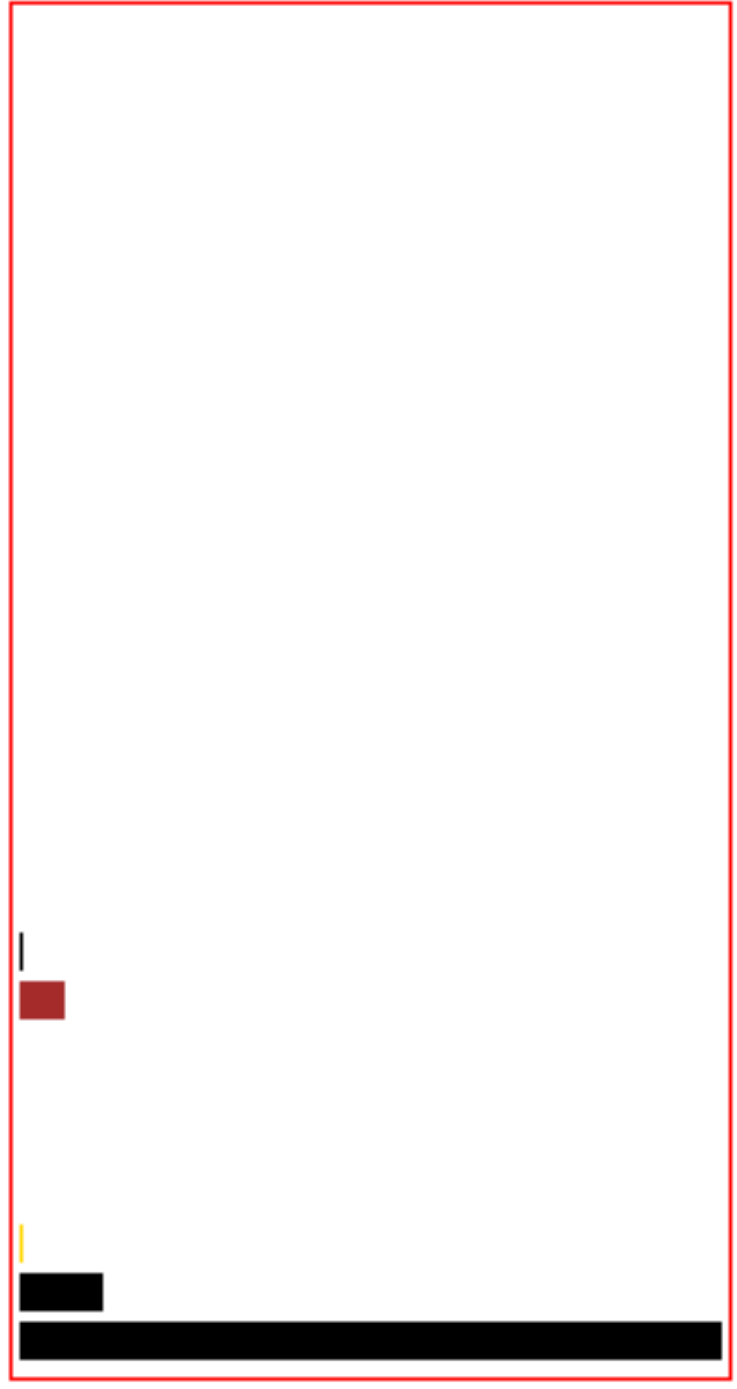}\\
4
\end{minipage}
\hfill
\begin{minipage}[c]{.11\linewidth}
\includegraphics[width=1\linewidth]{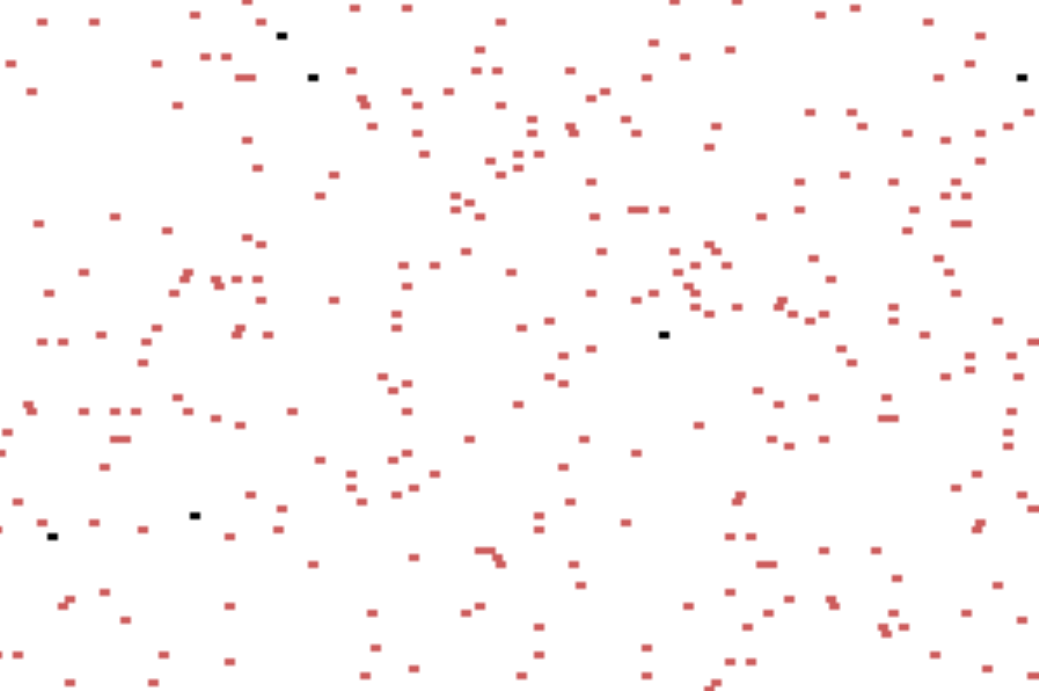}\\[2ex]
\includegraphics[width=1\linewidth]{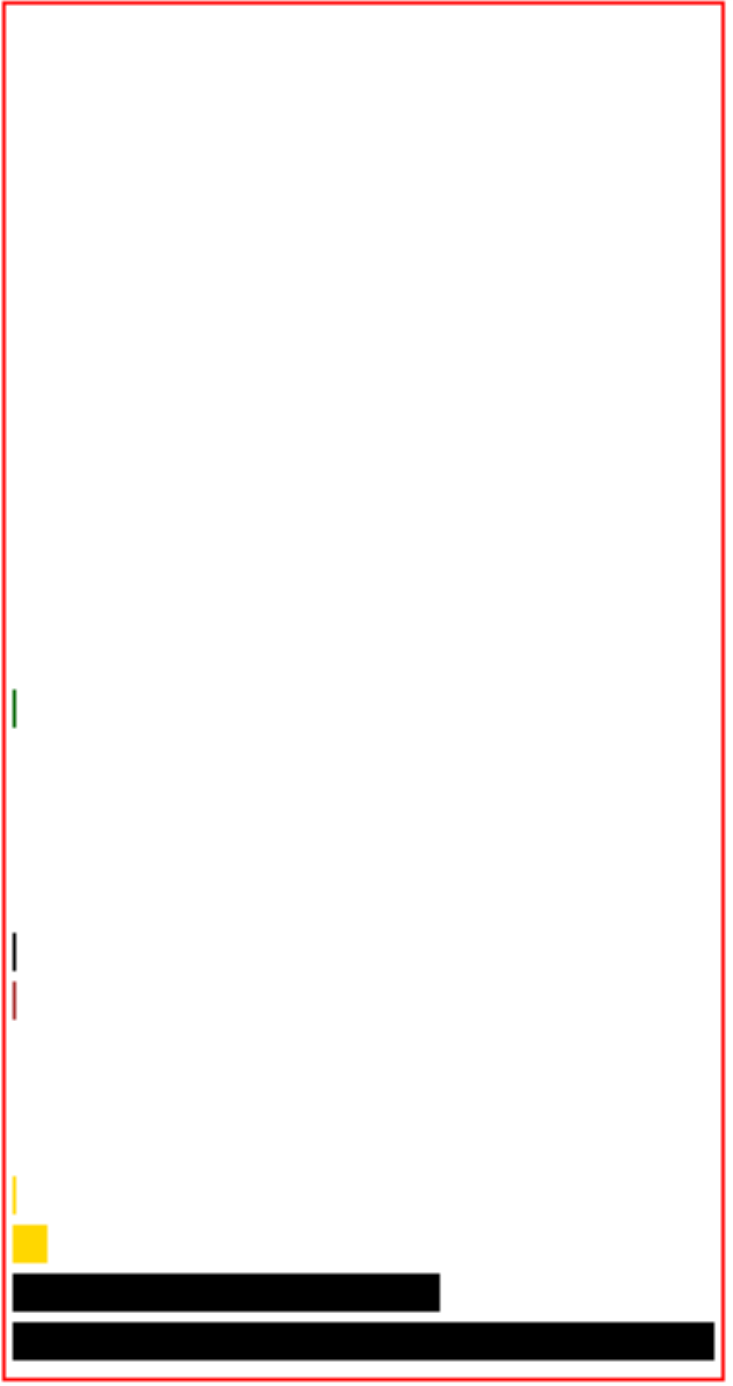}\\
5
\end{minipage}
\hfill
\begin{minipage}[c]{.11\linewidth}
\includegraphics[width=1\linewidth]{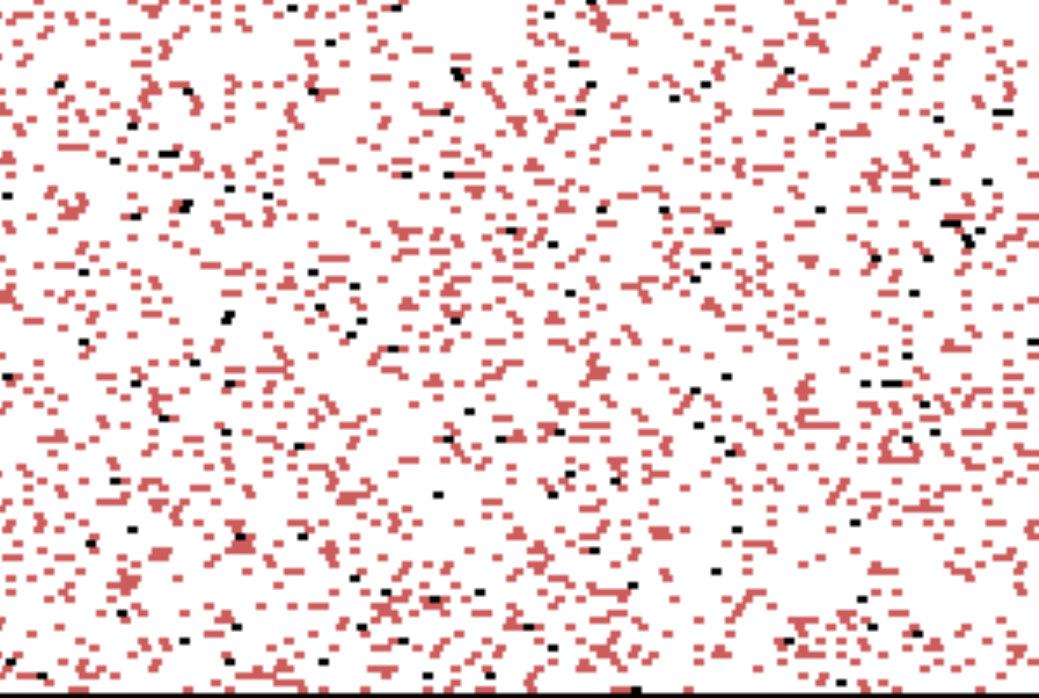}\\[2ex]
\includegraphics[width=1\linewidth]{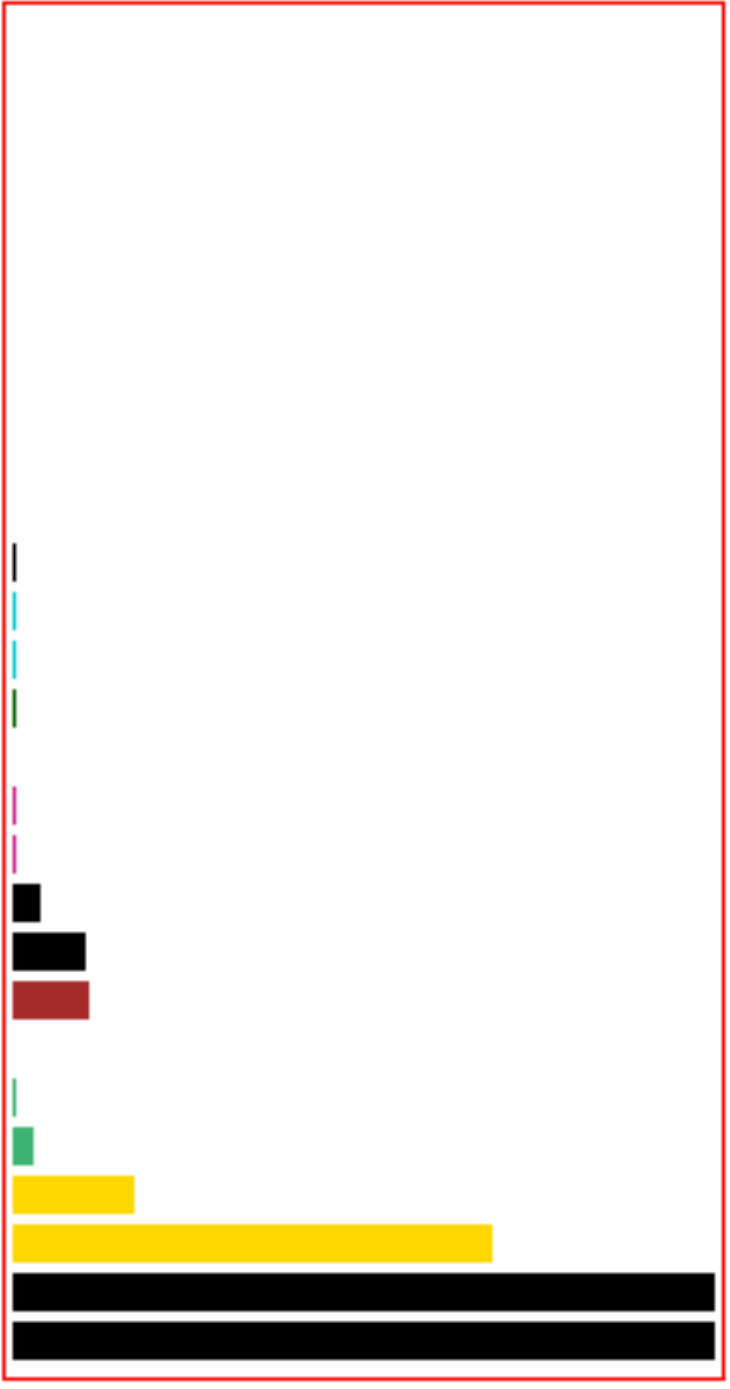}\\
6
\end{minipage}
\hfill
\begin{minipage}[c]{.11\linewidth}
\includegraphics[width=1\linewidth]{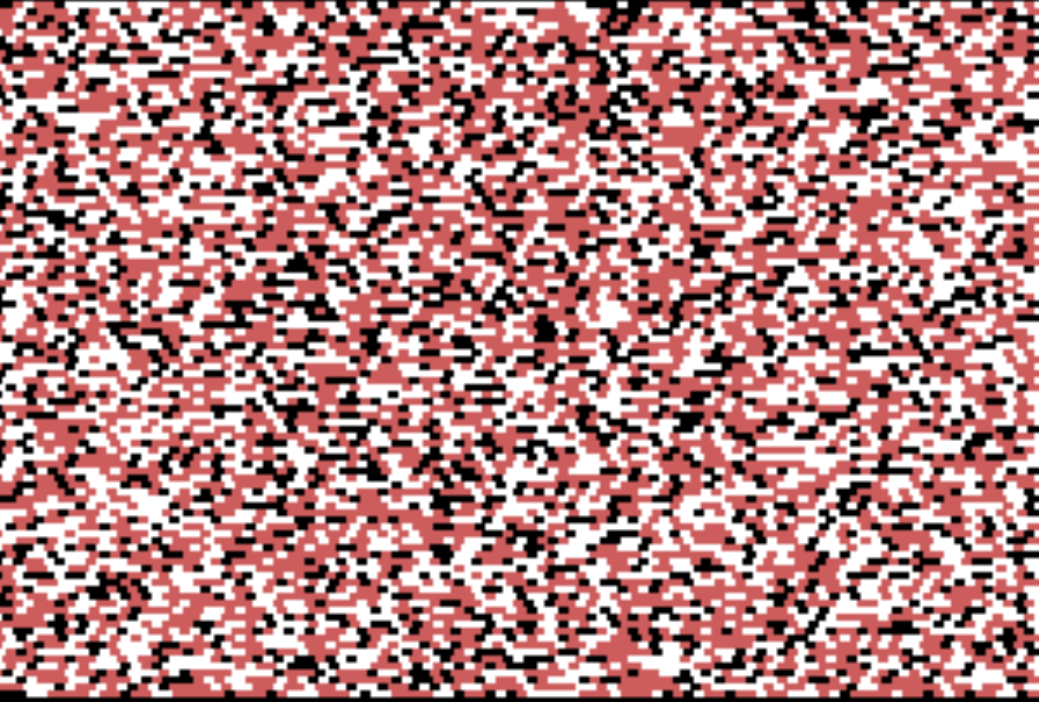}\\[2ex]
\includegraphics[width=1\linewidth]{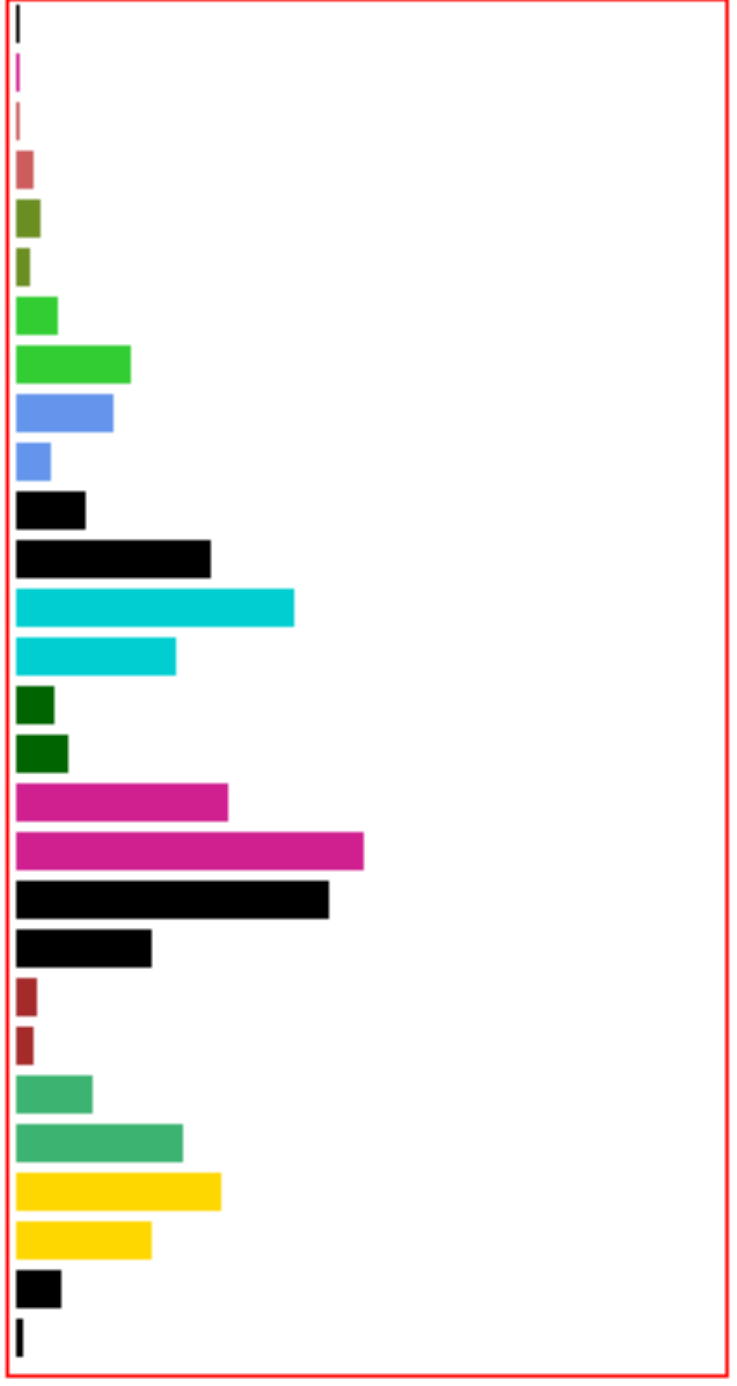}\\
7
\end{minipage}
\hfill
\begin{minipage}[c]{.11\linewidth}
\includegraphics[width=1\linewidth]{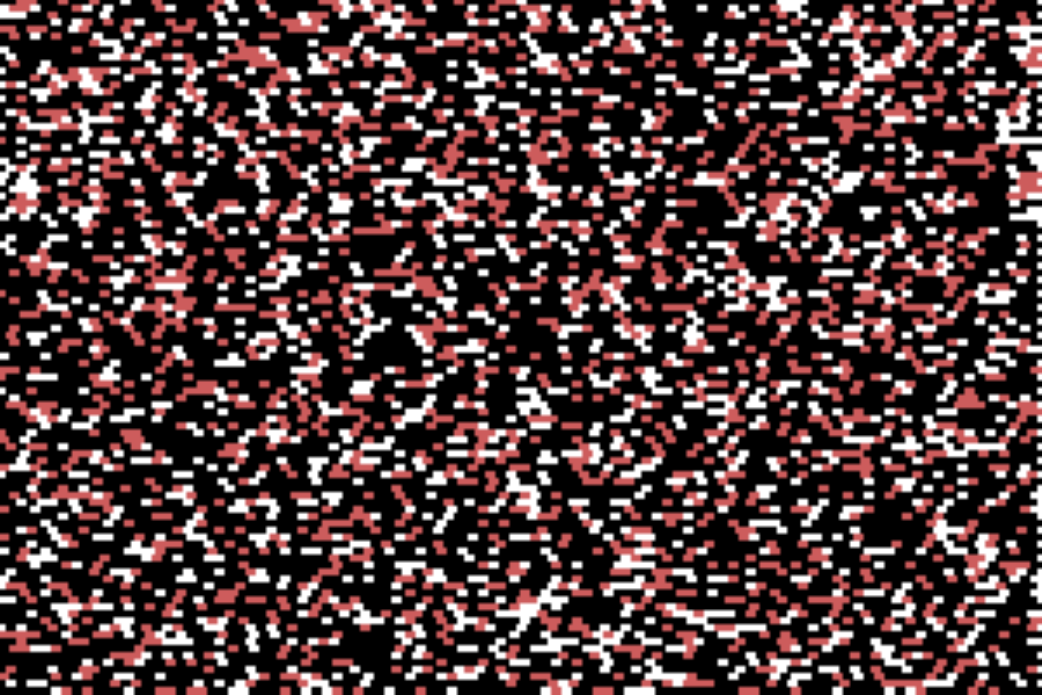}\\[2ex]
\includegraphics[width=1\linewidth]{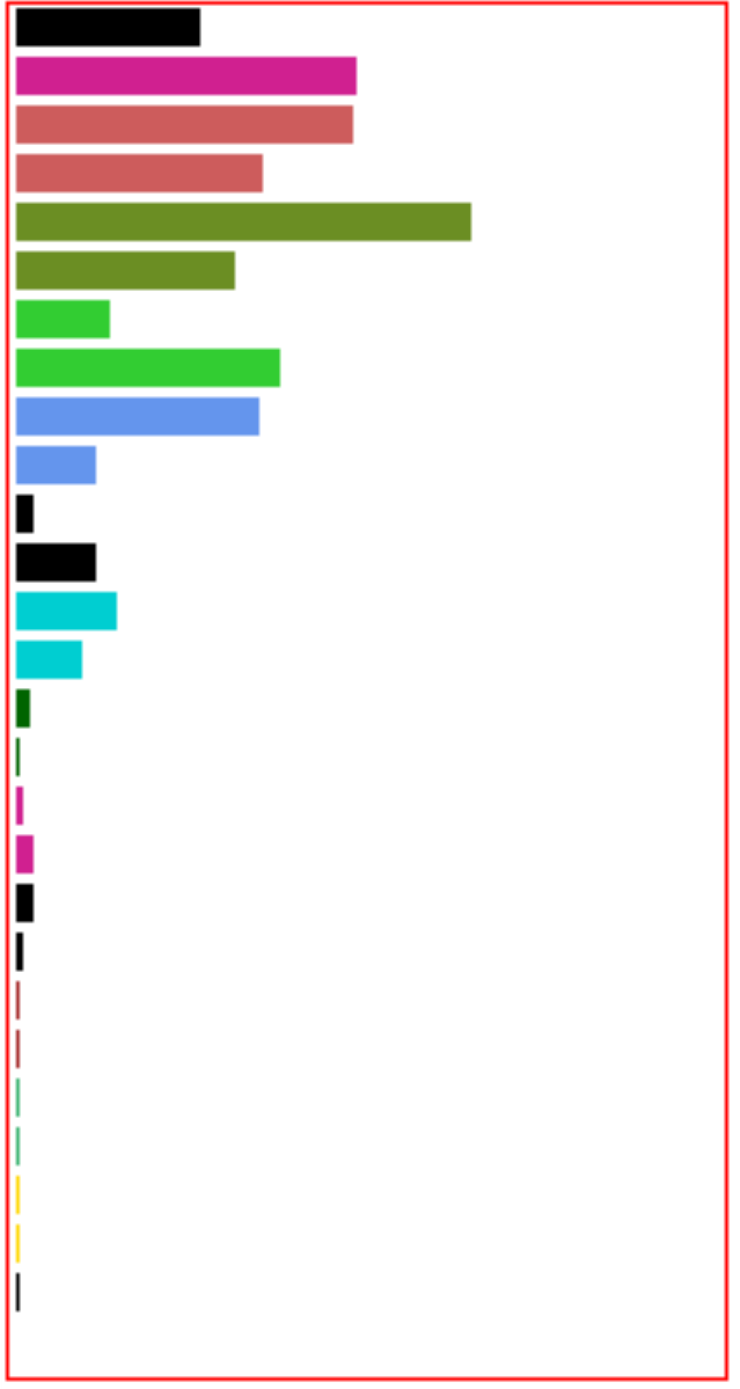}\\
repeat
\end{minipage}
\end{minipage}
\end{center}
\begin{center}
\begin{minipage}[c]{.95\linewidth}
{\LARGE $\searrow$}
\begin{minipage}[c]{.33\linewidth}
\includegraphics[width=1\linewidth]{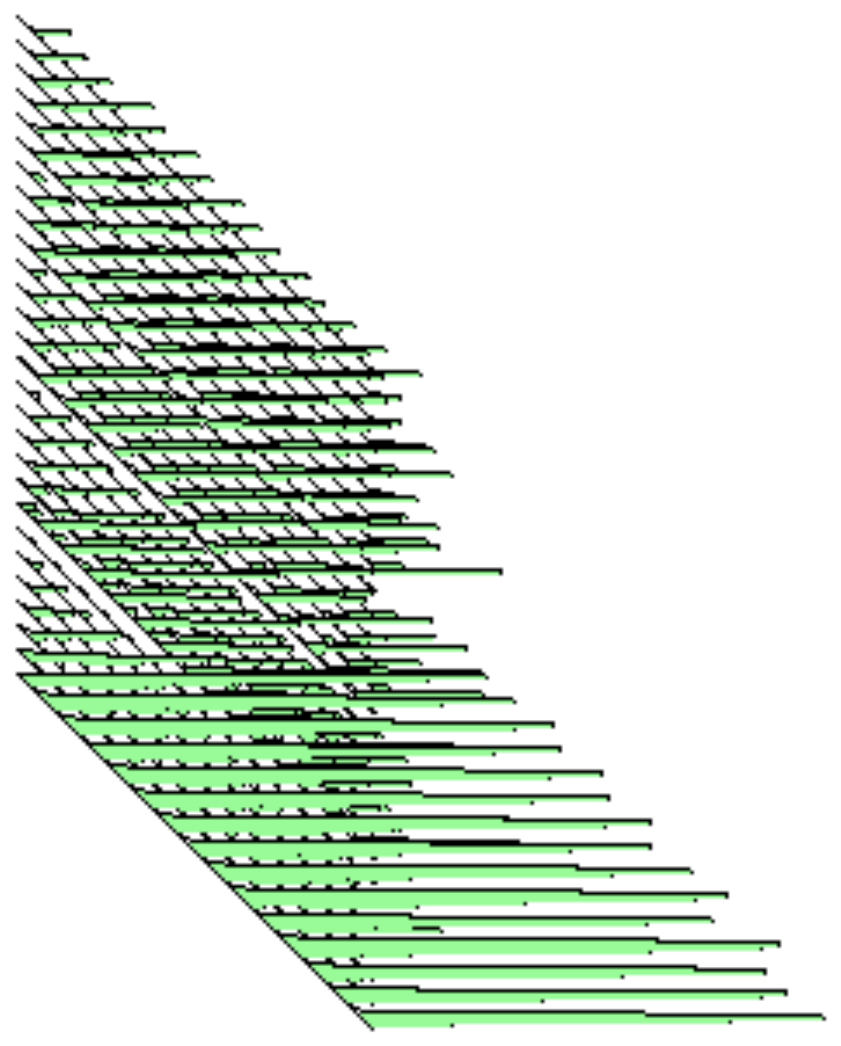}\\
(a)
\end{minipage}
\hfill
{\LARGE $\downarrow$}
\begin{minipage}[c]{.3\linewidth}
\includegraphics[width=1\linewidth]{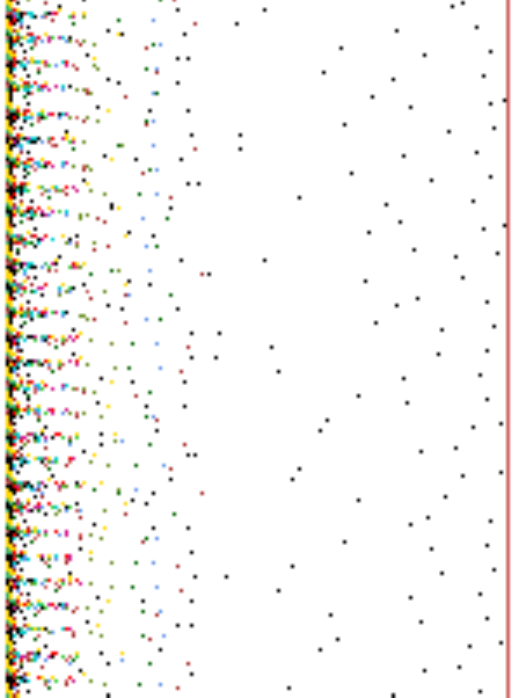}\\
(b)
\end{minipage}
\hfill
{\LARGE $\downarrow$}
\begin{minipage}[c]{.17\linewidth}
\fbox{\includegraphics[width=1\linewidth]{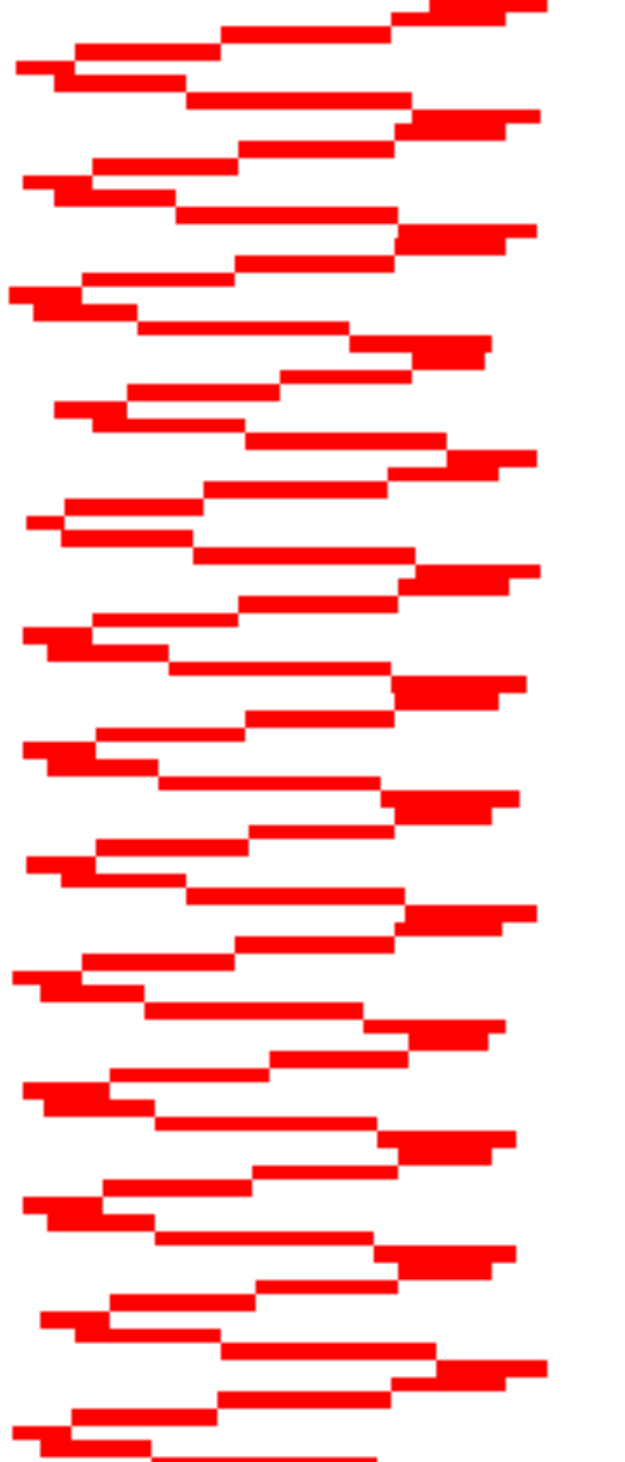}}\\
(c)
\end{minipage}
\end{minipage}
\end{center}
}
\vspace{-4ex}
\caption[input-histograms]
{\textsf{
\noindent Dynamic graphics in DDLab show up pulsing
in the $v3k6$ Beehive rule\cite{Wuensche05} (hex 0a0282816a0264) on
a 100$\times$100 hexagonal lattice with fully random wiring.
The period varies between 7 and 8 time-steps.\\ 
(top) A typical sequence of the space-time patterns displaying pulsing densities,
and related input-histograms, repeating on the 8th time-step. 
The horizontal bars represent the lookup-frequency of 28 neighborhoods, 
as in figure~\ref{kcode in vertical layout}(a). Below (arrows show time's direction):
(a) The input-histogram shown scrolling with time (z-axis).
(b) Plotting the input-histograms values (x-axis) for successive time-steps (y-axis).
(c) Input-entropy values (x-axis) plotted for successive time-steps (y-axis).
}}
\label{input-histograms}
\end{figure}
\clearpage

\section{The input-frequency and input-entropy}
\label{The input-frequency and input-entropy}

\noindent The input-frequency histogram tracks how frequently the
different entries in a rule-table are actually looked~up.
This is usually averaged over a moving window of $w$ time-steps\cite{Wuensche99}
to classify rules by the variability of input-entropy,
but to track pulsing dynamics we take the measures over each time-step.
The input-entropy is the Shannon entropy $H$ of this input-frequency histogram.
$H$, at time-step $t$, for one time-step ($w$=1), is given by 
$H^t = -\sum_{i=0}^{S-1}\left( {Q_{i}^{t}}/{n} \times
  log_{2}\left( {Q_{i}^{t}}/{n}\right)\right)$, 
where $Q_{i}^{t}$ is the lookup-frequency of neighborhood $i$ at time~$t$. $S$ is the
  rule-table size, and $n$ is the network size. The normalised Shannon entropy $H_N$ is
  a value between 0 and 1,  $H_N=H^t/log_{2}n$,
  which measures the heterogeneity of the histogram ---
  henceforth ``entropy'' will refer to $H_N$. 
Figure~\ref{input-histograms} shows how space-time patterns, their
input-frequency (histogram), and the input-entropy measures are tracked by
dynamic graphics in DDLab to show
up pulsing, taking the $v3k6$ Beehive rule\cite{Wuensche05} as an example.

\section{Pulsing case studies}
\label{Pulsing case studies}

\begin{figure}[htb]
\begin{center}
\textsf{\small
\begin{minipage}[c]{1\linewidth} 
\begin{minipage}[c]{.3\linewidth}
\includegraphics[width=1\linewidth]{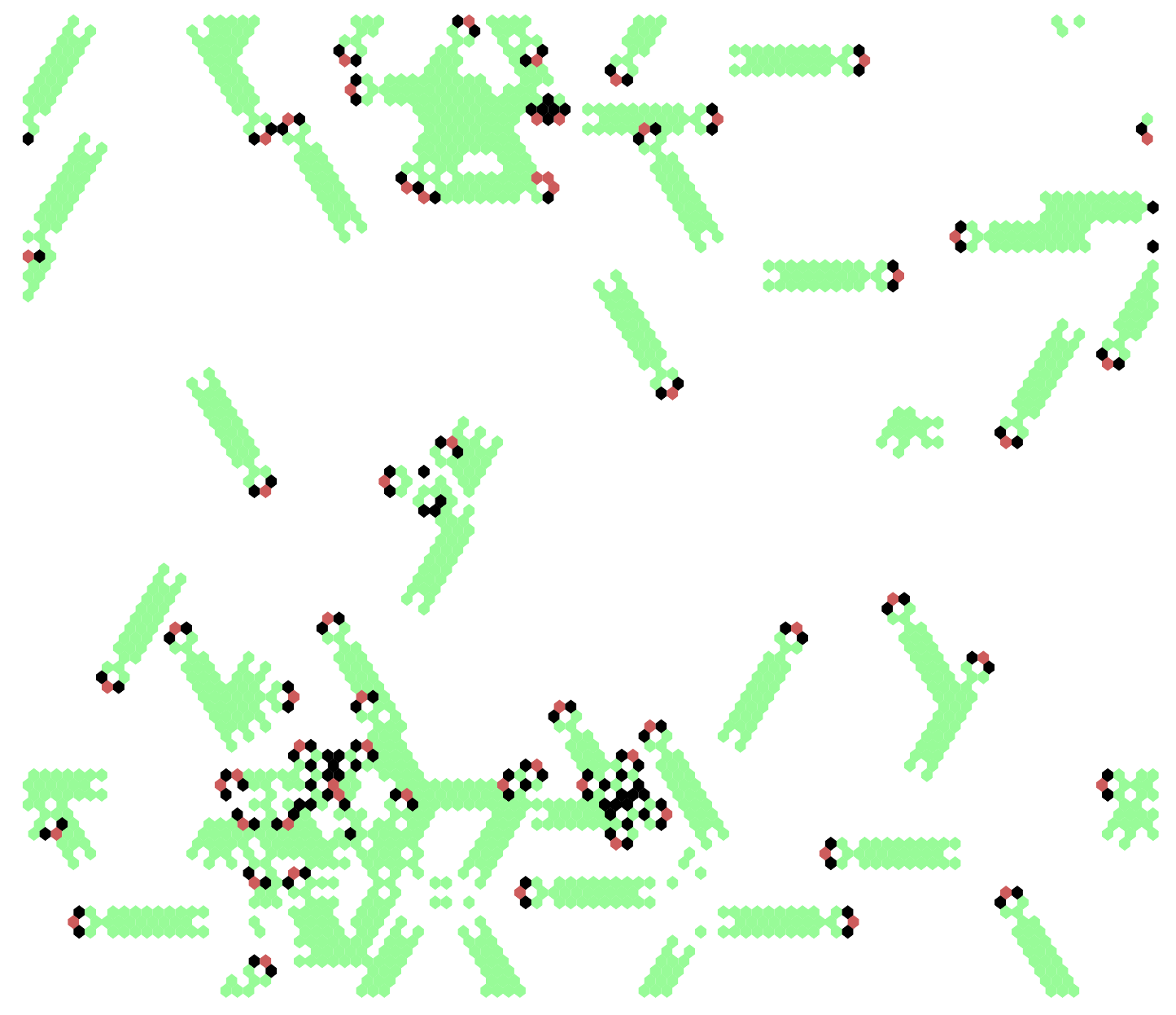}\\
(a)
\end{minipage}
\hfill
\begin{minipage}[c]{.3\linewidth}
\includegraphics[width=1\linewidth]{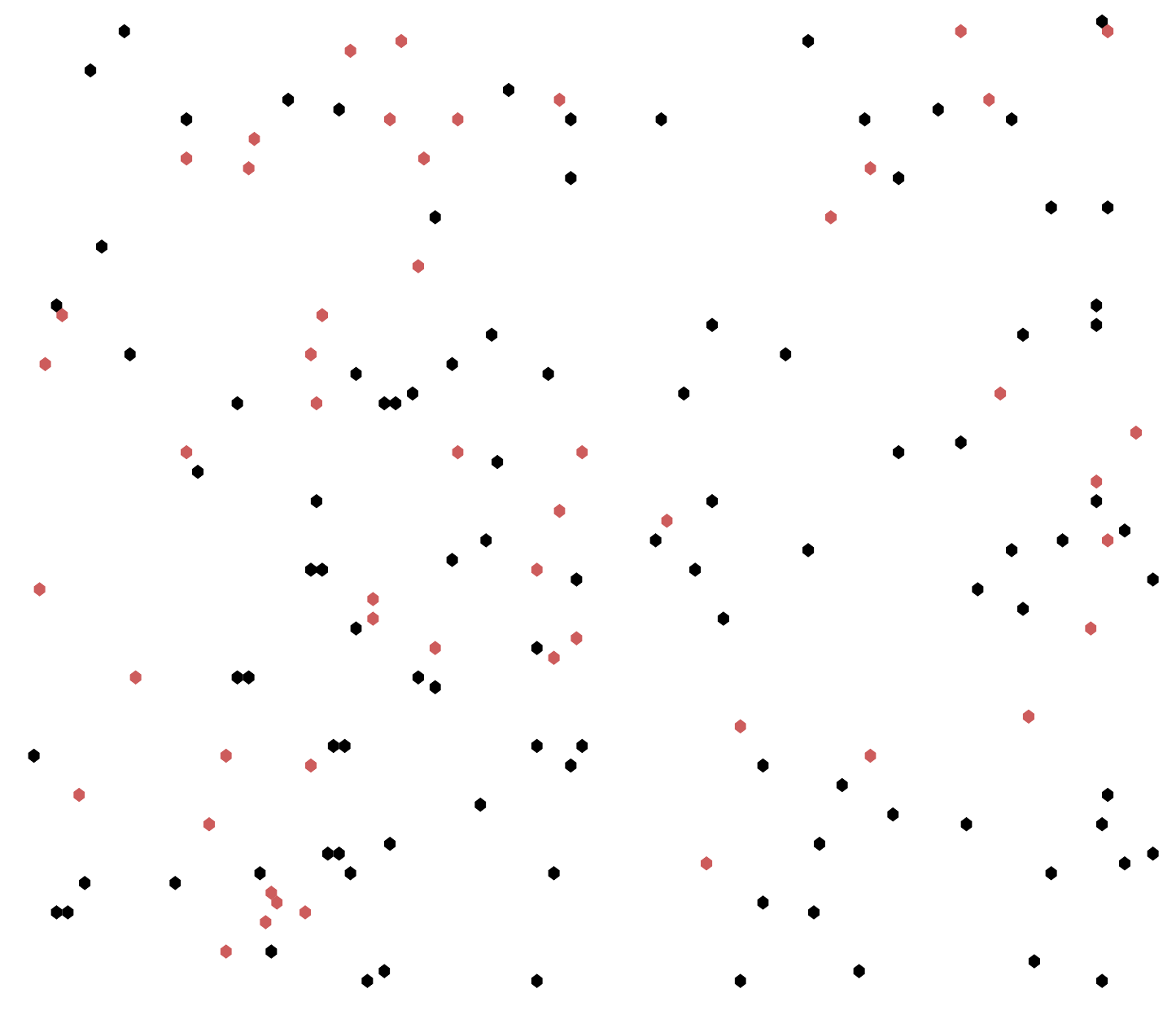}\\
(b)
\end{minipage}
\hfill
\begin{minipage}[c]{.3\linewidth}
\includegraphics[width=1\linewidth]{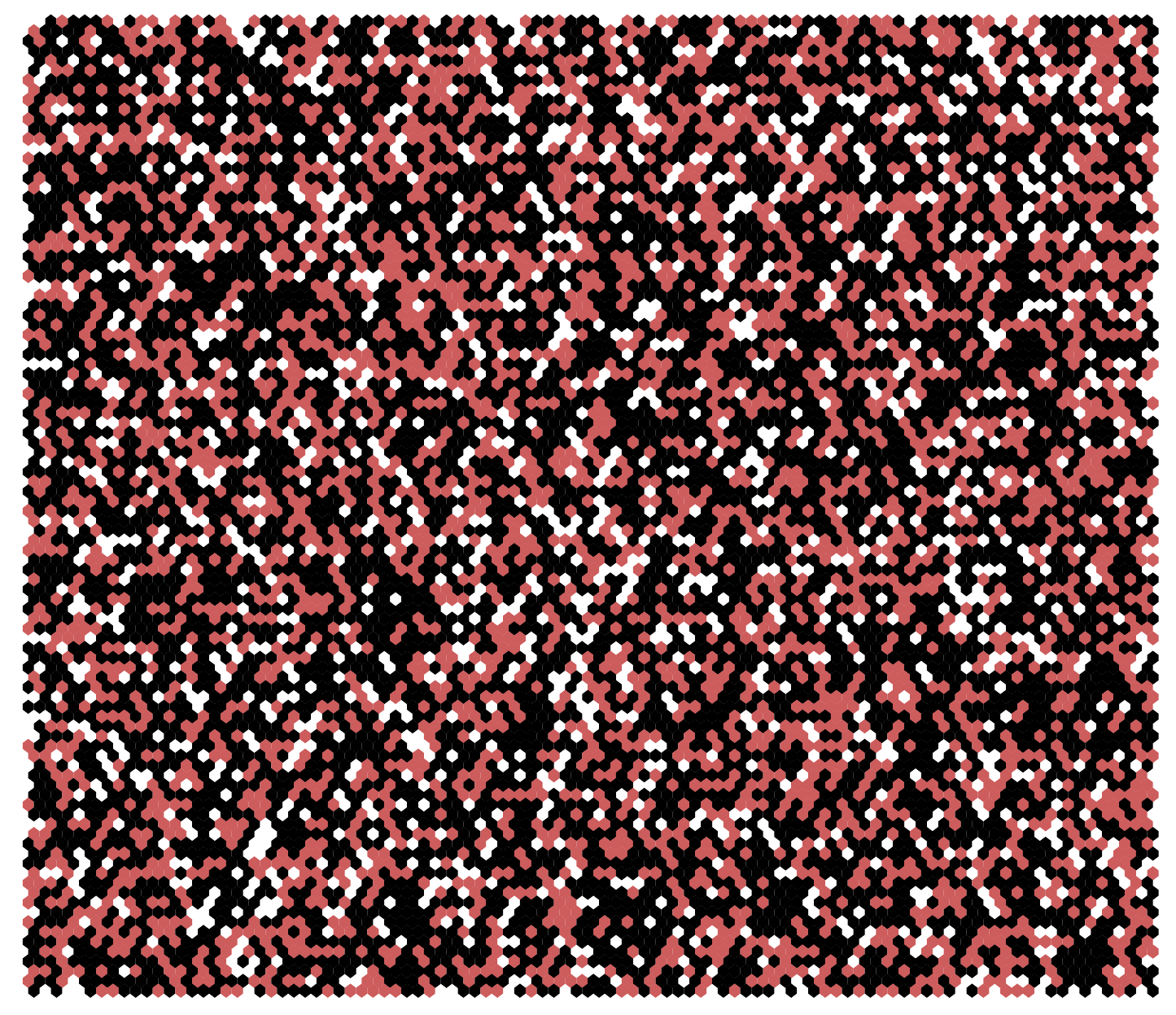}\\
\textsf{\small(c)}
\end{minipage}
\end{minipage}
}
\end{center}
\vspace{-1ex}
\textsf{\small
\noindent Space-time pattern snapshots,
(a) CA showing emergent gliders.
Randomised wiring results in disordered patterns,
(b) minimum density, and (c) maximum density.
}

\begin{center}
\textsf{\small
\begin{minipage}[c]{.95\linewidth} 
\begin{minipage}[c]{.40\linewidth} 
\includegraphics[width=1\linewidth]{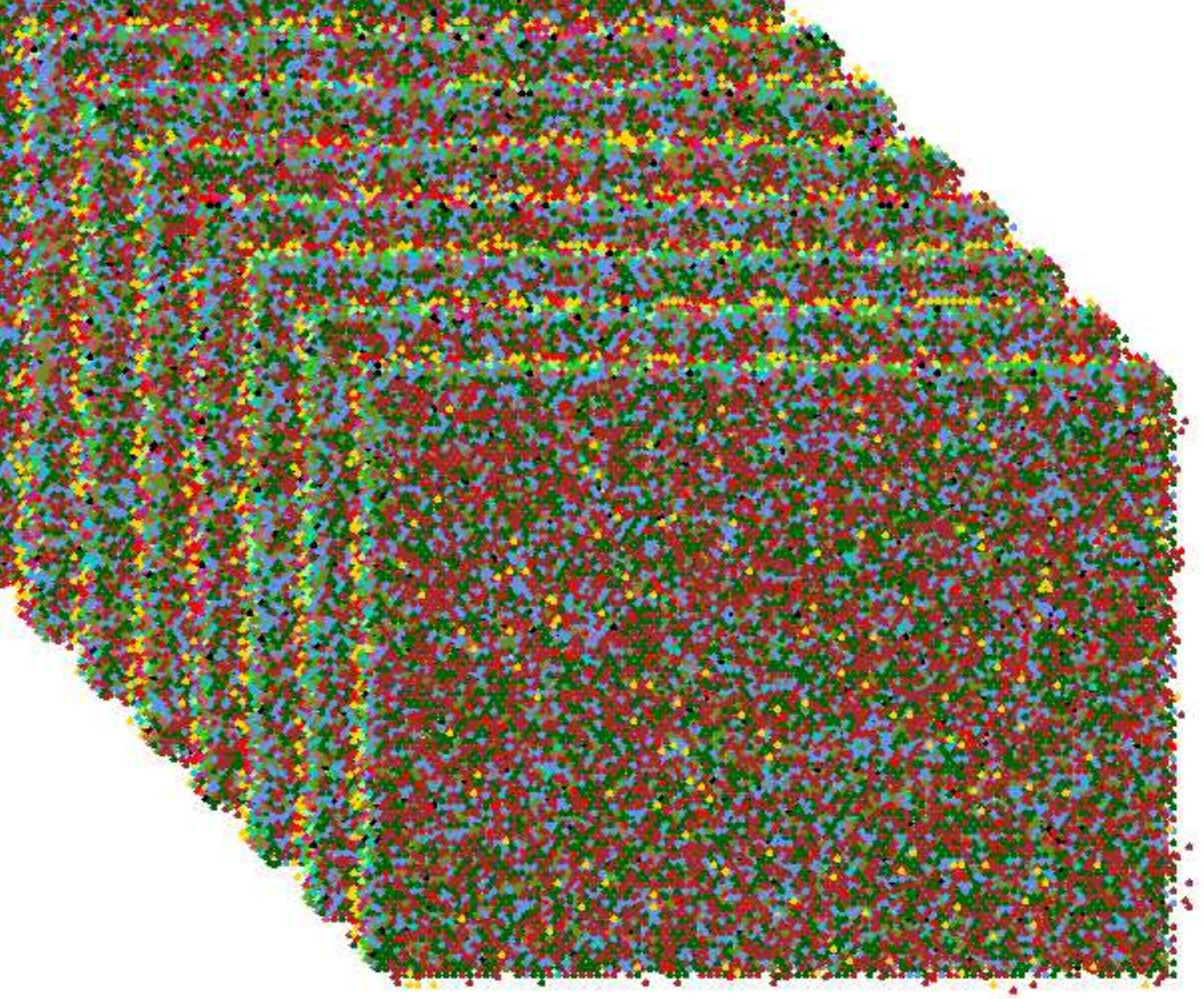}\\
(d)
\end{minipage}
\hfill
\begin{minipage}[c]{.33\linewidth} 
\includegraphics[width=1\linewidth]{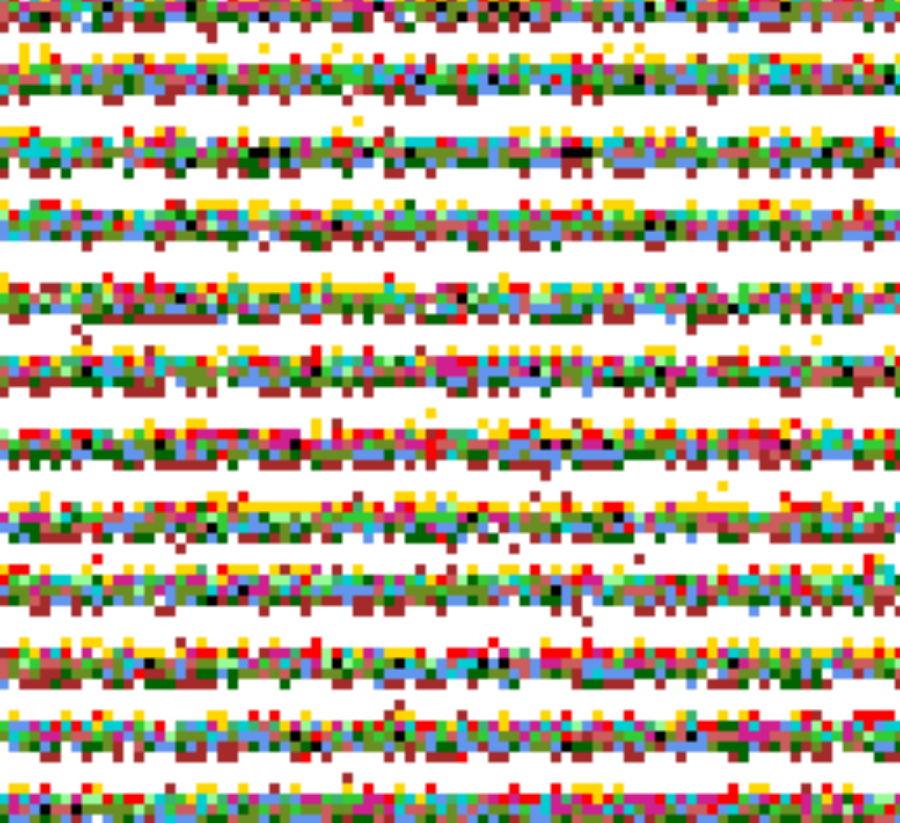}\\
(e)
\end{minipage}
\hfill 
\begin{minipage}[c]{.07\linewidth}
\fbox{\includegraphics[width=1\linewidth]{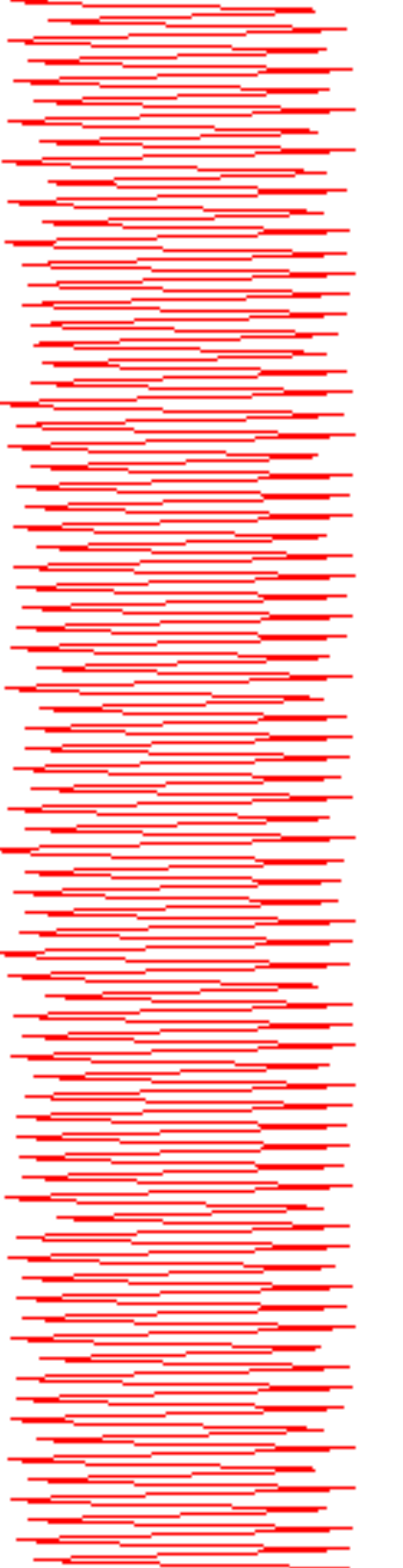}}\\[1ex]
(f1)
\end{minipage}
\end{minipage}
}
\end{center}
\vspace{-1ex}
\textsf{\small
\noindent Space-time patterns illustrating density oscillations.
(d) scrolling diagonally, the present moment is at the front
leaving a trail of time-steps behind. 
(e) a 1d segment, scrolling vertically with the most recent time-step at the bottom.
(f1) input-entropy oscillations with time (y-axis). 
}

\begin{center}
\textsf{\small
\begin{minipage}[c]{.95\linewidth} 
\begin{minipage}[c]{.15\linewidth}
\fbox{\includegraphics[width=.85\linewidth]{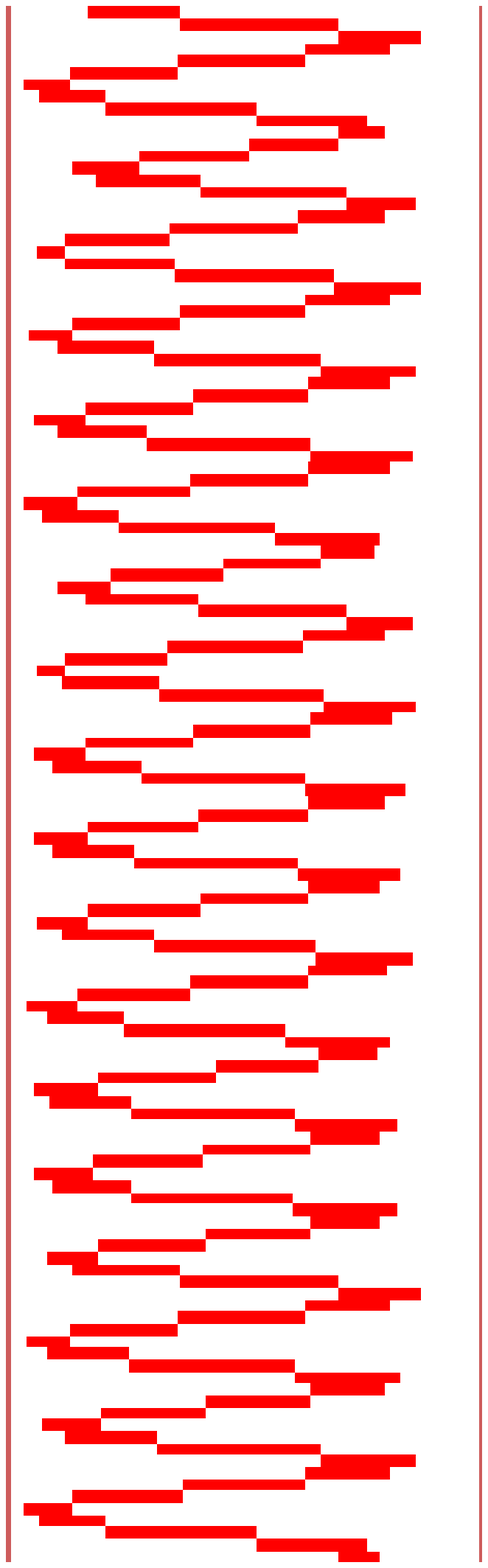}}\\[1ex]
(f2)
\end{minipage}
\hfill
\begin{minipage}[c]{.35\linewidth}
\includegraphics[width=1\linewidth]{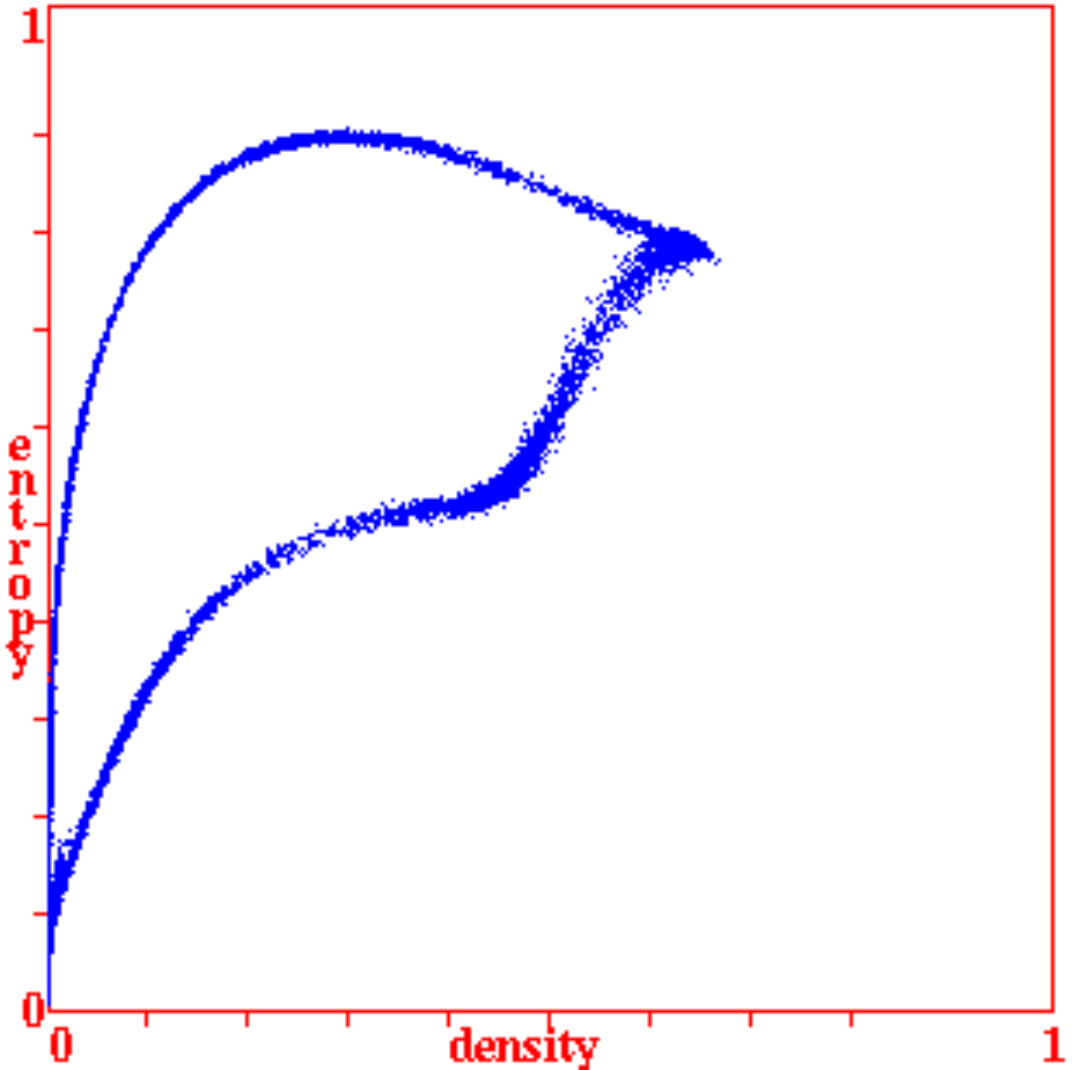}\\
(g)
\end{minipage}
\hfill
\begin{minipage}[c]{.35\linewidth}
\includegraphics[width=1\linewidth]{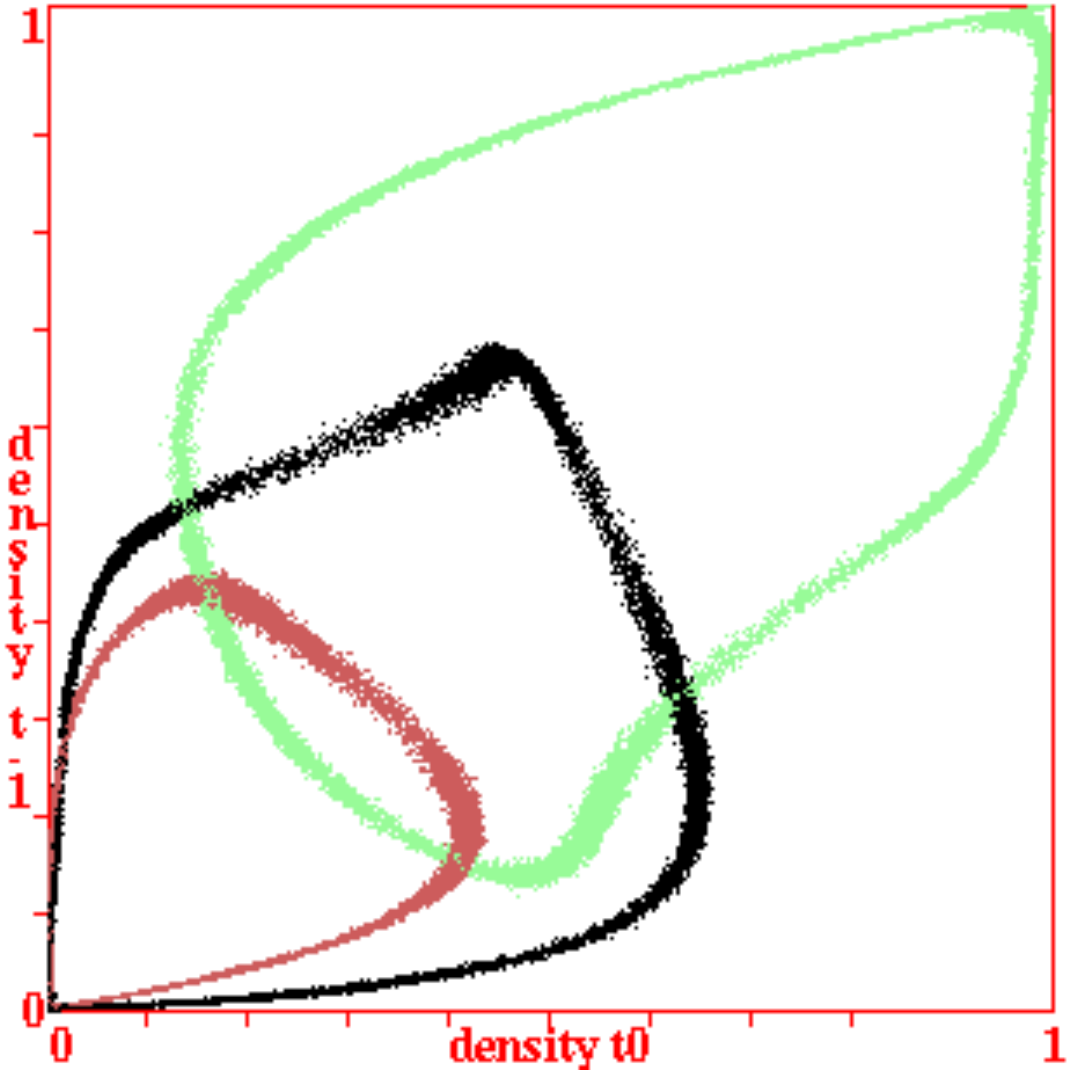}\\
(h)
\end{minipage}
\end{minipage}
}
\end{center}
\vspace{-1ex}
\textsf{\small
\noindent Time-plots of measures.
(f2) input-entropy oscillations with time (y-axis, stretched)\\
$wl$ = 7 or 8 time-steps, $wh\approx 0.8$.
(g) entropy-density scatter plot -- input-entropy (x-axis)
against the non-zero density (y-axis). (h) density return map scatter plot.}

\caption[Pulsing dynamics beehive rule]
{\textsf{
Pulsing dynamics for the $v3k6$ ``g2'' Beehive rule, (hex) 0a0282816a0264, 
on a 100x100 hexagonal lattice,
showing space-time patterns --- snapshots, scrolling, and time-plots of measures.
}}
\label{Pulsing dynamics beehive rule}
\end{figure}

\begin{figure}[htb]
\begin{center}
\textsf{\small
\begin{minipage}[c]{1\linewidth} 
\begin{minipage}[c]{.3\linewidth}
\includegraphics[width=1\linewidth]{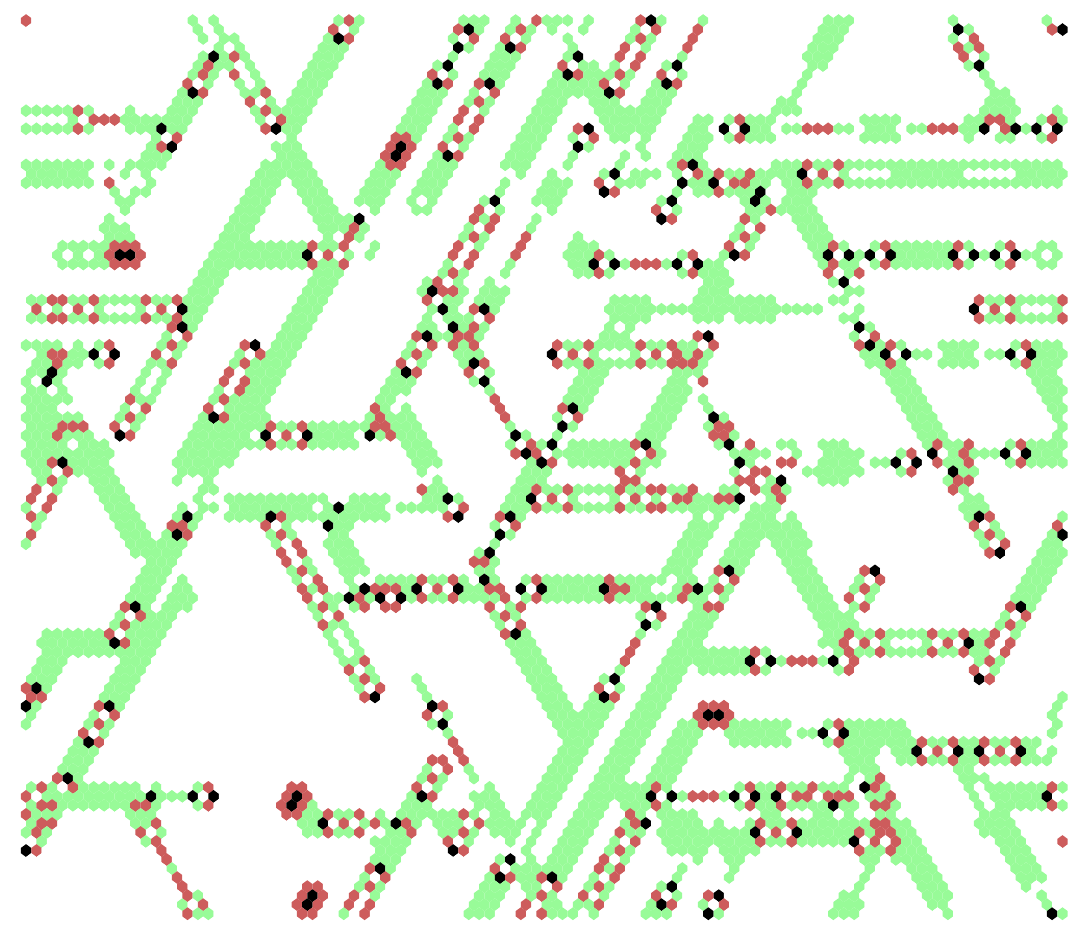}\\
(a)
\end{minipage}
\hfill
\begin{minipage}[c]{.3\linewidth}
\includegraphics[width=1\linewidth]{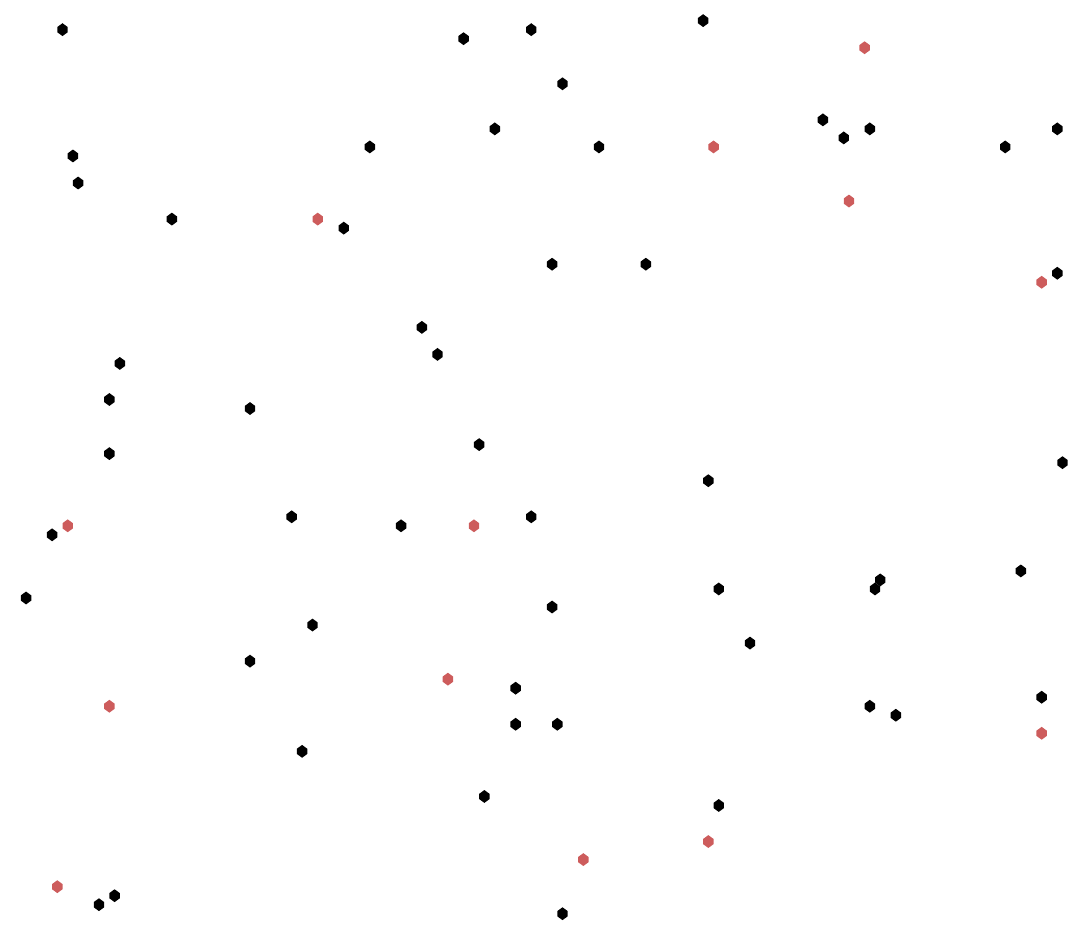}\\
(b)
\end{minipage}
\hfill
\begin{minipage}[c]{.3\linewidth}
\includegraphics[width=1\linewidth]{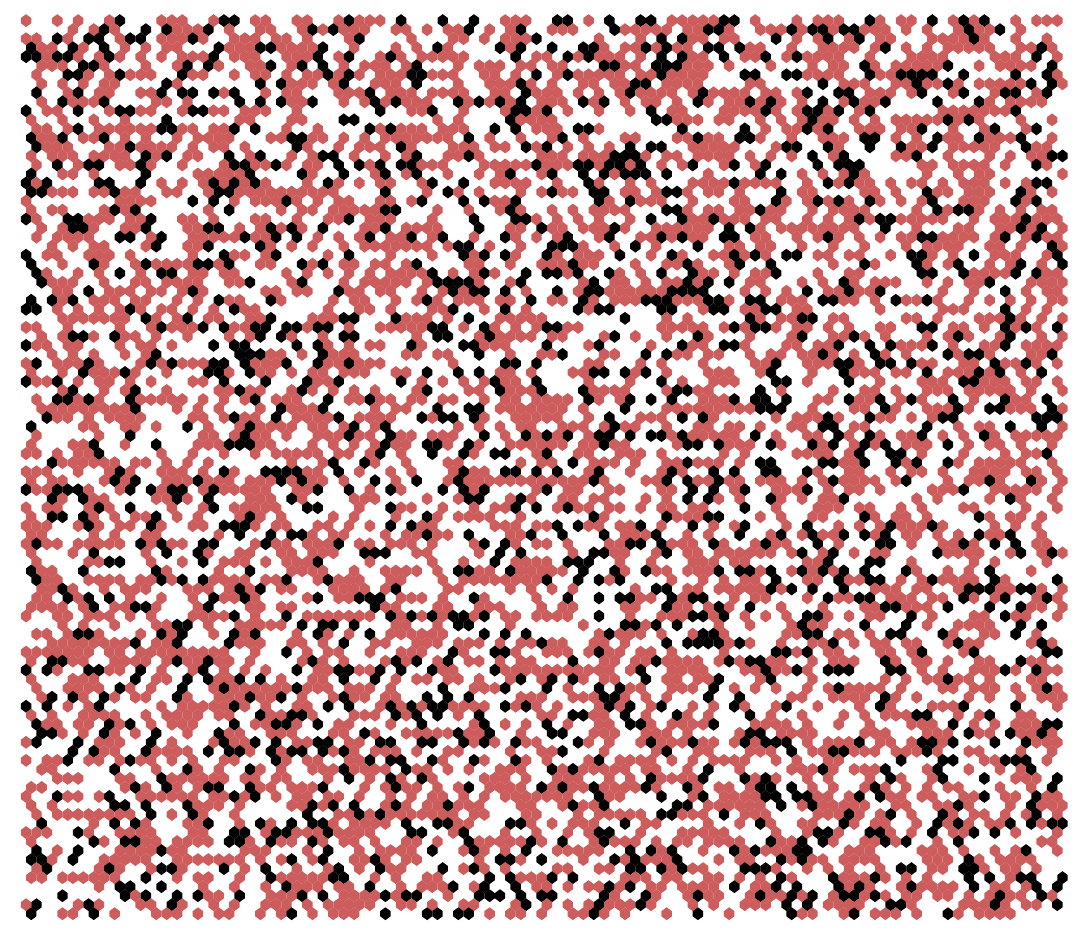}\\
(c)
\end{minipage}
\end{minipage}
}
\end{center}
\vspace{-1ex}
\textsf{\small
\noindent Space-time pattern snapshots,
(a) CA showing emergent gliders.
Randomised wiring results in disordered patterns,
(b) minimum density, and (c) maximum density.
}

\begin{center}
\textsf{\small
\begin{minipage}[c]{.95\linewidth} 
\begin{minipage}[c]{.37\linewidth} 
\includegraphics[width=1\linewidth]{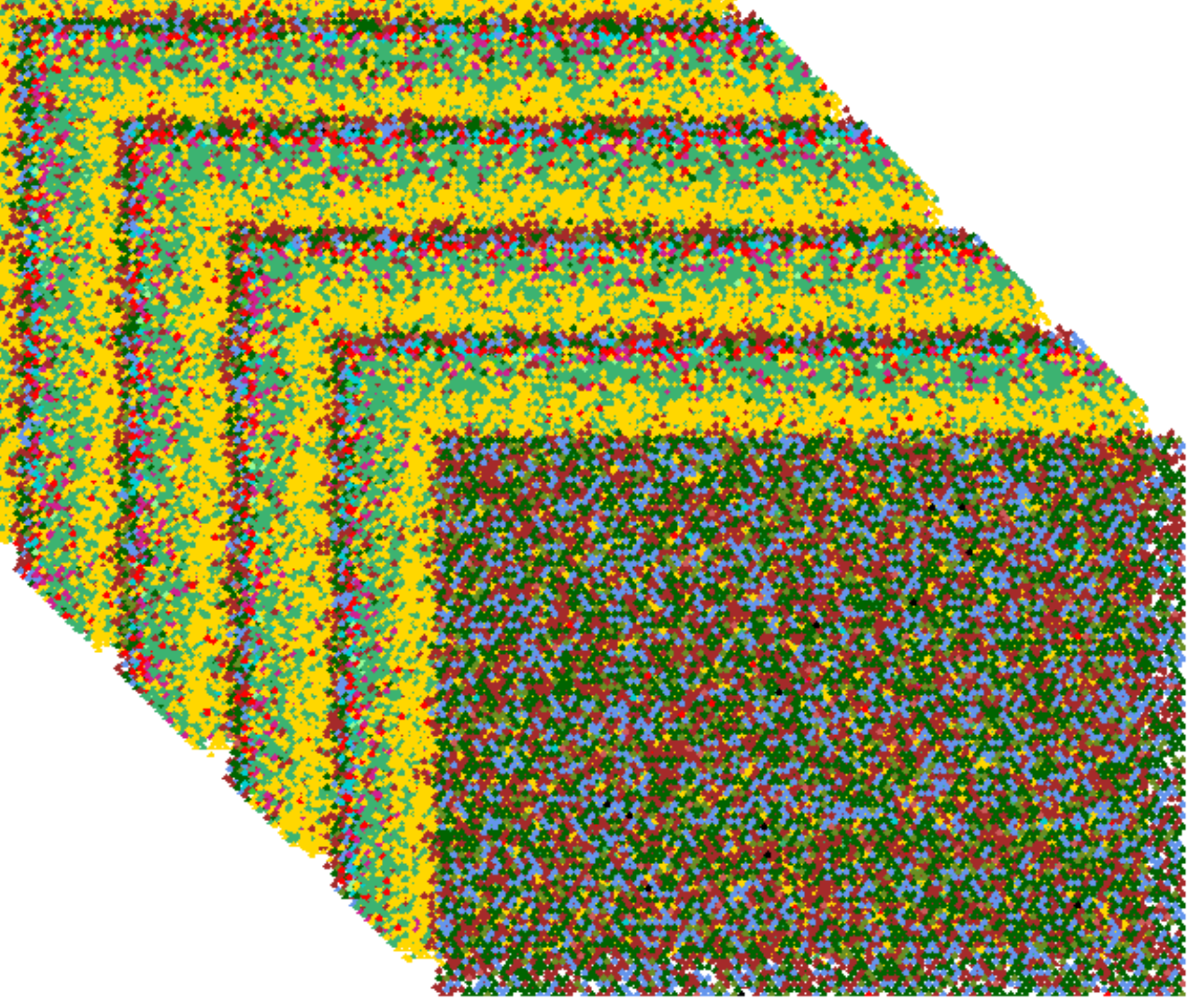}\\
(d)
\end{minipage}
\hfill
\begin{minipage}[c]{.33\linewidth} 
\includegraphics[width=1\linewidth,]{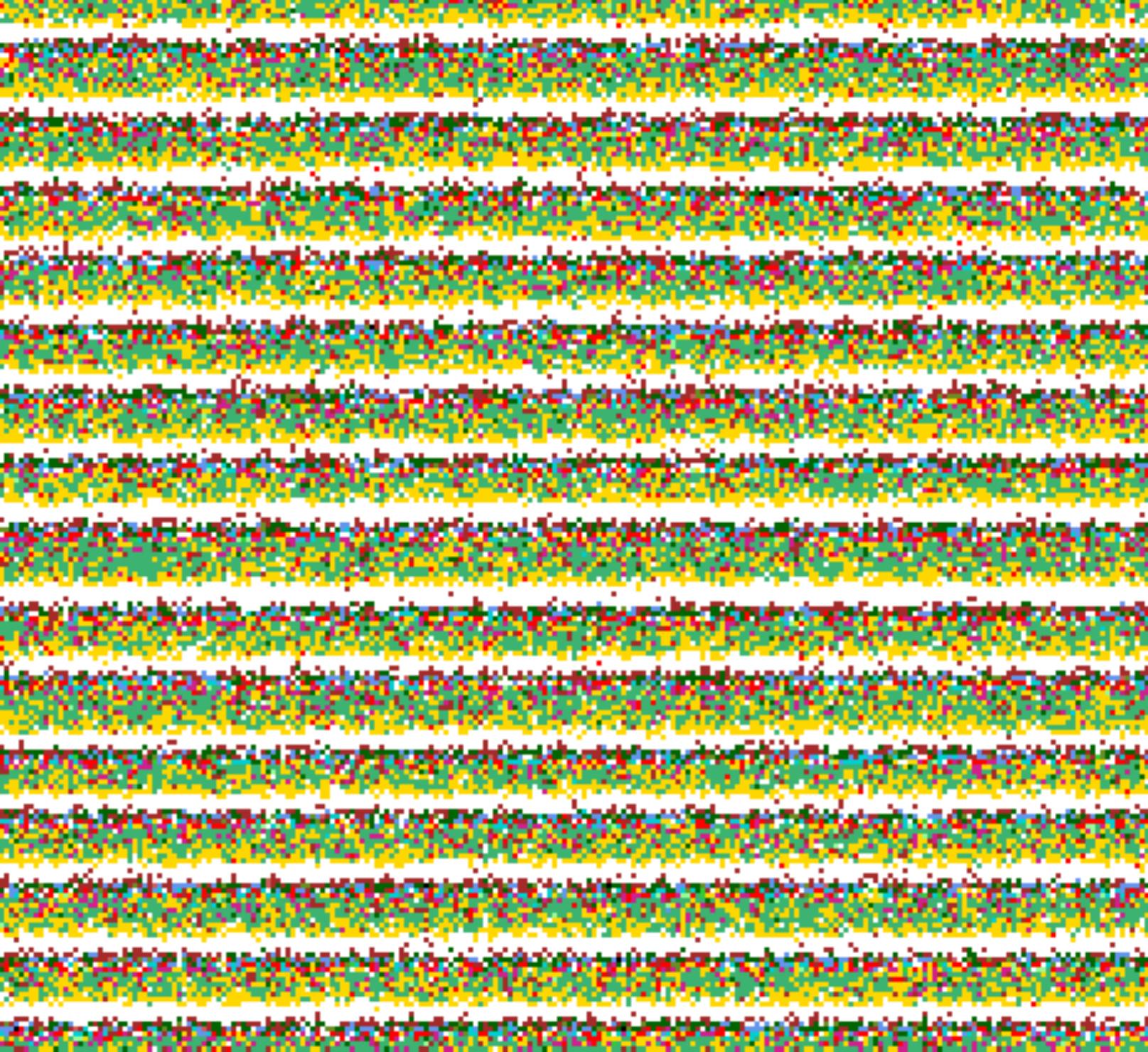}\\
(e)
\end{minipage}
\hfill 
\begin{minipage}[c]{.08\linewidth}
\fbox{\includegraphics[width=1\linewidth]{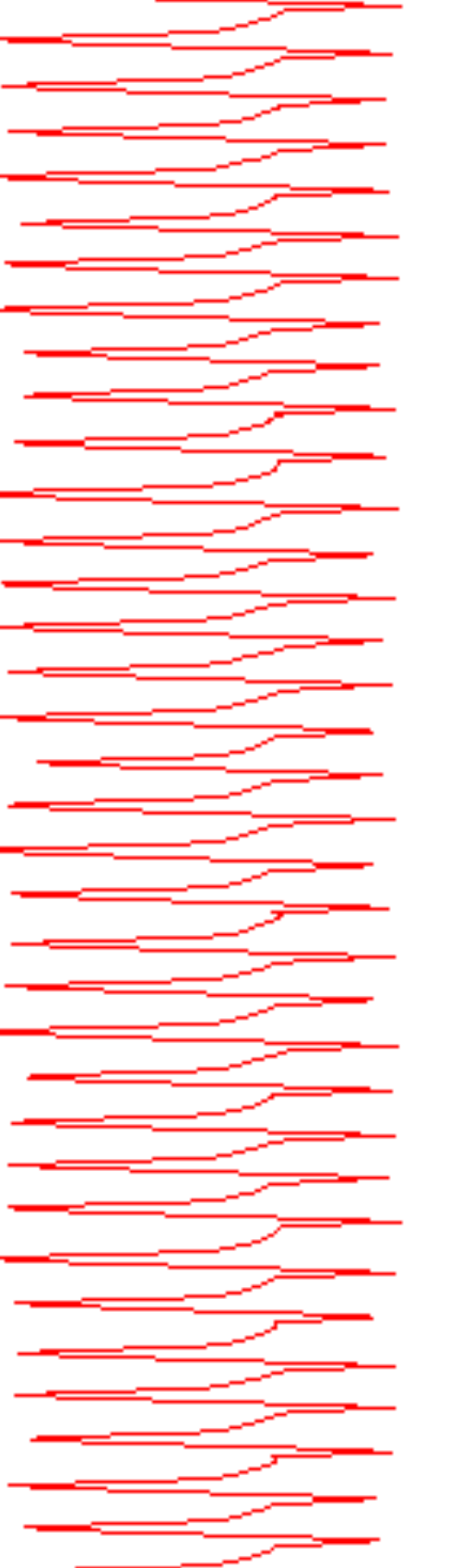}}\\[1ex]
(f1)
\end{minipage}
\end{minipage}
}
\end{center}
\vspace{-1ex}
\textsf{\small
\noindent Space-time patterns illustrating density oscillations.
(d) scrolling diagonally, the present moment is at the front
leaving a trail of time-steps behind. 
(e) a 1d segment, scrolling vertically with the most recent time-step at the bottom.
(f1) input-entropy oscillations with time (y-axis). 
}

\begin{center}
\textsf{\small
\begin{minipage}[c]{.95\linewidth} 
\begin{minipage}[c]{.15\linewidth}
\fbox{\includegraphics[width=.85\linewidth]{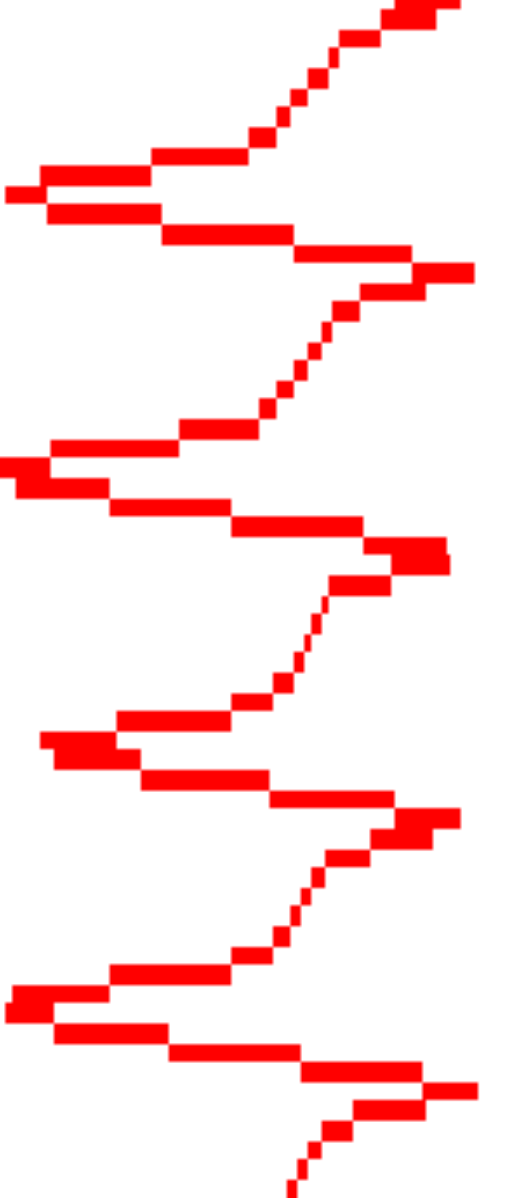}}\\[1ex]
(f2)
\end{minipage}
\hfill
\begin{minipage}[c]{.35\linewidth}
\includegraphics[width=1\linewidth]{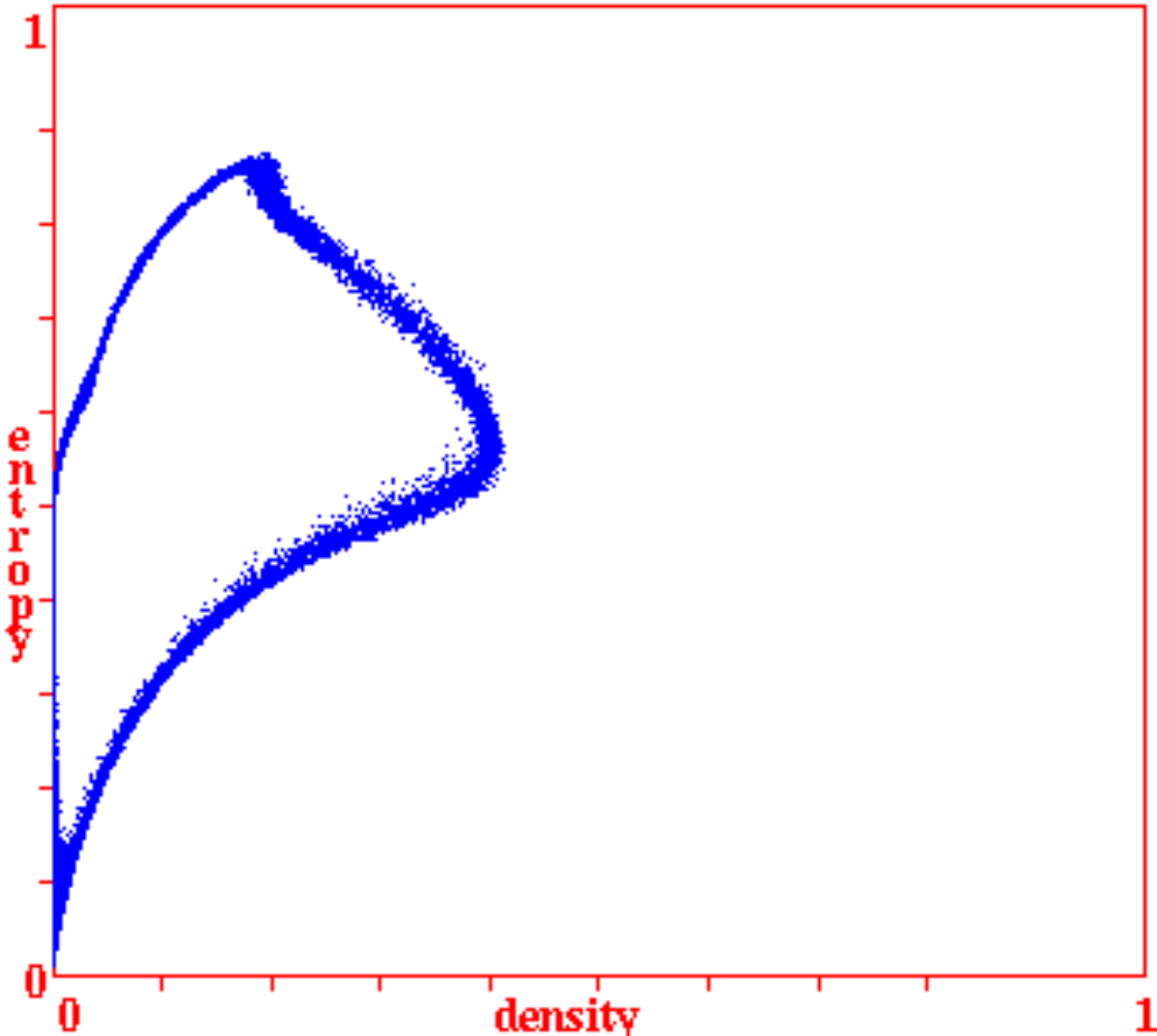}\\
(g)
\end{minipage}
\hfill
\begin{minipage}[c]{.35\linewidth}
\includegraphics[width=1\linewidth]{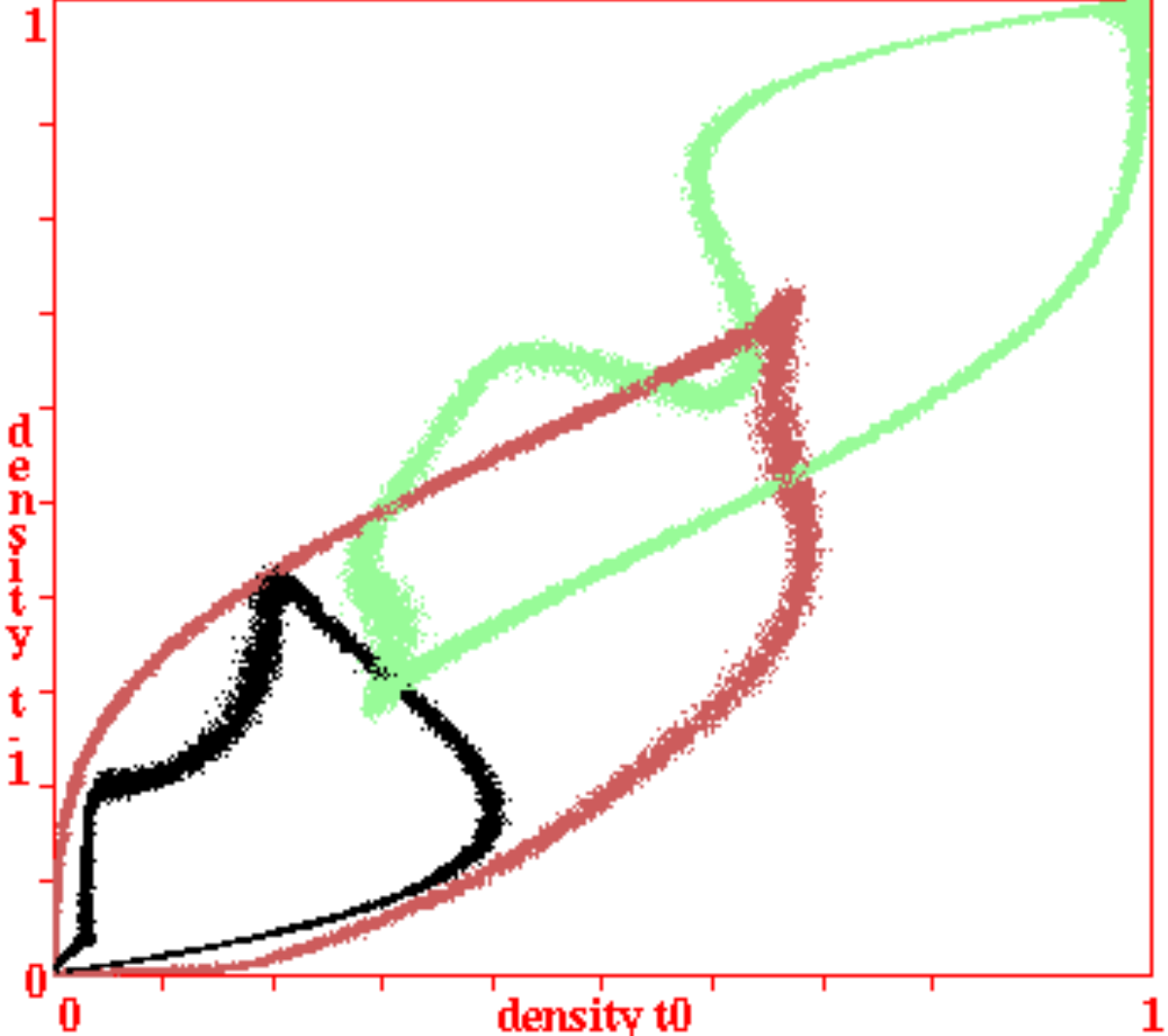}\\
(h)
\end{minipage}
\end{minipage}
}
\end{center}
\vspace{-1ex}
\textsf{\small
\noindent Time-plots of measures.
(f2) input-entropy oscillations with time (y-axis, stretched)\\
$wl$ 14 or 15 time-steps, $wh\approx 0.8$.
(g) entropy-density scatter plot -- input-entropy (x-axis)
against the non-zero density (y-axis). (h) density return map scatter plot.}

\caption[Pulsing dynamics g39]
{\textsf{
Pulsing dynamics for the $v3k6$ ``g39''rule, (hex) 0a184552558500(hex), 
on a 100x100 hexagonal lattice,
showing space-time patterns --- snapshots, scrolling, and time-plots of measures.
}}
\label{Pulsing dynamics k6-g39}
\end{figure}

\begin{figure}[htb]
\begin{center}
\textsf{\small
\begin{minipage}[c]{1\linewidth} 
\begin{minipage}[c]{.3\linewidth}
\includegraphics[width=1\linewidth]{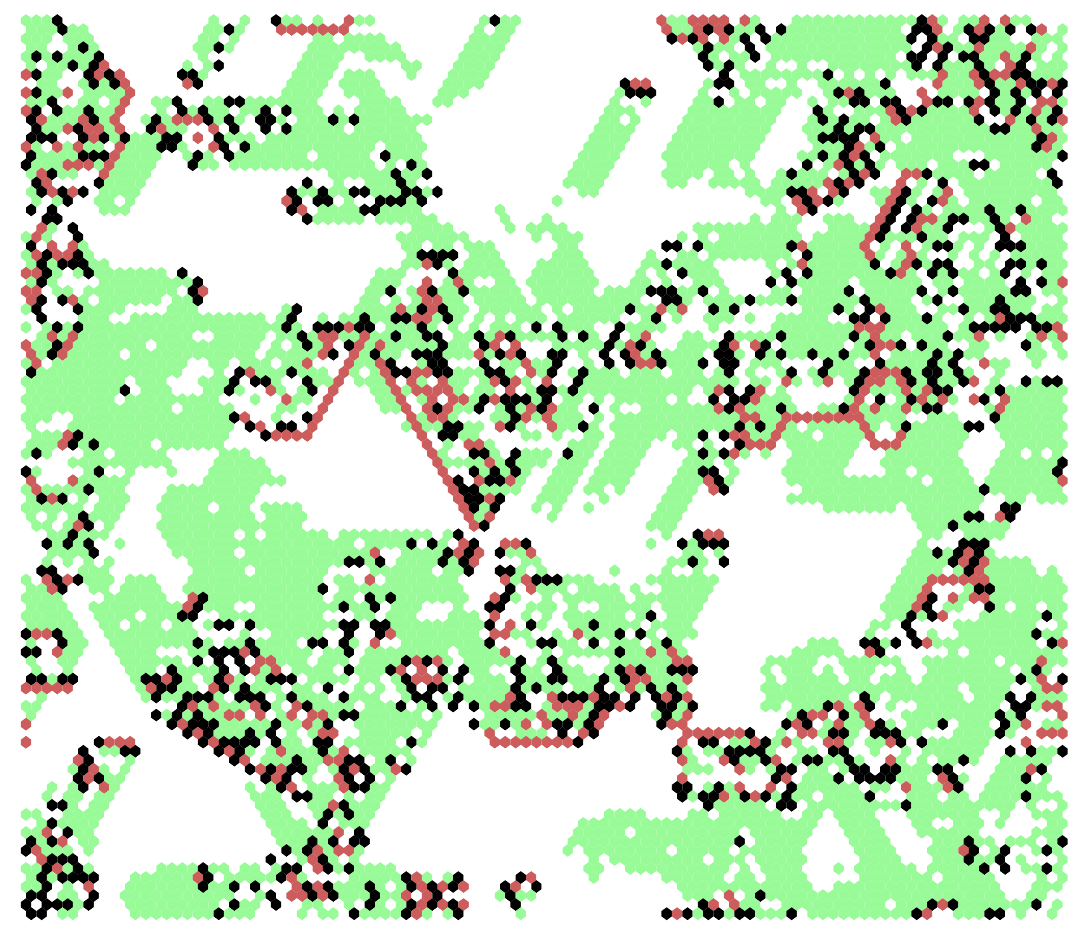}\\
(a)
\end{minipage}
\hfill
\begin{minipage}[c]{.3\linewidth}
\includegraphics[width=1\linewidth]{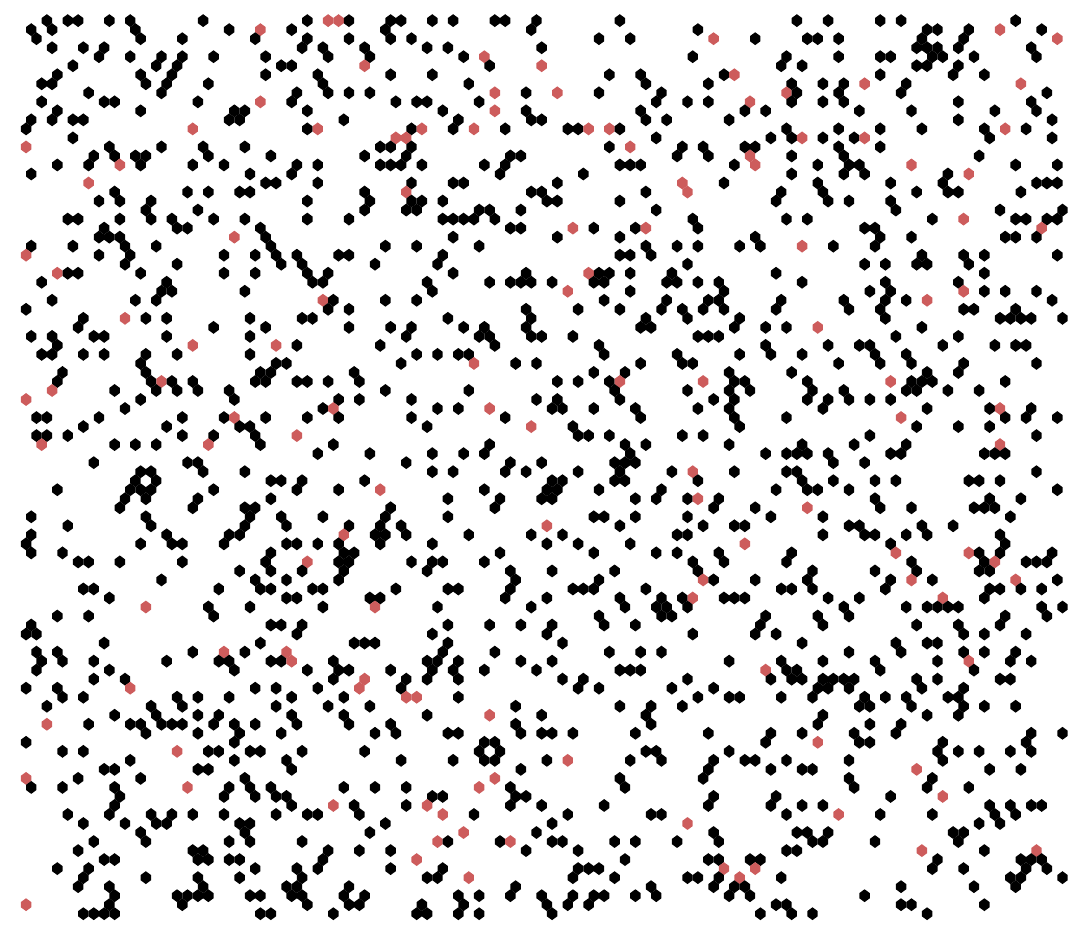}\\
(b)
\end{minipage}
\hfill
\begin{minipage}[c]{.3\linewidth}
\includegraphics[width=1\linewidth]{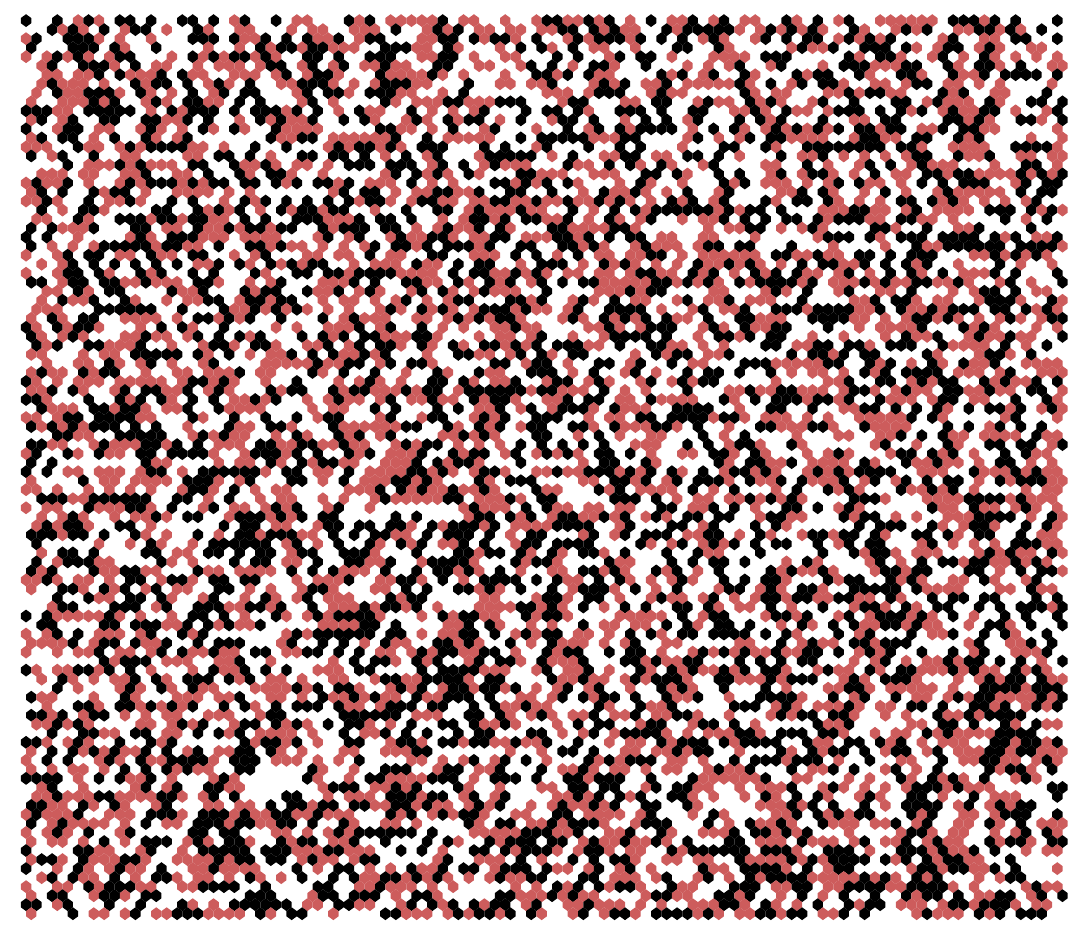}\\
(c)
\end{minipage}
\end{minipage}
}
\end{center}
\vspace{-1ex}
\textsf{\small
\noindent Space-time pattern snapshots,
(a) CA showing emergent gliders.
Randomised wiring results in disordered patterns,
(b) minimum density, and (c) maximum density.
}

\begin{center}
\textsf{\small
\begin{minipage}[c]{.95\linewidth} 
\begin{minipage}[c]{.37\linewidth} 
\includegraphics[width=1\linewidth]{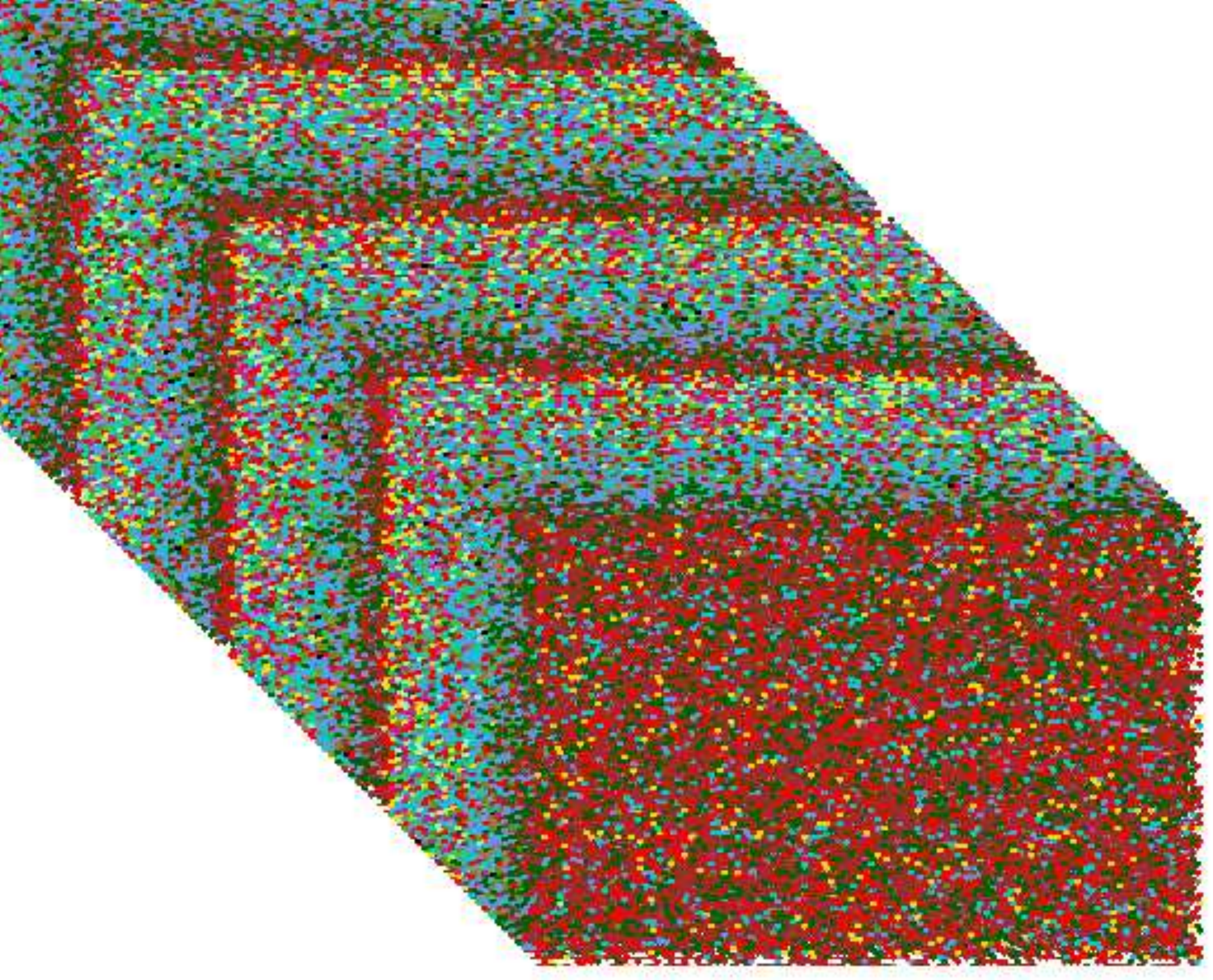}\\
(d)
\end{minipage}
\hfill
\begin{minipage}[c]{.33\linewidth} 
\includegraphics[width=1\linewidth]{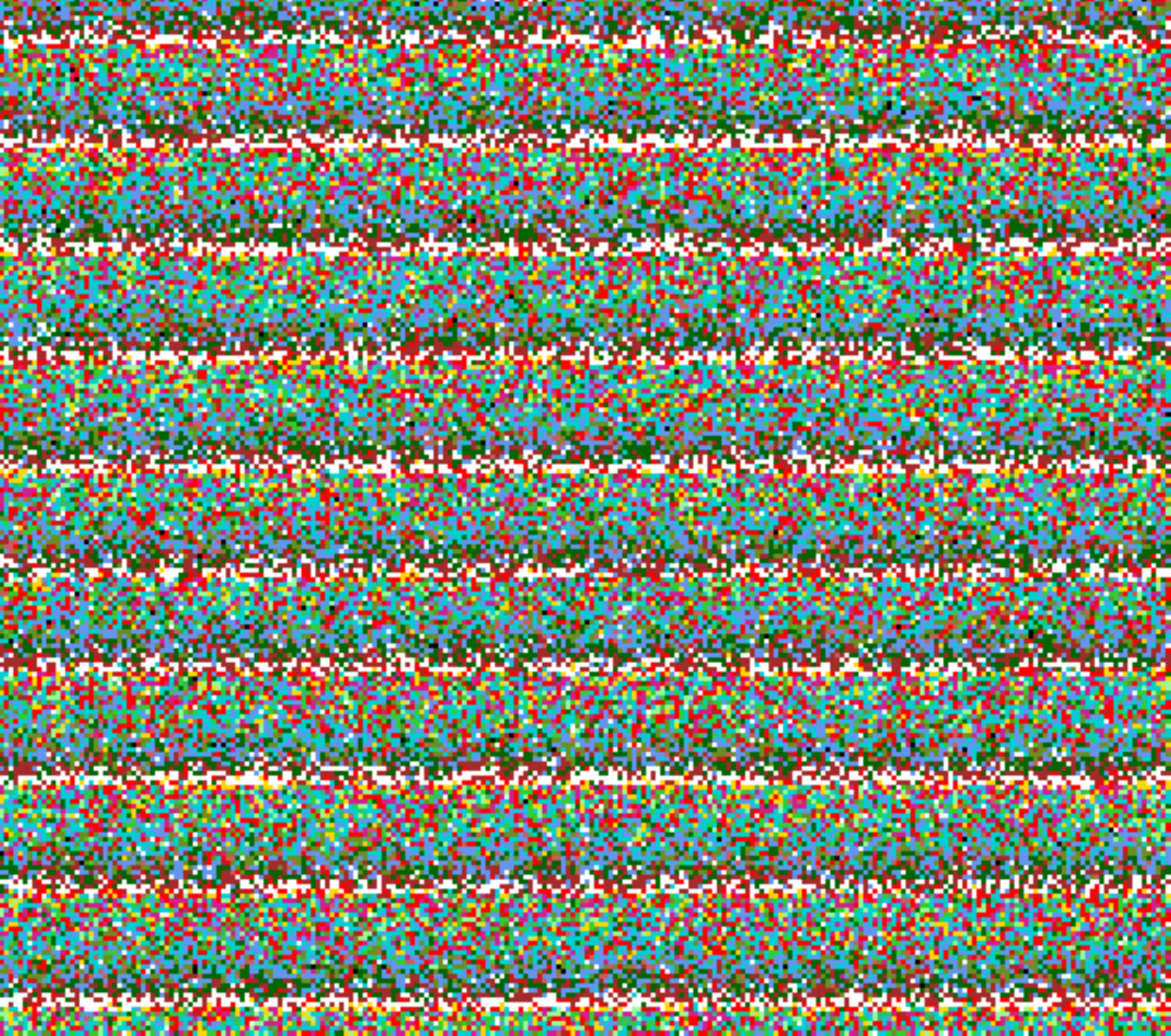}\\
(e)
\end{minipage}
\hfill 
\begin{minipage}[c]{.08\linewidth}
\fbox{\includegraphics[width=1\linewidth]{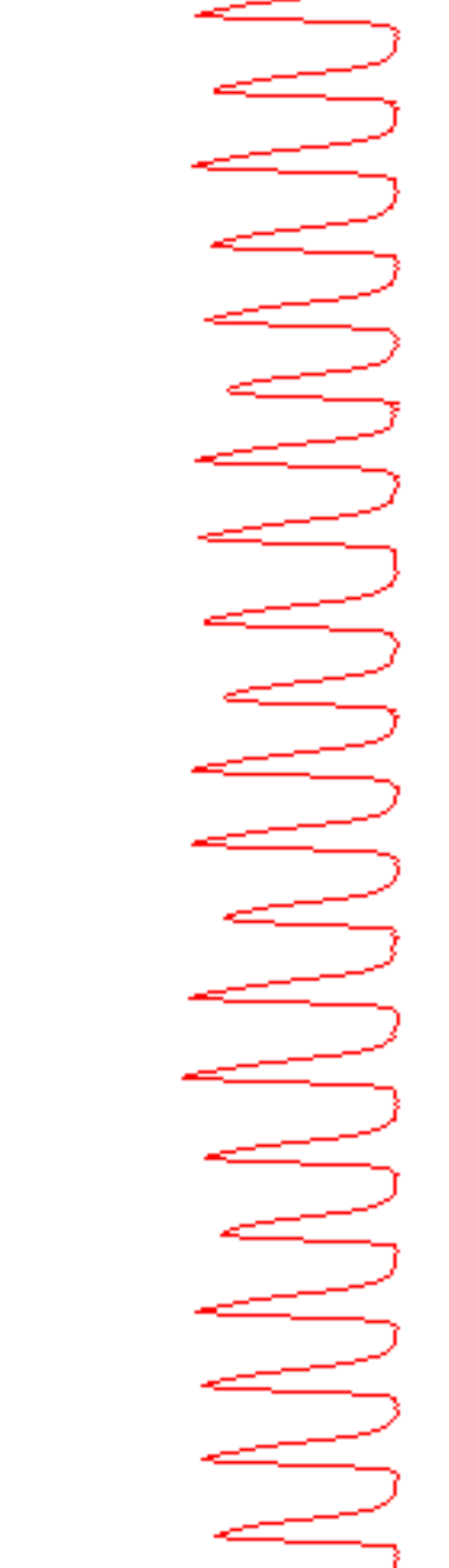}}\\[1ex]
(f1)
\end{minipage}
\end{minipage}
}
\end{center}
\vspace{-1ex}
\textsf{\small
\noindent Space-time patterns illustrating density oscillations.
(d) scrolling diagonally, the present moment is at the front
leaving a trail of time-steps behind. 
(e) a 1d segment, scrolling vertically with the most recent time-step at the bottom.
(f1) input-entropy oscillations with time (y-axis). 
}

\begin{center}
\textsf{\small
\begin{minipage}[c]{.95\linewidth} 
\begin{minipage}[c]{.15\linewidth}
\fbox{\includegraphics[width=.85\linewidth]{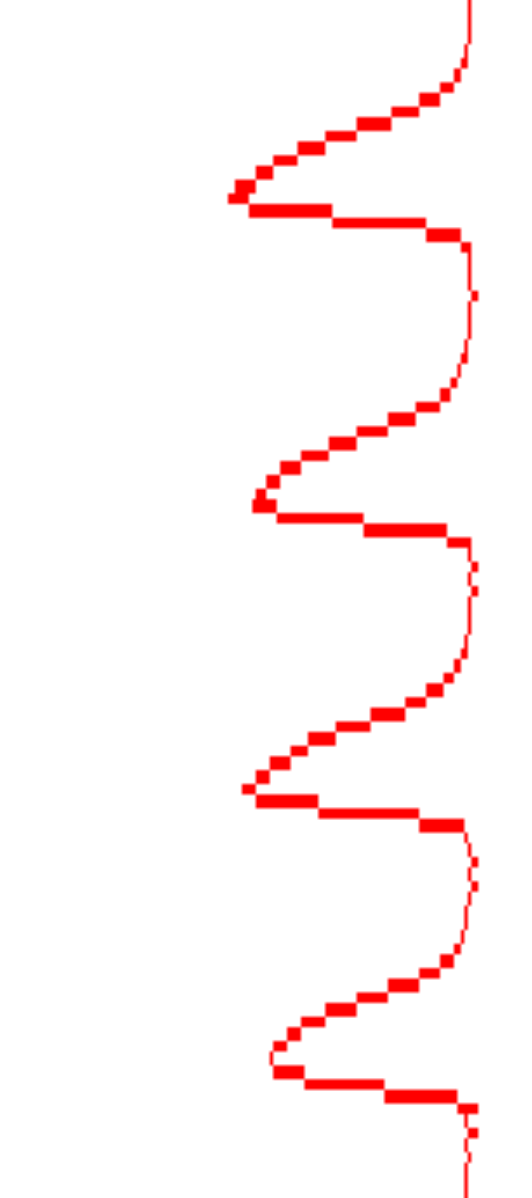}}\\[1ex]
(f2)
\end{minipage}
\hfill
\begin{minipage}[c]{.35\linewidth}
\includegraphics[width=1\linewidth]{figsR-arXiv/k6-g26-ed-100-33}\\
(g)
\end{minipage}
\hfill
\begin{minipage}[c]{.35\linewidth}
\includegraphics[width=1\linewidth]{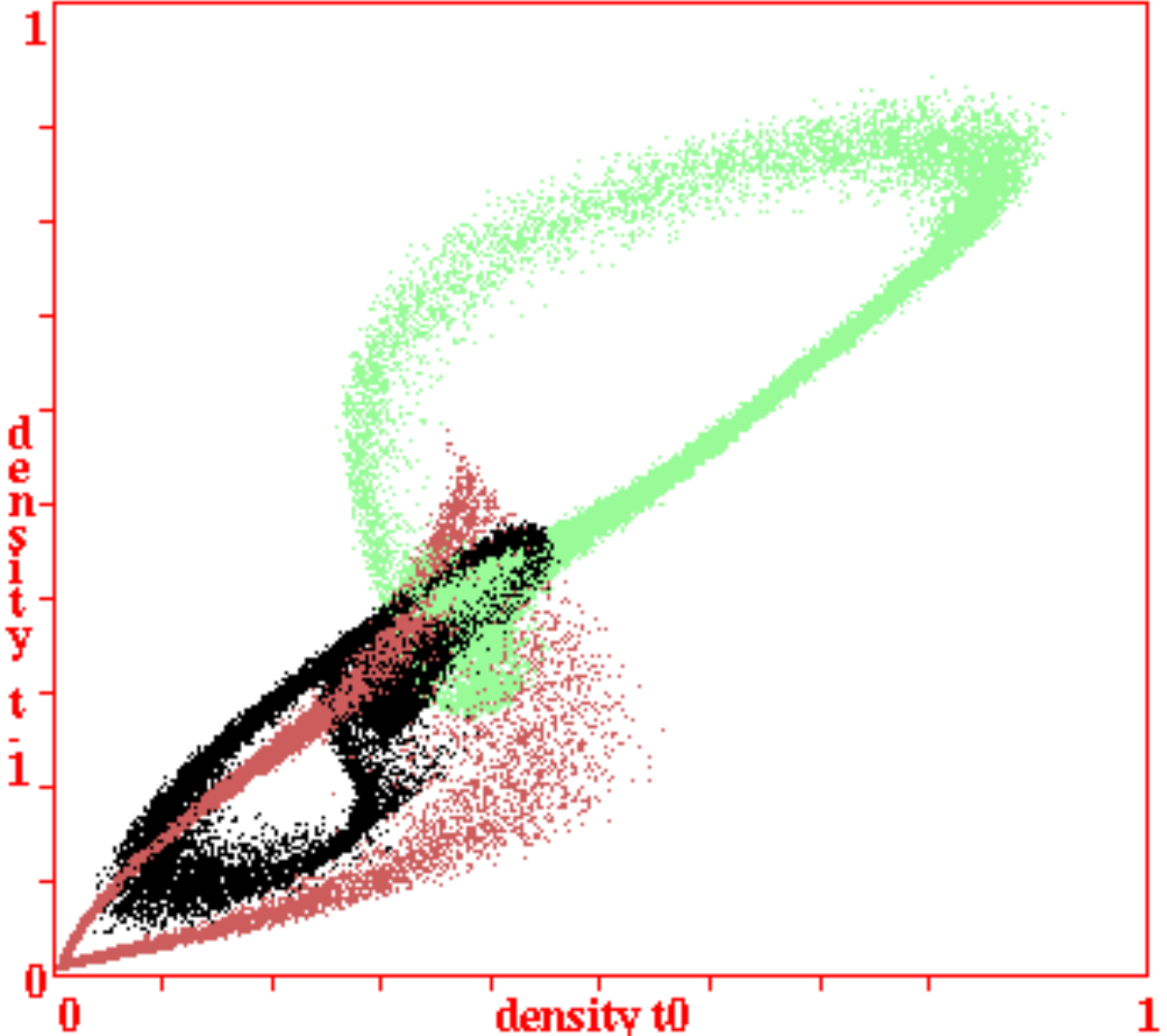}\\
(h)
\end{minipage}
\end{minipage}
}
\end{center}
\vspace{-1ex}
\textsf{\small
\noindent Time-plots of measures.
(f2) input-entropy oscillations with time (y-axis, stretched)\\
$wl\approx23$ time-steps, $wh\approx0.4$.
(g) entropy-density scatter plot -- input-entropy (x-axis)
against the non-zero density (y-axis). (h) density return map scatter plot.}

\caption[Pulsing dynamics g26]
{\textsf{
Pulsing dynamics for the $v3k6$ ``g26'' rule, (hex) 1000a121960214, 
on a 100x100 hexagonal lattice,
showing space-time patterns --- snapshots, scrolling, and time-plots of measures.
}}
\label{Pulsing dynamics k6-g26}
\end{figure}

\begin{figure}[htb]
\textsf{\small
\begin{center}
\begin{minipage}[c]{1\linewidth} 
\begin{minipage}[c]{.3\linewidth}
\includegraphics[width=1\linewidth]{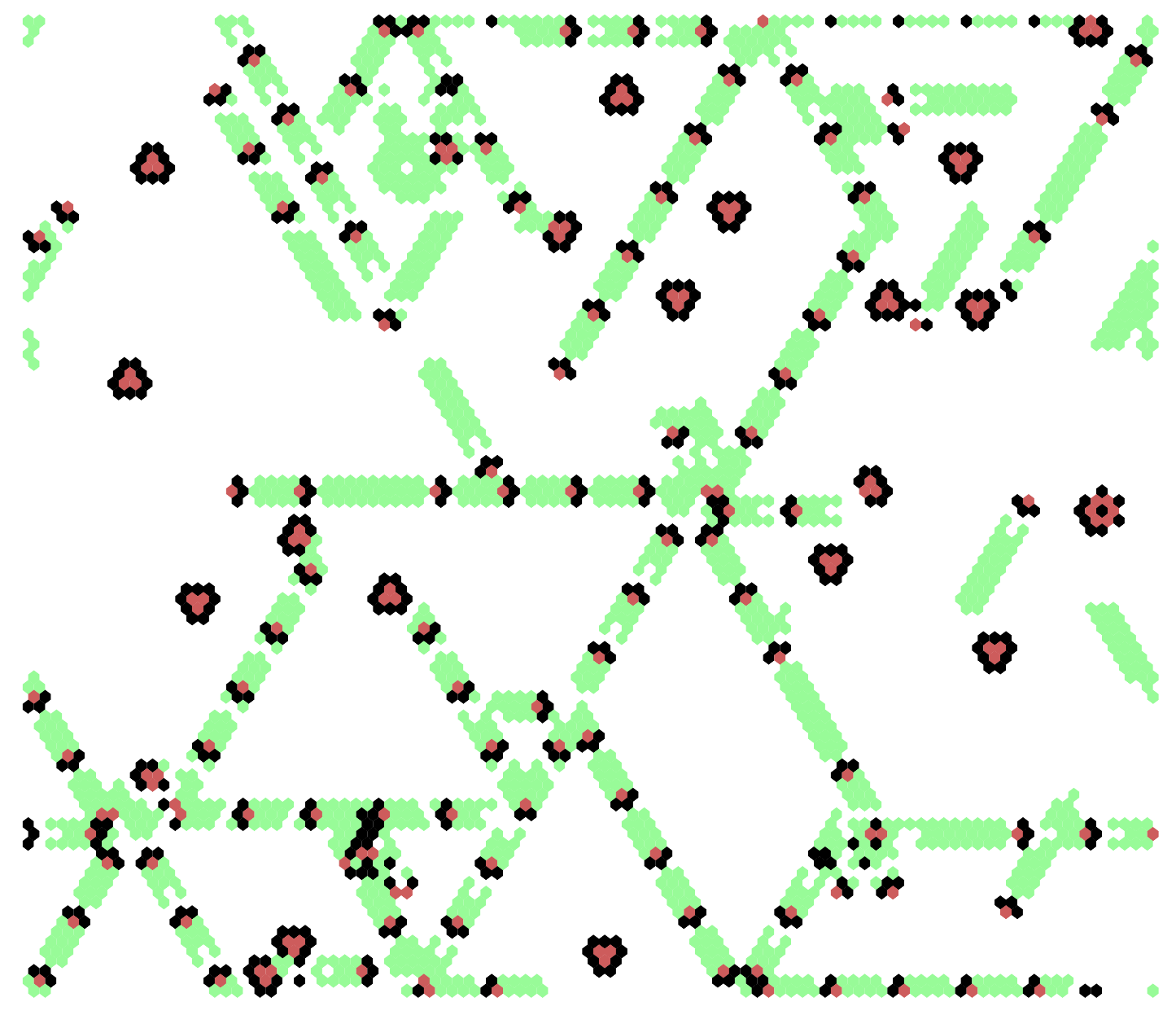}\\
(a)
\end{minipage}
\hfill
\begin{minipage}[c]{.3\linewidth}
\includegraphics[width=1\linewidth]{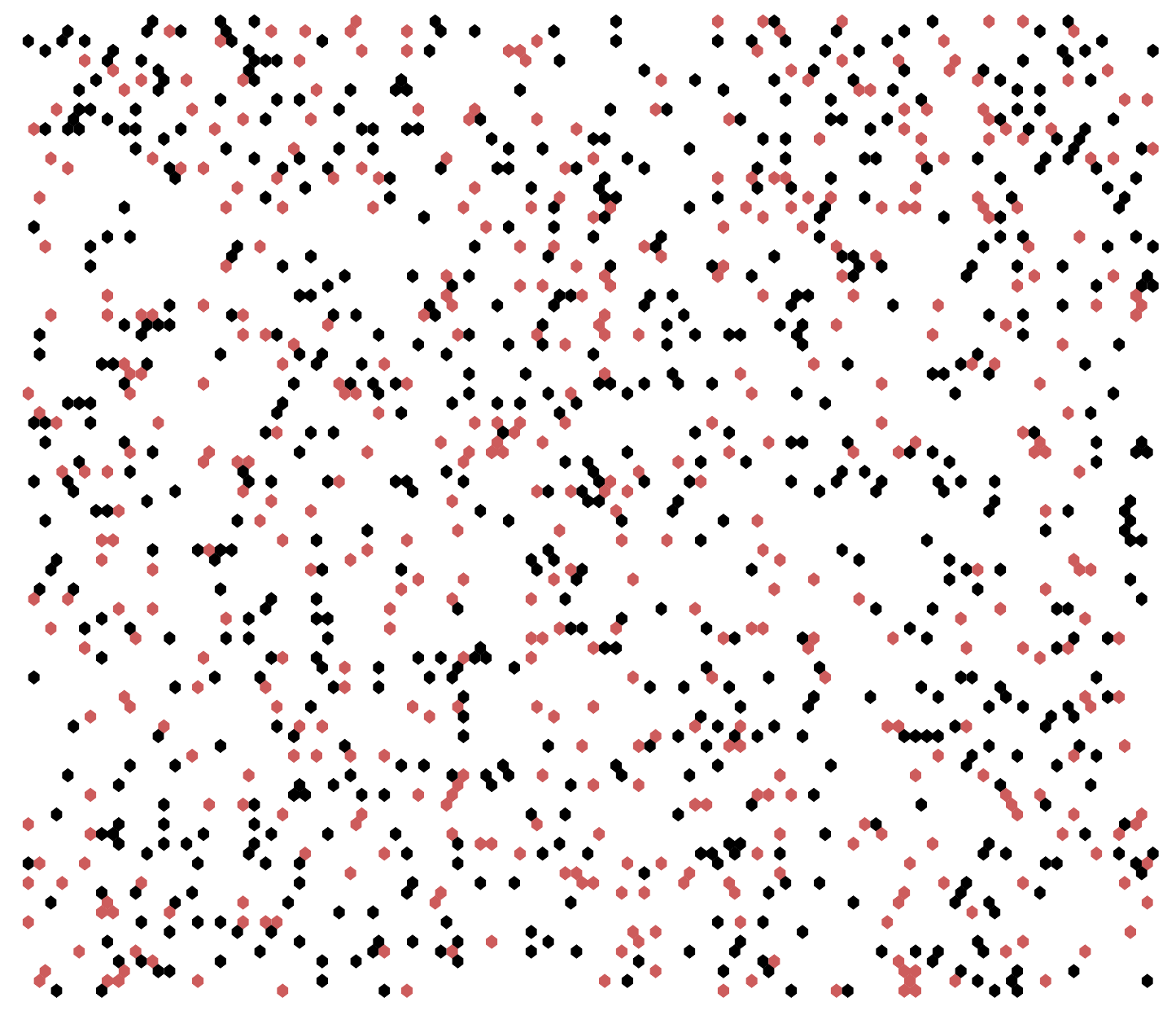}\\
(b)
\end{minipage}
\hfill
\begin{minipage}[c]{.3\linewidth}
\includegraphics[width=1\linewidth]{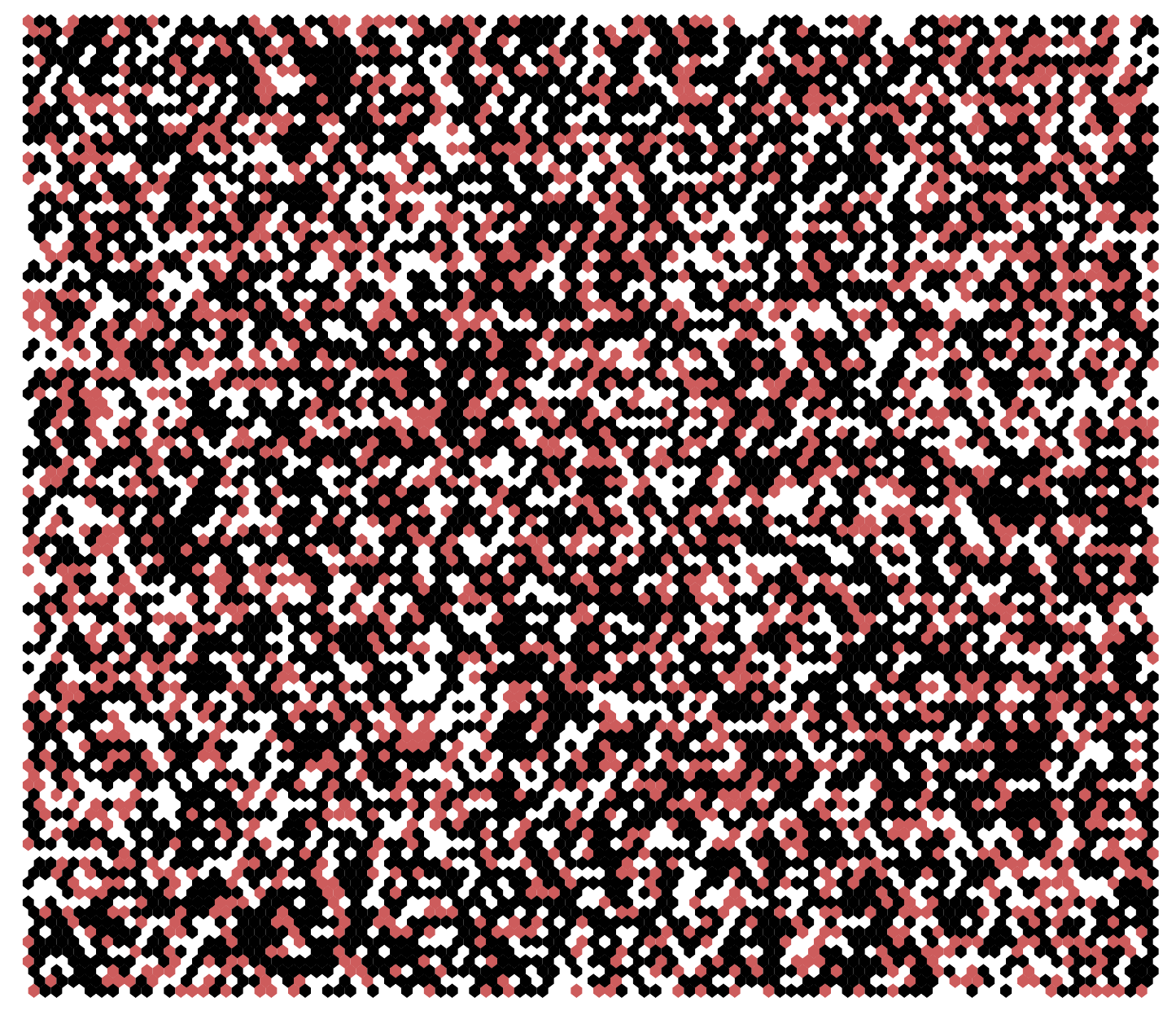}\\
(c)
\end{minipage}
\end{minipage}
\end{center}
}
\vspace{-1ex}
\textsf{\small
\noindent Space-time pattern snapshots.
(a) CA showing emergent gliders.
Randomising wiring results in disordered patterns, 
(b) minimum density, and (c) maximum density.
}

\begin{center}
\textsf{\small
\begin{minipage}[c]{.95\linewidth} 
\begin{minipage}[c]{.37\linewidth} 
\includegraphics[width=1\linewidth]{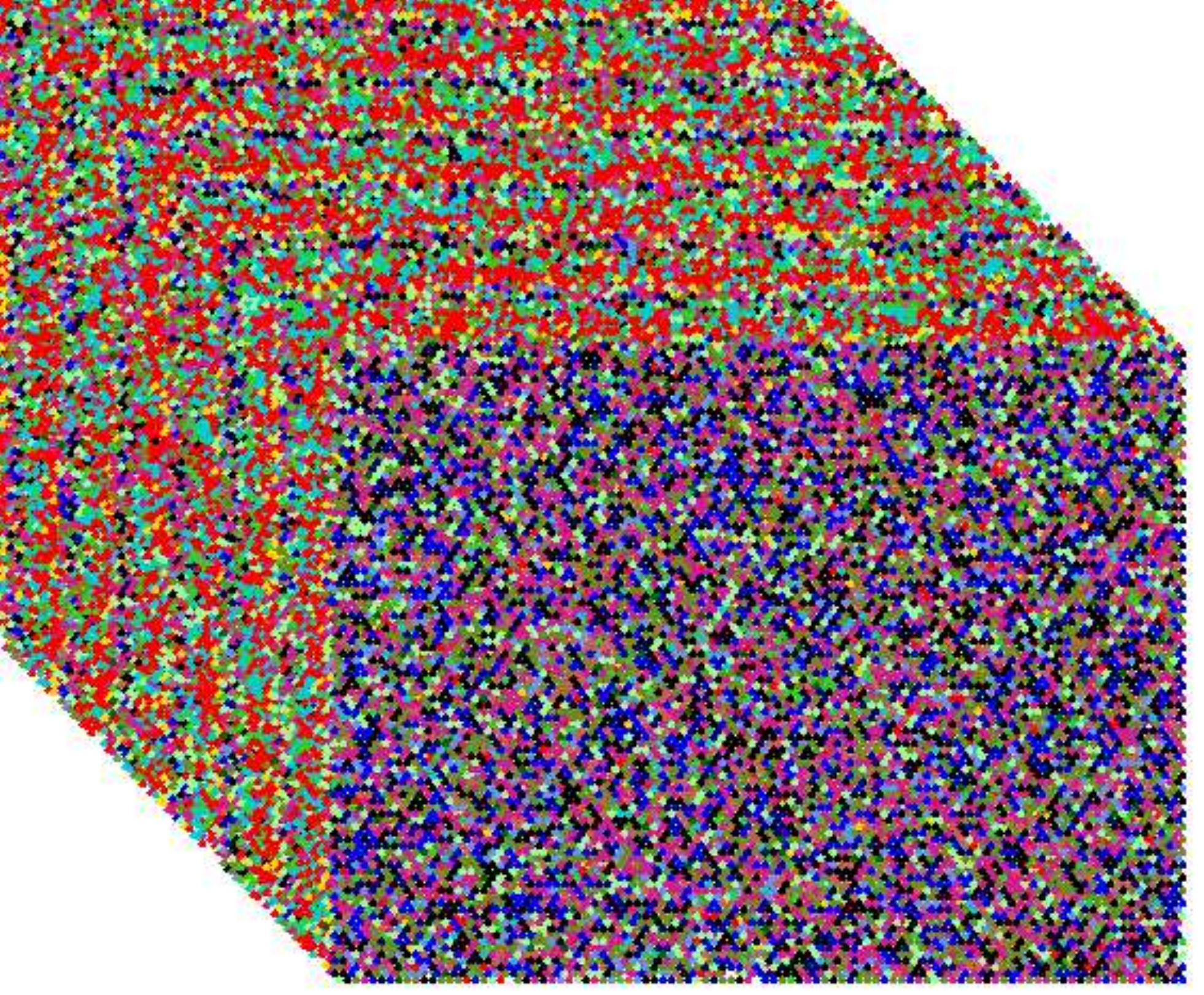}\\
(d)
\end{minipage}
\hfill
\begin{minipage}[c]{.33\linewidth} 
\includegraphics[width=1\linewidth]{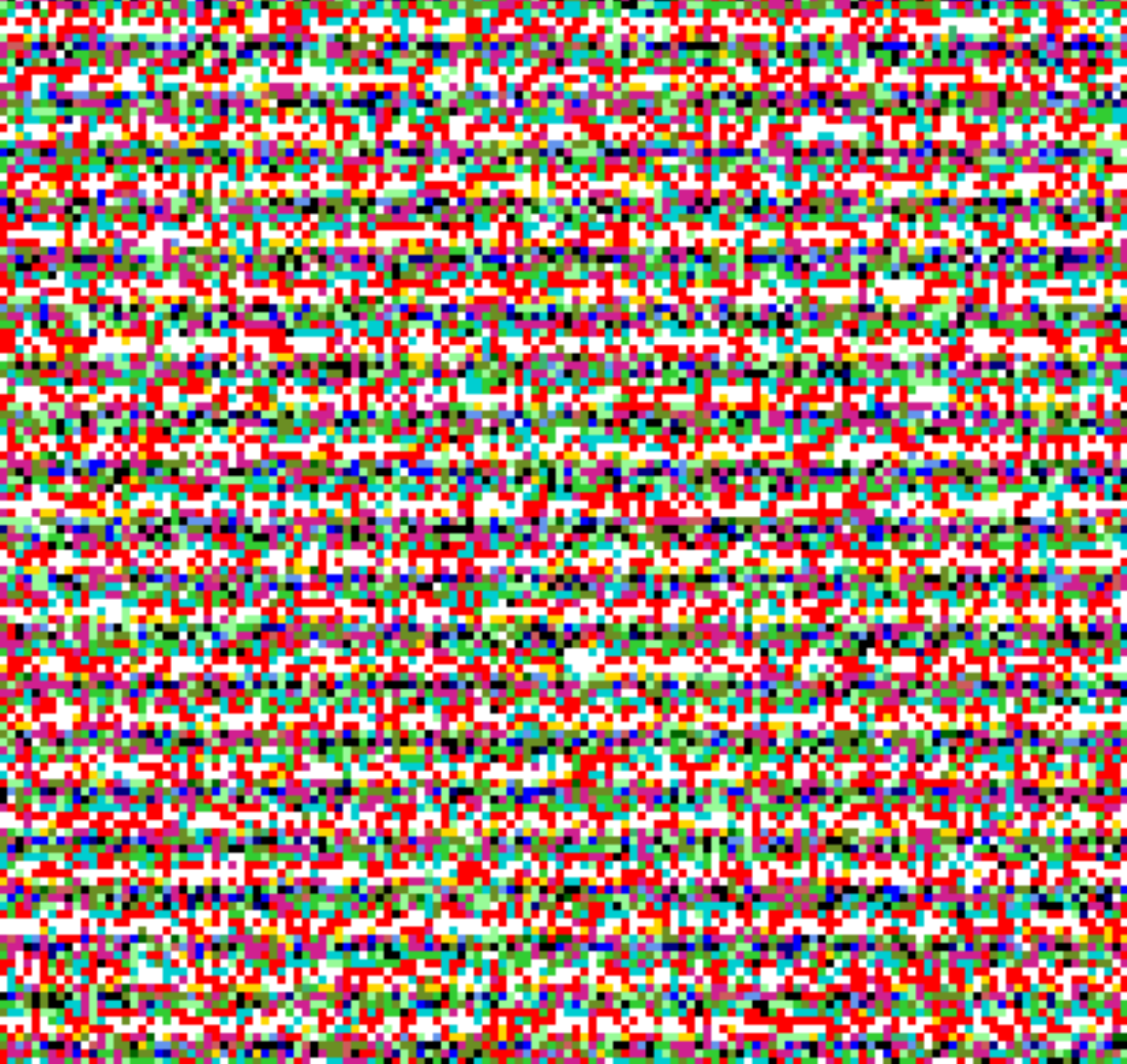}\\
(e)
\end{minipage}
\hfill 
\begin{minipage}[c]{.08\linewidth}
\fbox{\includegraphics[width=1\linewidth]{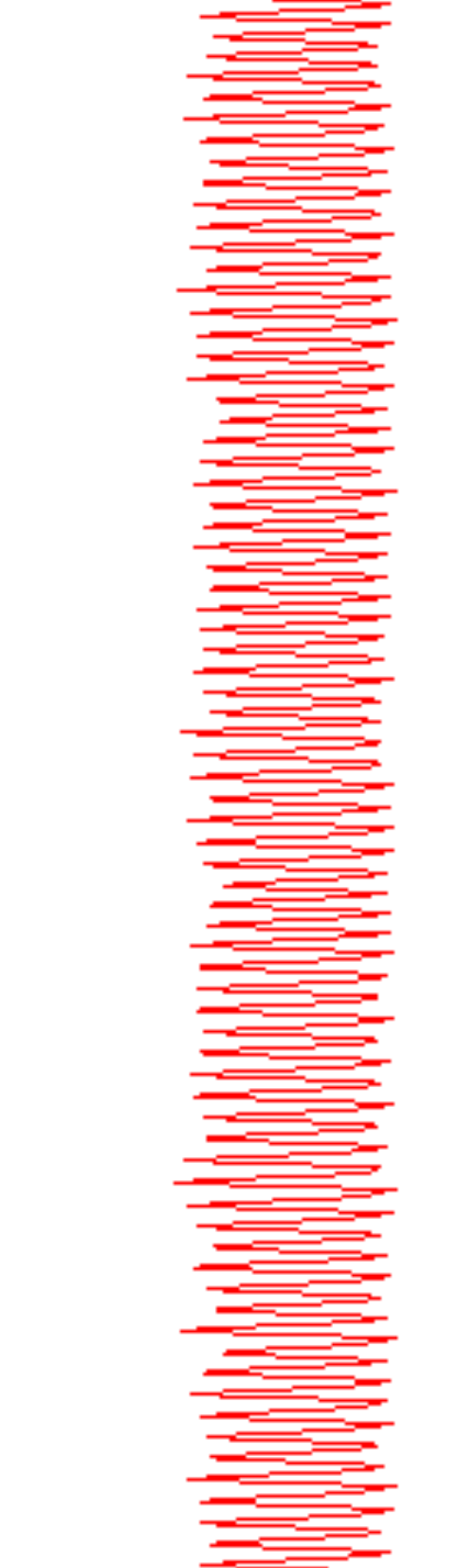}}\\[1ex]
(f1)
\end{minipage}
\end{minipage}
}
\end{center}
\vspace{-1ex}
\textsf{\small
\noindent Space-time patterns illustrating density oscillations.
(d) scrolling diagonally, the present moment is at the front
leaving a trail of time-steps behind. (e) a 1d segment, scrolling vertically with the most recent
time-step at the bottom. (f1) input-entropy oscillations with time (y-axis).
}
\begin{center}
\textsf{\small
\begin{minipage}[c]{.95\linewidth} 
\begin{minipage}[c]{.15\linewidth}
\fbox{\includegraphics[width=.85\linewidth]{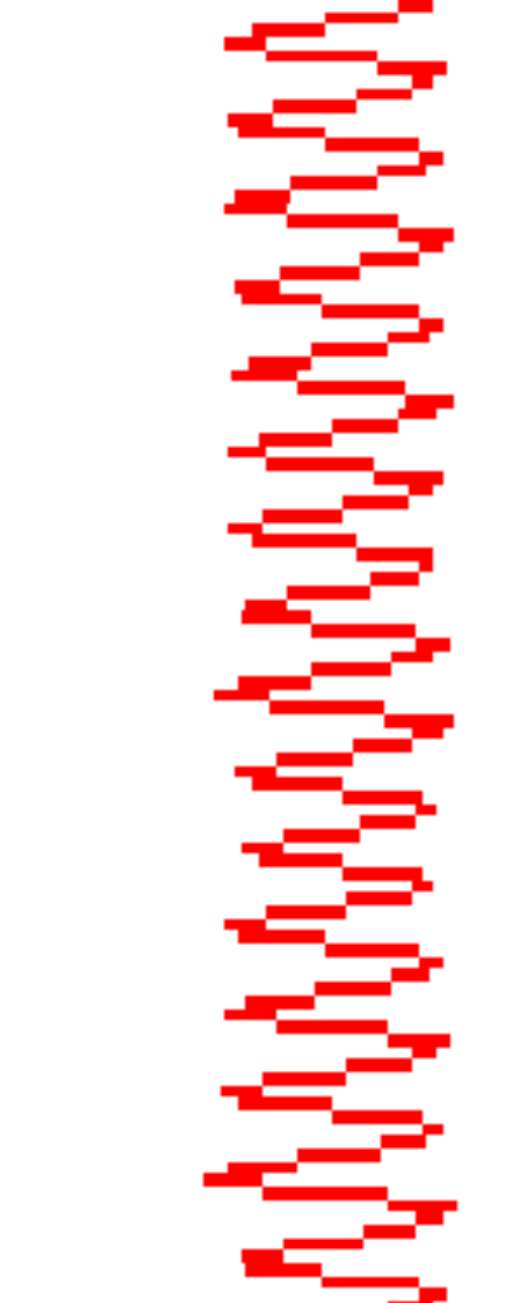}}\\[1ex]
(f2)
\end{minipage}
\hfill
\begin{minipage}[c]{.35\linewidth}
\includegraphics[width=1\linewidth]{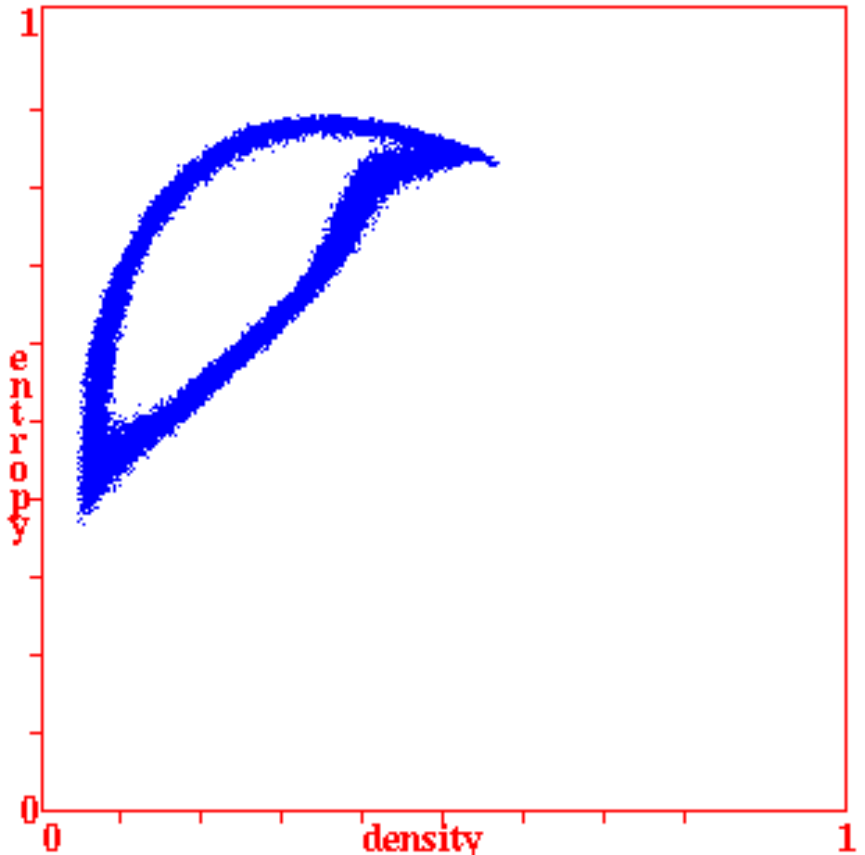}\\
(g)
\end{minipage}
\hfill
\begin{minipage}[c]{.35\linewidth}
\includegraphics[width=1\linewidth]{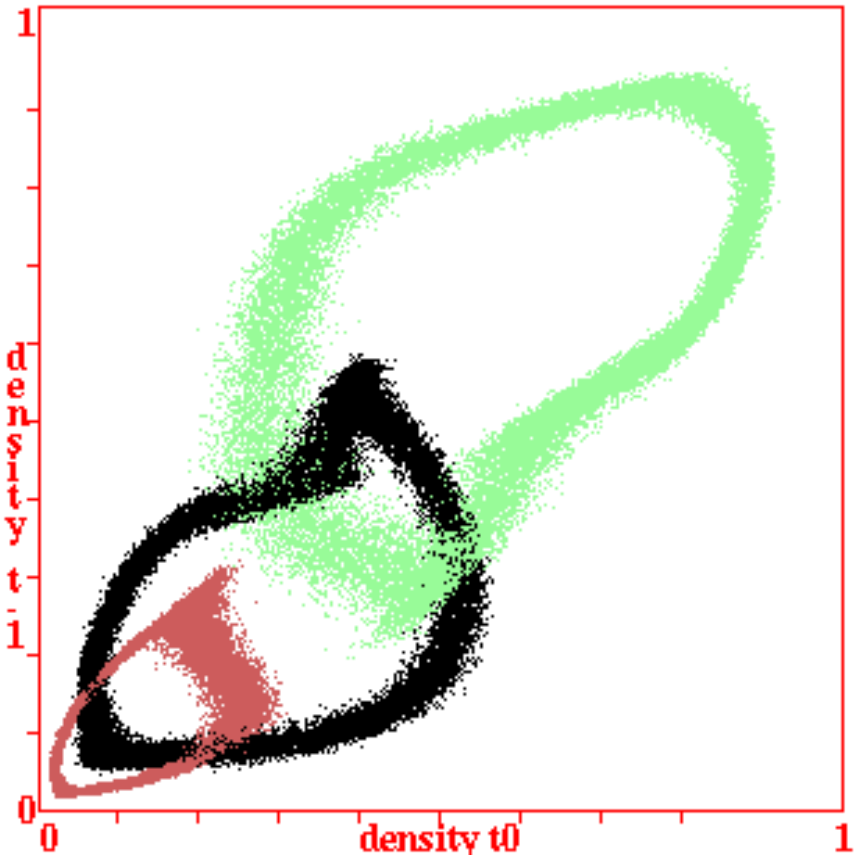}\\
(h)
\end{minipage}
\end{minipage}
}
\end{center}
\vspace{-1ex}
\textsf{\small
\noindent Time-plots of measures.
(f2) input-entropy oscillations with time (y-axis, stretched)\\
$wl$= 6 or 7 time-steps, $wh\approx0.4$.
(g) entropy-density scatter plot -- input-entropy (x-axis)
against the non-zero density (y-axis). (h) density return map scatter plot.}

\caption[Pulsing dynamics spiral rule]
{\textsf{
Pulsing dynamics for the $v3k7$ ``g1'' Spiral rule, (hex) 020609a2982a68aa64,
on a 100x100 hexagonal lattice,
showing space-time patterns --- snapshots, scrolling, and time-plots of measures.
}}
\label{Pulsing dynamics spiral rule}
\end{figure}

\begin{figure}[htb]
\begin{center}
\textsf{\small
\begin{minipage}[c]{1\linewidth} 
\begin{minipage}[c]{.3\linewidth}
\includegraphics[width=1\linewidth]{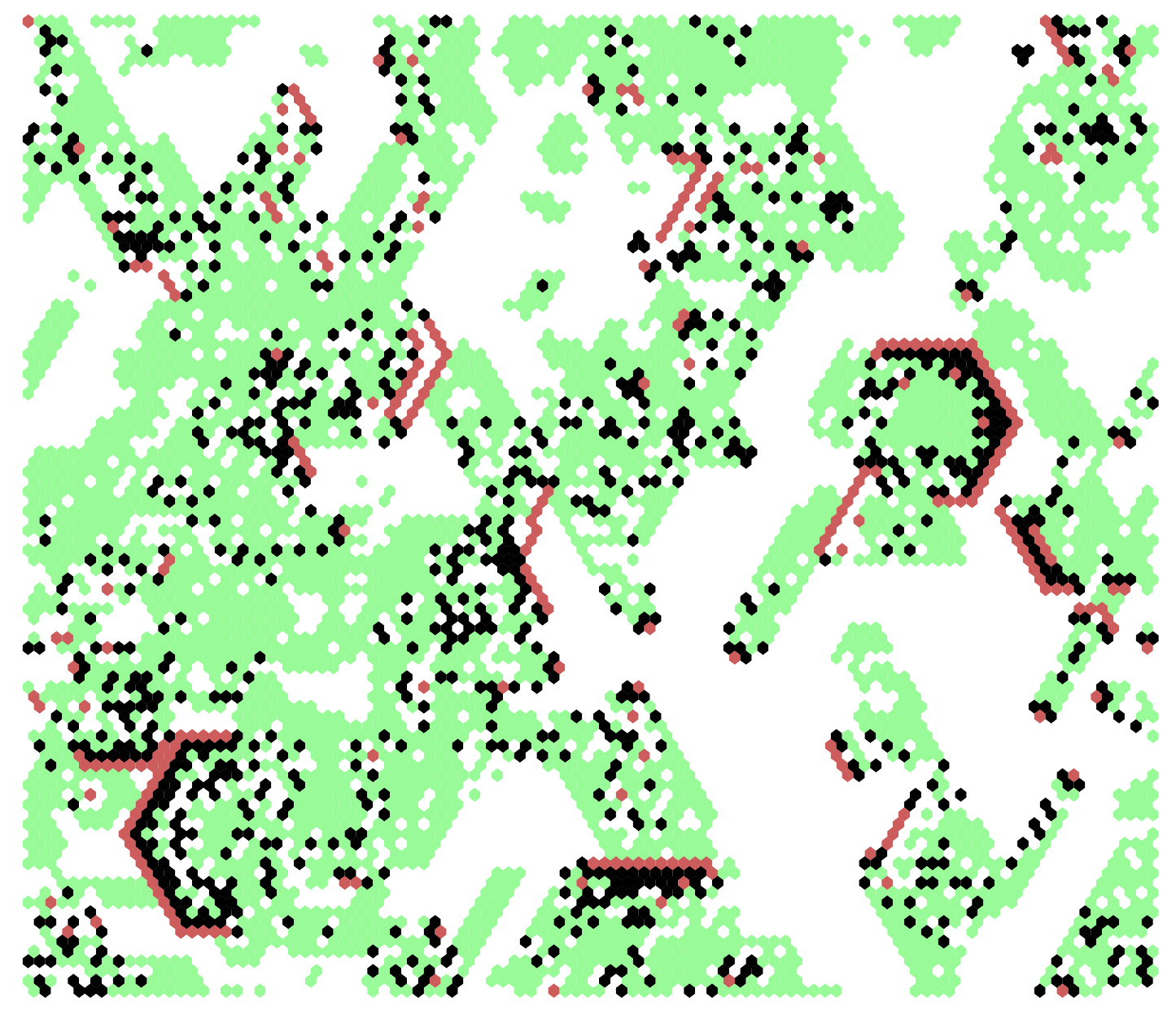}\\
(a)
\end{minipage}
\hfill
\begin{minipage}[c]{.3\linewidth}
\includegraphics[width=1\linewidth]{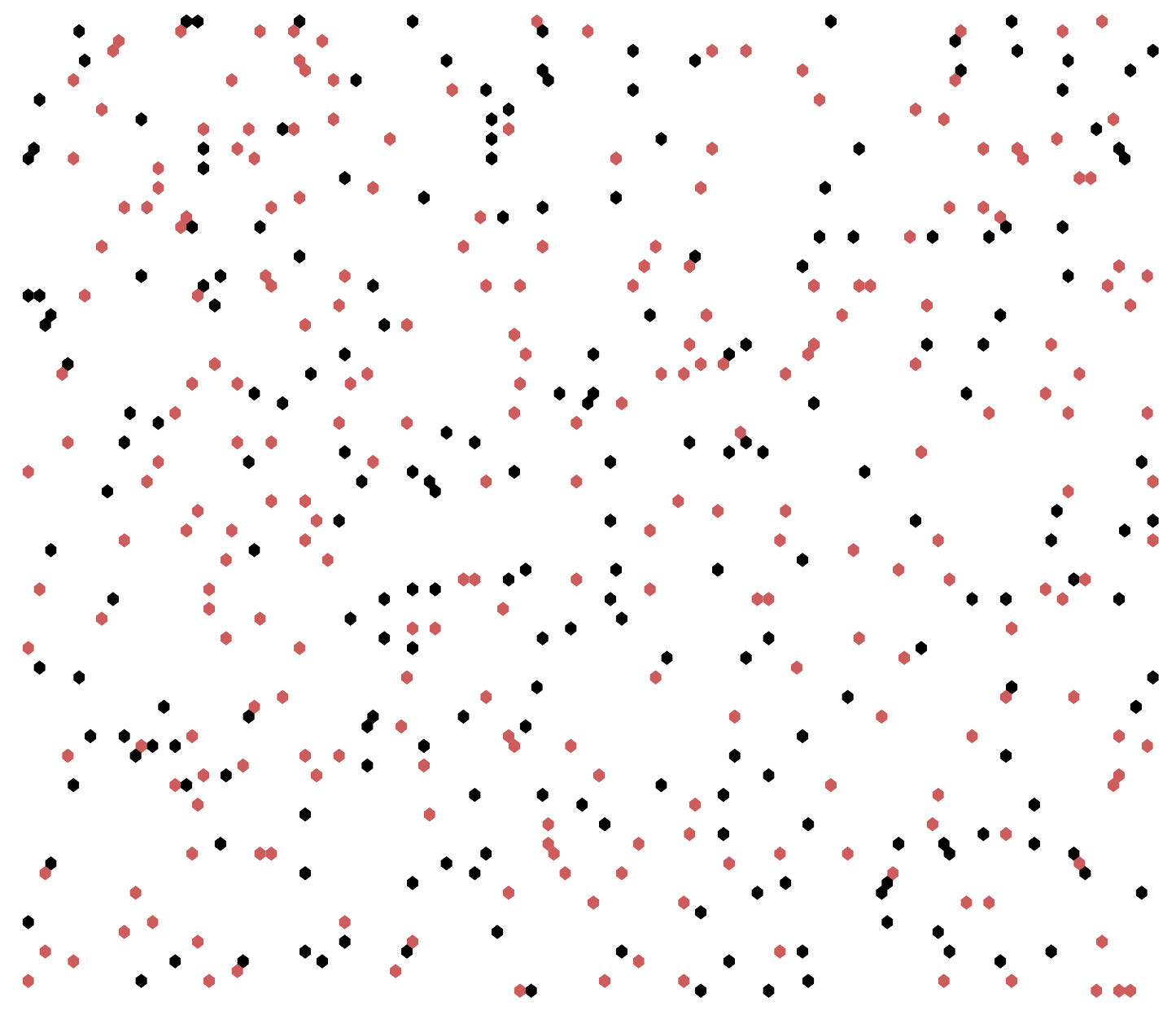}\\
(b)
\end{minipage}
\hfill
\begin{minipage}[c]{.3\linewidth}
\includegraphics[width=1\linewidth]{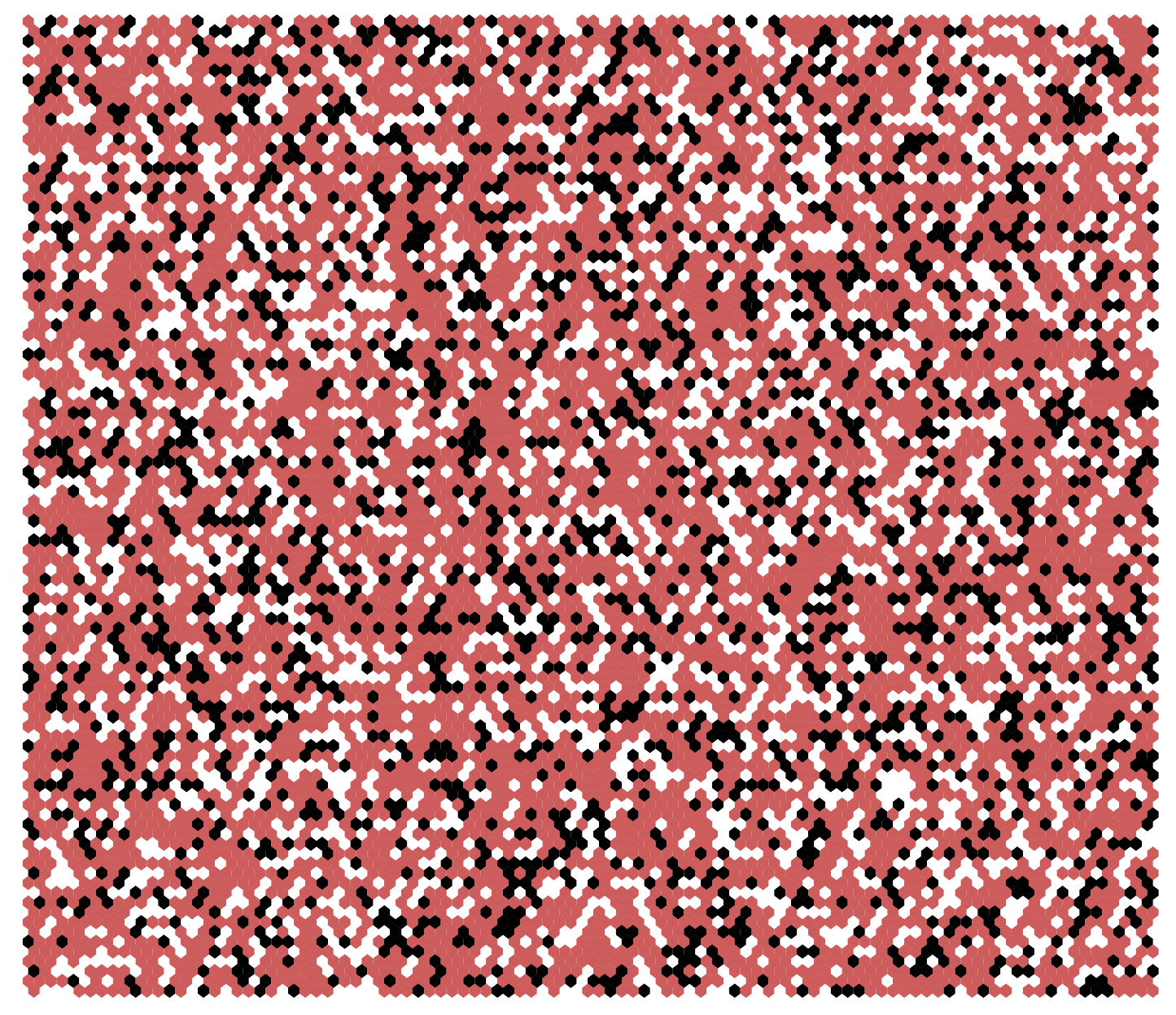}\\
(c)
\end{minipage}
\end{minipage}
}
\end{center}
\vspace{-1ex}
\textsf{\small
\noindent Space-time pattern snapshots.
(a) CA showing emergent gliders.
Randomised wiring results in disordered patterns, 
(b) minimum density, and (c) maximum density.
}

\begin{center}
\textsf{\small
\begin{minipage}[c]{.85\linewidth}
\begin{minipage}[c]{.43\linewidth} 
\includegraphics[width=1\linewidth]{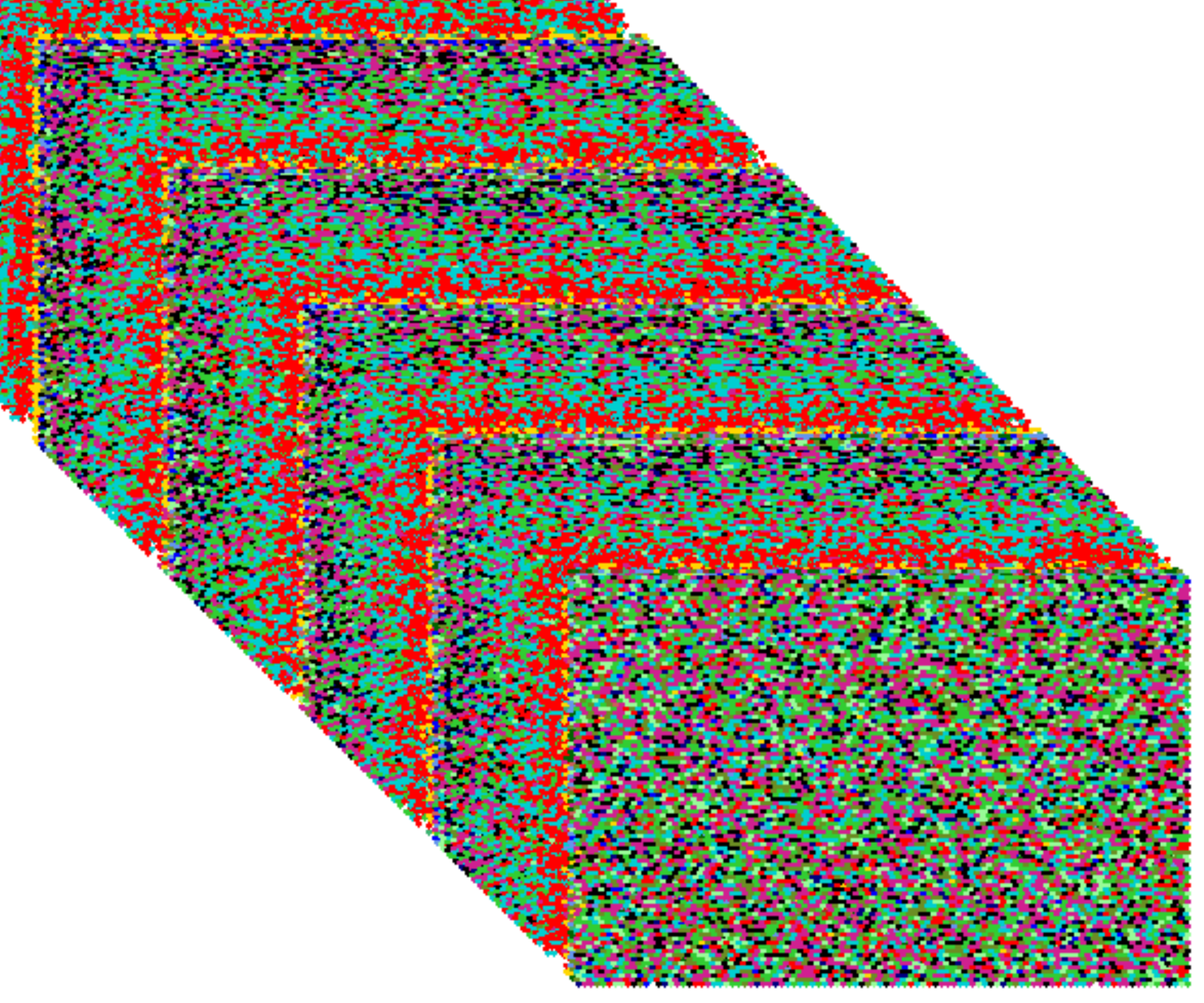}\\
(d)
\end{minipage}
\hfill
\begin{minipage}[c]{.36\linewidth} 
\includegraphics[width=1\linewidth]{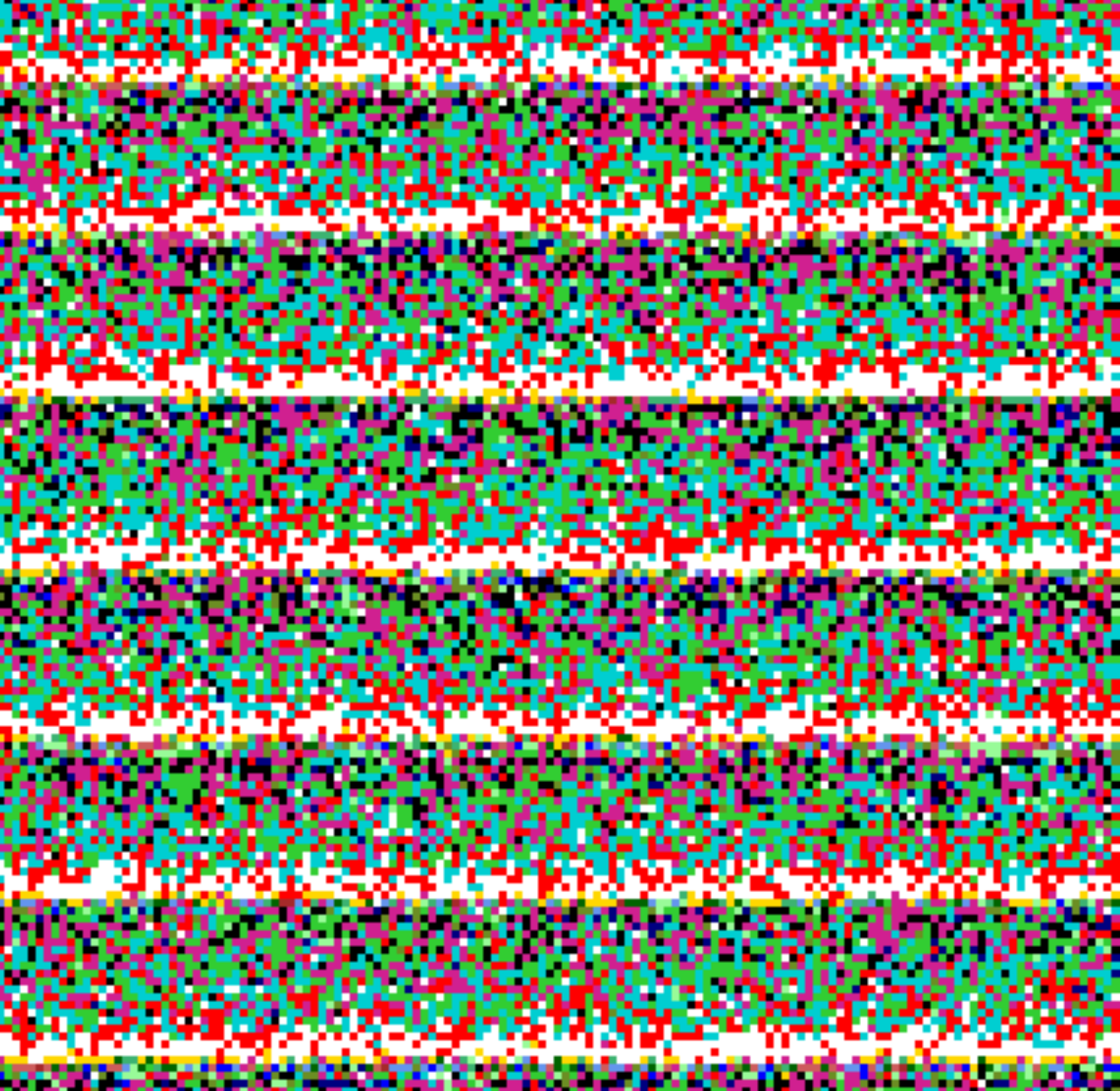}\\
(e)
\end{minipage}
\hfill 
\begin{minipage}[c]{.075\linewidth}
\fbox{\includegraphics[width=1\linewidth]{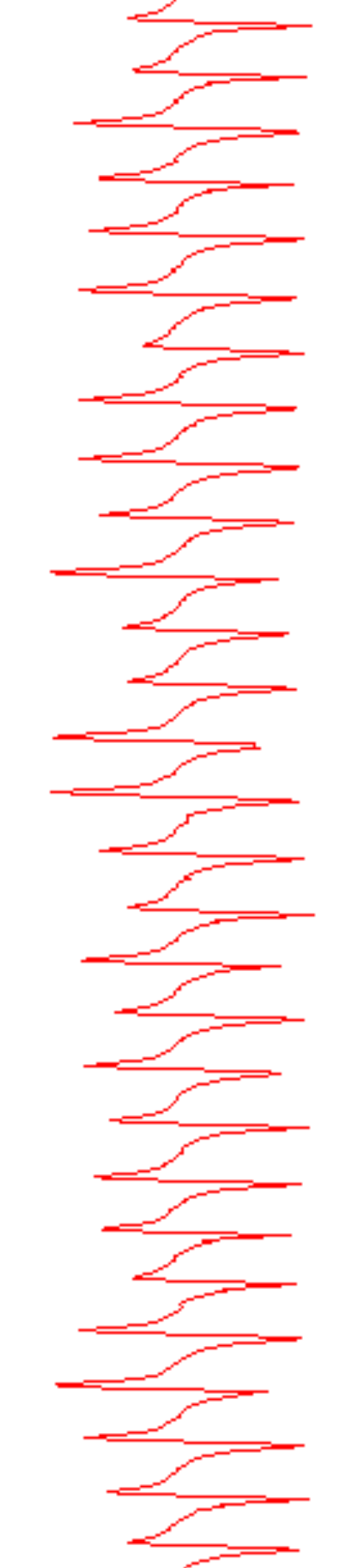}}\\[1ex]
(f1)
\end{minipage}
\end{minipage}
}
\end{center}
\vspace{-1ex}
\textsf{\small
\noindent Space-time patterns illustrating density oscillations. 
(d) scrolling diagonally, the present moment is at the front
leaving a trail of time-steps behind. (e) a 1d segment, scrolling vertically with the most recent
time-step at the bottom. (f1) input-entropy oscillations with time (y-axis).
}
\begin{center}
\textsf{\small
\begin{minipage}[c]{.95\linewidth} 
\begin{minipage}[c]{.083\linewidth}
\fbox{\includegraphics[width=.8\linewidth]{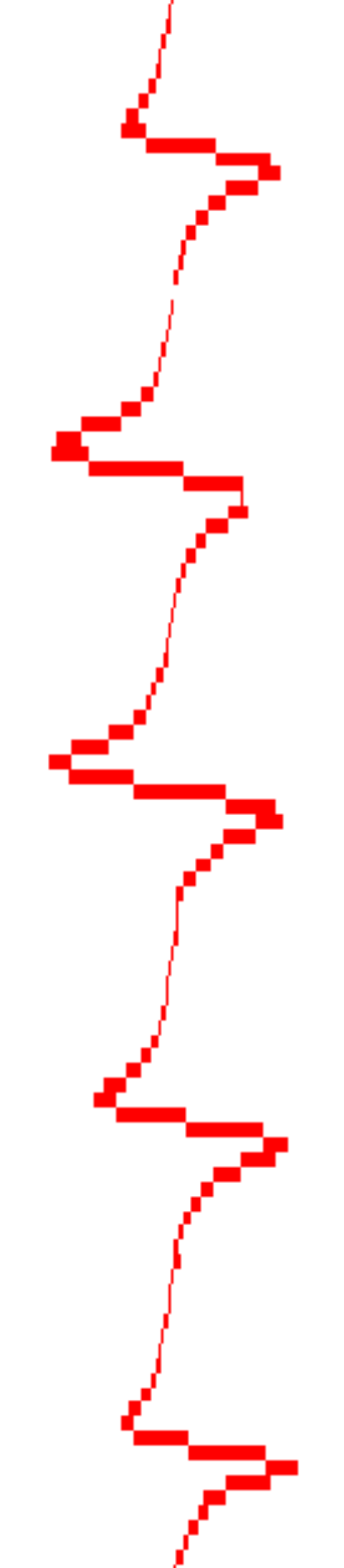}}\\[1ex]
(f2)
\end{minipage}
\hfill
\begin{minipage}[c]{.35\linewidth}
\includegraphics[width=1\linewidth]{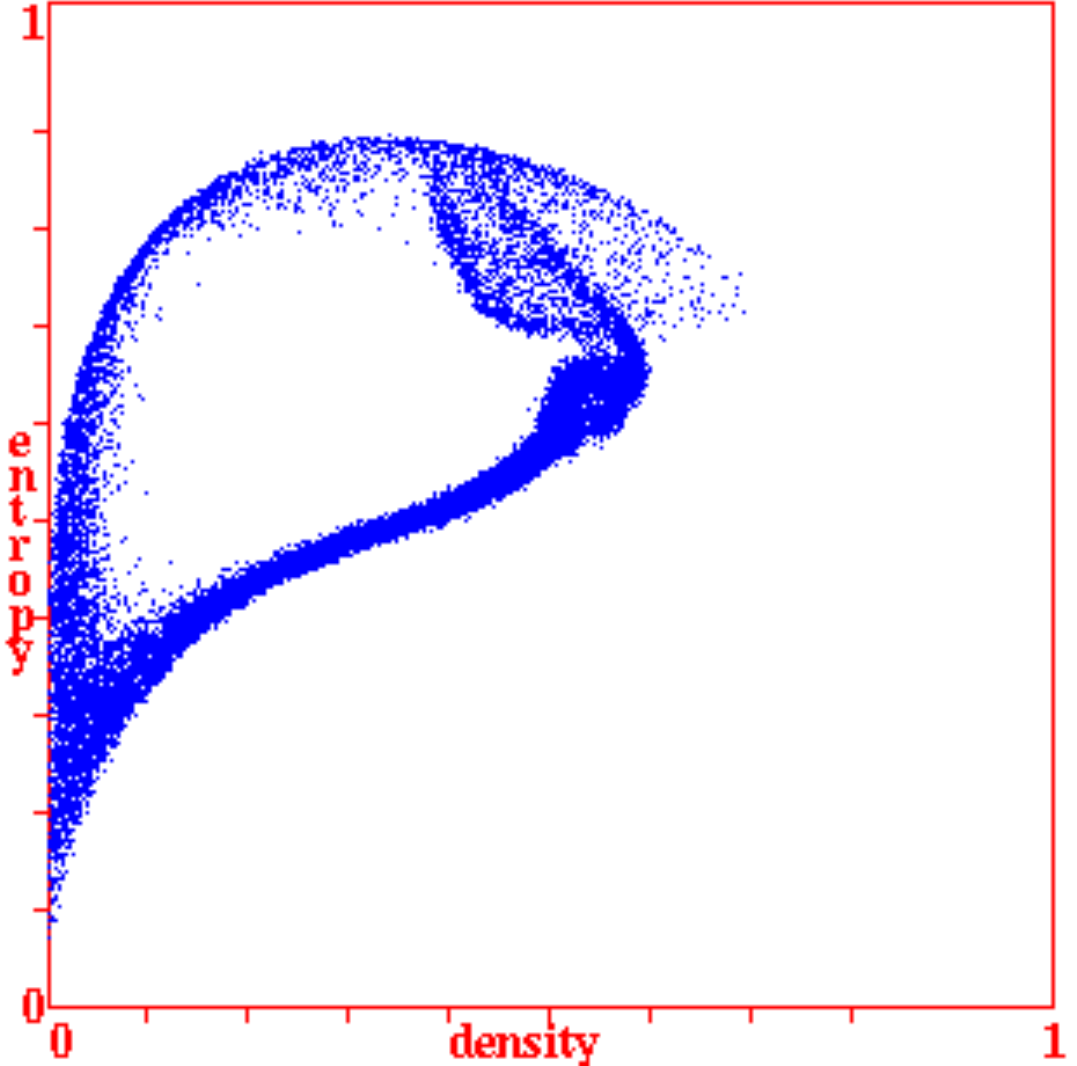}\\
(g)
\end{minipage}
\hfill
\begin{minipage}[c]{.35\linewidth}
\includegraphics[width=1\linewidth]{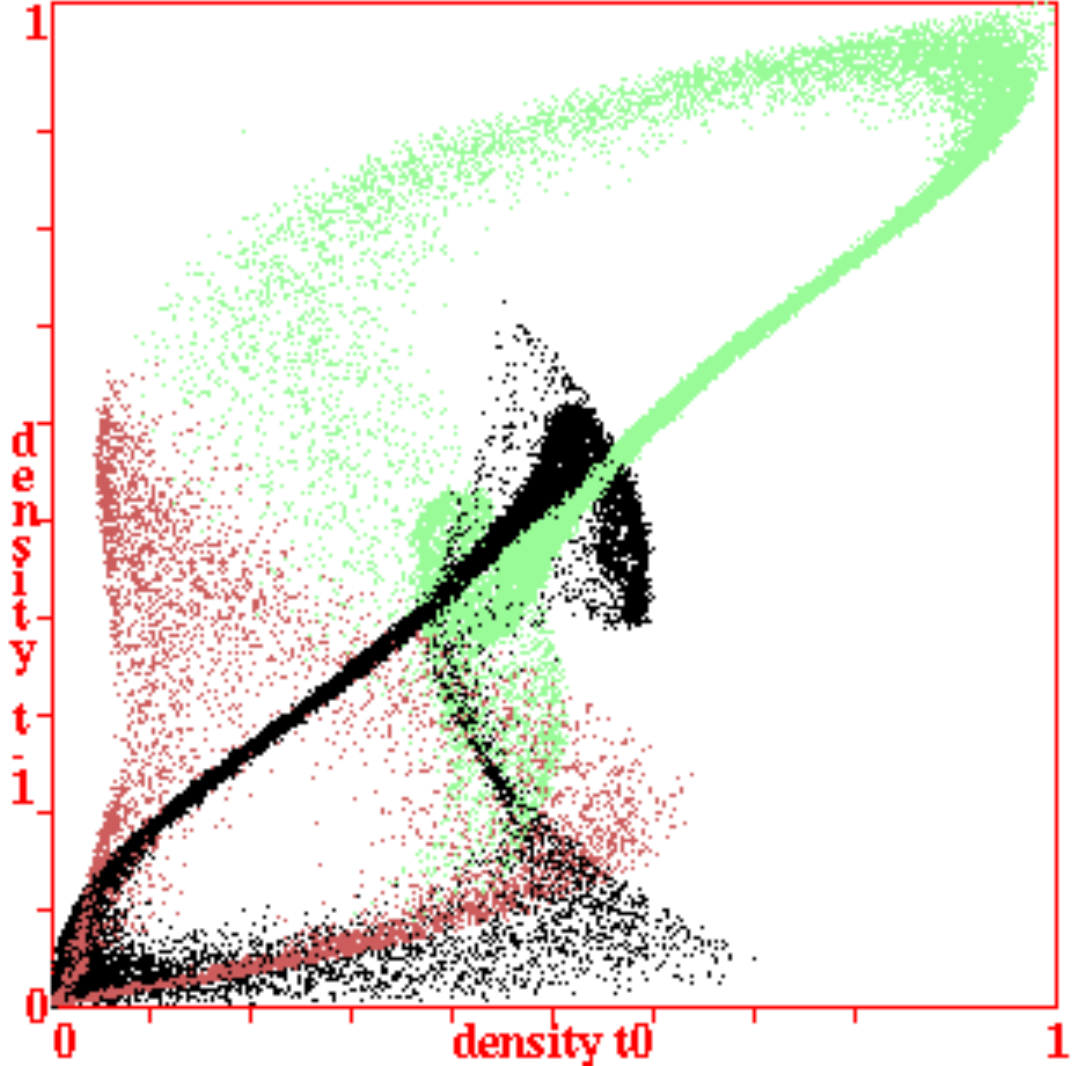}\\
\textsf{(h)}
\end{minipage}
\end{minipage}
}
\end{center}
\vspace{-1ex}
\textsf{\small
\noindent Time-plots of measures.
(f2) input-entropy oscillations with time (y-axis, stretched)\\
$wl\approx$ 21 time-steps, $wh\approx0.55$.
(g) entropy-density scatter plot -- input-entropy (x-axis)
against the non-zero density (y-axis). (h) density return map scatter plot.}

\caption[Pulsing dynamics g3]
{\textsf{
Pulsing dynamics for the $v3k7$ ``g3'' rule, (hex) 622984288a08086a94,
on a 100x100 hexagonal lattice,
showing space-time patterns --- snapshots, scrolling, and time-plots of measures.
}}
\label{Pulsing dynamics g3 rule}
\end{figure}

\begin{figure}[htb]
\begin{center}
\textsf{\small
\begin{minipage}[c]{1\linewidth} 
\begin{minipage}[c]{.3\linewidth}
\includegraphics[width=1\linewidth]{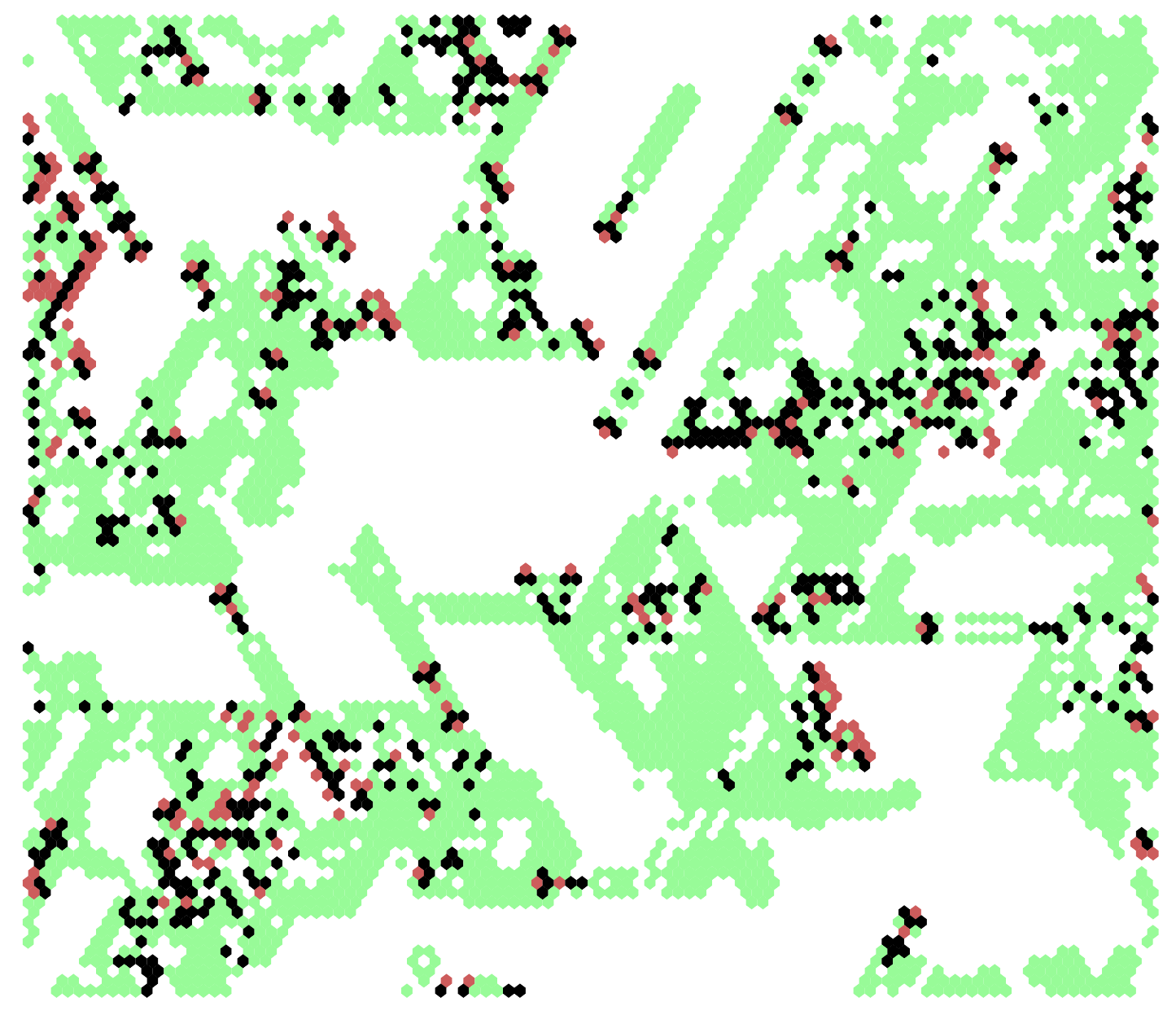}\\
(a)
\end{minipage}
\hfill
\begin{minipage}[c]{.3\linewidth}
\includegraphics[width=1\linewidth]{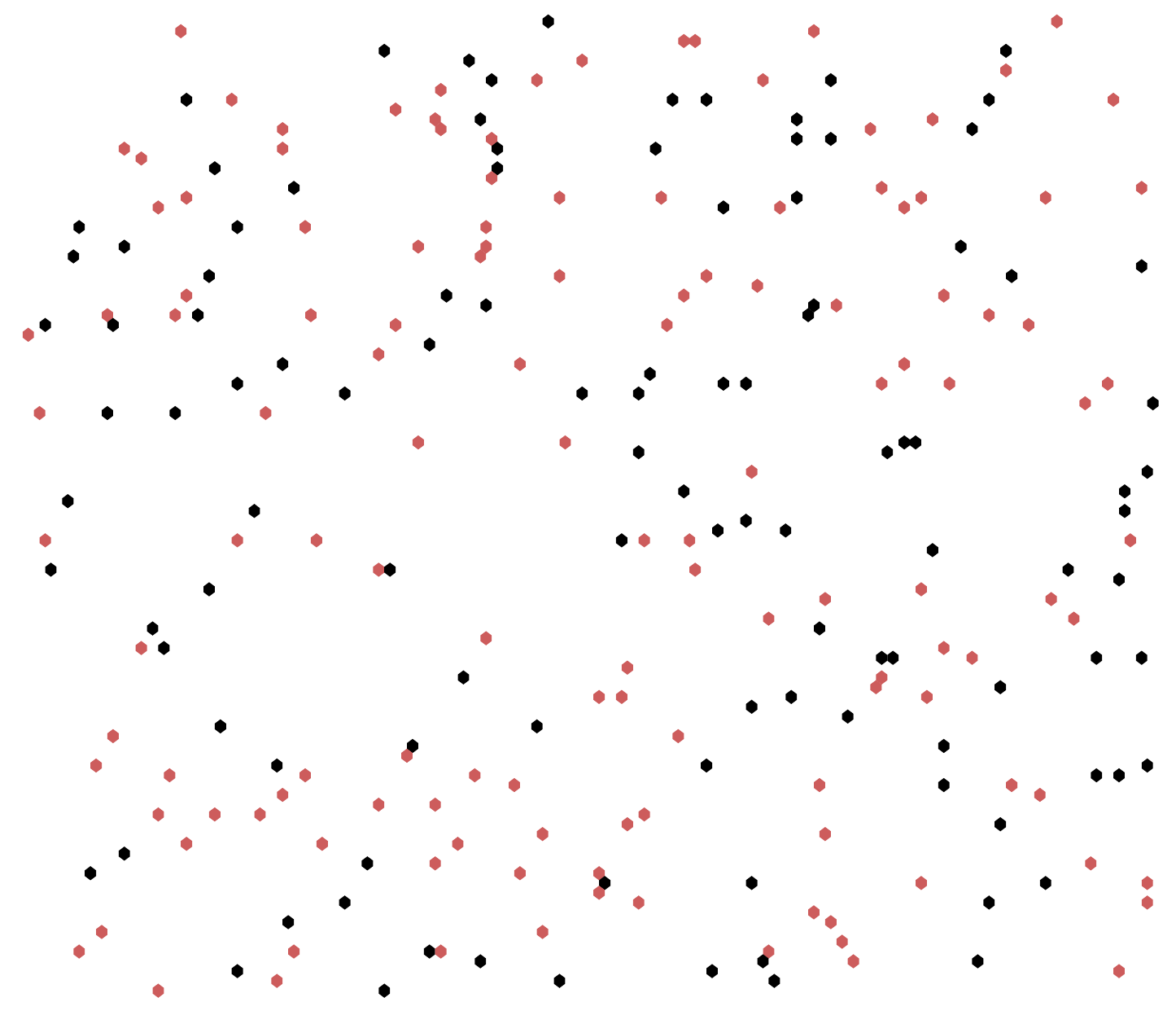}\\
(b)
\end{minipage}
\hfill
\begin{minipage}[c]{.3\linewidth}
\includegraphics[width=1\linewidth]{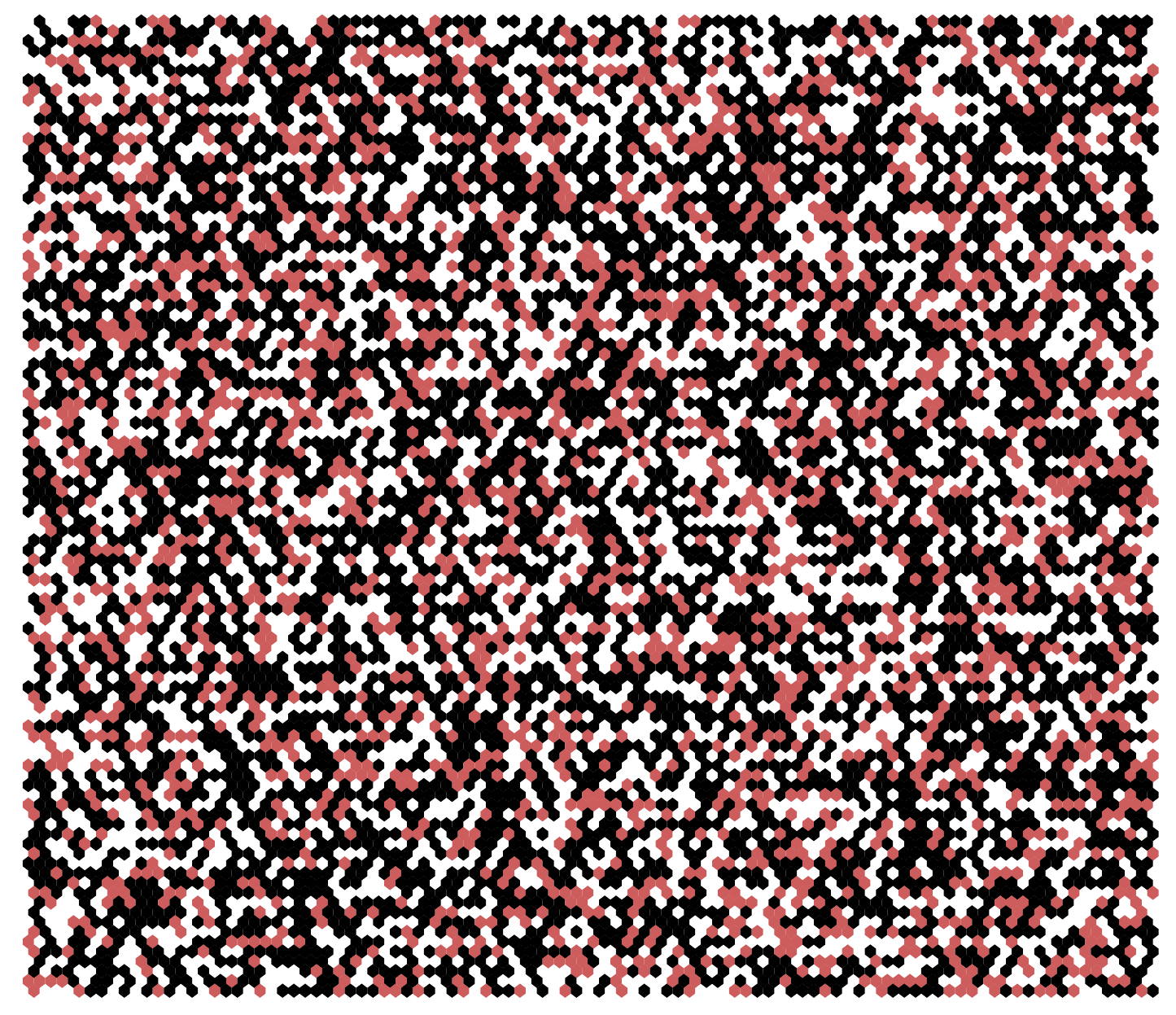}\\
(c)
\end{minipage}
\end{minipage}
}
\end{center}
\vspace{-1ex}
\textsf{\small
\noindent Space-time pattern snapshots.
(a) CA showing emergent gliders.
Randomised wiring results in disordered patterns, 
(b) minimum density, and (c) maximum density.
}

\begin{center}
\textsf{\small
\begin{minipage}[c]{.85\linewidth}
\begin{minipage}[c]{.38\linewidth} 
\includegraphics[width=1\linewidth]{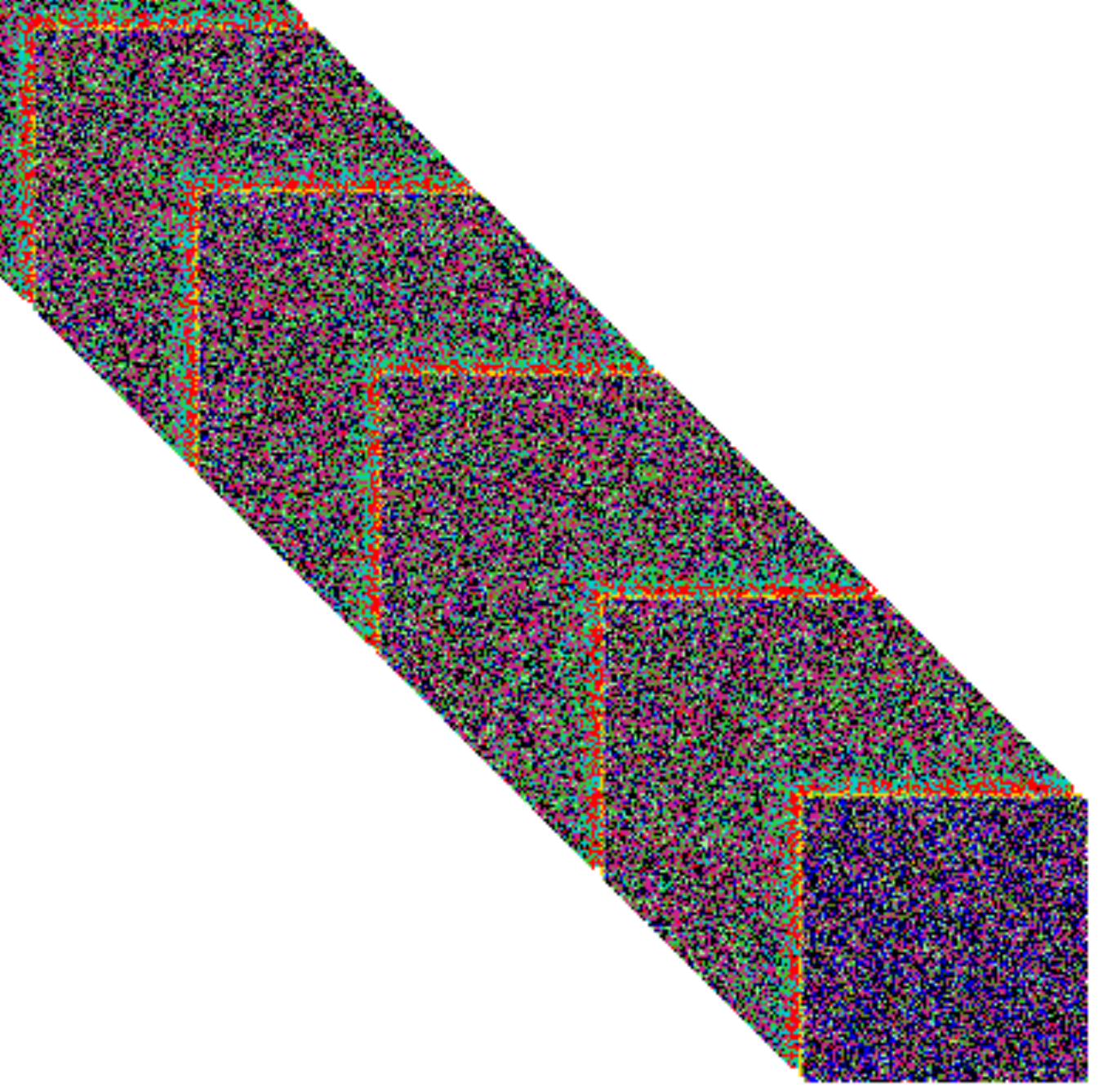}\\
(d)
\end{minipage}
\hfill
\begin{minipage}[c]{.38\linewidth} 
\includegraphics[width=1\linewidth]{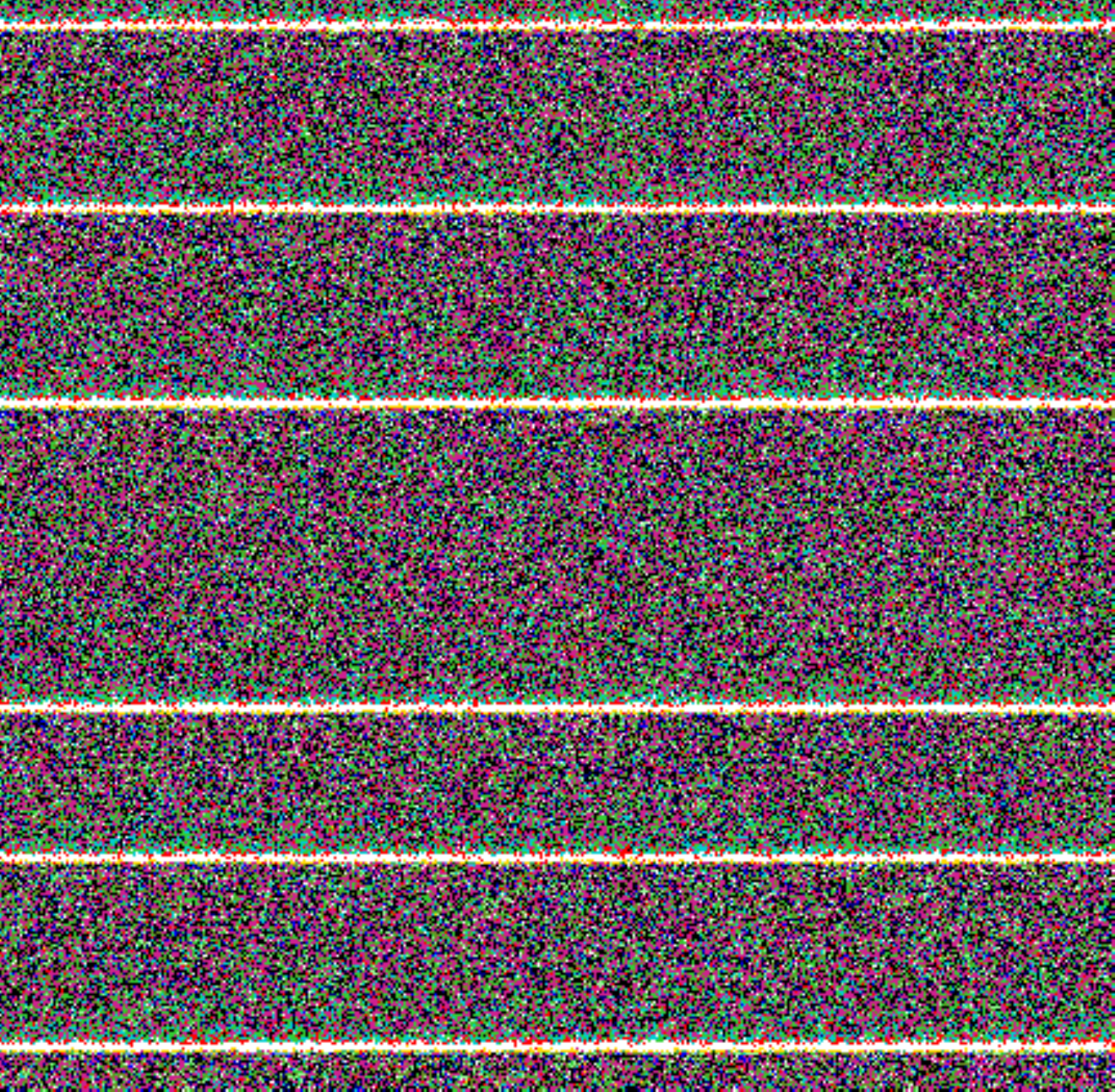}\\
(e)
\end{minipage}
\hfill 
\begin{minipage}[c]{.075\linewidth}
\fbox{\includegraphics[width=1\linewidth]{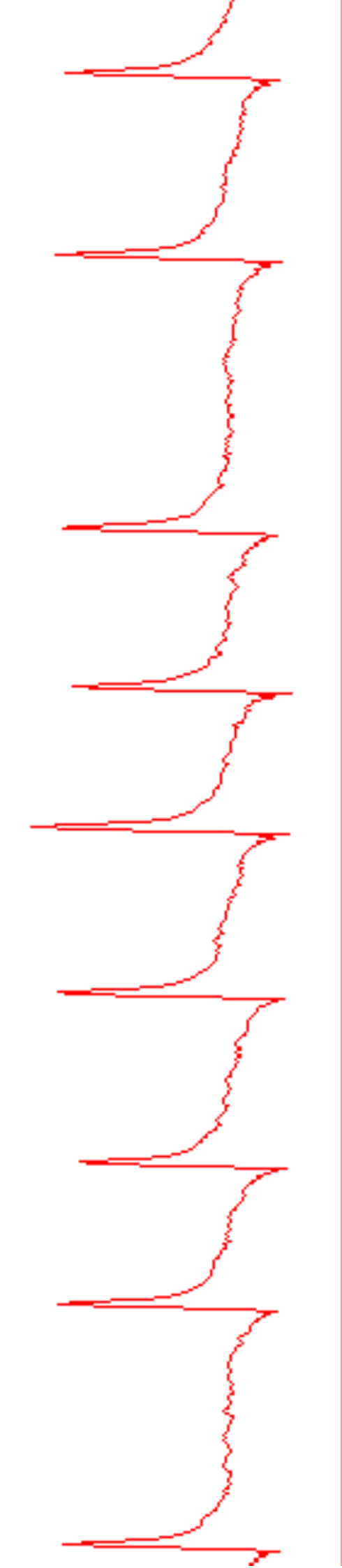}}\\[1ex]
(f1)
\end{minipage}
\end{minipage}
}
\end{center}
\vspace{-1ex}
\textsf{\small
\noindent Space-time patterns illustrating density oscillations.
(d) scrolling diagonally, the present moment is at the front
leaving a trail of time-steps behind. (e) a 1d segment, scrolling vertically with the most recent
time-step at the bottom. (f1) input-entropy oscillations with time (y-axis).
}
\begin{center}
\textsf{\small
\begin{minipage}[c]{.95\linewidth} 
\begin{minipage}[c]{.16\linewidth}
\fbox{\includegraphics[width=.85\linewidth]{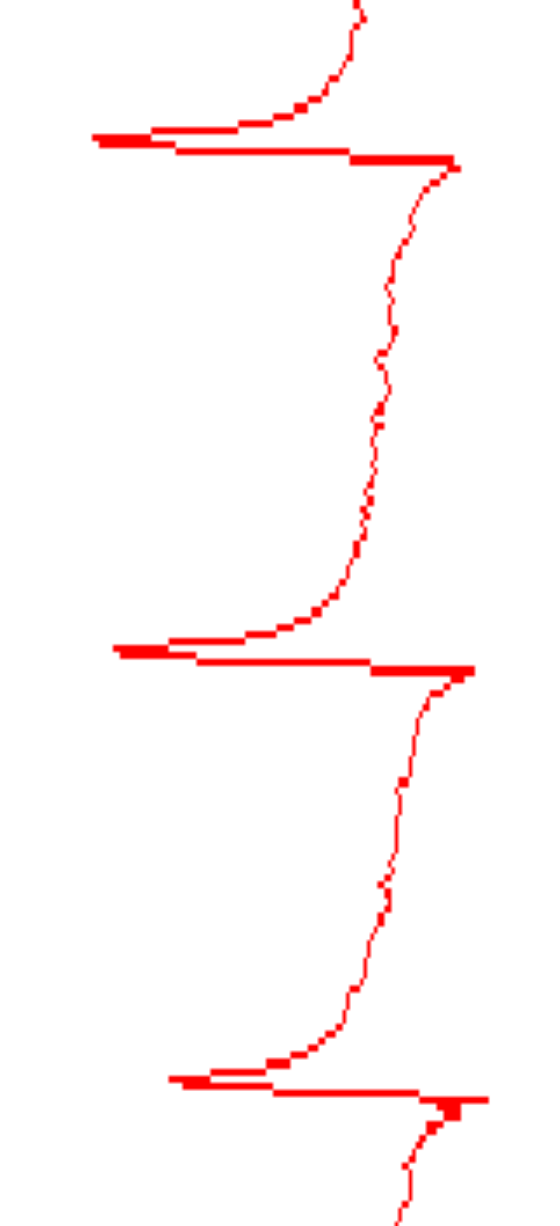}}\\[1ex]
(f2)
\end{minipage}
\hfill
\begin{minipage}[c]{.35\linewidth}
\includegraphics[width=1\linewidth]{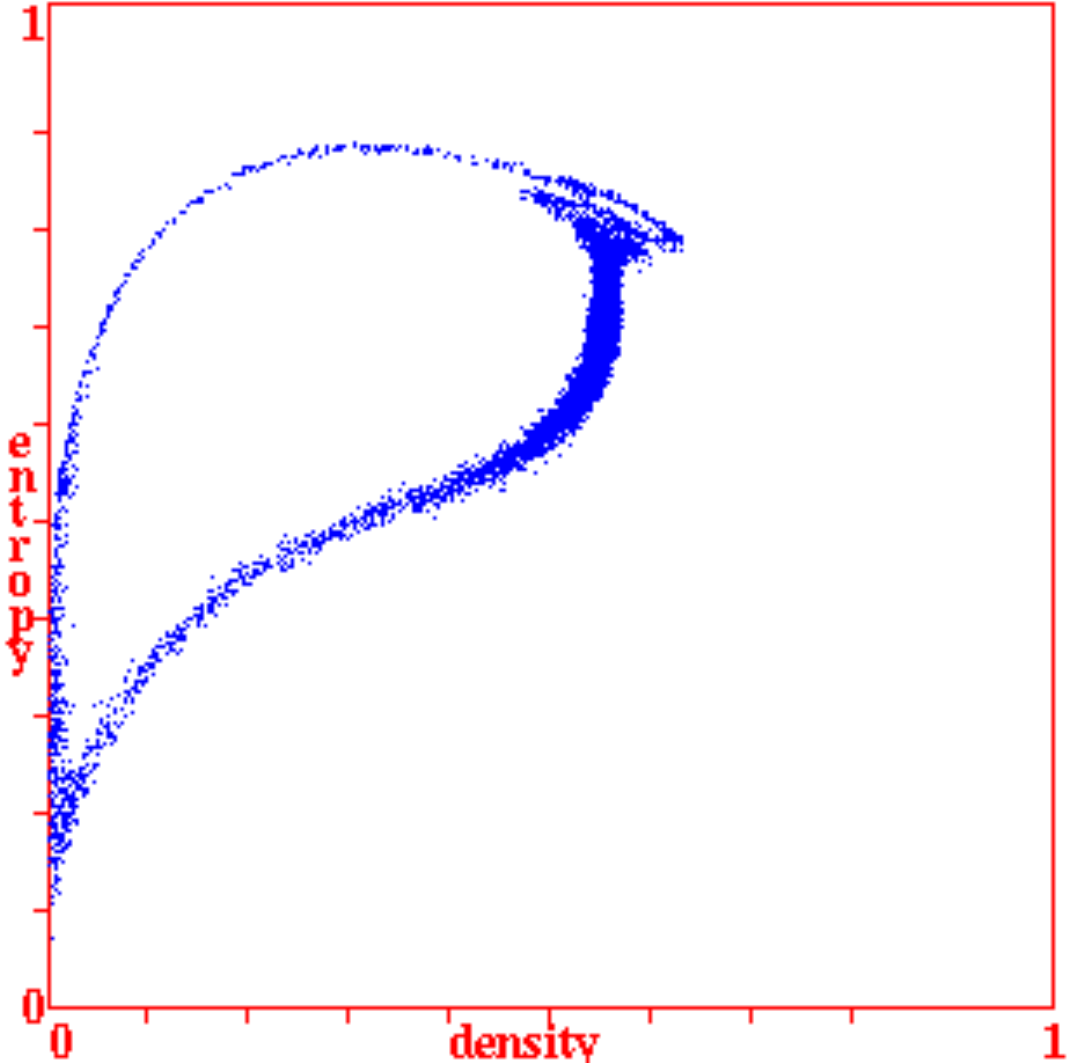}\\
(g)
\end{minipage}
\hfill
\begin{minipage}[c]{.35\linewidth}
\includegraphics[width=1\linewidth,]{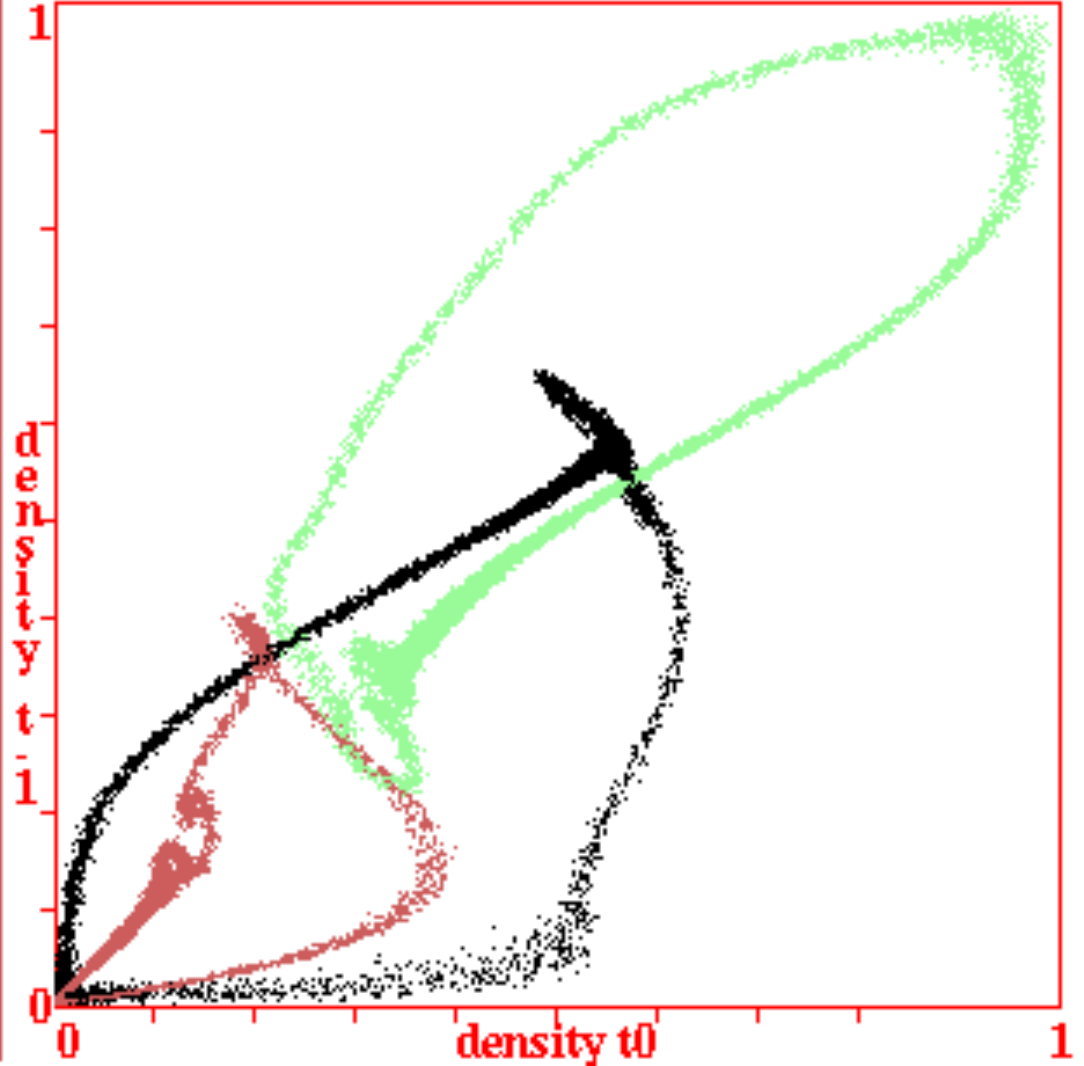}\\
\textsf{(h)}
\end{minipage}
\end{minipage}
}
\end{center}
\vspace{-1ex}
\textsf{\small
\noindent Time-plots of measures.
(f2) input-entropy oscillations with time (y-axis, stretched)\\
diverse $wl$ between 52 and 122 time-steps (average $wl$$\approx$ 82), $wh\approx0.6$.
(g) entropy-density scatter plot -- input-entropy (x-axis)
against the non-zero density (y-axis). (h) density return map scatter plot.}

\caption[Pulsing dynamics f82]
{\textsf{
Pulsing dynamics for the $v3k7$ ``g35'' rule, (hex) 806a22a29a12182a84,
on a 100x100 hexagonal lattice,
showing space-time patterns --- snapshots, scrolling, and time-plots of measures.
}}
\label{Pulsing dynamics f82 rule}
\end{figure}

\noindent We present six case studies of the CA pulsing model, based on
glider rule samples assembled previously\cite{Wuensche2016}, three for $k$=6 
(figures~\ref{Pulsing dynamics beehive rule}-\ref{Pulsing dynamics k6-g26}), 
and three for $k$=7, (figures~\ref{Pulsing dynamics spiral rule}-\ref{Pulsing dynamics f82 rule}),
selected for a variety of waveform profiles. Surprisingly, wave-lengths ($wl$) are very diverse,
with average $wl$ varying between 6 and 82 time-steps.
Two well documented rules are included, the $v3k6$ 
Beehive rule\cite{Adamatzky&Wuensche&Cosello2006,Wuensche05,beehivewebpage}, 
and the $v3k7$ Spiral rule\cite{Wuensche&Adamatzky2006,Adamatzky&Wuensche2006,spiralwebpage}. 

Each case study examines pulsing dynamics on a 2D 100$\times$100
hexagonal lattice.  Random wiring is unconstrained (giving the
``RW-waveform'') where the $k$ inputs to each cell are independently
assigned at random without bias. This makes the 2d geometry
irrelevant --- it is retained for convenience.

The results are displayed graphically as follows:

\begin{itemize}

\item (a) A typical snapshot before the wiring was randomised,
of the CA with its emergent gliders, with 10 time-step green trails.

\item (b, c) Two snapshots after the wiring was randomised,
showing the now disordered pattern (b) at its minimum and (c) maximum density
of non-zero values (2=black, 1=red, 0=white). 

\item (d, e) Space-time patterns showing evidence of pulsing,
with cells colored according to lookup instead of value,
following the histogram colors in figure~\ref{input-histograms}.
(d) 2D space-time patterns scrolling diagonally, with the
latest time-step at the front, leaving a trail of time-steps behind.
(e) A stretch of the space-times pattern transformed to 1D, scrolling vertically,
with the present moment at the bottom, leaving a trail of time-steps behind.

\item (f1) The input-entropy plotted
for each time-step, showing the pulsing waveform. 
(f2) A stretched or magnified version of this plot,
noting the wavelength $wl$ and wave-height $wh$.

\item (g) The entropy-density scatter plot --- input-entropy (x-axis)
against the non-zero density (y-axis), plotted as blue dots,
for about 33000 time-steps.

\item (h) The density return-map scatter plot ---
the density of each value  at $t_0$ (x-axis) 
against its density at $t_1$, plotted as colored dots (2=black, 1=red, 0=green),
for about 33000 time-steps.
\end{itemize}

Case study experiments (confirmed for many other glider rules)
give the following general results: Any random initial state, 
(within reason\footnote{Initial states with some of the 3-values missing,
or with very low/high density may not converge on the waveform.}) 
will initiate a transient that rapidly converges on the waveform, 
which is impervious to reasonable noise. 
With fully random wiring (without bias),
changing the actual wiring makes no difference to the waveform, neither
does re-randomising at each time-step.  The scatter plots,
both input-entropy and the density return-map,
show unique signatures, which have the characteristics of chaotic strange
attractors in the context of deterministic discrete dynamical
systems --- sensitivity to initial conditions evolving towards a
compact global attracting set, local instability but globally stability.
Varying the network size also preserves the waveform --- 
the signatures are diffused for small sizes,
becoming more focused as the size of the network increases
(figure~\ref{scatter plot 50x50, 200x200}), and this would continue
towards infinity.
Reducing the network size, however,
increases the probability of reaching a uniform value attractor, such as
all zeros, where the system would stop.

\enlargethispage{4ex}
The ``RW-waveform'' results of these
case studies will serve as a base of comparison with
the other wiring biases investigated: CA with freed wires, 
localised random wiring (and with freed wires), and the equivalent in 3D.

\begin{figure}[htb]
\begin{center}
\textsf{\small
\begin{minipage}[c]{.8\linewidth}
\begin{minipage}[c]{.3\linewidth}
\includegraphics[width=1\linewidth]{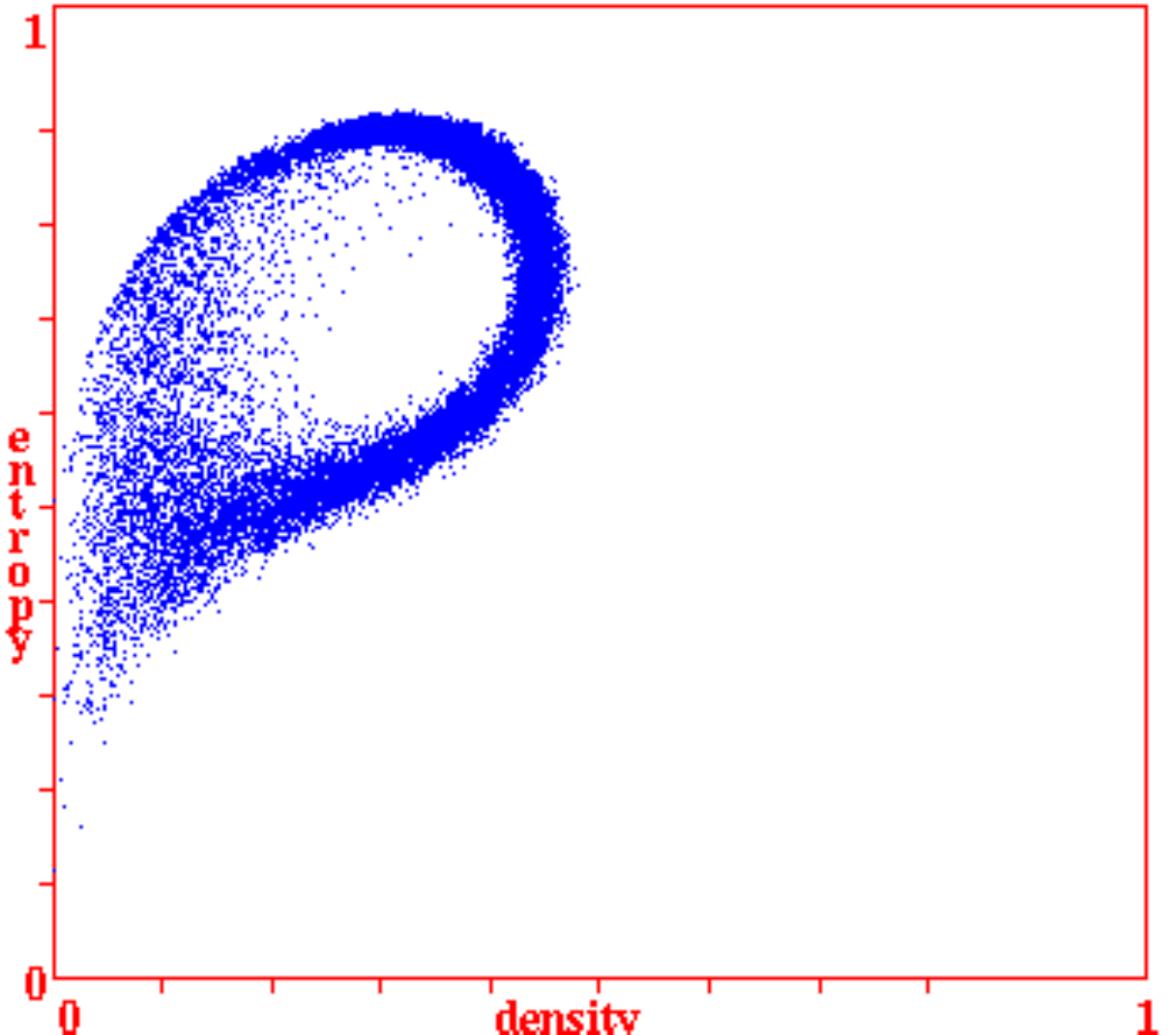}\\[-1ex]
50$\times$50
\end{minipage}
\hfill
\begin{minipage}[c]{.3\linewidth}
\includegraphics[width=1\linewidth]{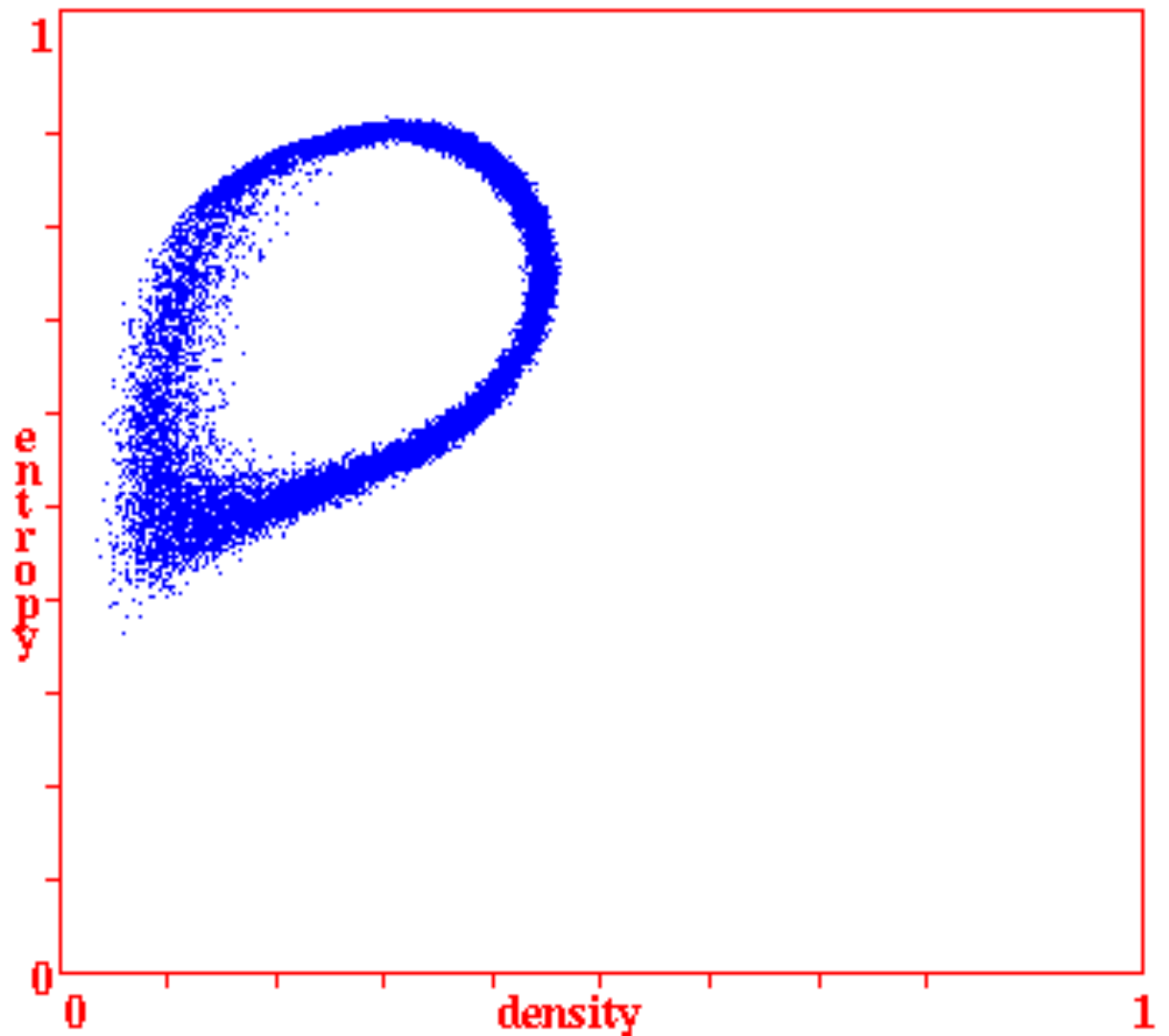}\\[-1ex]
100$\times$100
\end{minipage}
\hfill
\begin{minipage}[c]{.3\linewidth}
\includegraphics[width=1\linewidth]{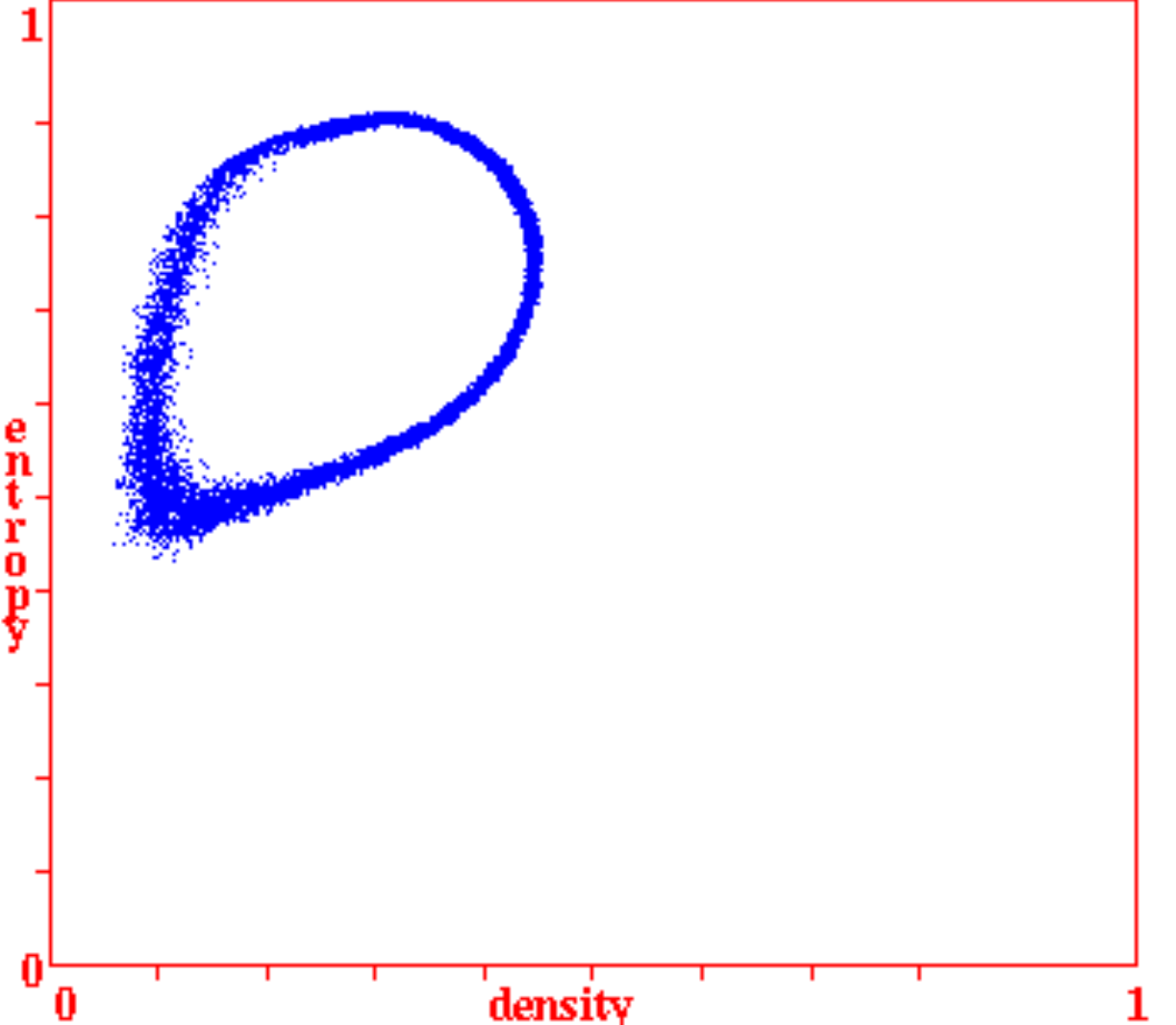}\\[-1ex]
200$\times$200
\end{minipage}
\end{minipage}\\[1ex]
Entropy-density scatter plots,
$v3k6$ ``g26'' rule, see figure~\ref{Pulsing dynamics k6-g26}(g).
\\[3ex]
\begin{minipage}[c]{.8\linewidth}
\begin{minipage}[c]{.3\linewidth}
\includegraphics[width=1\linewidth]{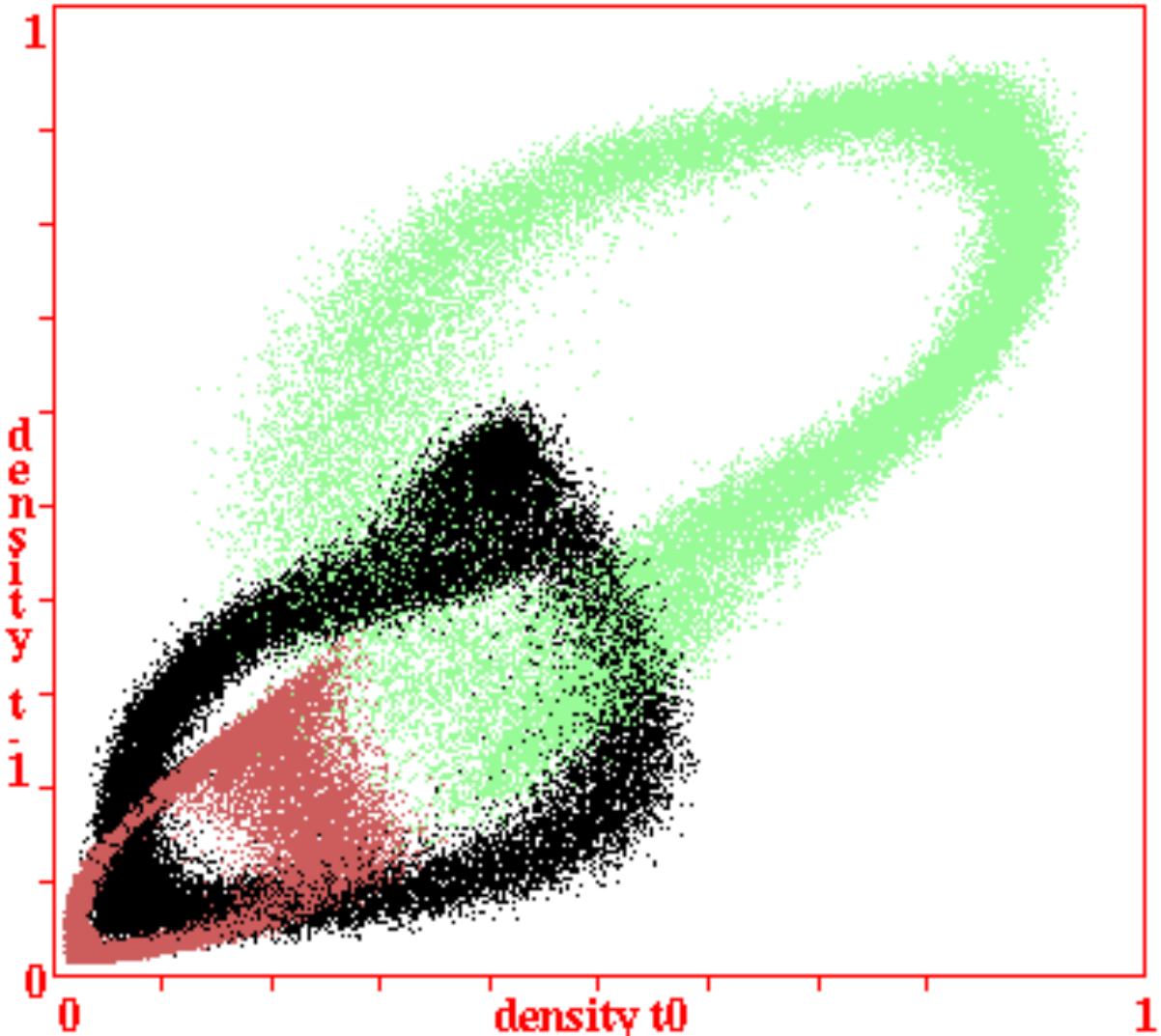}\\[-1ex]
50$\times$50
\end{minipage}
\hfill
\begin{minipage}[c]{.3\linewidth}
\includegraphics[width=1\linewidth]{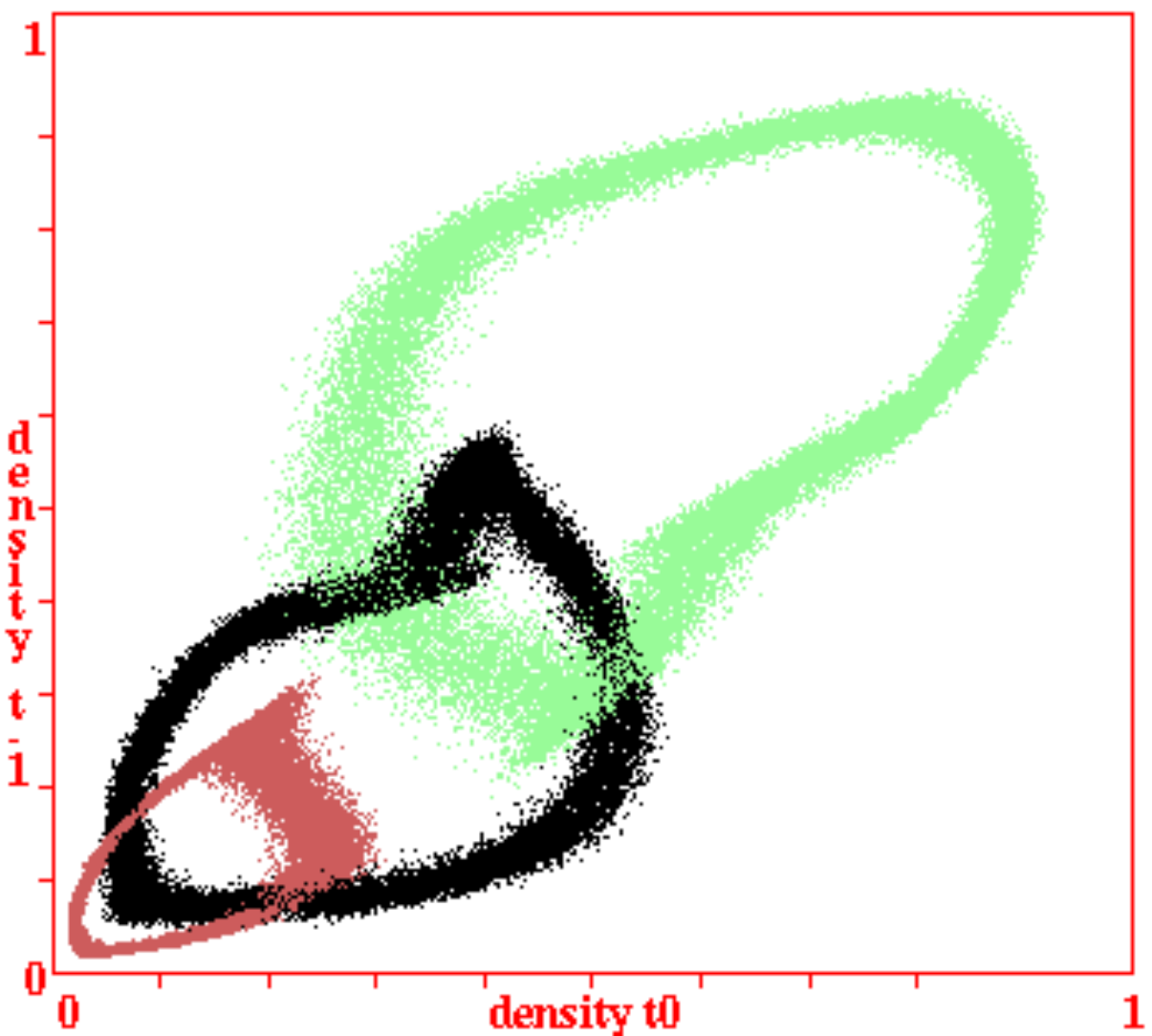}\\[-1ex]
100$\times$100
\end{minipage}
\hfill
\begin{minipage}[c]{.3\linewidth}
\includegraphics[width=1\linewidth]{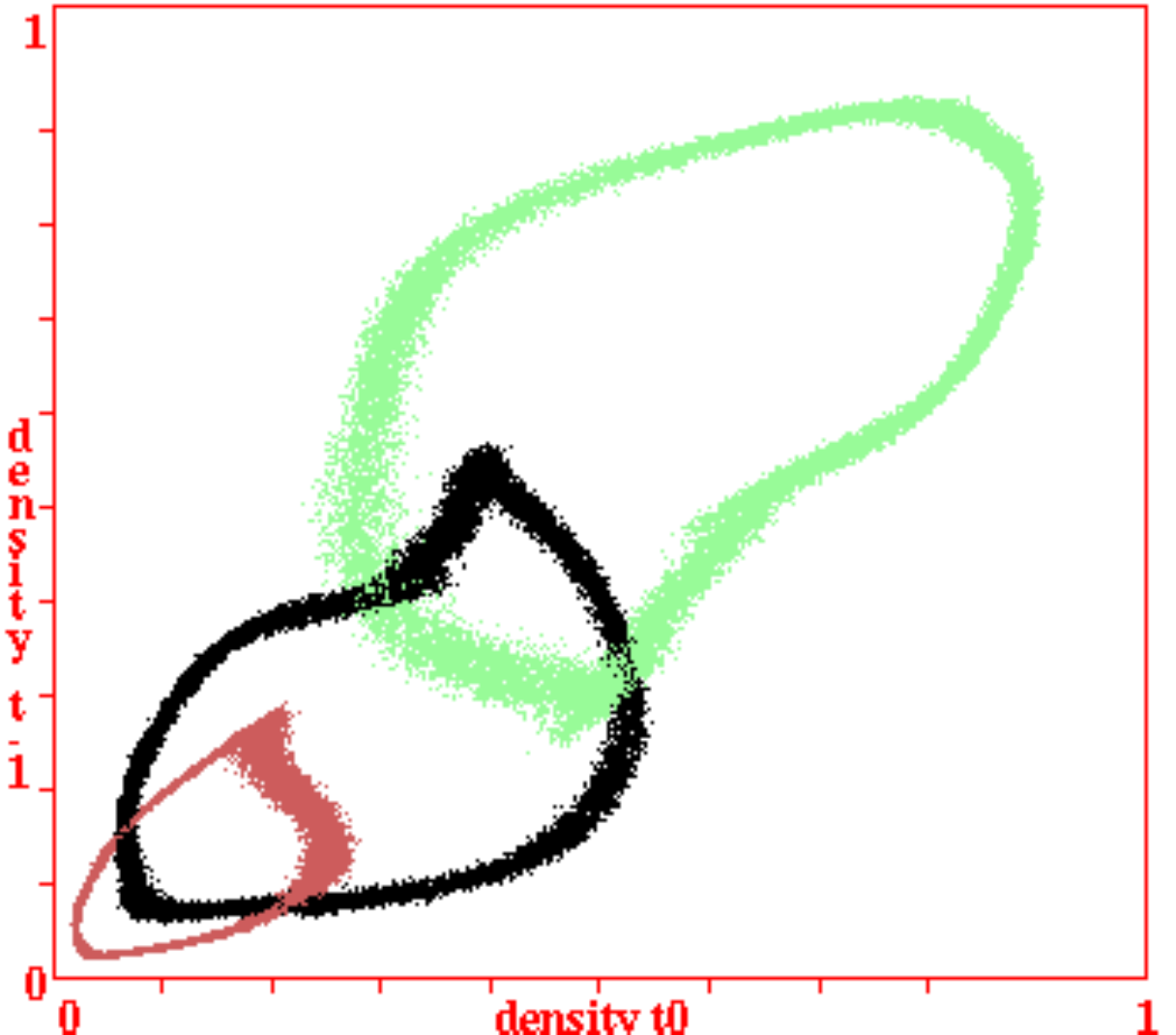}\\[-1ex]
200$\times$200
\end{minipage}
\end{minipage}\\[1ex]
Density return-map scatter plots,
$v3k7$ ``g1'' Spiral rule, see figure~\ref{Pulsing dynamics spiral rule}(h).
}
\end{center}
\vspace{-3ex}
\caption[scatter plot 50x50, 200x200]
{\textsf{
Scatter plots (for about 33000 time-steps) become focused as the size
of the nework increases, but the underlying signature is preserved.
}}
\label{scatter plot 50x50, 200x200}
\end{figure}
\clearpage

\clearpage

\clearpage

\clearpage


\clearpage

\clearpage

\clearpage

\section{Freeing one wire from CA neighborhoods}
\label{Freeing one wire from CA neighboroods}
\vspace{-3ex}
\begin{figure}[htb]
\begin{center}
\textsf{\small
\begin{minipage}[c]{.95\linewidth} 
\begin{minipage}[c]{.12\linewidth}
\fbox{\includegraphics[width=1\linewidth]{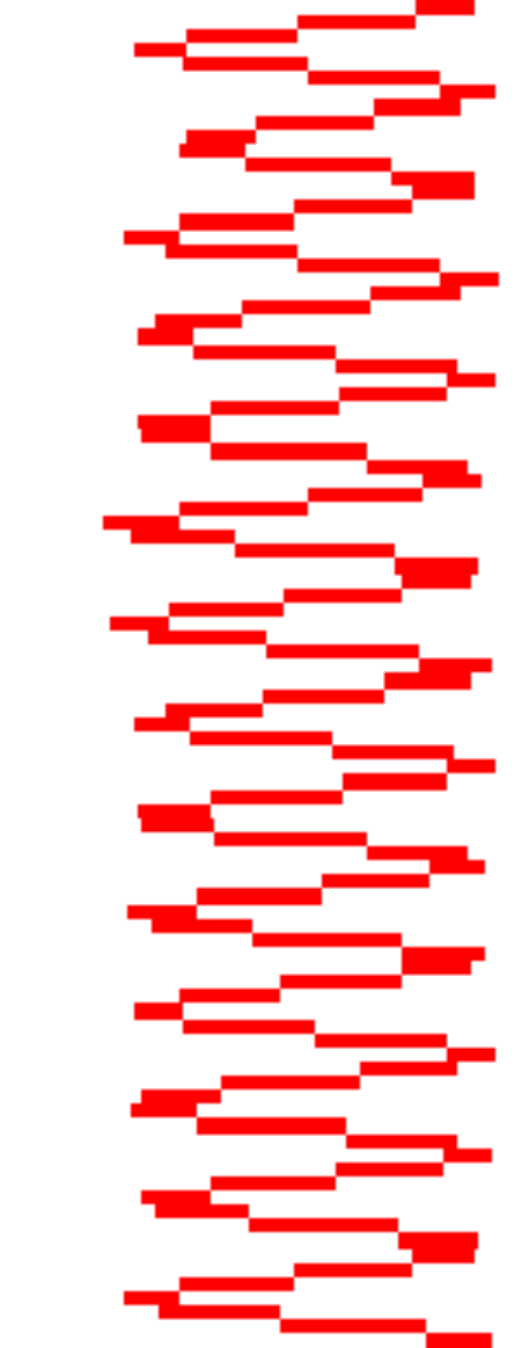}}\\[1ex]
(f2)~$wl$$\approx$6
\end{minipage}
\hfill
\begin{minipage}[c]{.35\linewidth}
\includegraphics[width=1\linewidth]{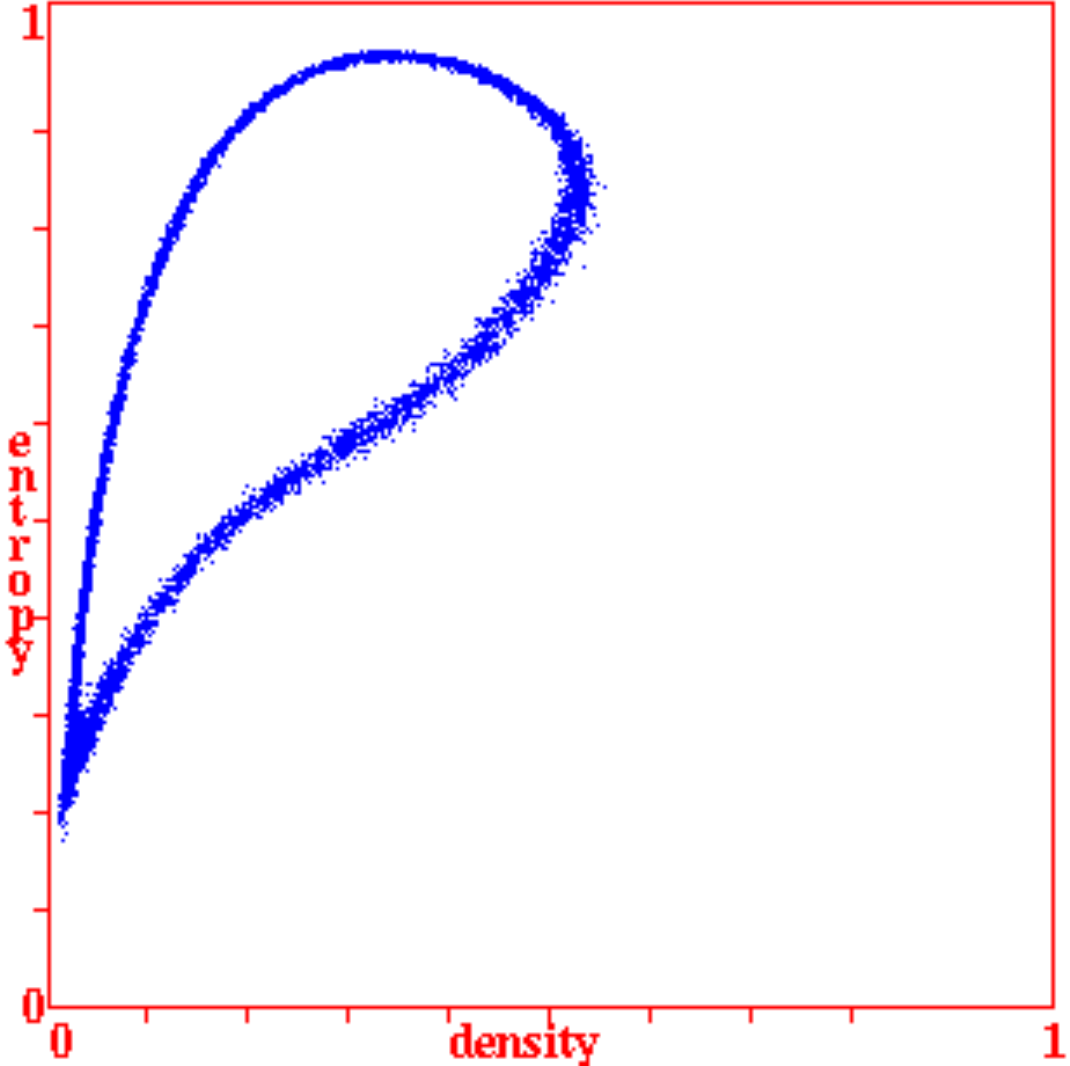}\\
(g) $v3k6$ ``g2, beehive''
\end{minipage}
\hfill
\begin{minipage}[c]{.35\linewidth}
\includegraphics[width=1\linewidth]{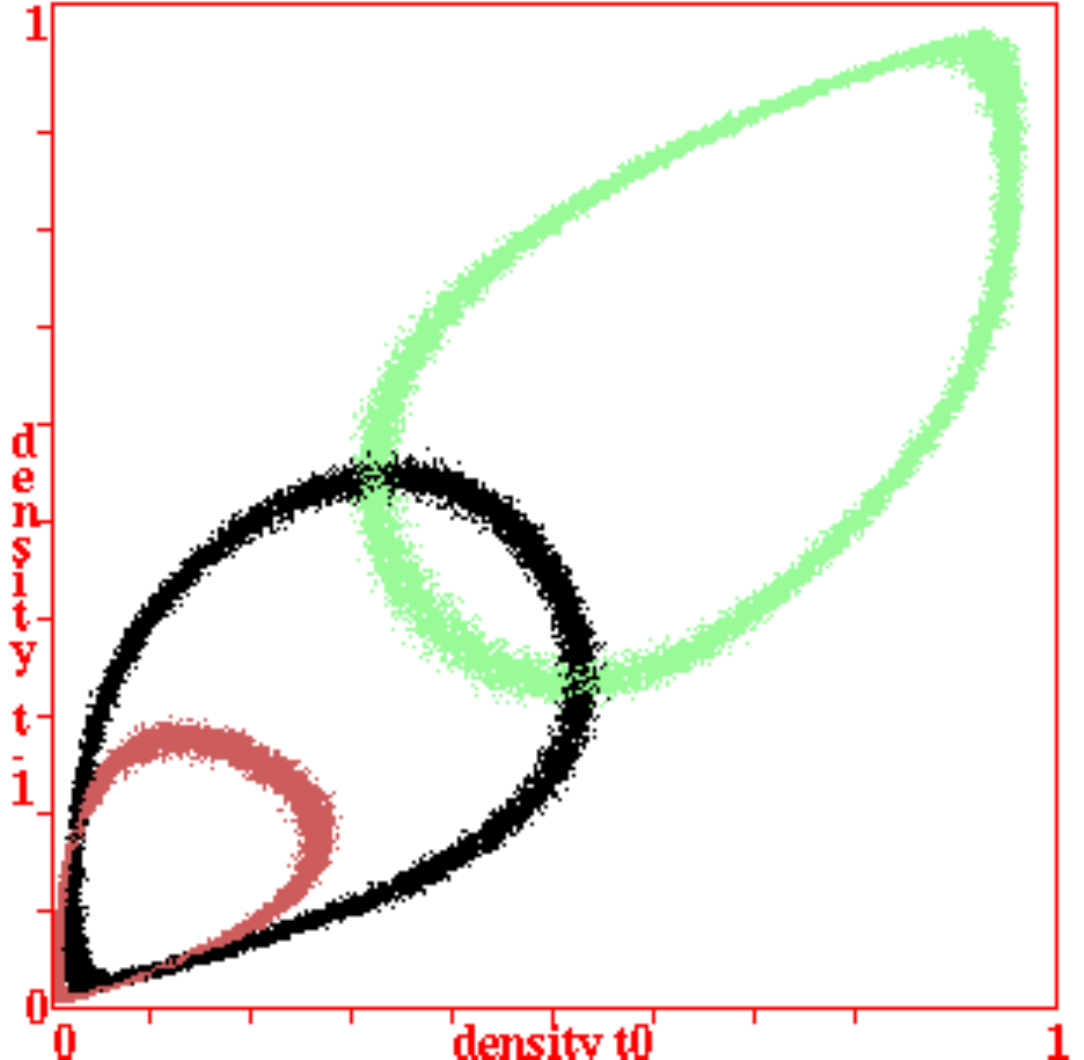}\\
(h)
\end{minipage}
\end{minipage}
}
\end{center}
\begin{center}
\textsf{\small
\begin{minipage}[c]{.95\linewidth} 
\begin{minipage}[c]{.12\linewidth}
\fbox{\includegraphics[width=1\linewidth]{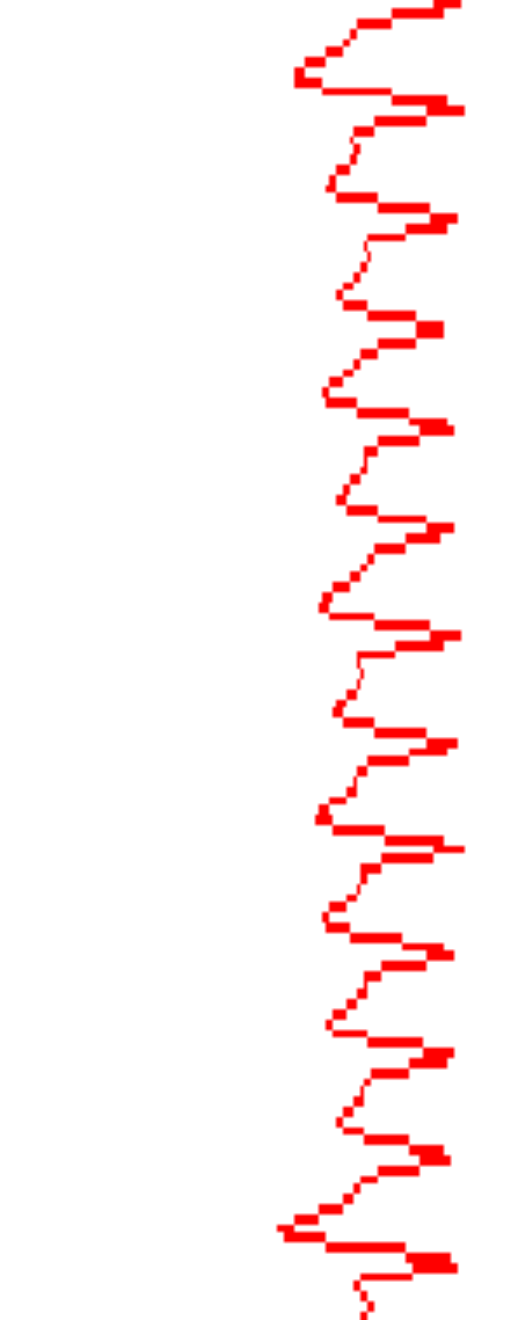}}\\[1ex]
(f2)~$wl$$\approx$10
\end{minipage}
\hfill
\begin{minipage}[c]{.35\linewidth}
\includegraphics[width=1\linewidth]{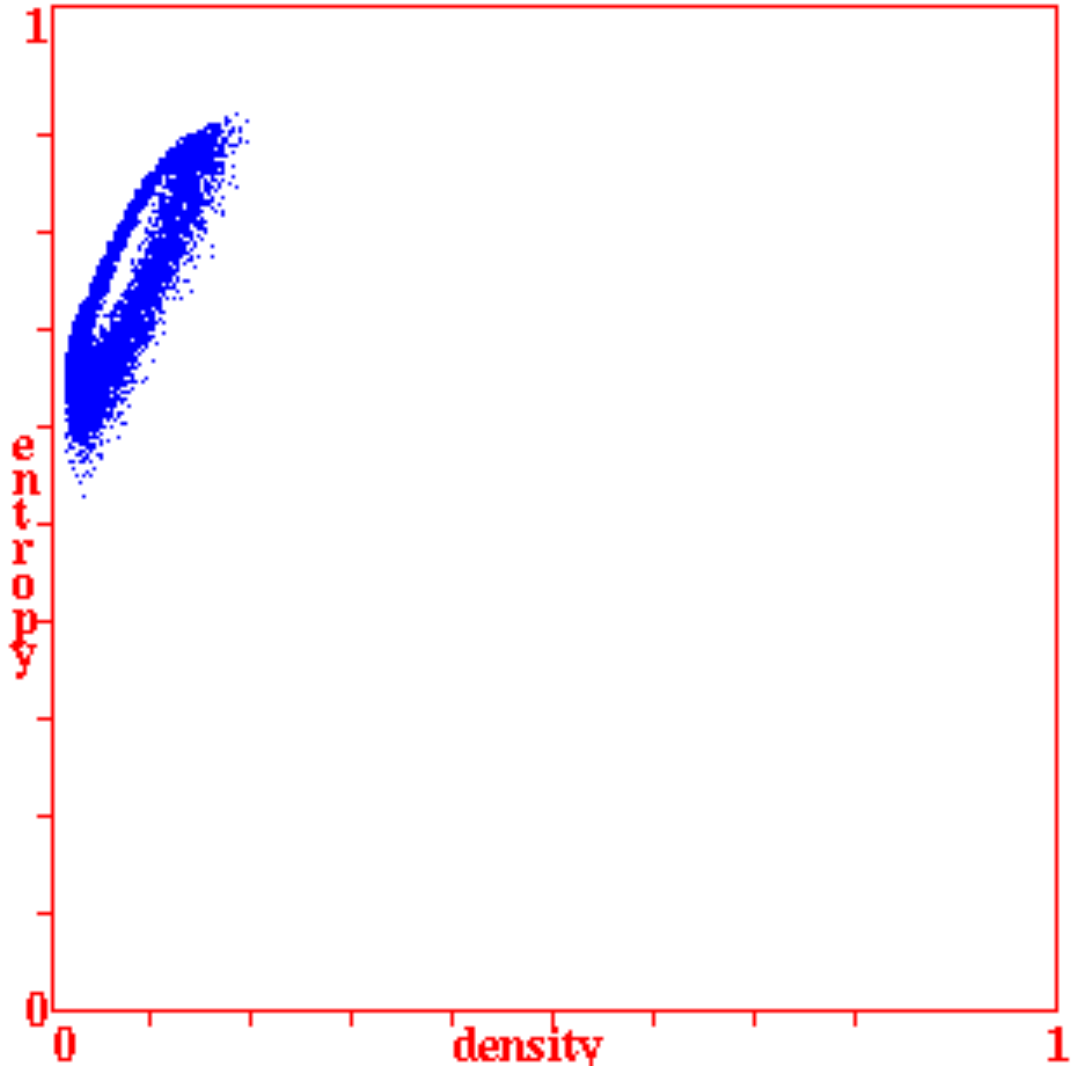}\\
(g) $v3k6$ ``g39'' 
\end{minipage}
\hfill
\begin{minipage}[c]{.35\linewidth}
\includegraphics[width=1\linewidth]{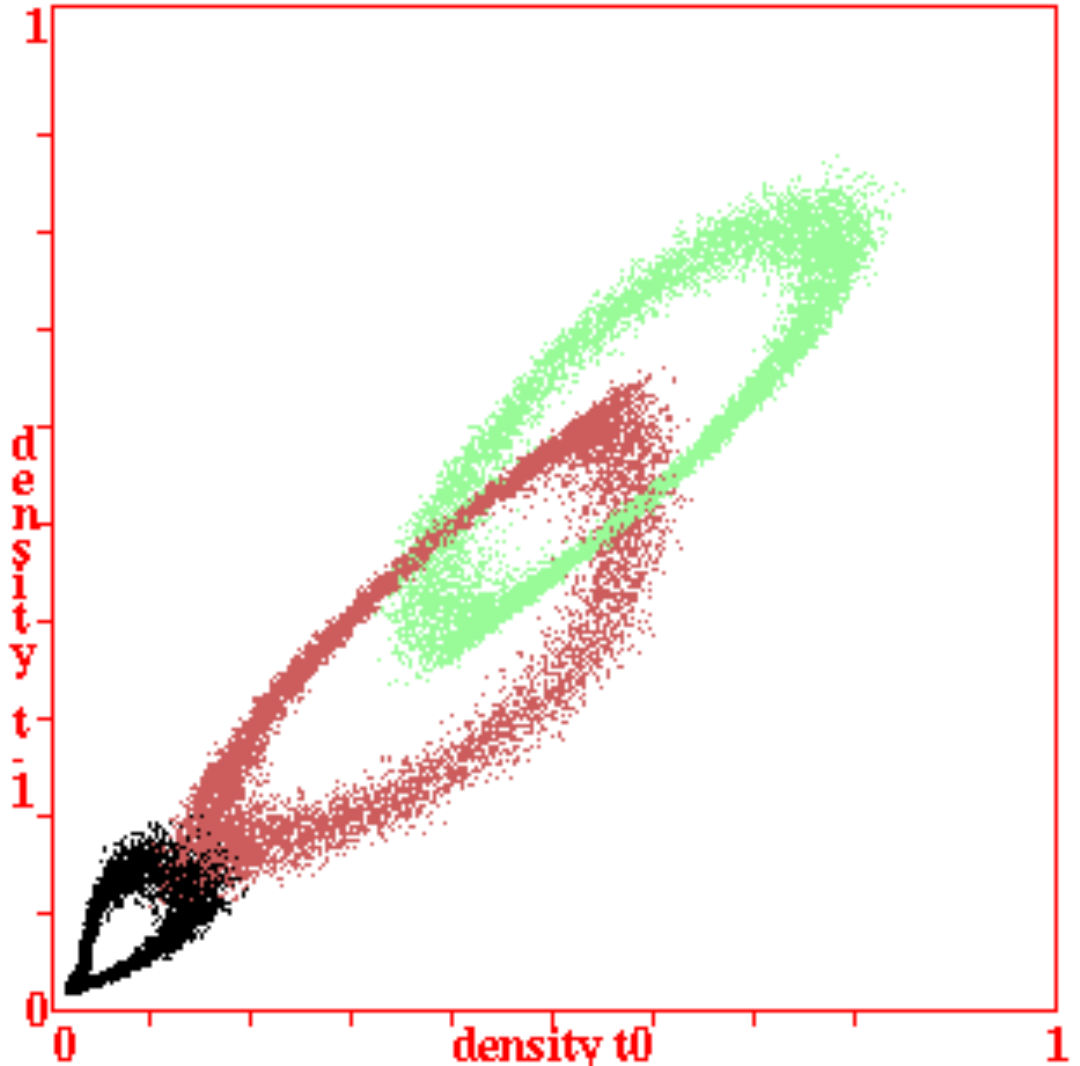}\\
(h)
\end{minipage}
\end{minipage}
}
\end{center}
\begin{center}
\textsf{\small
\begin{minipage}[c]{.95\linewidth} 
\begin{minipage}[c]{.12\linewidth}
\fbox{\includegraphics[width=1\linewidth]{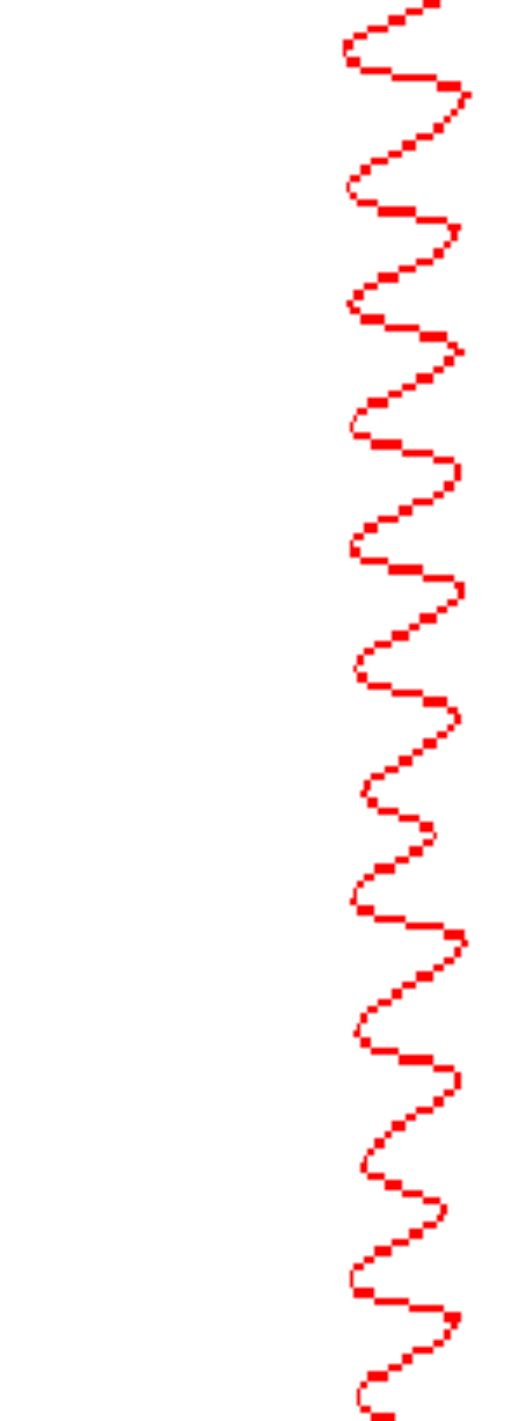}}\\[1ex]
(f2)~$wl$$\approx$15
\end{minipage}
\hfill
\begin{minipage}[c]{.35\linewidth}
\includegraphics[width=1\linewidth]{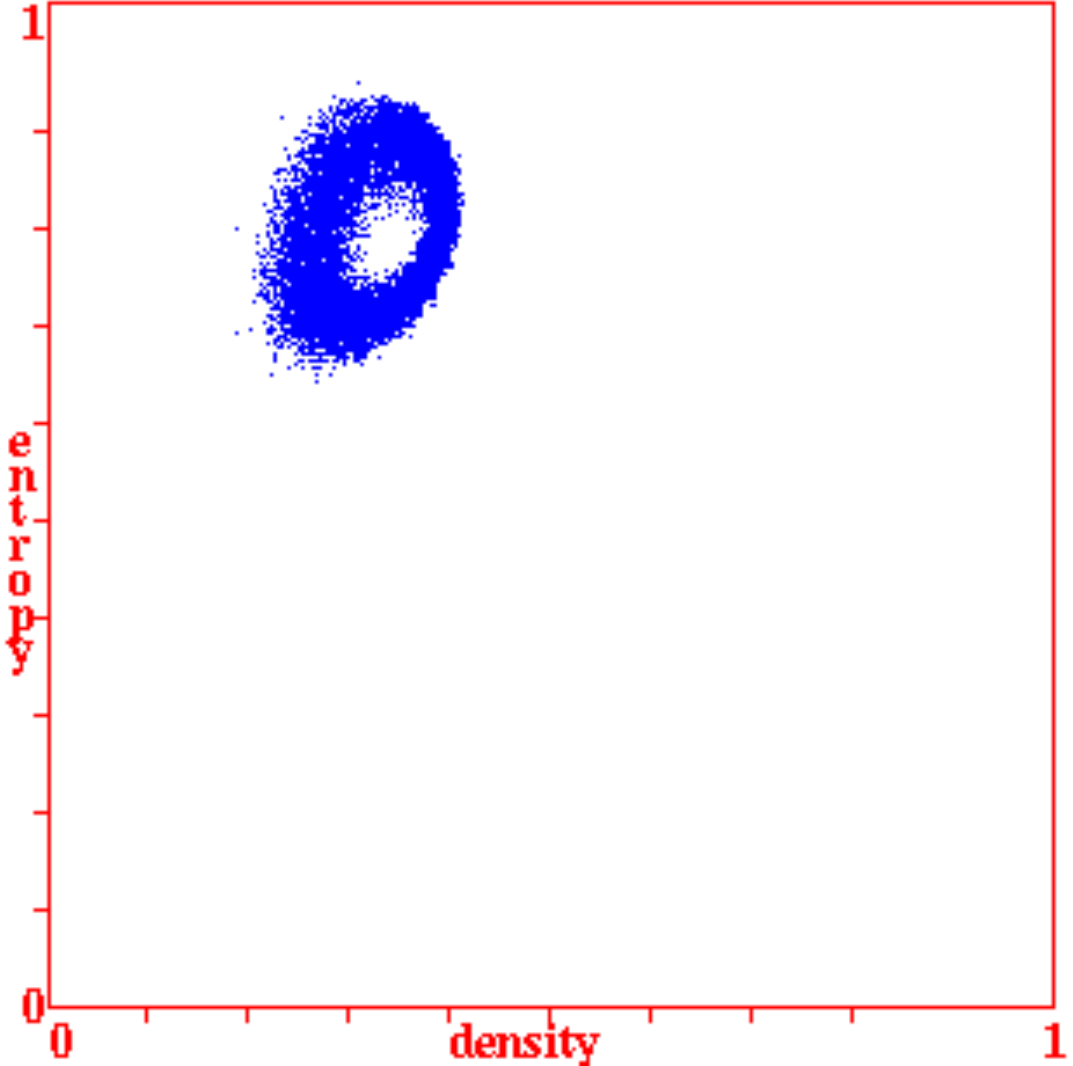}\\
(g)  $v3k6$ ``g26''
\end{minipage}
\hfill
\begin{minipage}[c]{.35\linewidth}
\includegraphics[width=1\linewidth]{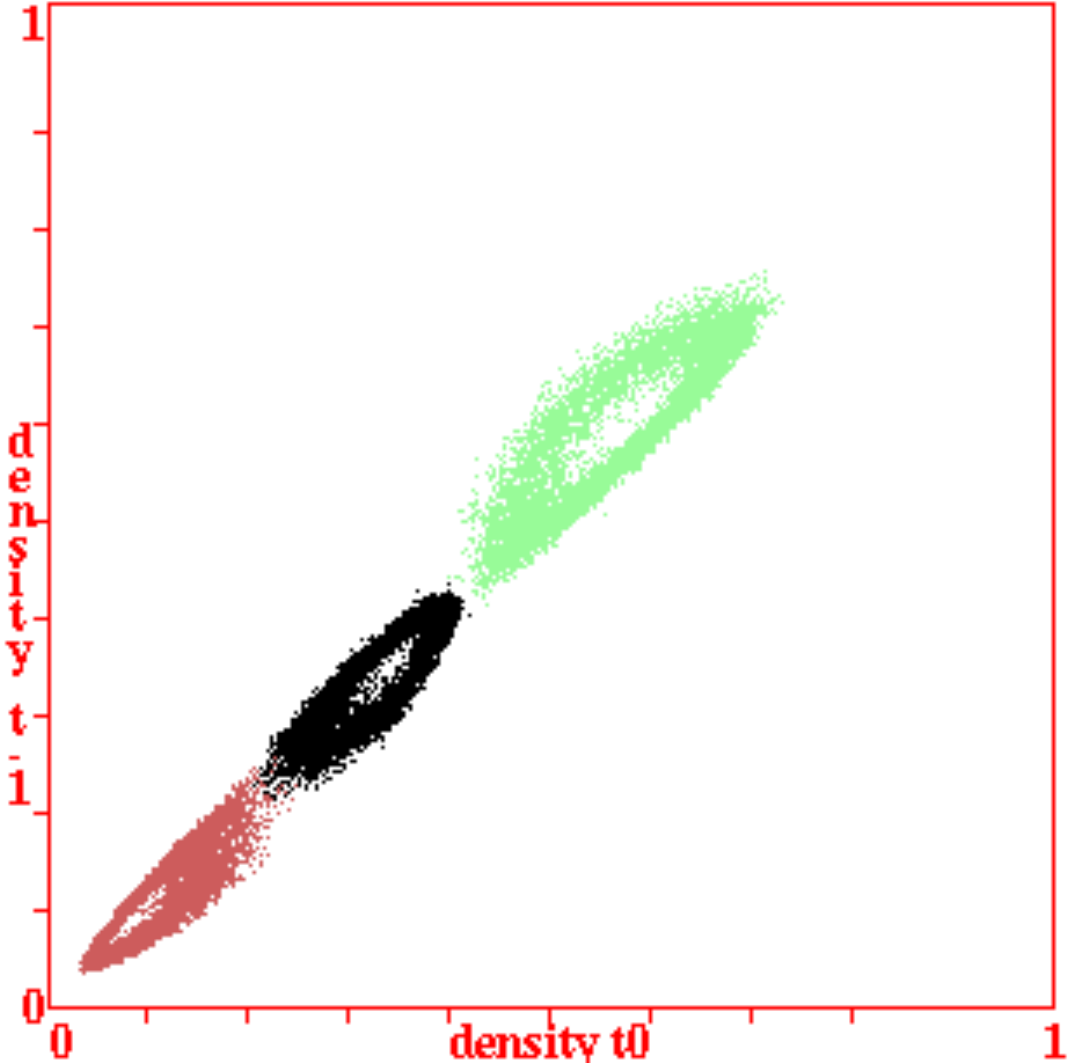}\\
(h)
\end{minipage}
\end{minipage}
}
\end{center}
\vspace{-3ex}
\caption[k6CA freeing one wire]
{\textsf{
Pulsing measures for 2D CA with one free wire, for the three $v3k6$ case study rules  
in figures~\ref{Pulsing dynamics beehive rule},
\ref{Pulsing dynamics k6-g39} and \ref{Pulsing dynamics k6-g26}.
(f2) input-entropy/time plot, (g) entropy-density scatter plot, (h) density return map scatter plot, 
for a 100x100 hexagonal lattice.
}}
\label{k6CA freeing one wire}
\end{figure}

\begin{figure}[htb]
\begin{center}
\textsf{\small
\begin{minipage}[c]{.95\linewidth} 
\begin{minipage}[c]{.12\linewidth}
\fbox{\includegraphics[width=1\linewidth]{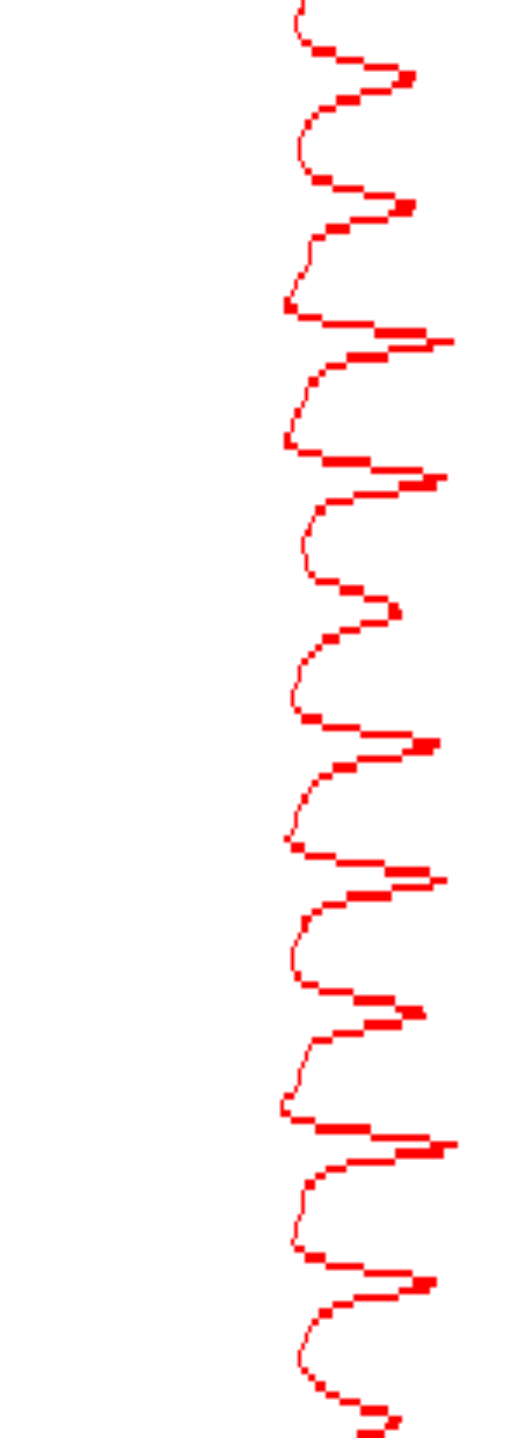}}\\[1ex]
(f2)~$wl$$\approx$16
\end{minipage}
\hfill
\begin{minipage}[c]{.35\linewidth}
\includegraphics[width=1\linewidth]{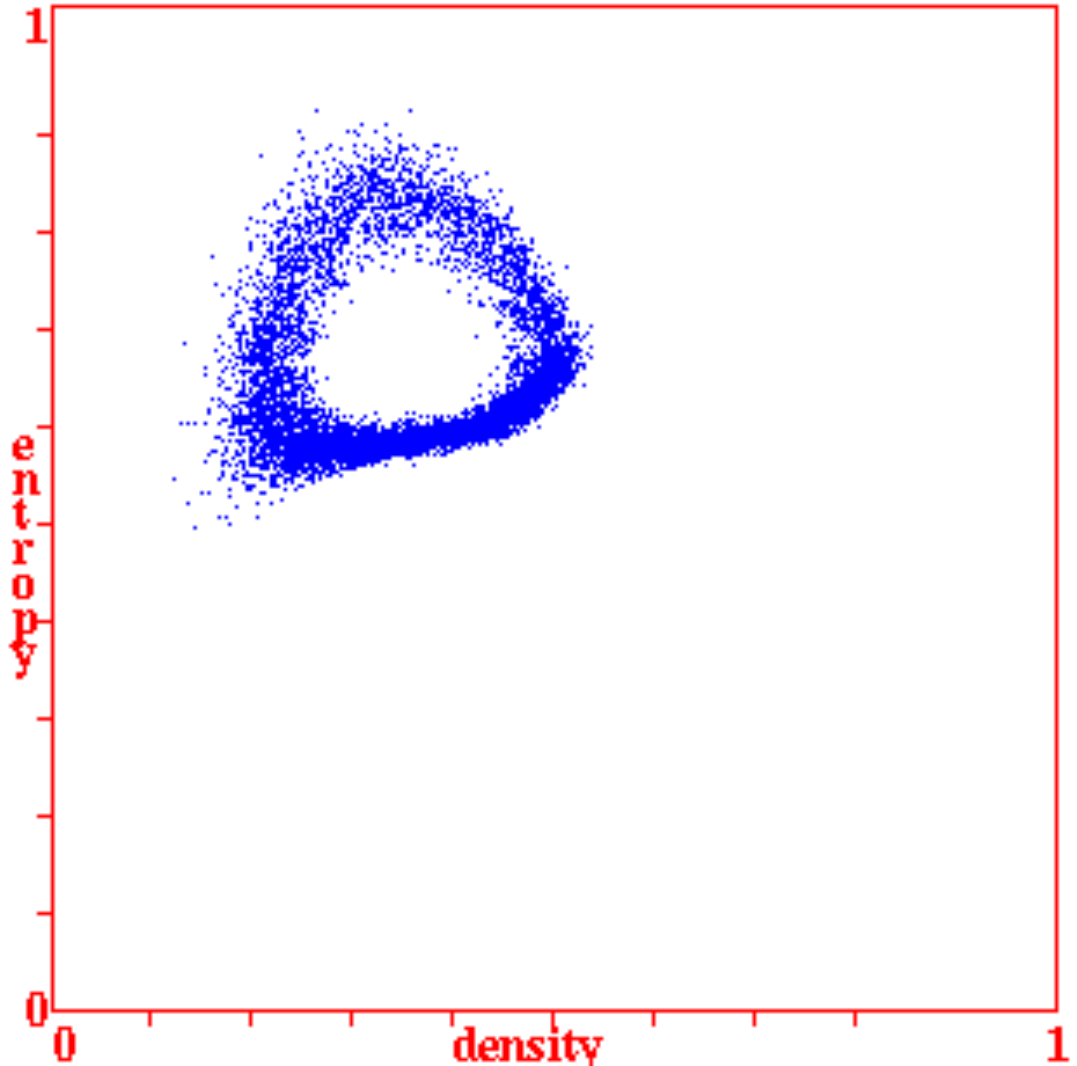}\\
(g) $v3k7$ ``g3''
\end{minipage}
\hfill
\begin{minipage}[c]{.35\linewidth}
\includegraphics[width=1\linewidth]{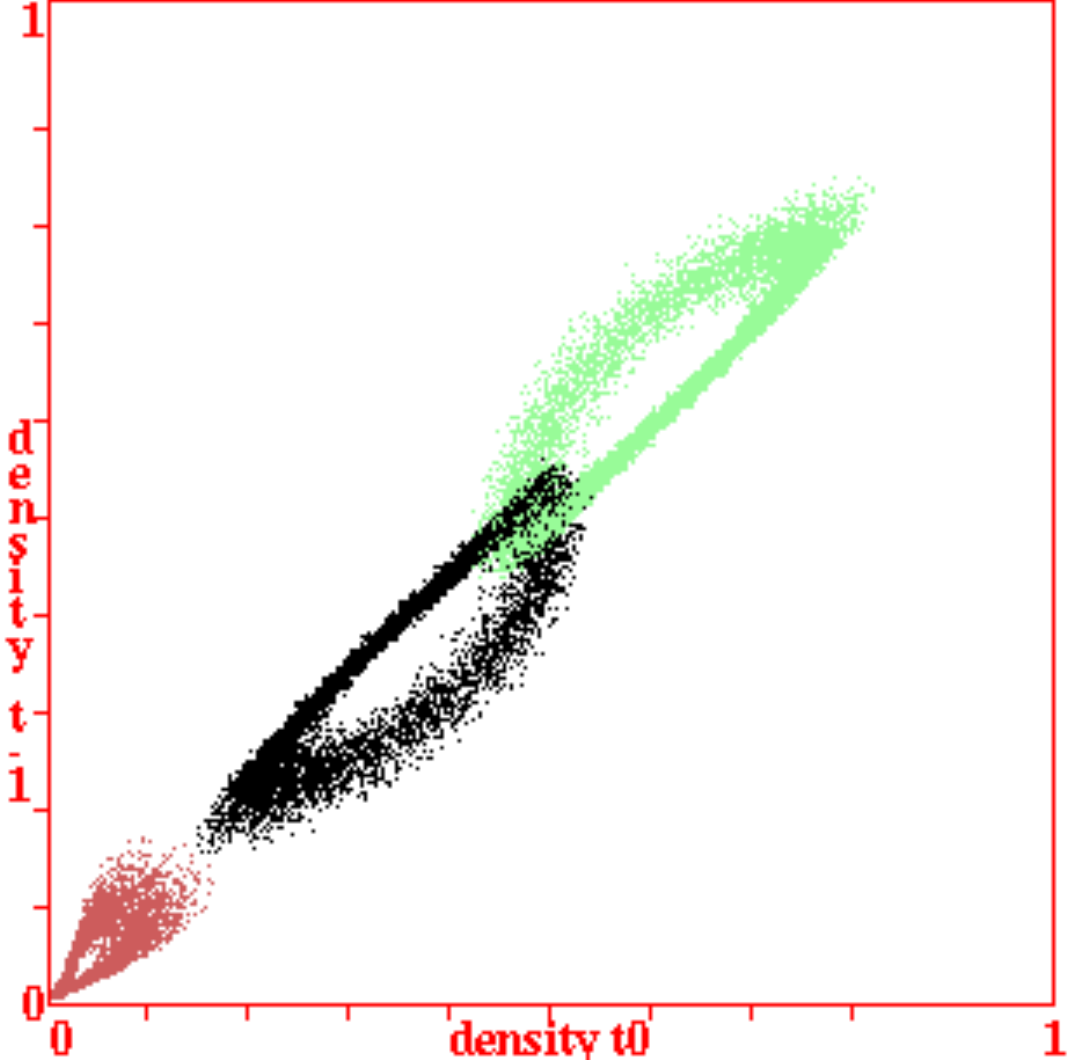}\\
(h)
\end{minipage}
\end{minipage}
}
\end{center}
\vspace{-3ex}
\caption[k7CA freeing one wire]
{\textsf{
Pulsing measures for 2D CA with one free wire for $v3k7$ --- only one case study rule,
``g3'' from figure~\ref{Pulsing dynamics g3 rule} showed significant pulsing.
(f2) input-entropy/time plot, (g) entropy-density scatter plot, (h) density return map scatter plot, 
for a 100x100 hexagonal lattice.
}}
\label{k7CA freeing one wire}
\vspace{-3ex}
\end{figure}

\noindent If one wire is released from each neighborhood in the 2D CA,
and freely connected anywhere
in the lattice, glider dynamics is destroyed and we may begin to see pulsing.
Freeing one wire results in significant pulsing in all three $k$=6 rules in our
case study (figure~\ref{k6CA freeing one wire}), 
and is also probable in other $k$=6 glider rules.
For $k$=7,  pulsing is less probable because a smaller proportion of the
neighbrhood is randomised --- only one rule from the case study gave distinct pulsing
(figure~\ref{k7CA freeing one wire}).
If two wires are freed, pulsing is highly probable for both $k$=6 and  $k$=7,
and with more free wires pulsing properties approach the RW-waveform.
The waveform is unaffected by re-randomising at each time-step.

\section{Localised random wiring}
\label{Localised random wiring}
\vspace{-4ex}

\begin{figure}[h]
\begin{center}
\begin{minipage}[c]{.8\linewidth}
\begin{minipage}[c]{.45\linewidth}
\includegraphics[width=1\linewidth,bb=20 89 379 445, clip=]{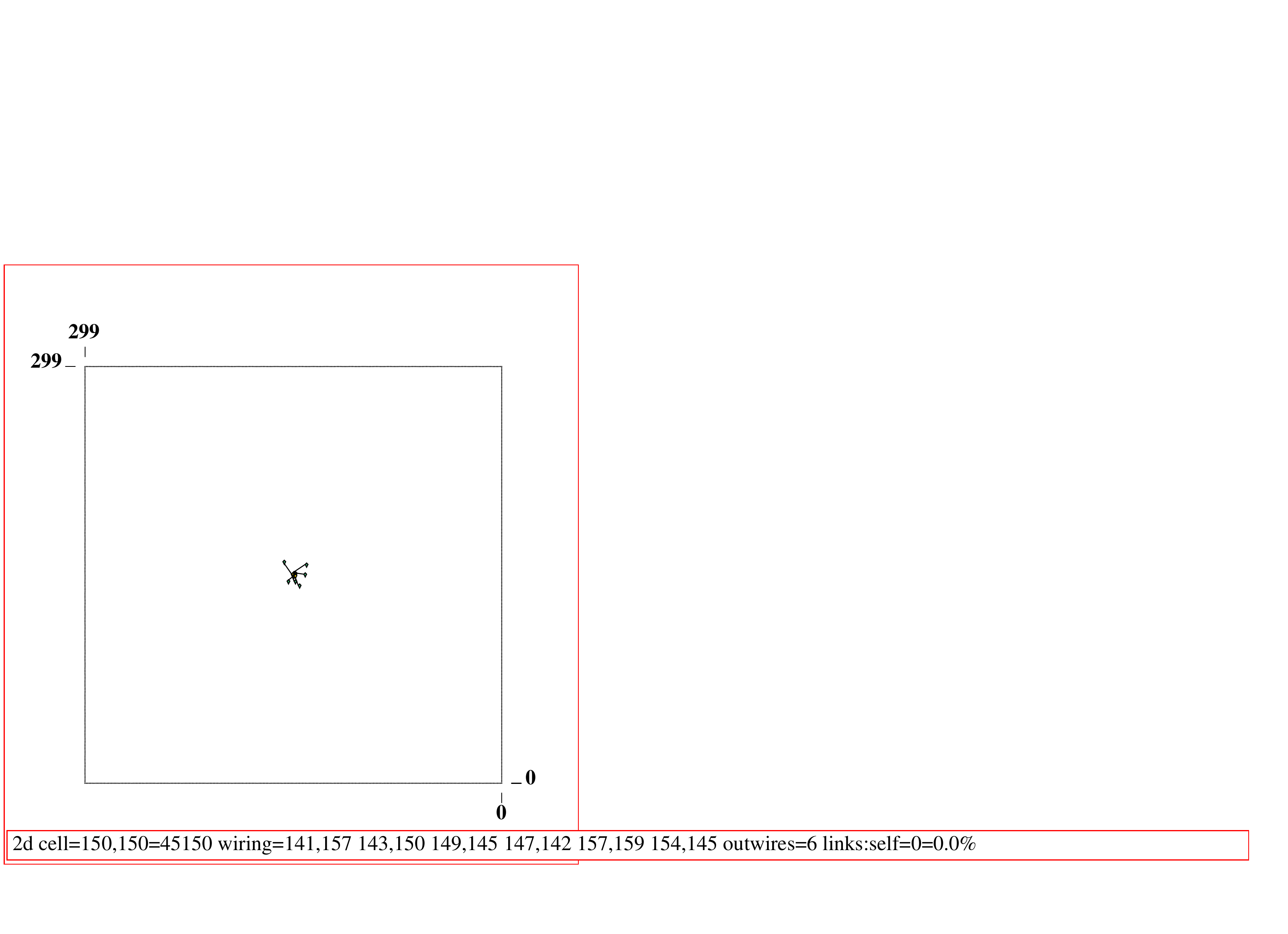}
\end{minipage}
\hfill
\begin{minipage}[c]{.45\linewidth}
\includegraphics[width=1\linewidth,bb=170 199 456 486, clip=]{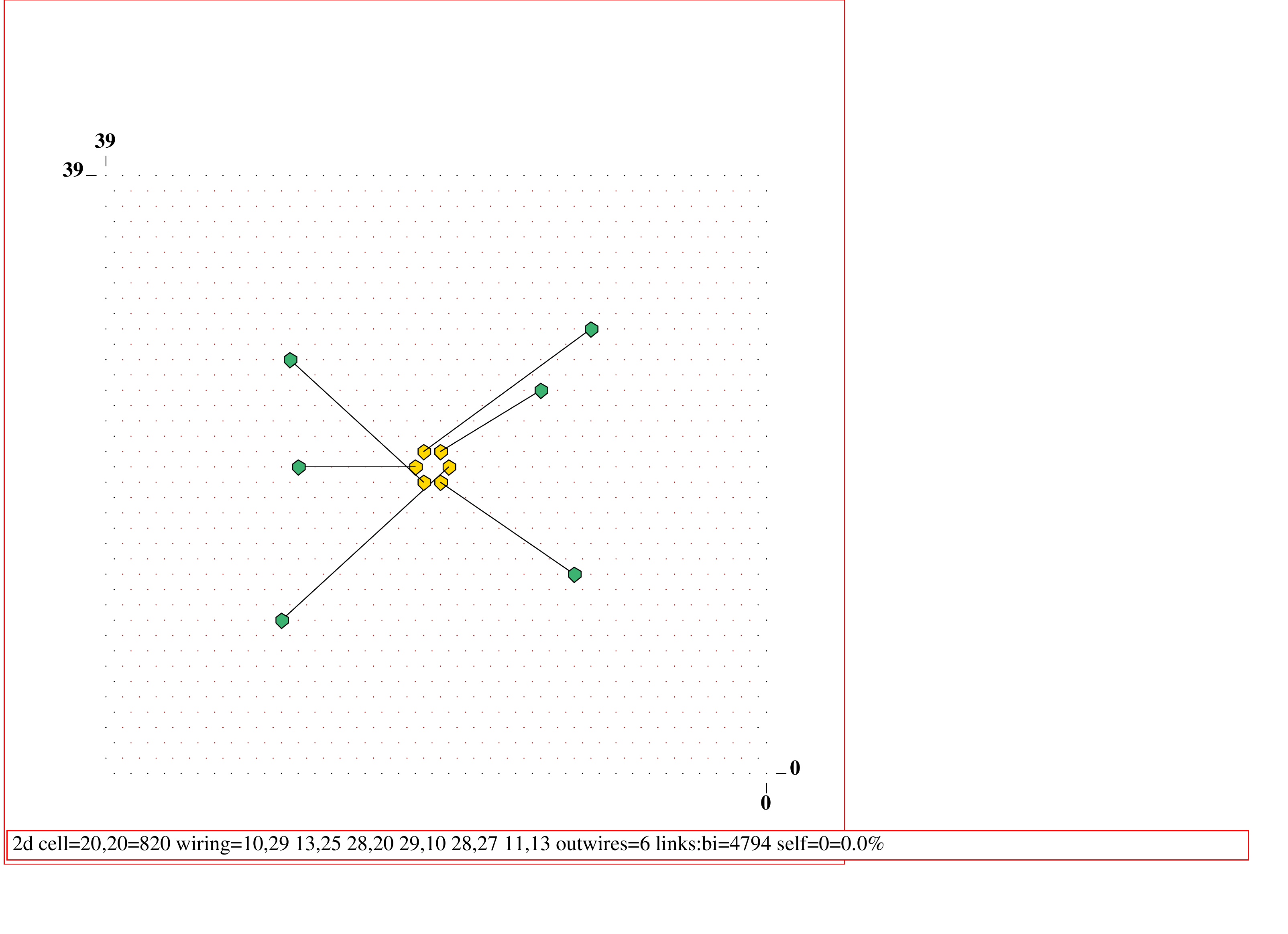}
\end{minipage}
\end{minipage}
\end{center}
\vspace{-4ex}
\caption[Localised random wiring 300x300]
{\textsf{
(Left) Random wiring confined within
20$\times$20 local zones within a 300$\times$300 hexagonal lattice ($f=$20/300$\approx$0.066),
(Right) shows a detail.
}}
\label{Localised random wiring 300x300}
\end{figure}

\begin{figure}[htb]
\begin{center}
\begin{minipage}[c]{1\linewidth}
\begin{minipage}[c]{.3\linewidth}
\includegraphics[width=1\linewidth]{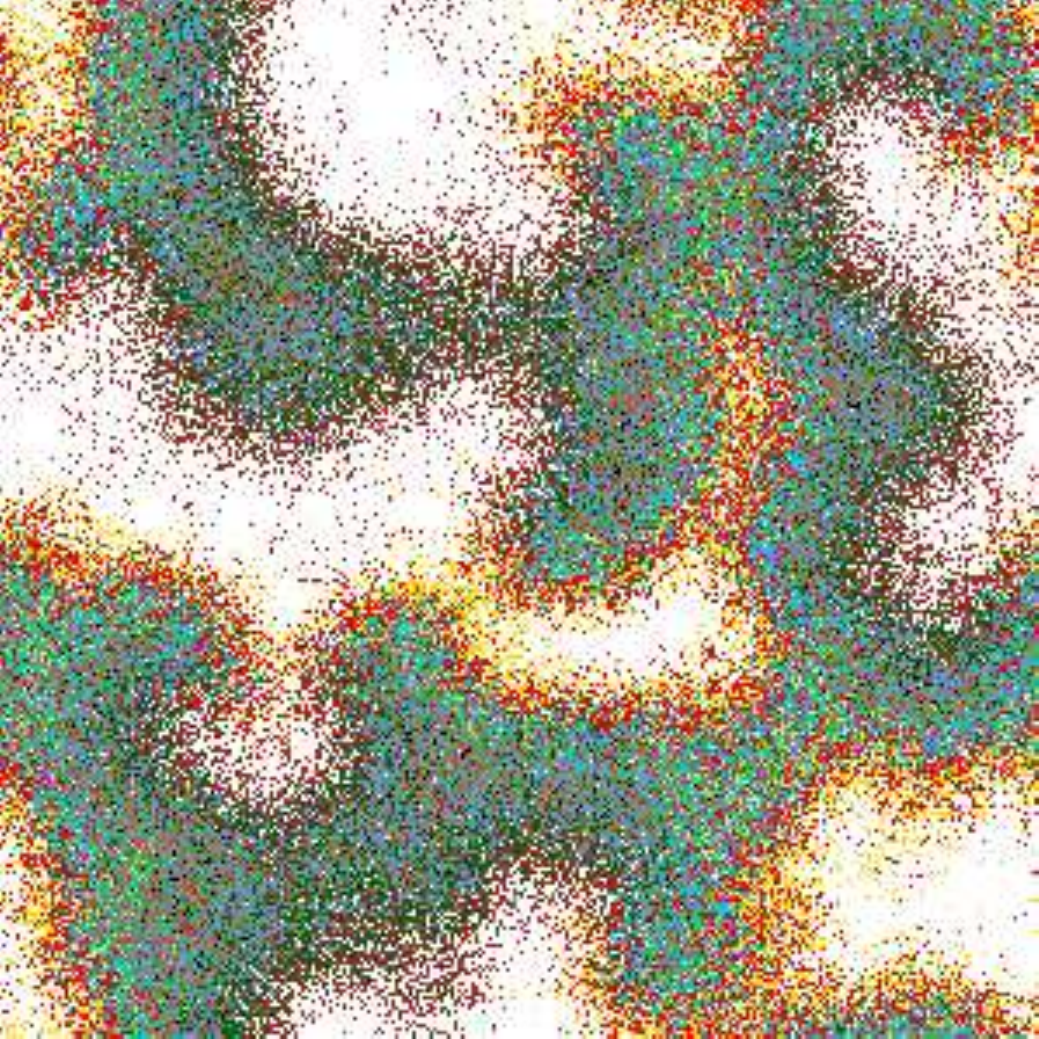}\\
\textsf{\small$v3k6$ ``g2''}
\end{minipage}
\hfill
\begin{minipage}[c]{.3\linewidth}
\includegraphics[width=1\linewidth]{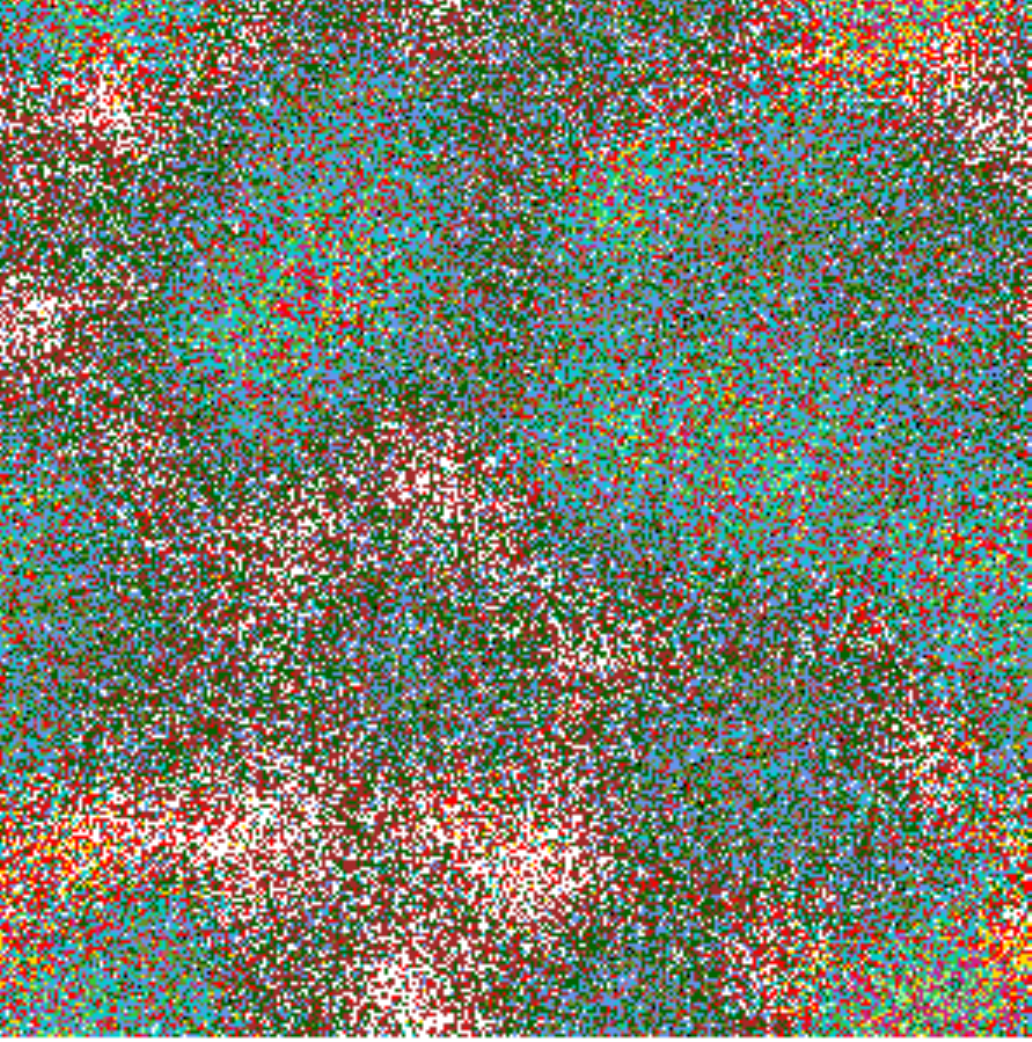}\\
$v3k6$ ``g39''
\end{minipage}
\hfill
\begin{minipage}[c]{.3\linewidth}
\includegraphics[width=1\linewidth]{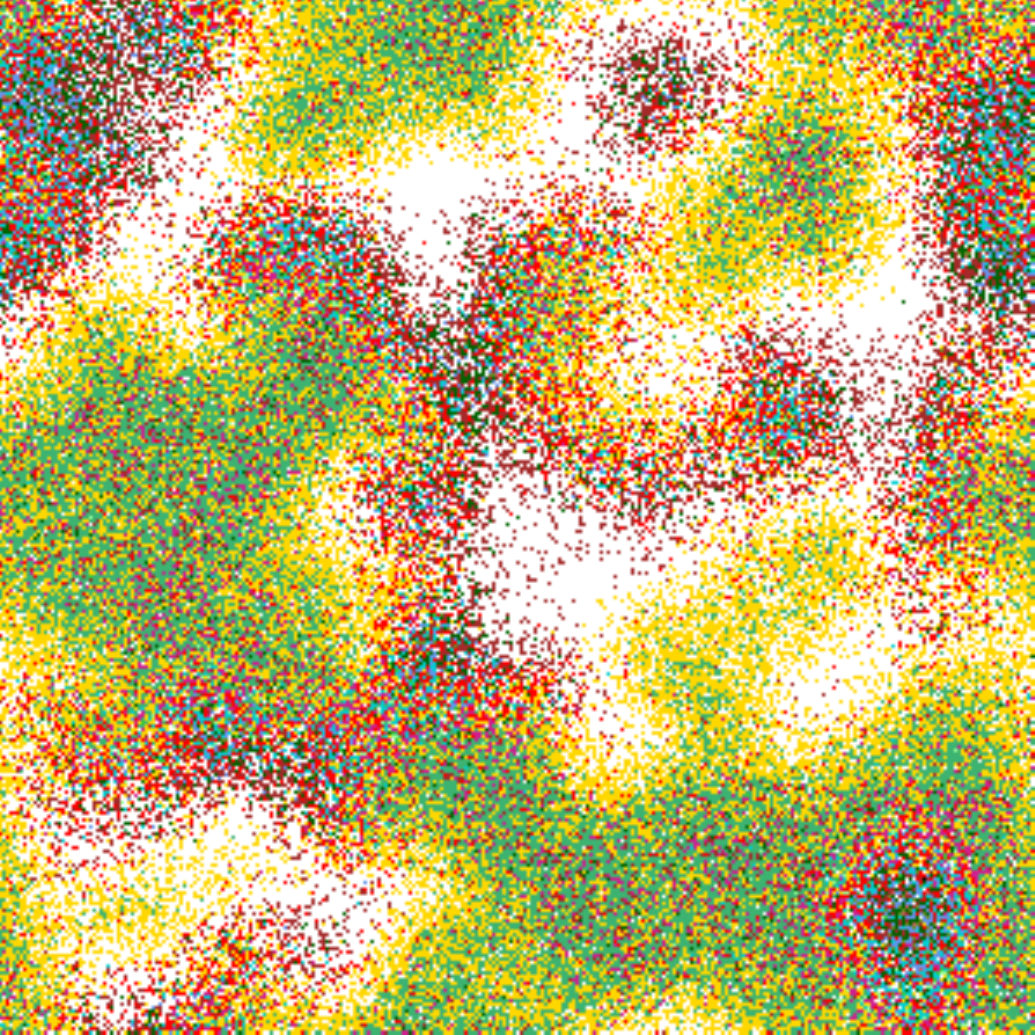}\\
\textsf{\small$v3k6$ ``g26''}
\end{minipage}
\end{minipage}\\[2ex]
\begin{minipage}[c]{1\linewidth}
\begin{minipage}[c]{.3\linewidth}
\includegraphics[width=1\linewidth]{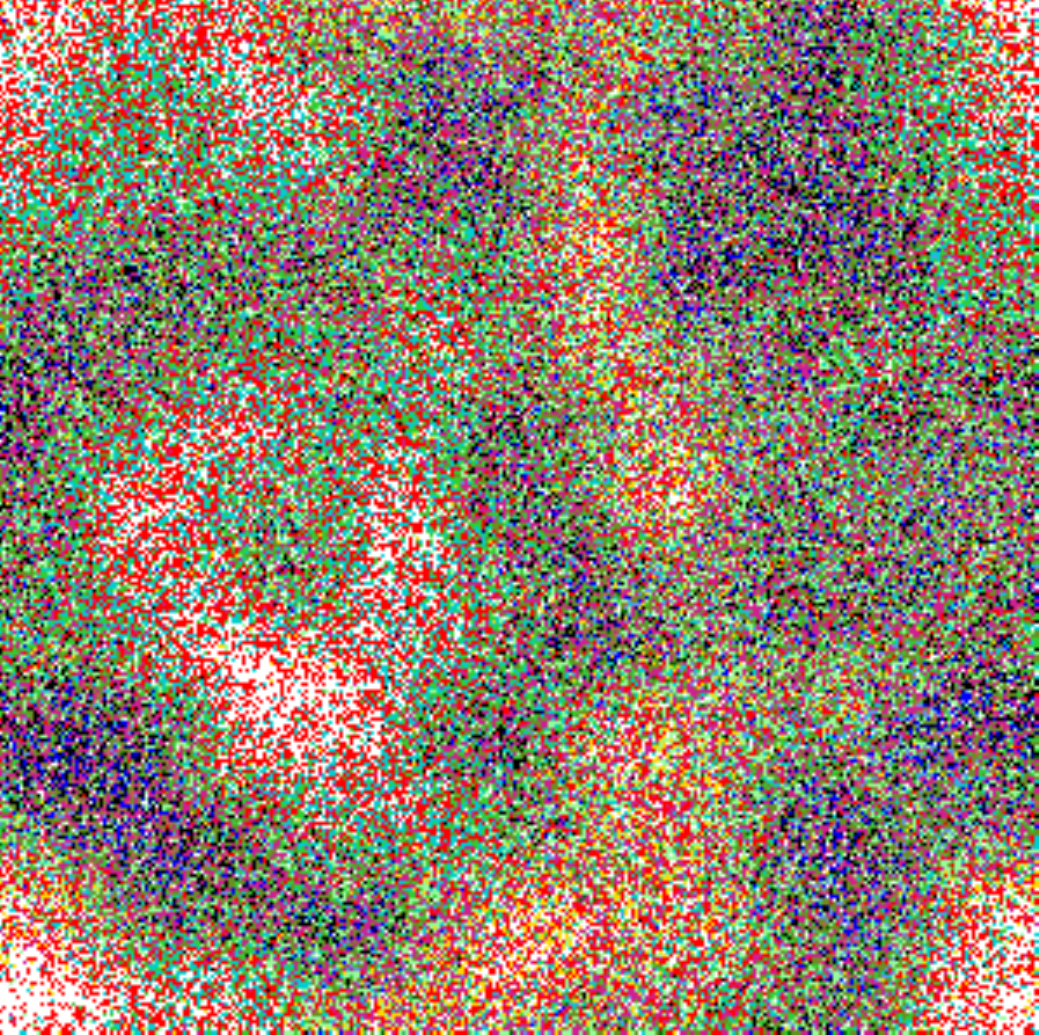}\\
$v3k7$ ``g1''
\end{minipage}
\hfill
\begin{minipage}[c]{.3\linewidth}
\includegraphics[width=1\linewidth]{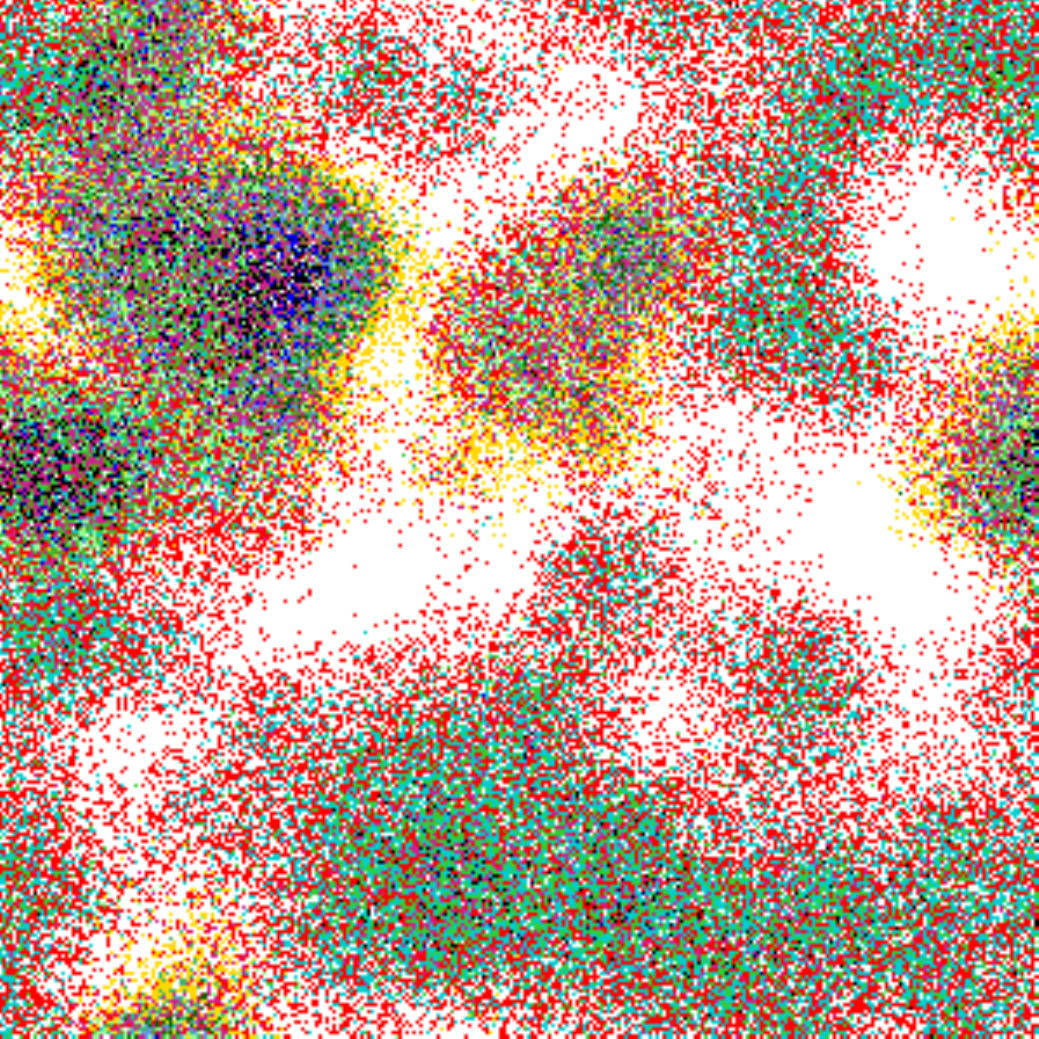}\\
\textsf{\small$v3k7$ ``g3''}
\end{minipage}
\hfill
\begin{minipage}[c]{.3\linewidth}
\includegraphics[width=1\linewidth]{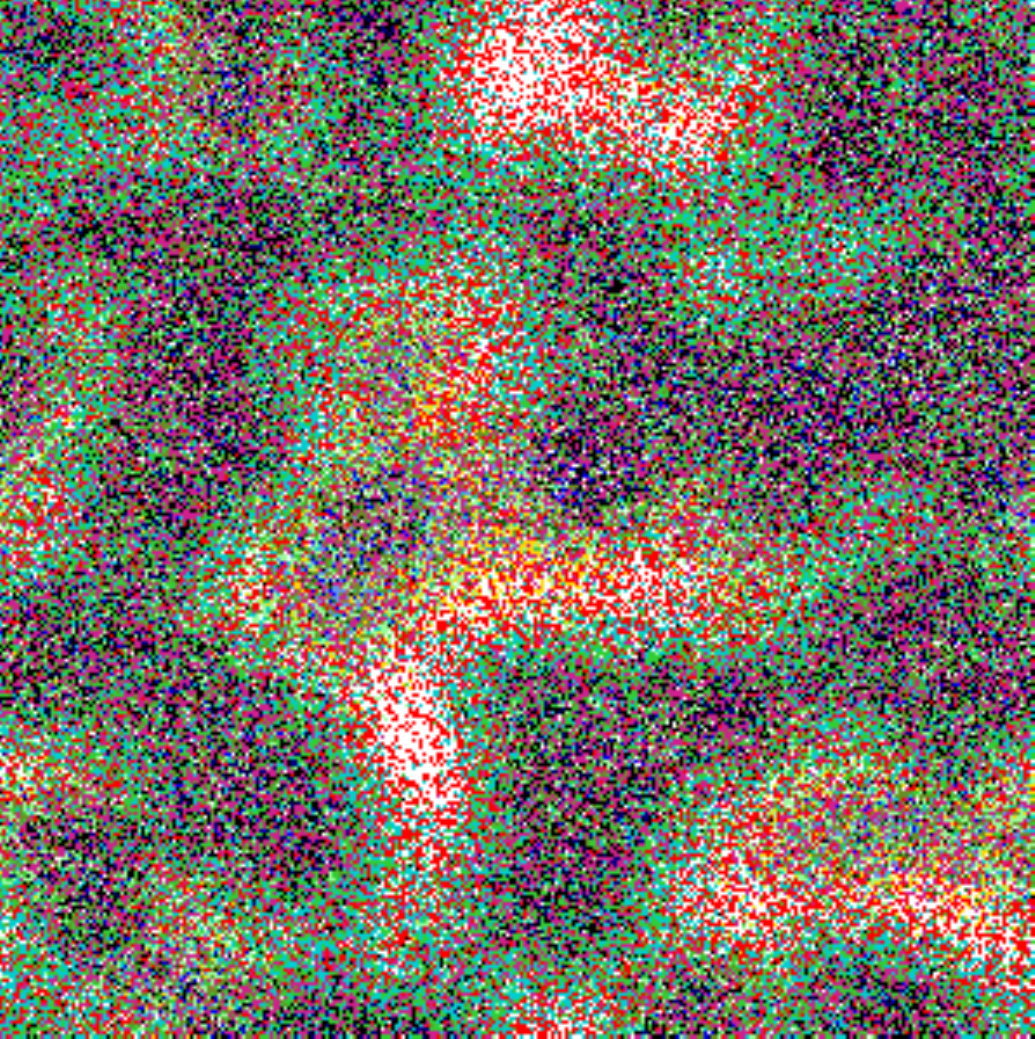}\\
\textsf{\small$v3k7$ ``g35''}
\end{minipage}
\end{minipage}
\end{center}
\vspace{-3ex}
\caption[spiral waves]
{\textsf{
Waves of density emerge when random wiring is localised
within 20$\times$20 zones in a 300$\times$300 hexagonal lattice
(figure~\ref{Localised random wiring 300x300}).
Typical pattern snapshots are shown, with cells colored according to
neighborhood lookup instead of value.\\
(Top row) $v3k6$ rules as in 
figures~\ref{Pulsing dynamics beehive rule}-\ref{Pulsing dynamics k6-g26}.
(Bottom row) $v3k7$ rules as in 
figures~\ref{Pulsing dynamics spiral rule}-\ref{Pulsing dynamics f82 rule}.
Any initial state will set off similar dynamics.
Overall entropy pulsing is still apparent, and also patchy and spiral
waves of density.
}}
\label{spiral waves}
\end{figure}

\noindent How does the dynamics play out if
random wiring is confined within a local zone relative to each
target cell? --- as in figure~\ref{Localised random wiring 300x300}, which makes the
2D geometry of the lattice significant, whereas with fully random wiring
the geometry loses significance.
Experiment shows that as the local zone diameter $d$ is reduced
relative to the network diameter $D$ --- the reach of random wiring ---
overall pulsing, though still apparent, turns into patchy waves of density.
At some threshold (of the fraction $f$=$d/D$) the stability and shape of the waveform will start
to deform relative to the RW-waveform,
and eventually break down, a process that could be interpreted as a type of phase transition,
though a proper description will require further research and analysis.  
Preliminary results show that the threshold $f_T$ is independent of
network size, but varies according to the rule. For $v3k7$ rules in 
figures~\ref{Pulsing dynamics spiral rule}, 
\ref{Pulsing dynamics g3 rule}, \ref{Pulsing dynamics f82 rule},
$f_T\approx$ 0.08, 0.05, 0.24, respectively. 
Above this relatively low threshold the pulsing waveform is robust.

As the local zone is further reduced, spiral density waves, reminiscent
of reaction-diffusion, can emerge in a large enough system
(figure~\ref{spiral waves}) with local pulsing as waves sweep across
a local area. Re-randomising the wiring at each time-step makes
no significant difference to the general behaviour. 

\section{Freeing one wire from localised neighborhoods}
\label{Localised --- but freeing one wire}

\noindent Freeing just one wire from the 20$\times$20
localised random zone in section~\ref{Localised random wiring}, allowing
it to connect anywhere, restores pulsing behaviour, but with a patchy
distribution of values. Experiment confirms this applies to 
all the rules in the pulsing case studies in
section~\ref{Pulsing case studies} --\ref{Pulsing dynamics f82 rule}. 
The waveforms are still recognisable when compared to the RW-waveforms,
including the entropy-density and density return-map scatter plot
signatures.
For example, the $v3k7$ ``g3'' rule waveform
(figure~\ref{one remote wire, 300x300 stp}) can be compared with 
its RW-waveform in figure~\ref{Pulsing dynamics g3 rule}.
\clearpage

\begin{figure}[htb]
\begin{center}
\textsf{\small
\begin{minipage}[c]{.95\linewidth} 
\begin{minipage}[c]{.12\linewidth}
\fbox{\includegraphics[width=1\linewidth]{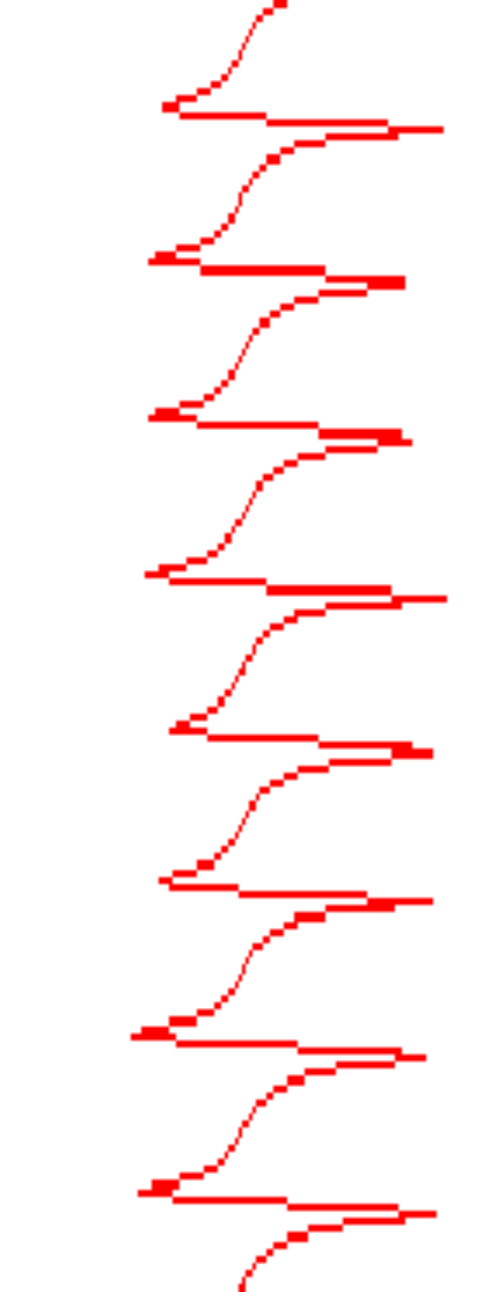}}\\[1ex]
(f2)~$wl$$\approx$21
\end{minipage}
\hfill
\begin{minipage}[c]{.35\linewidth}
\includegraphics[width=1\linewidth]{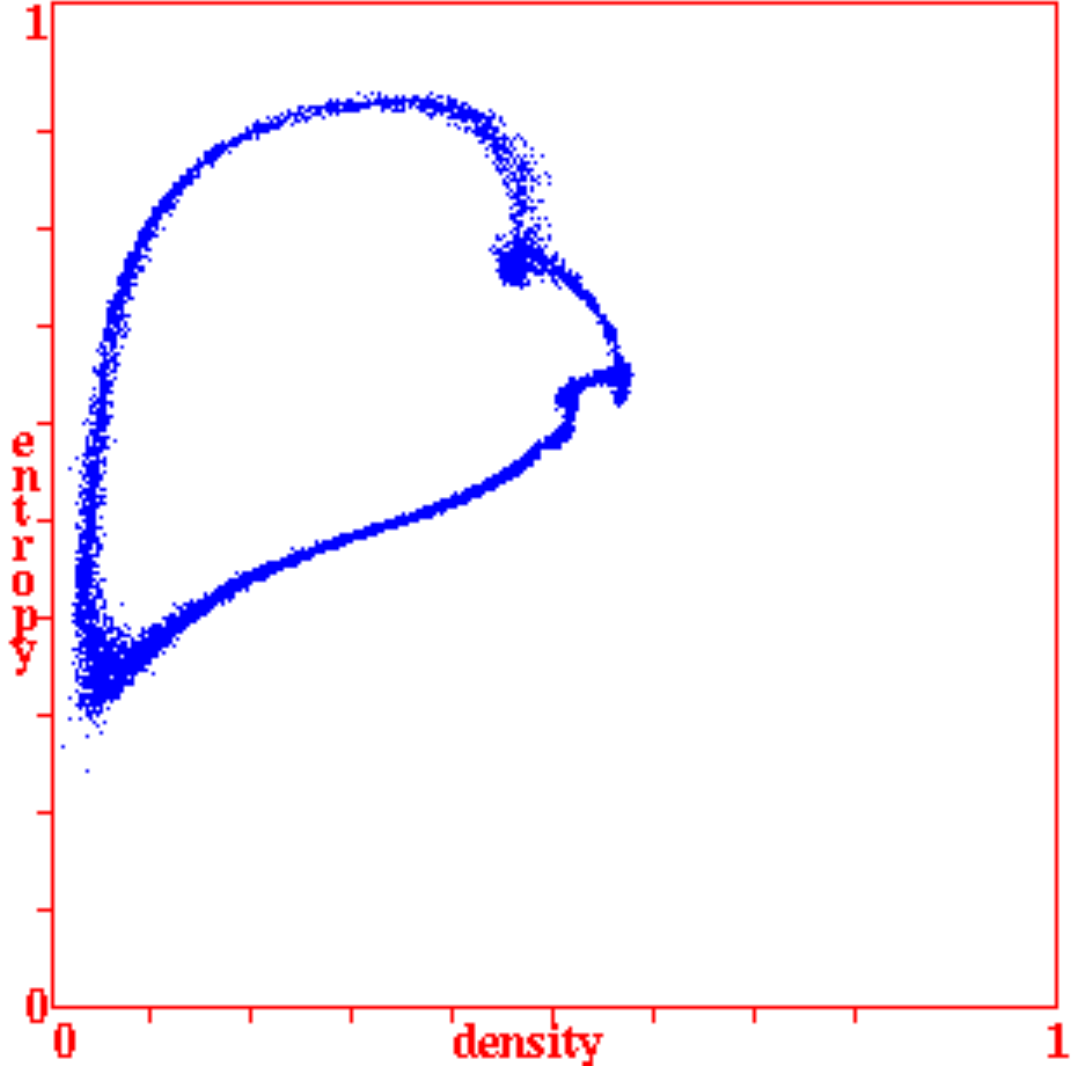}\\
(g) $v3k7$ ``g3''
\end{minipage}
\hfill
\begin{minipage}[c]{.35\linewidth}
\includegraphics[width=1\linewidth]{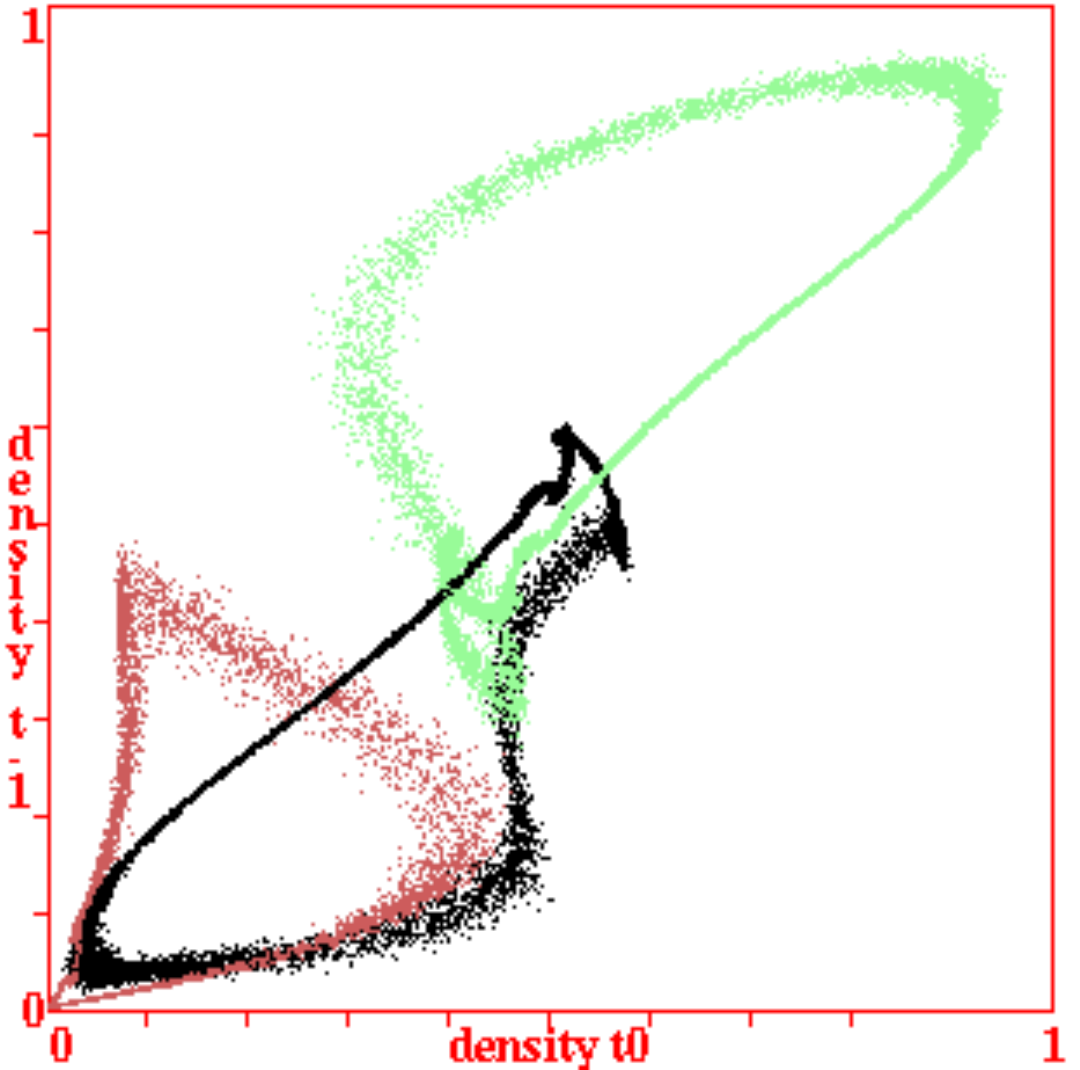}\\
(h)
\end{minipage}
\end{minipage}
\begin{minipage}[c]{.7\linewidth}
\end{minipage}\\[2ex]
\begin{minipage}[c]{.7\linewidth}
\begin{minipage}[c]{.43\linewidth}
\includegraphics[width=1\linewidth]{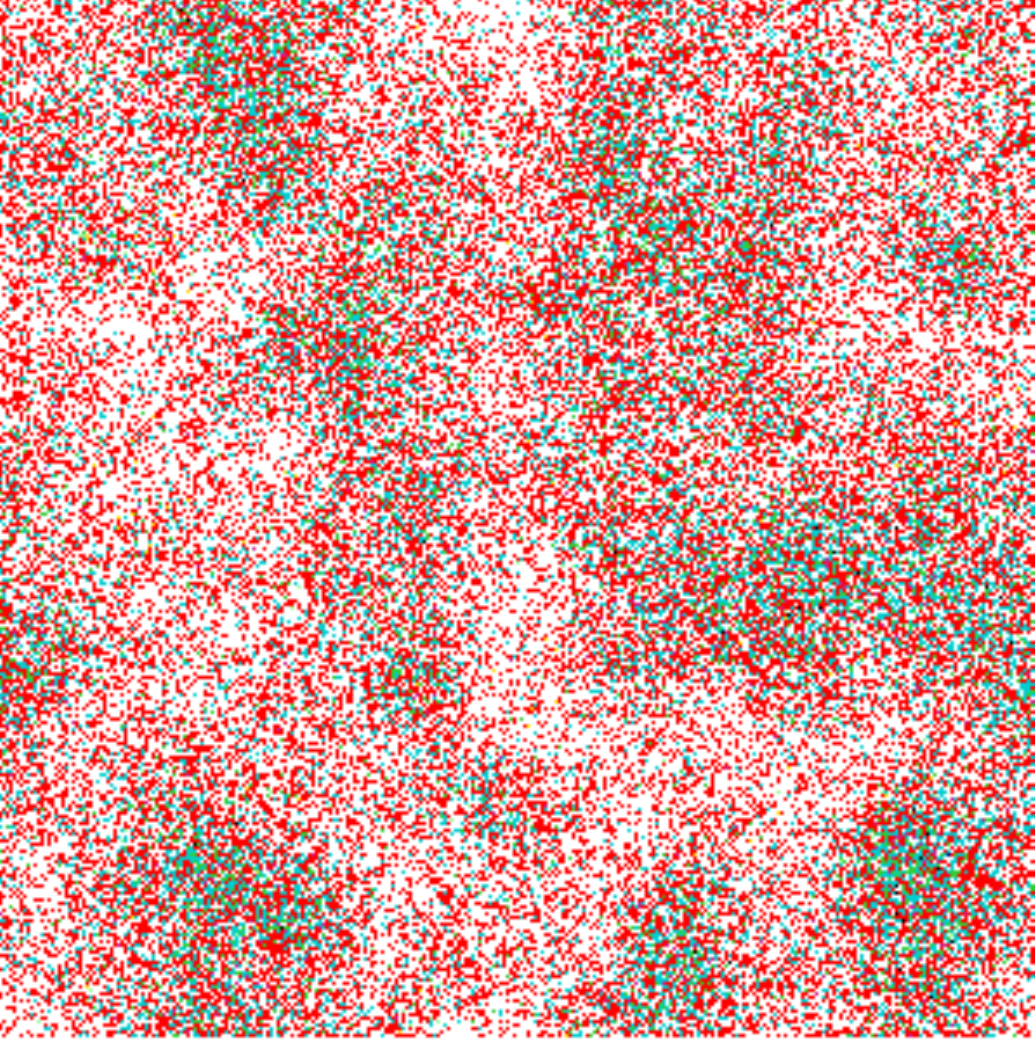}\\
$v3k7$ ``g3-min''
\end{minipage}
\hfill
\begin{minipage}[c]{.43\linewidth}
\includegraphics[width=1\linewidth]{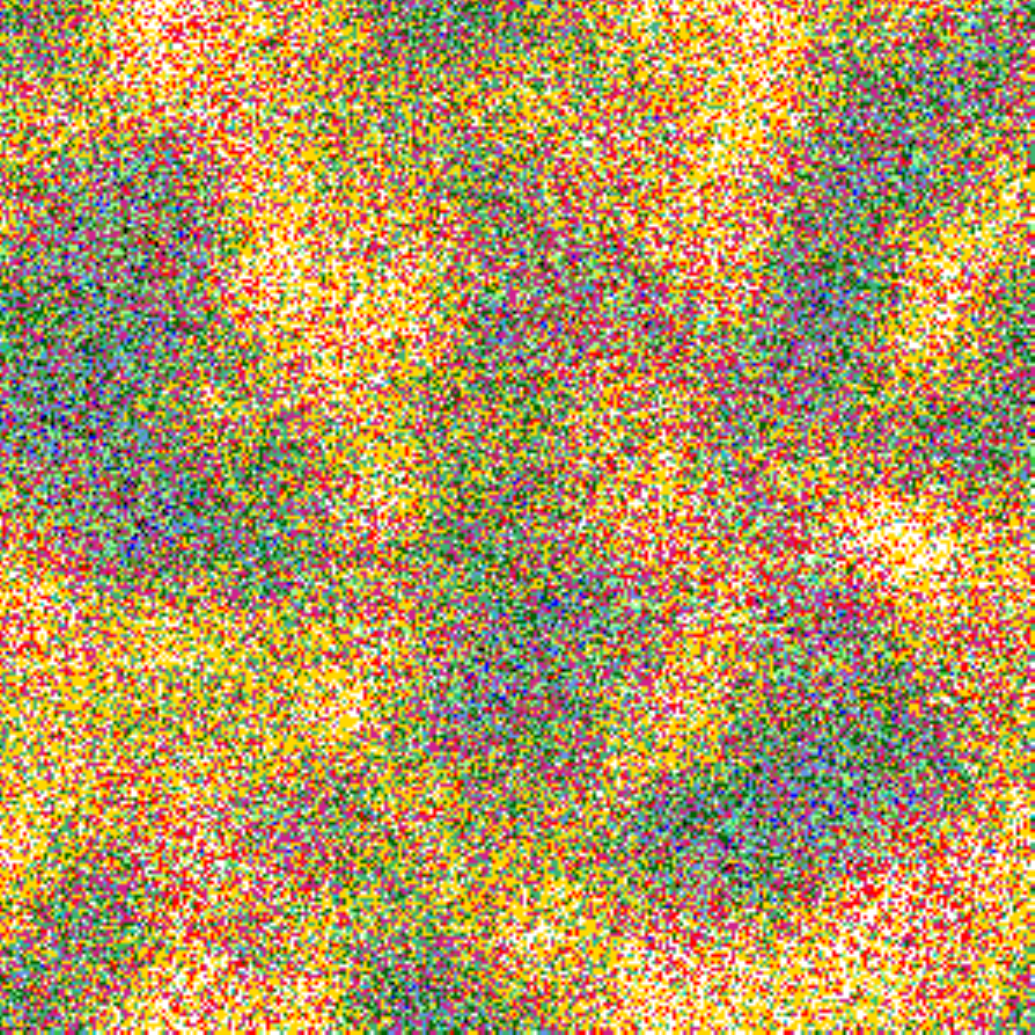}\\
$v3k7$ ``g3-max''
\end{minipage}
\end{minipage}\\[2ex]
}
\end{center}
\vspace{-2ex}
\caption[one remote wire, 300x300 stp]
{\textsf{
300$\times$300 2D hexagonal lattice with random wiring confined within
20$\times$20 local zones, but one wire freed,
rule $v3k7$ ``g3'' (from figure~\ref{Pulsing dynamics g3 rule}).\\
(Top) Pulsing measures: (f2) input-entropy/time plot, (g) entropy-density scatter plot, 
(h) density return map scatter plot,
with a strong similarity to the RW-waveform.
(Bottom) Pulsing patterns at the extremes of input-entropy, with
cells colored according to lookup instead of value.
}}
\label{one remote wire, 300x300 stp}
\end{figure}
\vspace{-2ex}

\section{3D systems}
\label{3D systems}
\enlargethispage{5ex}

\begin{figure}[htb]
\begin{minipage}[c]{.17\linewidth}
\includegraphics[width=1\linewidth,bb=189 103 301 215,clip=]{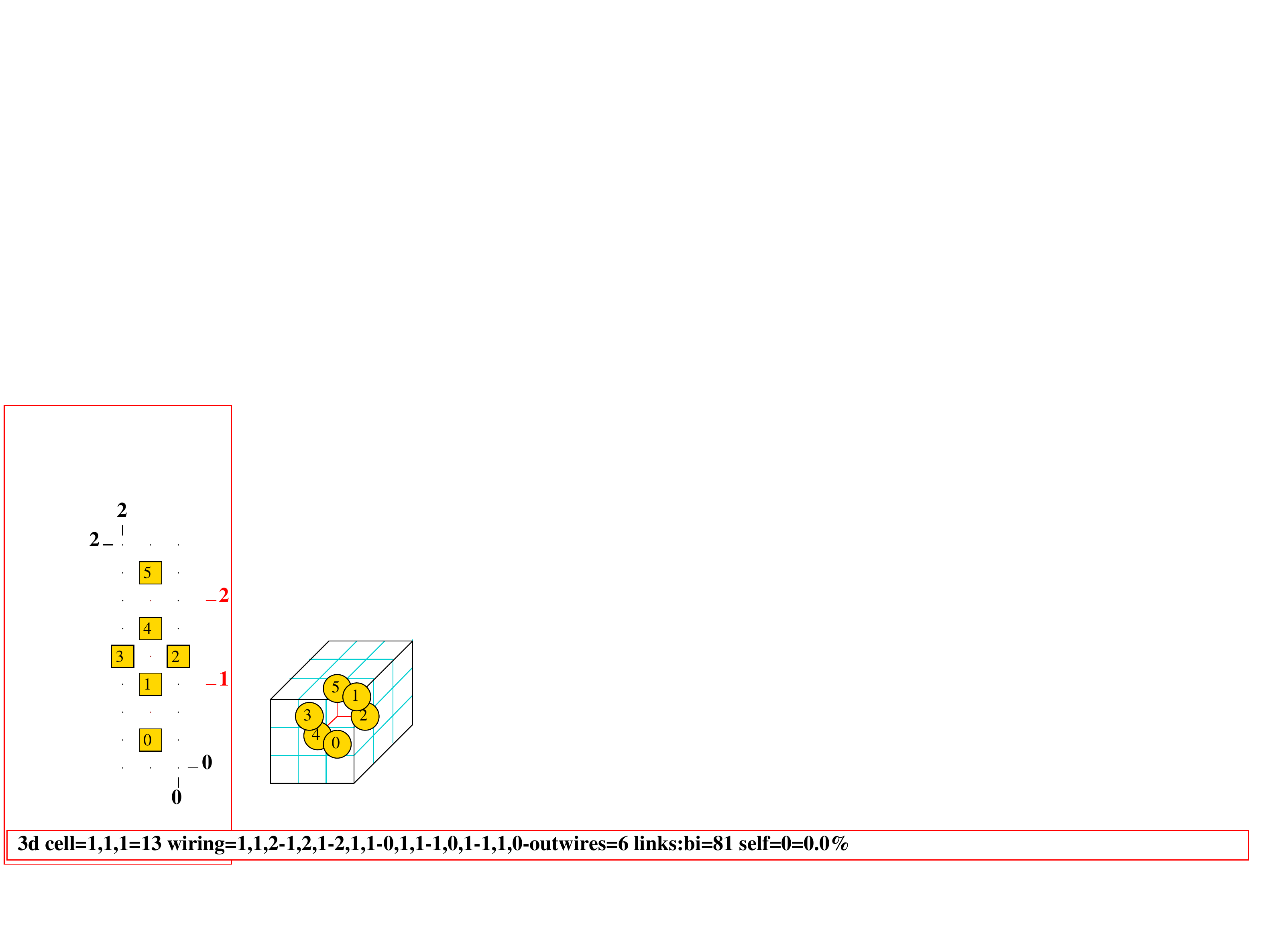}\\
\includegraphics[width=1\linewidth,bb=189 103 301 215,clip=]{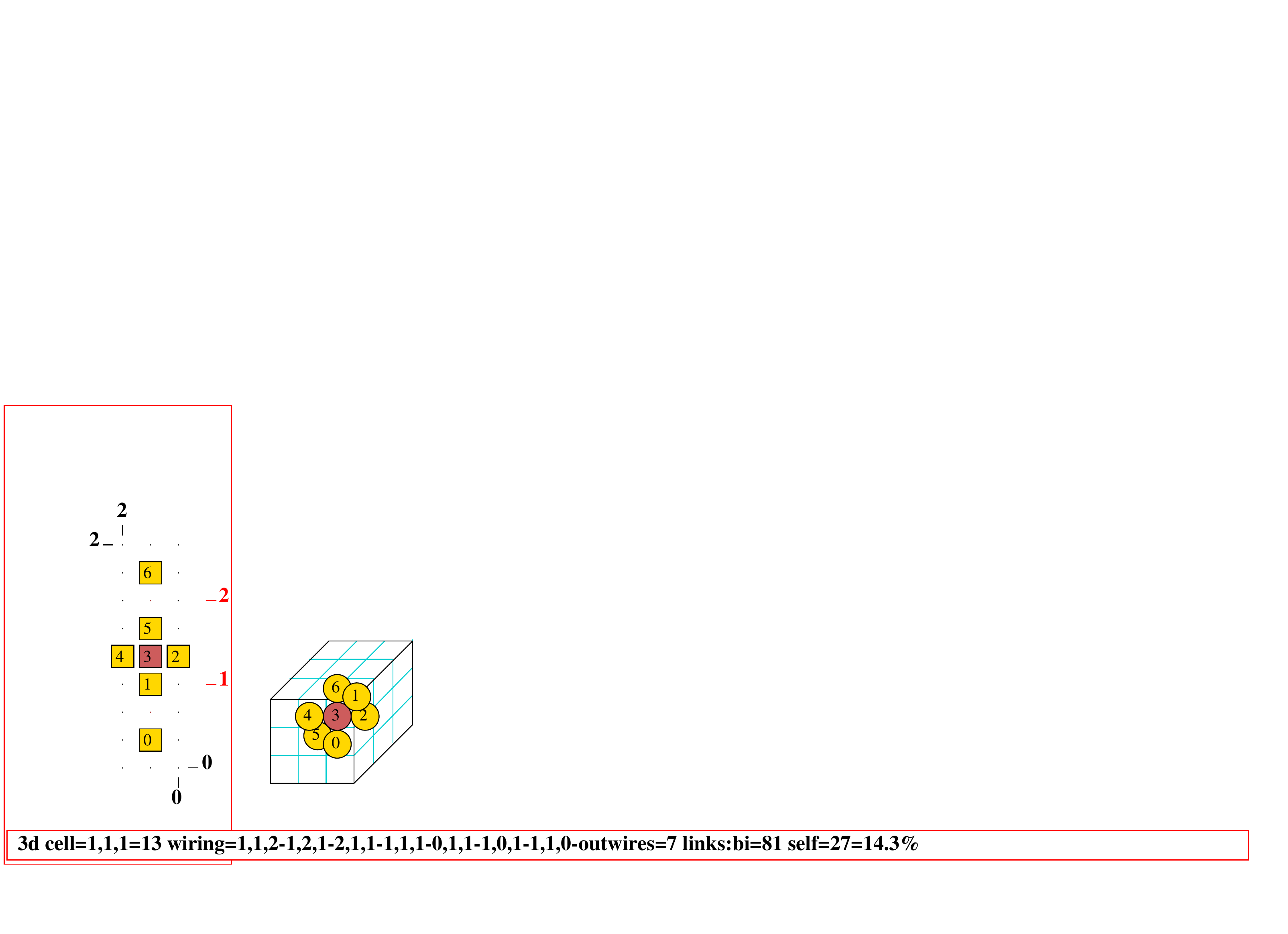}\\
(a)
\end{minipage}
\hfill
\begin{minipage}[c]{.35\linewidth}
\includegraphics[width=1\linewidth,bb=155 104 625 574,clip=]{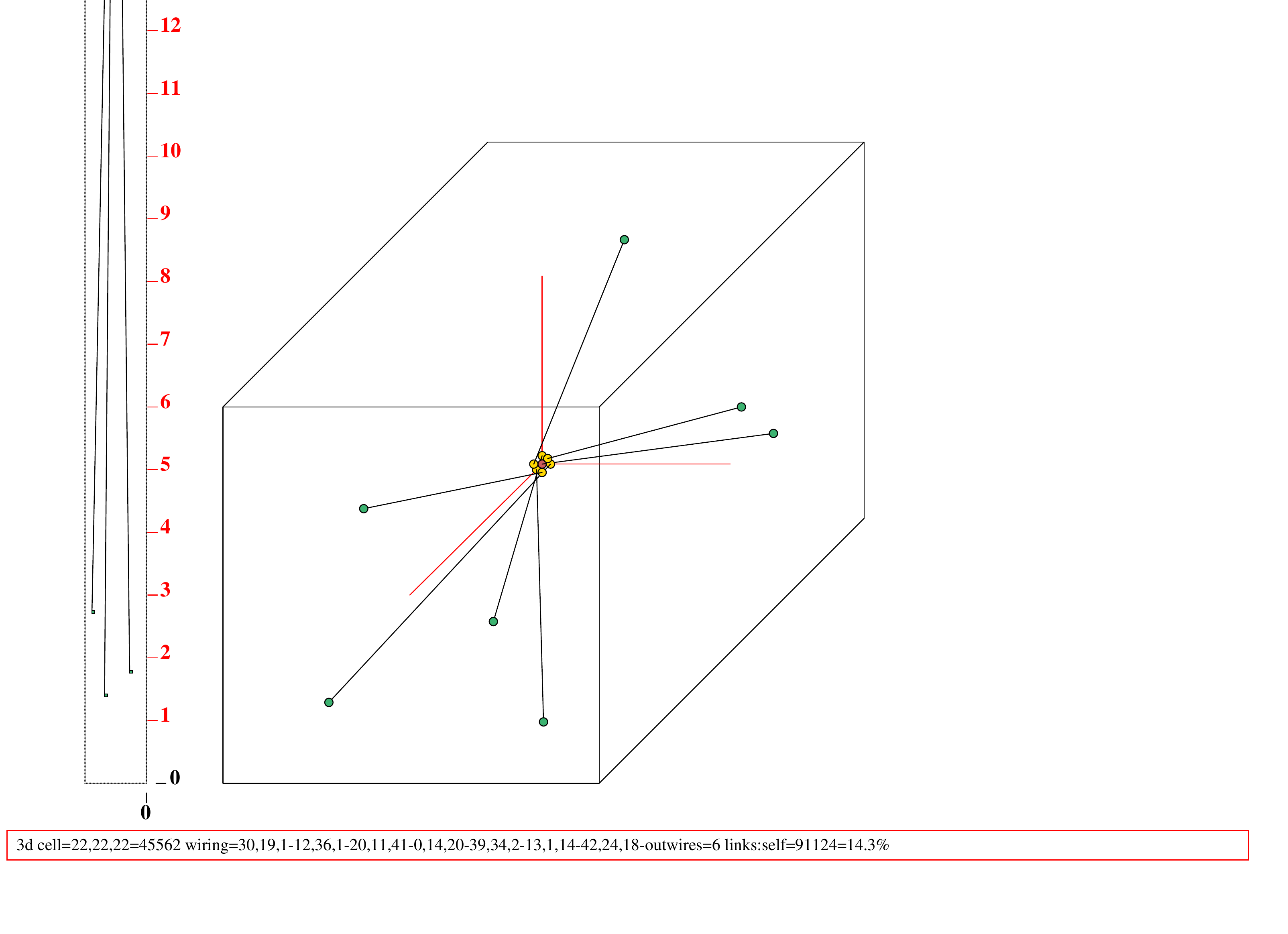}\\
(b)
\end{minipage}
\hfill
\begin{minipage}[c]{.35\linewidth}
\includegraphics[width=1\linewidth,bb=155 104 625 574,clip=]{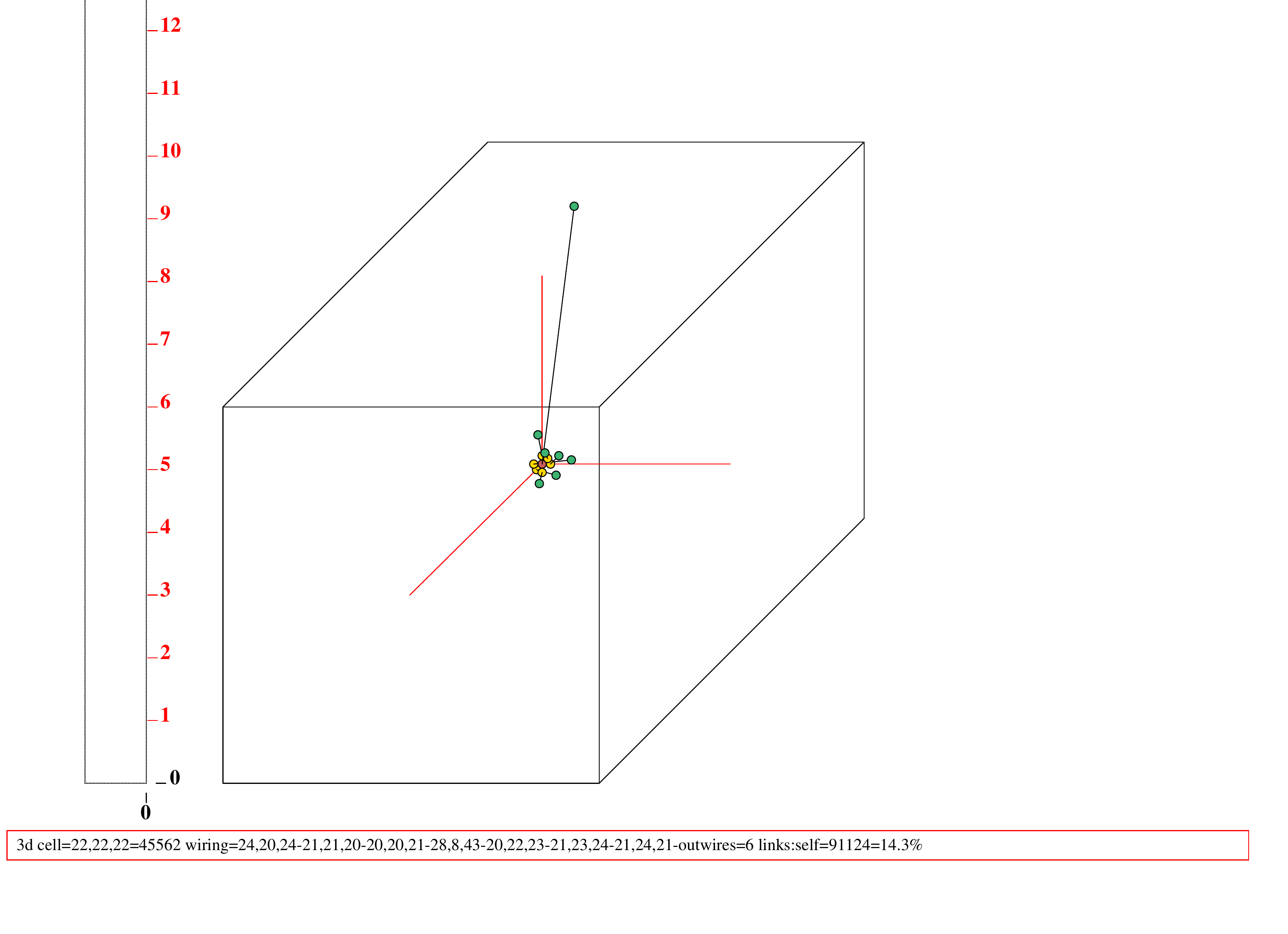}\\
(c)
\end{minipage}
\vspace{-2ex}
\caption[3D wiring]
{\textsf{
(a) 3D neighborhoods, $k$=6 and $k$=7. (b) 3D  45$\times$45$\times$45 lattice
with unrestrained random wiring. (c) Random wiring restrained in a 5$\times$5$\times$5
local zone, but with one wire freed.
}}
\label{3D wiring}
\end{figure}
\clearpage

\begin{figure}[htb]
\begin{center}
\textsf{\small
\begin{minipage}[c]{.95\linewidth}
\begin{minipage}[c]{.12\linewidth}
\fbox{\includegraphics[width=1\linewidth]{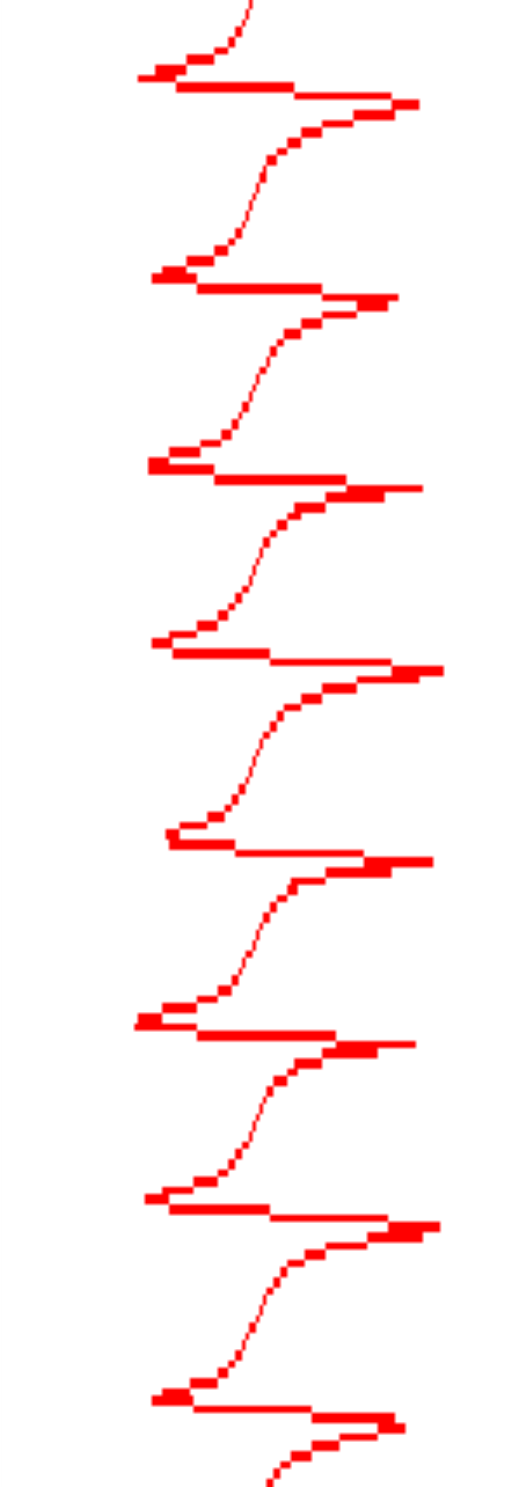}}\\
(f2)~$wl$$\approx$21
\end{minipage}
\hfill
\begin{minipage}[c]{.35\linewidth}
\includegraphics[width=1\linewidth]{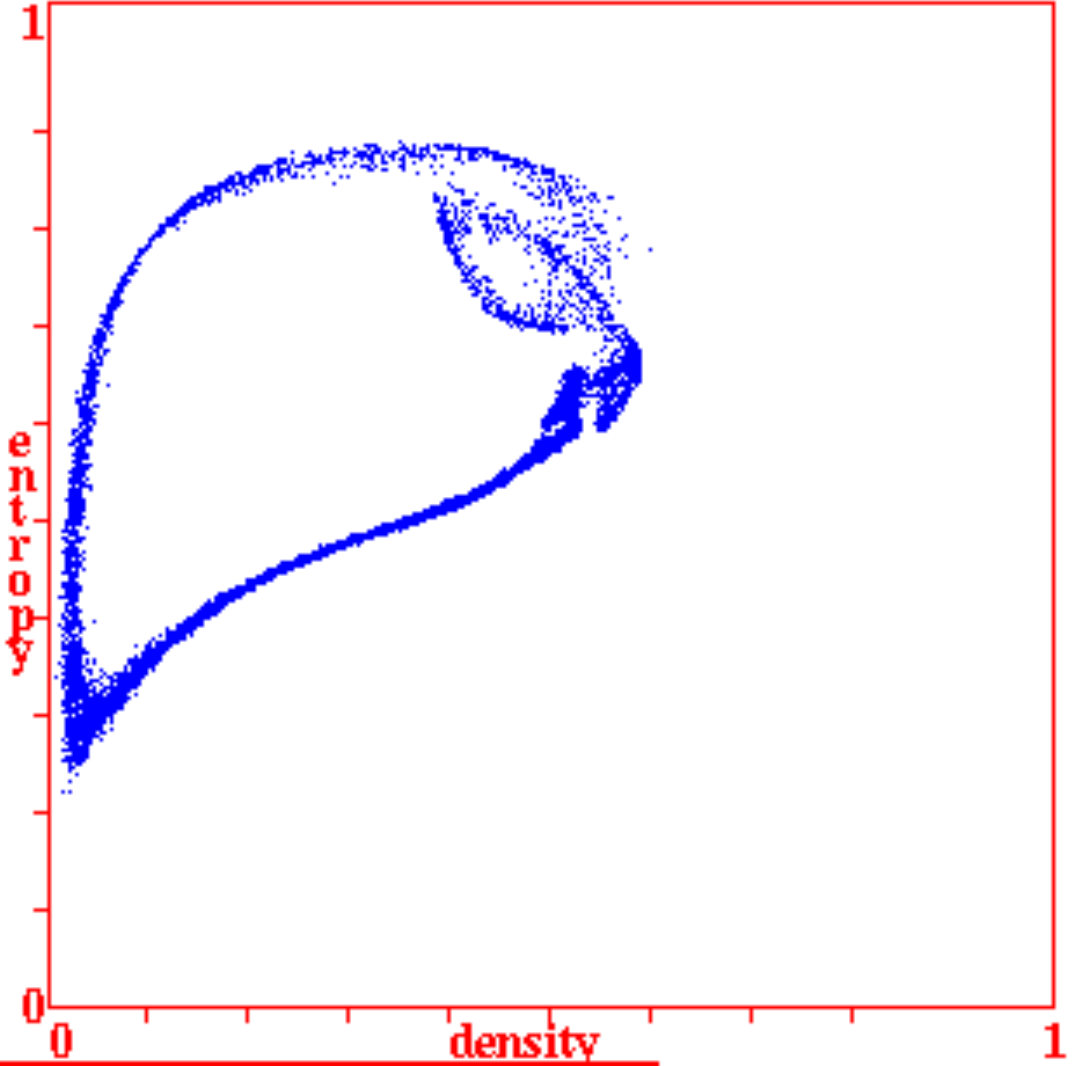}\\
(g)
\end{minipage}
\hfill
\begin{minipage}[c]{.35\linewidth}
\includegraphics[width=1\linewidth]{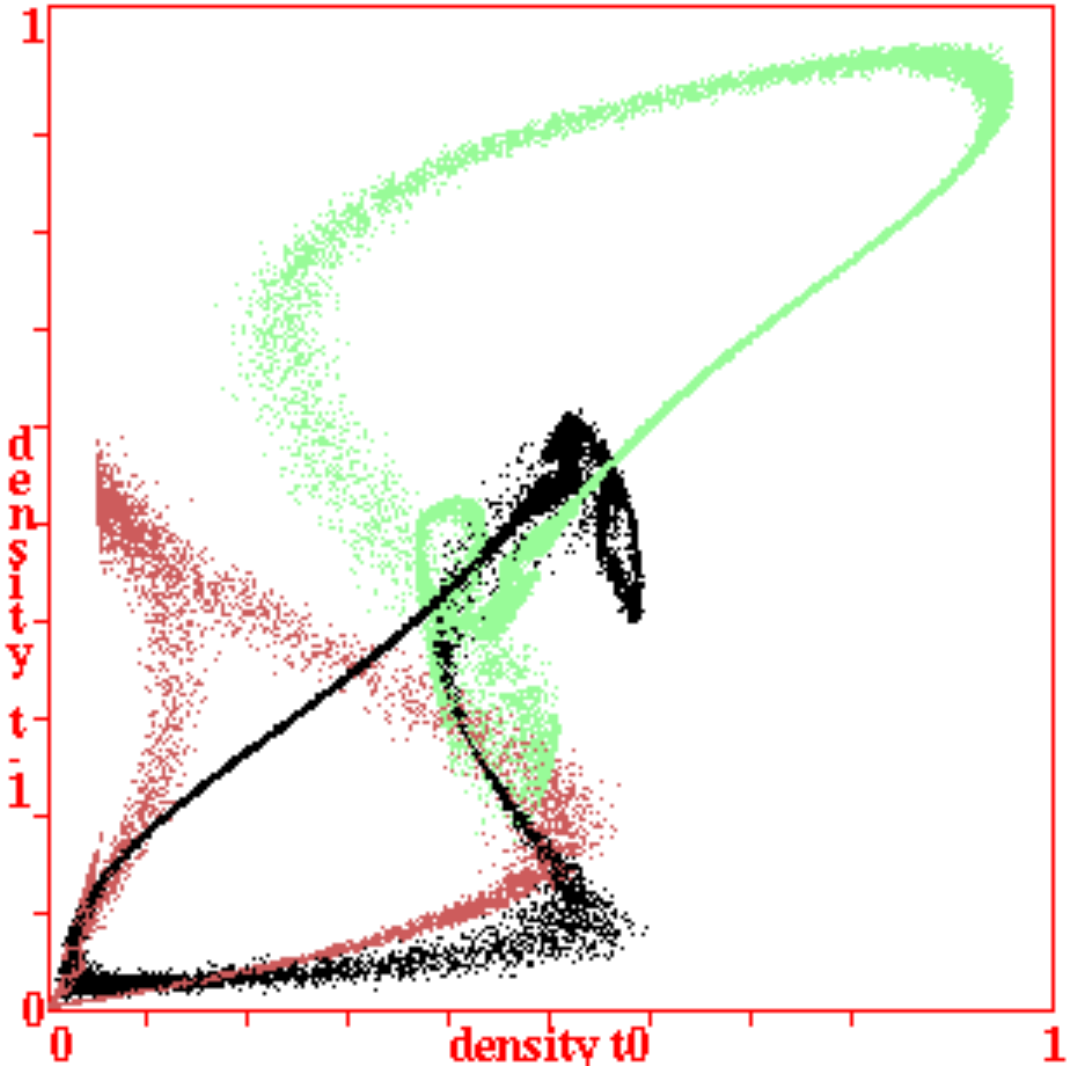}\\
(h)
\end{minipage}
\end{minipage}\\[2ex]
\begin{minipage}[c]{.95\linewidth}
\begin{minipage}[c]{.12\linewidth}
\fbox{\includegraphics[width=1\linewidth]{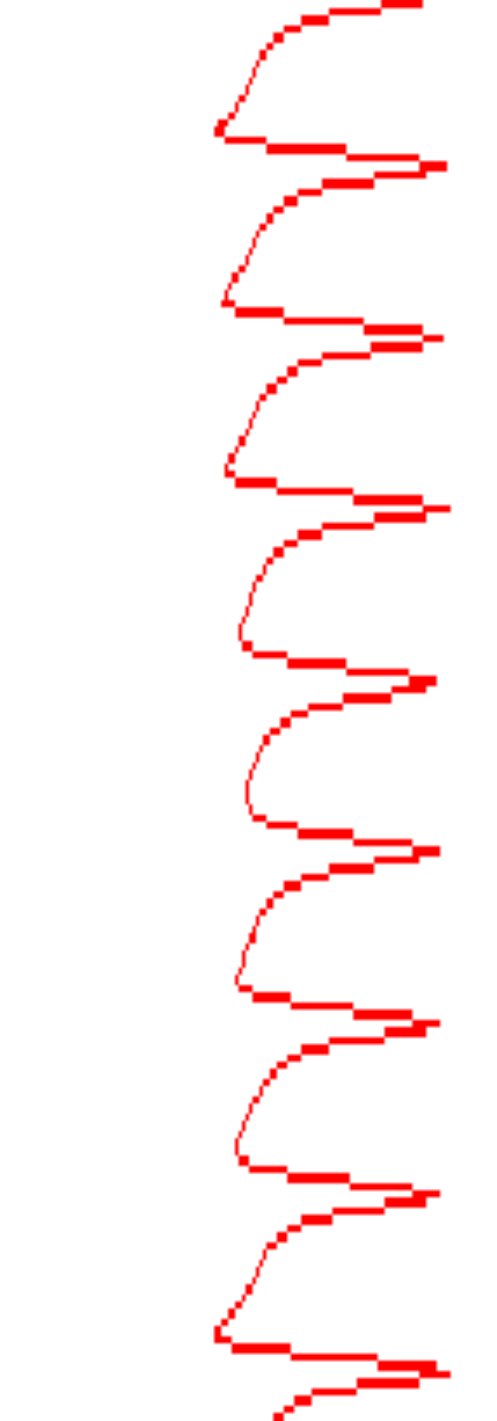}}\\
(f2)~$wl$$\approx$20
\end{minipage}
\hfill
\begin{minipage}[c]{.35\linewidth}
\includegraphics[width=1\linewidth]{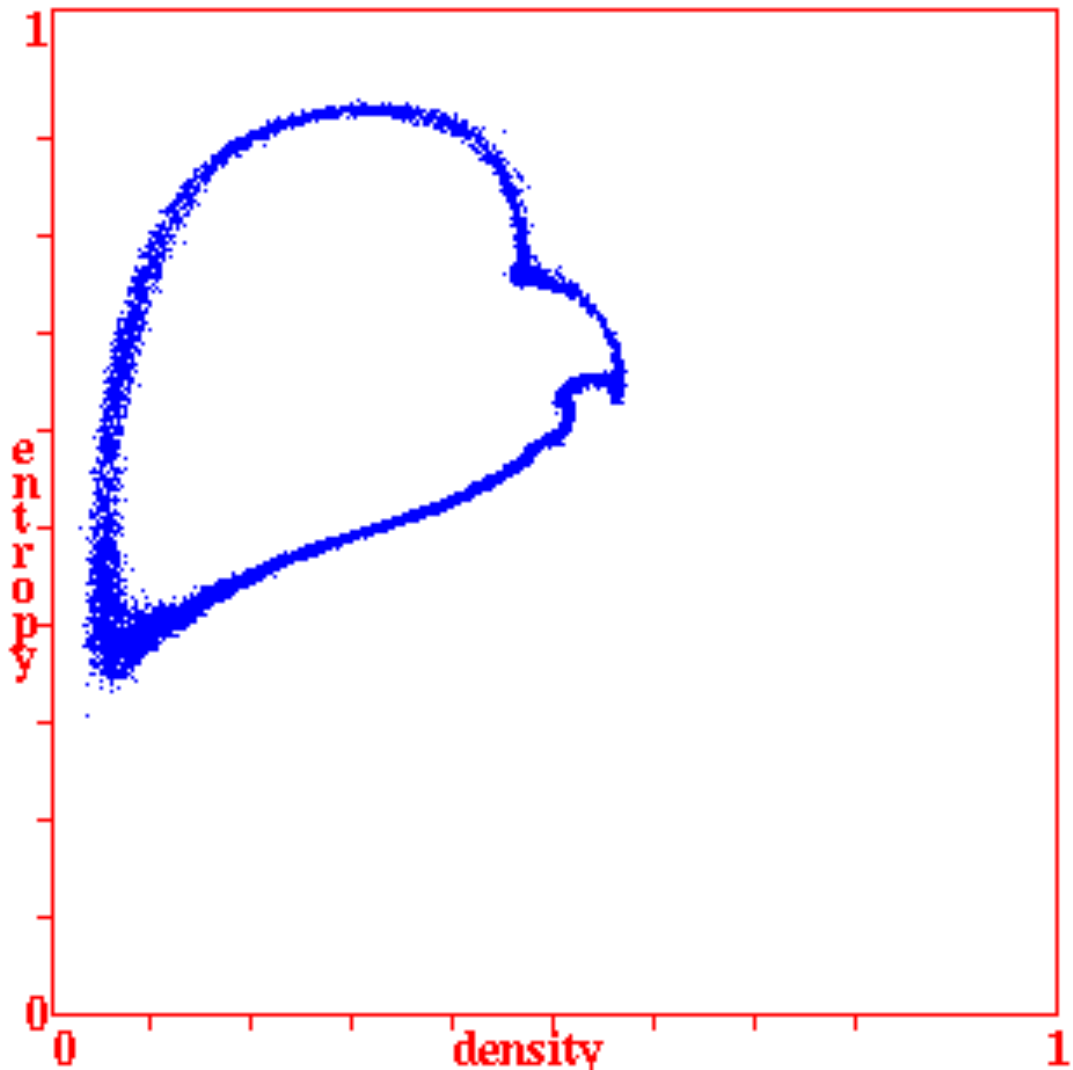}\\
(g)
\end{minipage}
\hfill
\begin{minipage}[c]{.35\linewidth}
\includegraphics[width=1\linewidth]{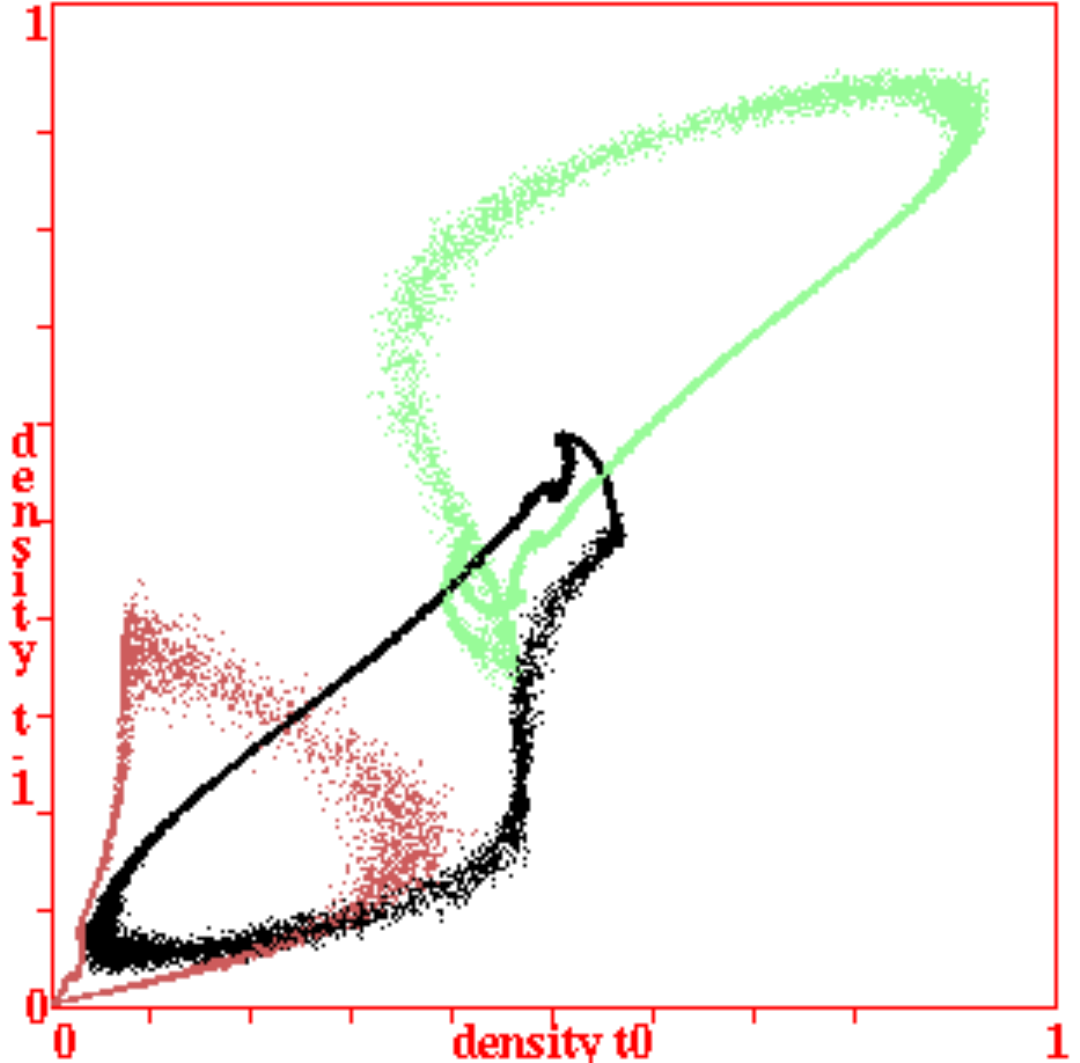}\\
(h)
\end{minipage}
\end{minipage}
}
\end{center}
\vspace{-2ex}
\caption[3D time-plots k7 g3]
{\textsf{
45$\times$45$\times$45 3D lattice with random wiring,
rule $v3k7$ ``g3'' ---  compare with the RW-waveform in figure~\ref{Pulsing dynamics g3 rule}.
Pulsing measures: (f2) input-entropy/time plot, (g) entropy-density scatter plot, 
(h) density return map scatter plot. 
(Top) Unconstrained random wiring gives the same RW-waveform.
(Bottom) Confined within 5$\times$5$\times$5 local zones but with one wire freed,
the waveform is similar to the RW-waveform.
}}
\label{3D time-plots k7 g3}
\vspace{-3ex}
\end{figure}

\noindent 3D $k$=6 and $k$=7 CA (figure~\ref{3D wiring})
were implemented in a 3D cubic lattice 45$\times$45$\times$45 with periodic boundaries.
We found that all the rules in the case study (section~\ref{Pulsing case studies}) exhibit
3D glider behaviour, and fully random wiring gave the same pulsing waveforms as 2D  
(figure~\ref{3D time-plots k7 g3} Top) --- not surprisingly because the shape
of the neighborhood is not significant for an isotropic rule.

When random wiring is confined within 5$\times$5$\times$5 local zones, pulsing is still evident
with strange attractor signatures deformed (figure~\ref{3D g3 45x45, 5x5, scatter plots}). 
However, when one wire is freed the signatures revert closer to RW-waveforms
(figure~\ref{3D time-plots k7 g3} Bottom).

\begin{figure}[htb]
\begin{center}
\begin{minipage}[c]{.95\linewidth} 
\begin{minipage}[c]{.12\linewidth}
\fbox{\includegraphics[width=1\linewidth]{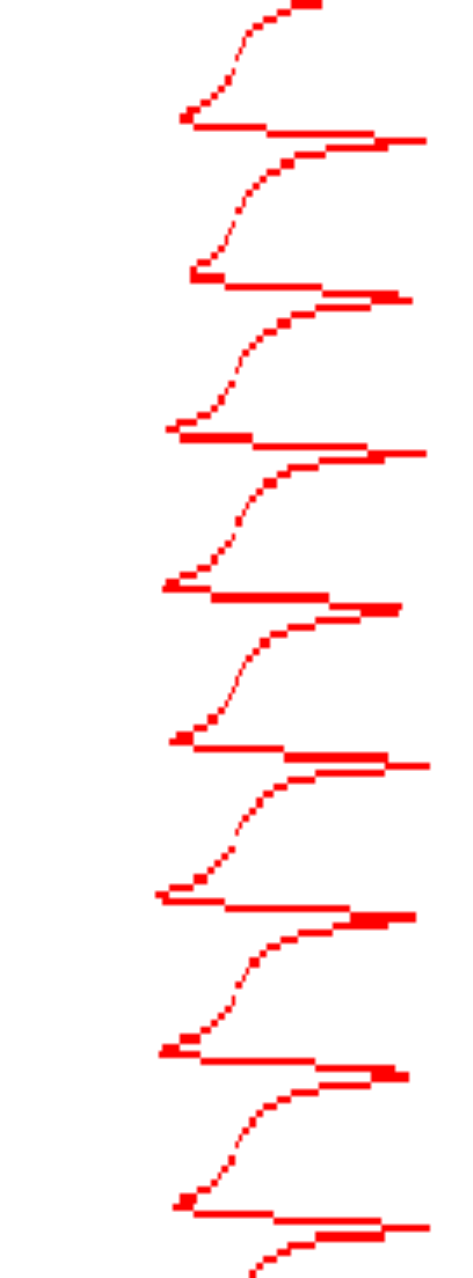}}\\[1ex]
(f2)~$wl$$\approx$21
\end{minipage}
\hfill
\begin{minipage}[c]{.35\linewidth}
\includegraphics[width=1\linewidth]{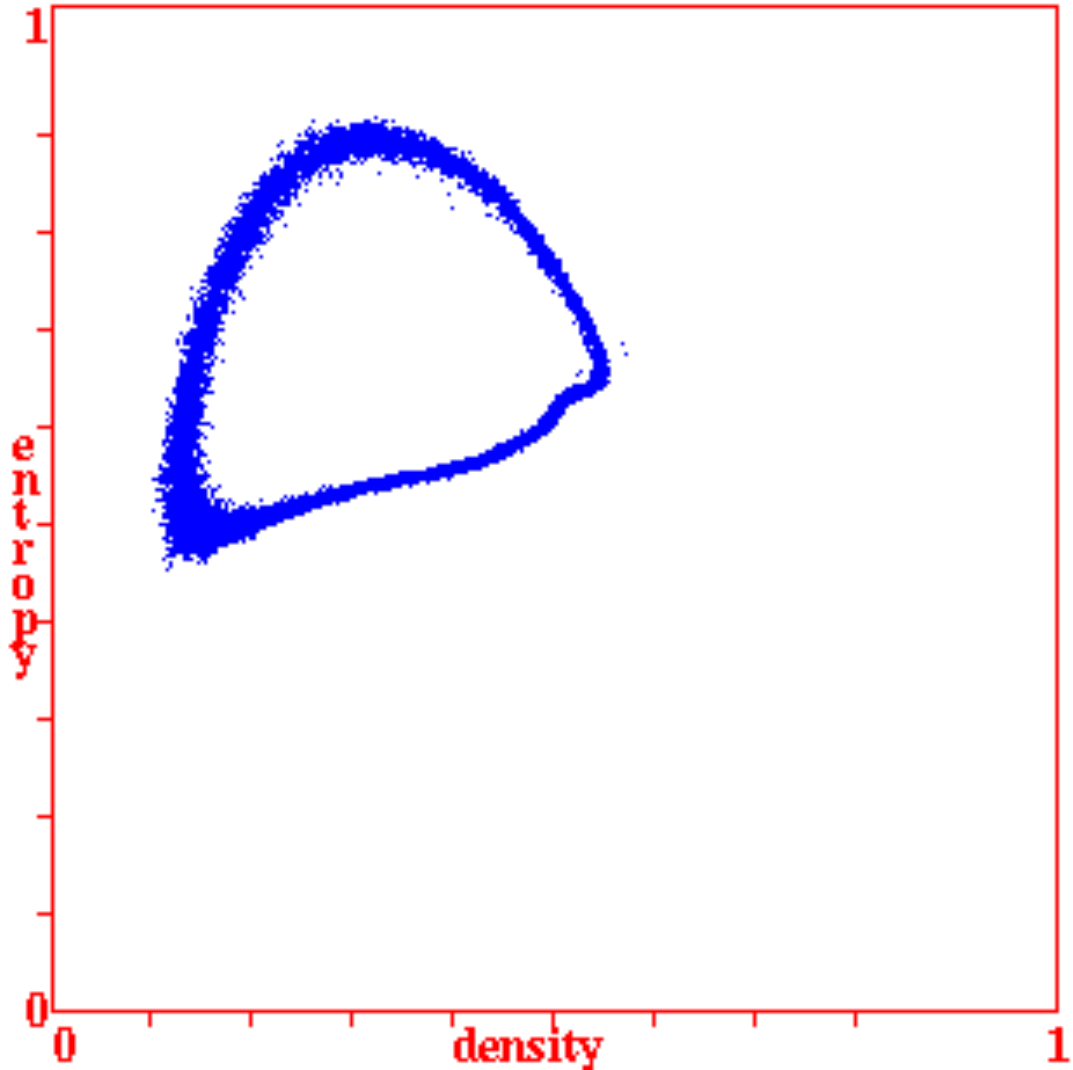}\\
(g) $v3k7$ ``g3''cp -p 
\end{minipage}
\hfill
\begin{minipage}[c]{.35\linewidth}
\includegraphics[width=1\linewidth]{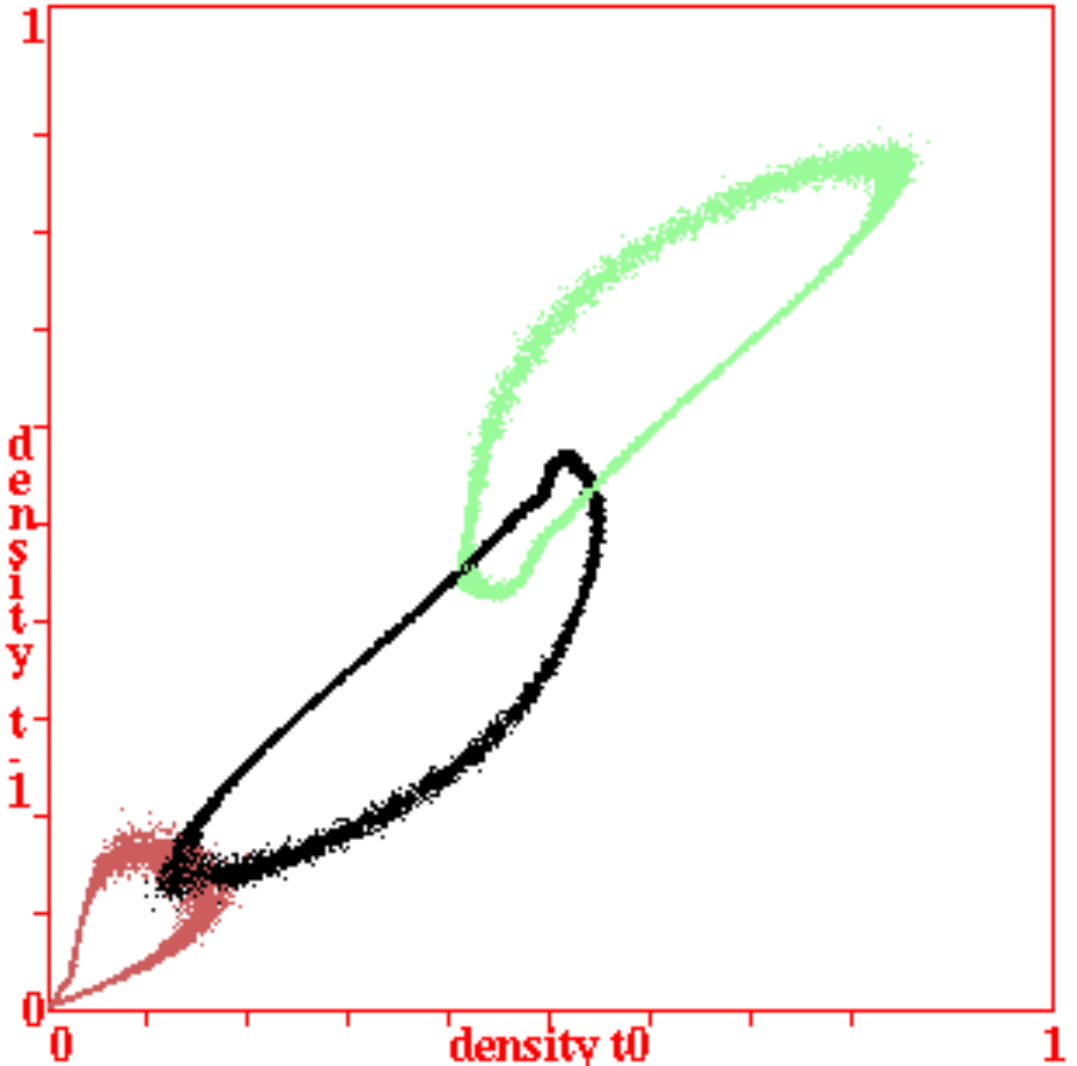}\\
(h)
\end{minipage}
\end{minipage}\\[2ex]
\begin{minipage}[c]{.8\linewidth}
\begin{minipage}[c]{.43\linewidth}
\includegraphics[width=1\linewidth]{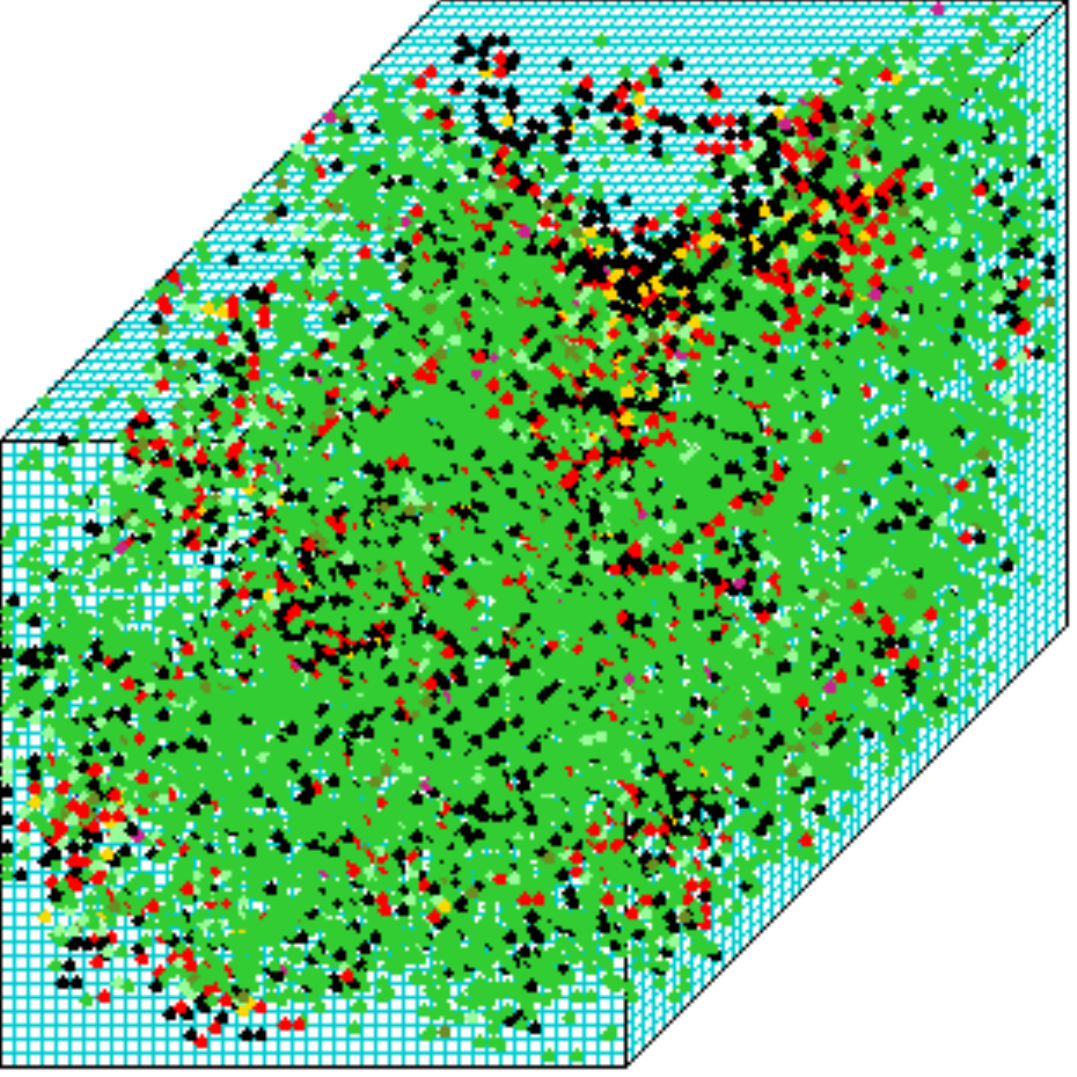}\\
$v3k7$ ``g3-min''
\end{minipage}
\hfill
\begin{minipage}[c]{.43\linewidth}
\includegraphics[width=1\linewidth]{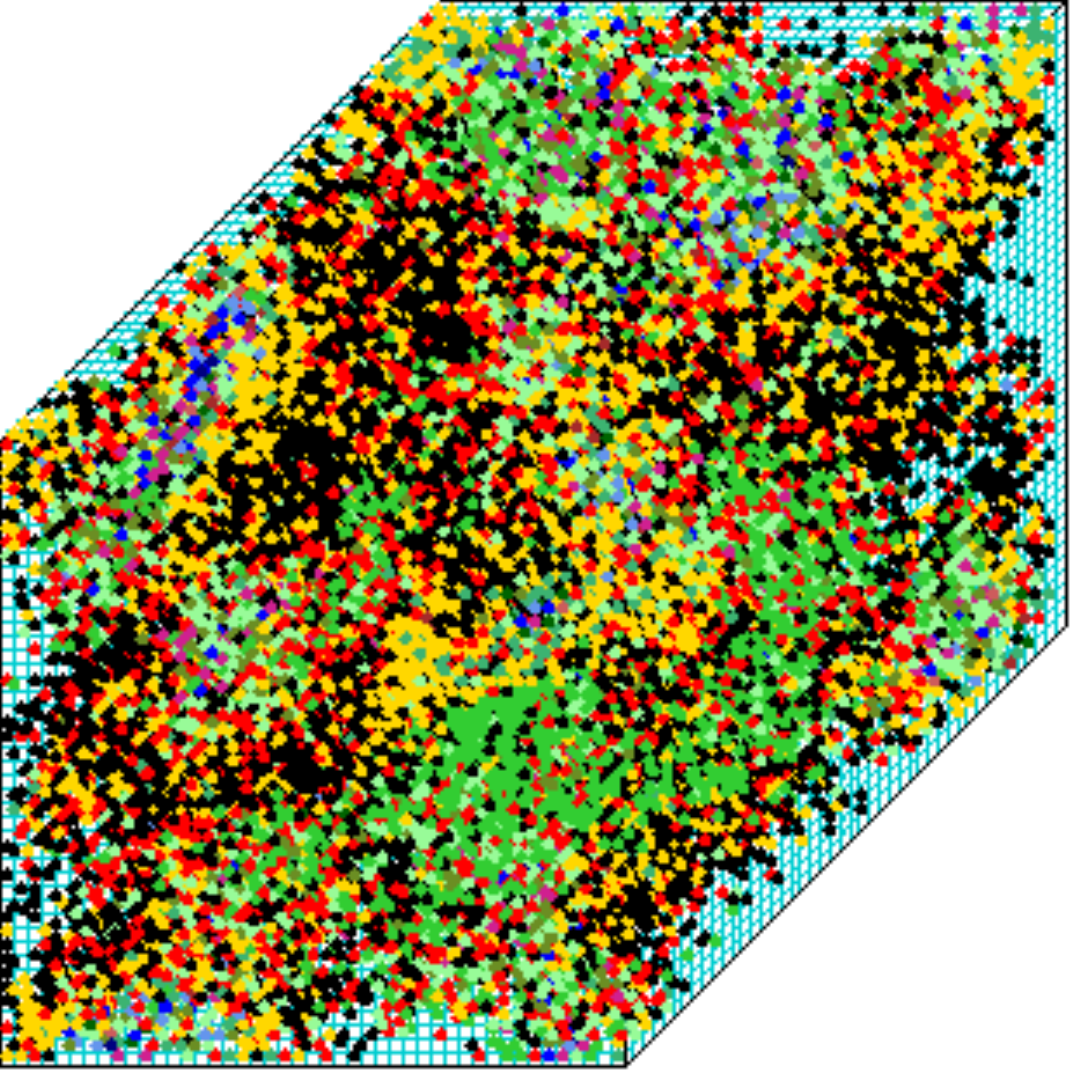}\\
$v3k7$ ``g3-max''
\end{minipage}
\end{minipage}
\end{center}
\vspace{-2ex}
\caption[3D g3 45x45, 5x5, scatter plots]
{\textsf{
45$\times$45$\times$45 3D lattice with random wiring confined within
5$\times$5$\times$5 local zones, rule $v3k7$
``g3'' (compare with figures~\ref{3D time-plots k7 g3} and \ref{Pulsing dynamics g3 rule}).
(Top) Pulsing measures (f2) input-entropy/time plot, (g)
entropy-density scatter plot, (h) density return map scatter plot ---
distorted compared to the RW-waveform.  (Bottom) Pulsing patterns at
the extremes of input-entropy showing patchy
waves of density --- cells colored according to lookup
instead of value.
}}
\label{3D g3 45x45, 5x5, scatter plots}
\end{figure}

\section{k-totalistic rules as reaction-diffusion systems}
\label{k-totalistic rules as reaction-diffusion systems}

\noindent An explanation of glider dynamics in k-totalistic rules
can be based on Adamatsky's reinterpretation of the
$k$=6 Beehive rule\cite{Adamatzky&Wuensche&Cosello2006}, and
the $k$=7 Spiral rule\cite{Wuensche&Adamatzky2006,Adamatzky&Wuensche2006},
as discrete models of reaction-diffusion systems with
inhibitor/activator reagents in a chemical medium.  The three CA
values are seen as: A=1 (Activator), I=2 (Inhibitor), and S=0
(Substrate). The three reagents perform a sort of non-linear feedback
dance, suppressing and catalysing each other at critical
concentrations. The analysis accounts for the movement of a glider's
head and following tail, but could also apply to the randomly wired
system seen as a neural network with three states: 1=(Activator, Firing),
2=(Refractory), 0=(Ready to Fire).  When wiring is randomized, it seems that
feedback becomes distributed, giving global pulsing instead of driving a
glider.

\enlargethispage{2ex}
In glider CA, gliders and their interactions quickly dominate the
dynamics, and thus the frequency of neighborhood lookup in the
rule-table.  The neighborhoods responsible for the background
``domain'' are the most frequent, followed by neighborhoods that drive
gliders, other (stable) structures, and those invloved in
collisions. The remaining neighborhoods rarely appear in an evolved
system and can be regarded as wild-cards in the rule-table ---
mutations of these have little or no effect\cite{beehivewebpage,spiralwebpage}.

\begin{figure}[h]
\textsf{\small
\begin{center}
\begin{minipage}[c]{.5\linewidth}
\begin{minipage}[c]{.45\linewidth}
\includegraphics[width=1\linewidth]{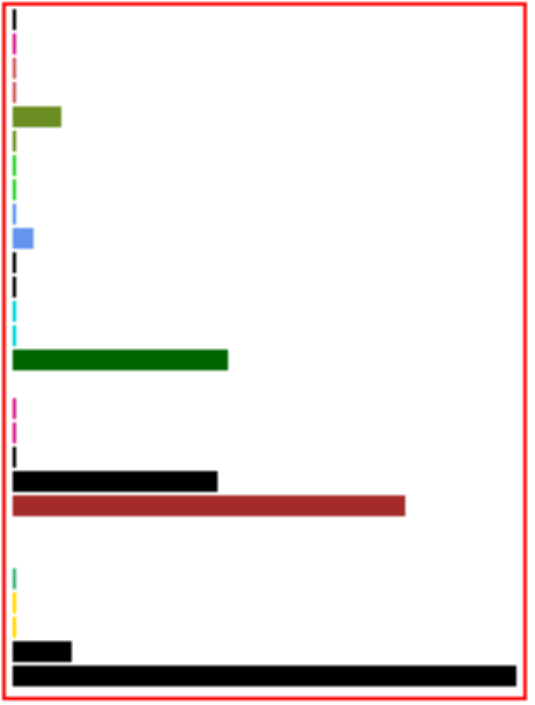}\\[-1ex]
k6g2-CA Beehive
\end{minipage}
\hfill
\begin{minipage}[c]{.45\linewidth}
\includegraphics[width=1\linewidth]{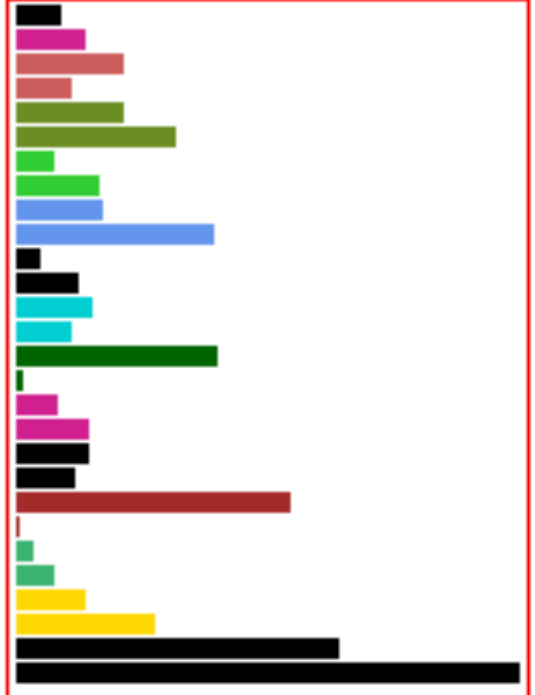}\\[-1ex]
k6g2-RW Beehive
\end{minipage}\\[2ex]
\begin{minipage}[c]{.45\linewidth}
\includegraphics[width=1\linewidth]{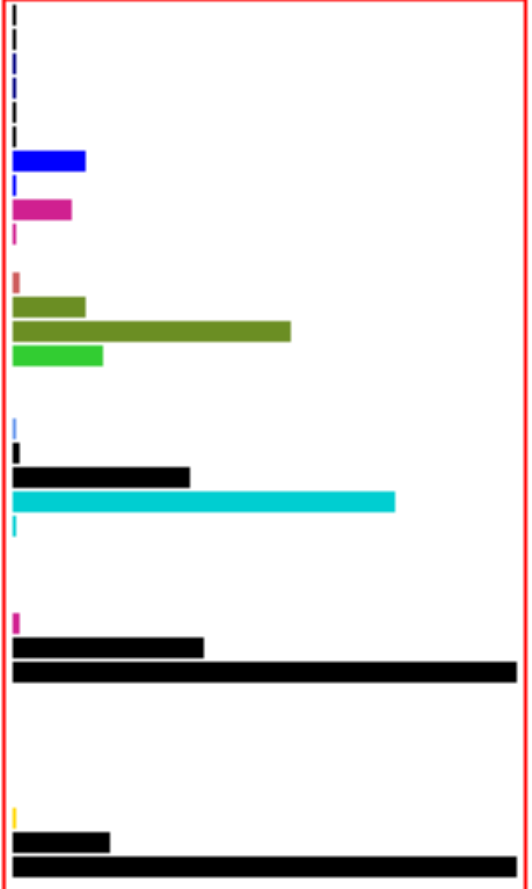}\\[-1ex]
k7g1-CA Spiral
\end{minipage}
\hfill
\begin{minipage}[c]{.45\linewidth}
\includegraphics[width=1\linewidth]{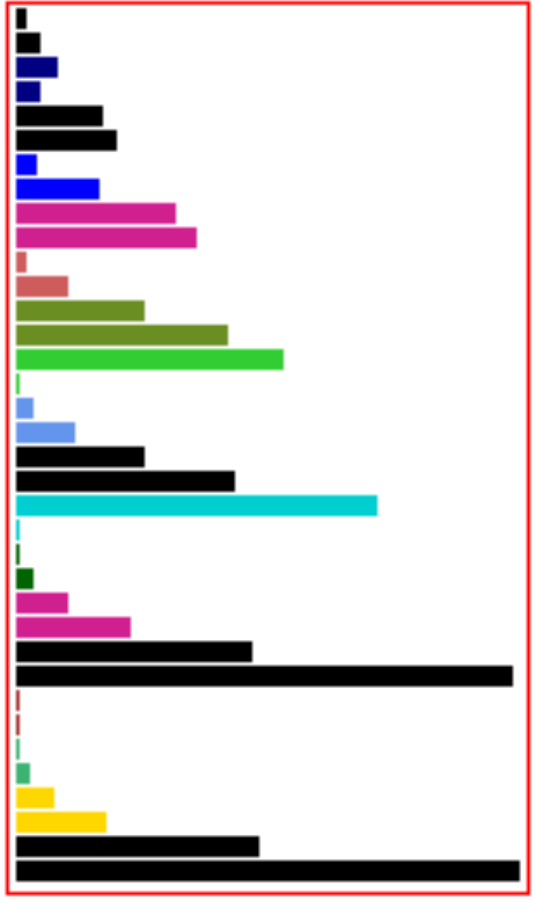}\\[-1ex]
k7g1-RW Spiral
\end{minipage}\\
\end{minipage}
\end{center}}
\vspace{-3ex}
\caption[lookup-histograms k6 k7 CA/RW]
{\textsf{
Lookup-histograms averaged over 100 time-steps,
for $k$6 g2 Beehive, and $k$7 g1 Spiral rules, for a 100$\times$100 lattice.
(Left) CA, (Right) Random Wiring, showing a correlation in neighborhooh frequency.
}}
\label{lookup-histograms k6 k7 CA/RW}
\end{figure}

This is captured by the lookup-frequency histograms
(figure~\ref{lookup-histograms k6 k7 CA/RW}) for the Beehive and
Spiral rules\cite{Wuensche05,Wuensche&Adamatzky2006},
averaged over 100 time-steps, where the CA histogram
highlights the background domain and glider dynamics, and the wild-cards
by gaps or reduced values.  The histogram for random wiring has a less
pronounced distribution, but there is a significant correlation with
the CA histogram, showing that the feedback between the three values
is at play globally.  Histograms for the other rules studied confirm
these results.

\section{Asynchronous and noisy updating}
\label{Asynchronous and noisy updating}

\begin{figure}[htb]
\begin{center}
\textsf{\small
\begin{minipage}[c]{.95\linewidth} 
\begin{minipage}[c]{.13\linewidth}
\fbox{\includegraphics[width=.8\linewidth]{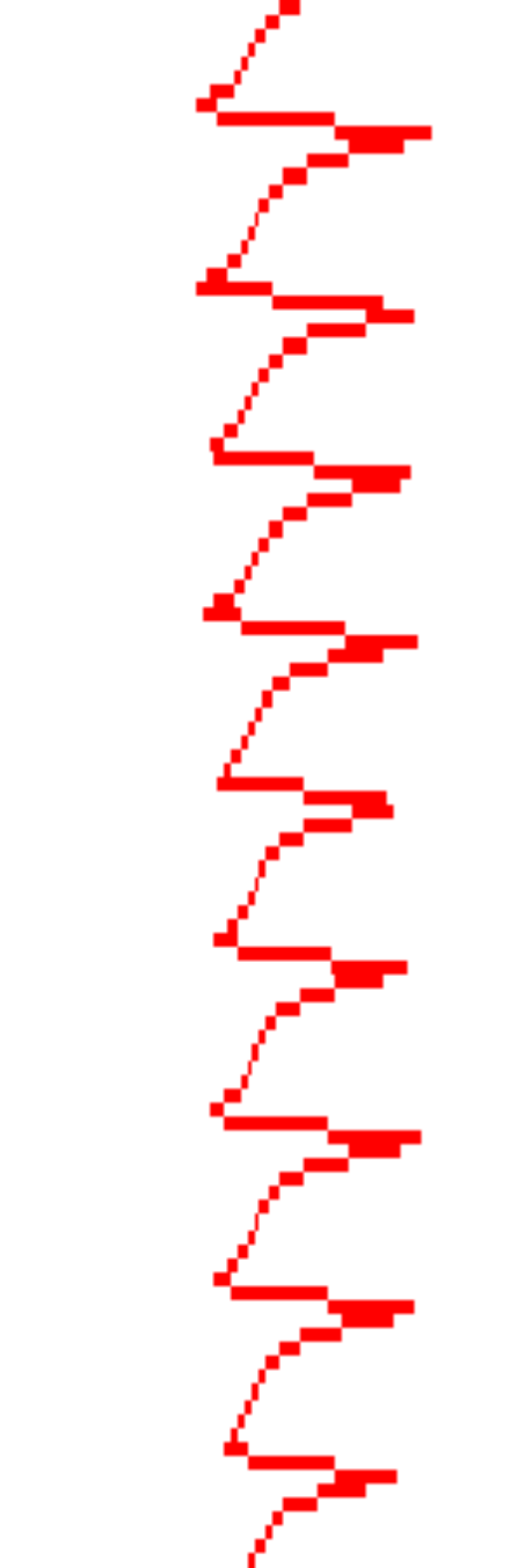}}\\[1ex]
(f2)
\end{minipage}
\hfill
\begin{minipage}[c]{.35\linewidth}
\includegraphics[width=1\linewidth]{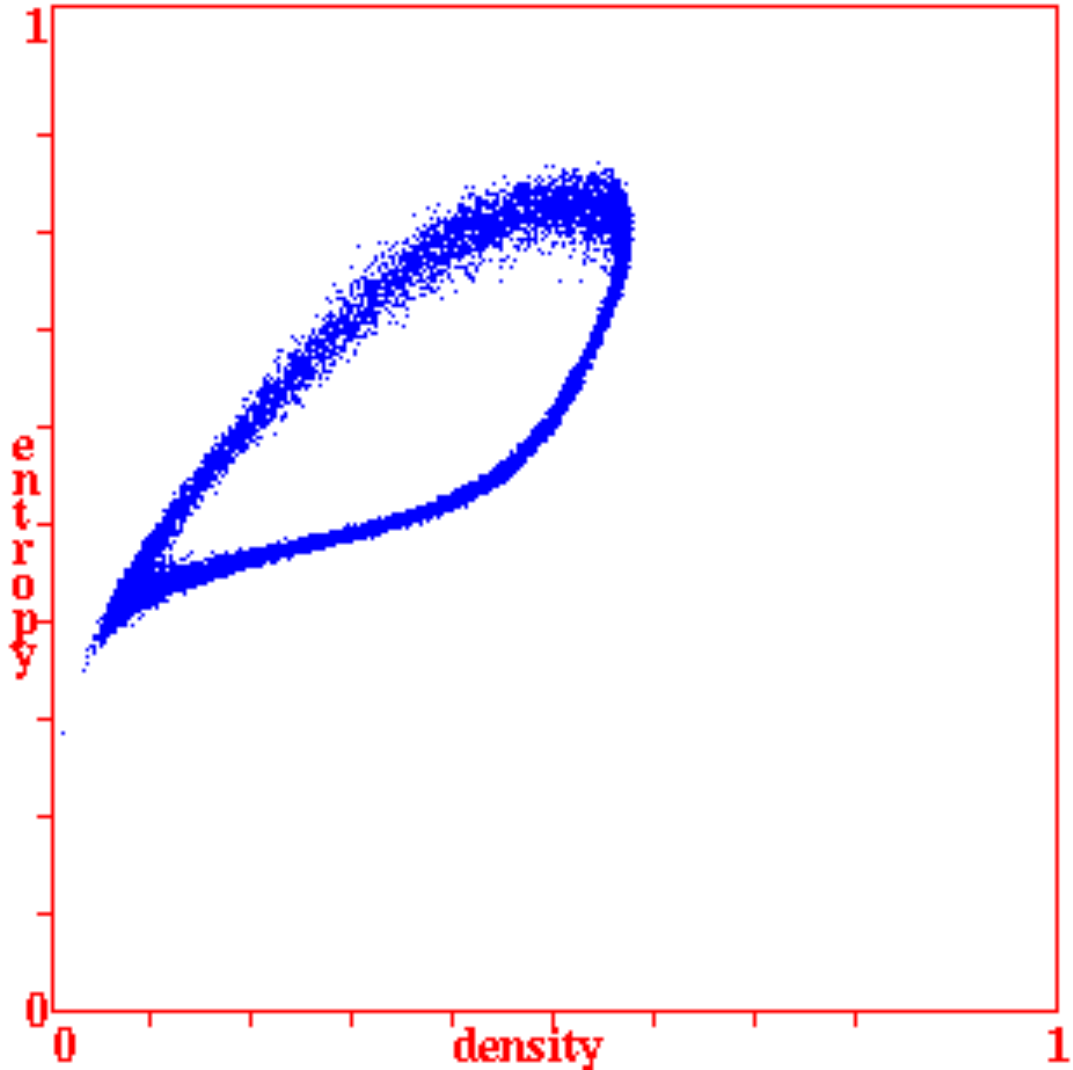}\\
(g)
\end{minipage}
\hfill
\begin{minipage}[c]{.35\linewidth}
\includegraphics[width=1\linewidth]{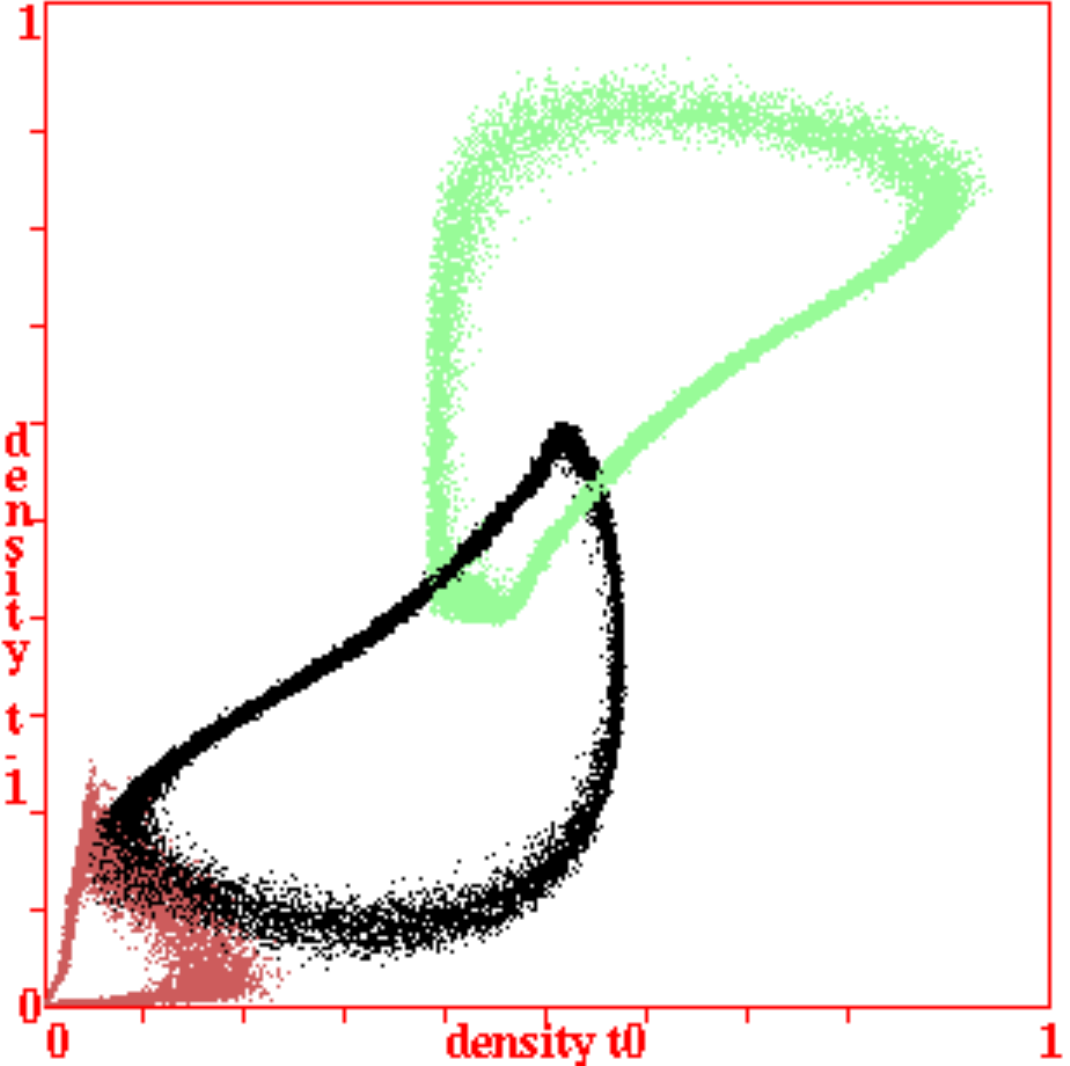}\\
\textsf{(h)}
\end{minipage}
\end{minipage}
}
\end{center}
\vspace{-1ex}
\textsf{\small
\noindent Time-plots of measures.
(f2) input-entropy oscillations with time (y-axis, stretched).
(g) entropy-density scatter plot -- input-entropy (x-axis)
against the non-zero density (y-axis). (h) density return map scatter plot.}
\vspace{-2ex}
\caption[Pulsing dynamics g3]
{\textsf{
Sequential updating in a random order, re-randomised at each time-step, showing
pulsing measures taken at time-step intervals, which can be compared with
figure~\ref{Pulsing dynamics g3 rule}.  $v3k7$ ``g3'' rule, (hex) 622984288a08086a94,
on a 200$\times$200 hexagonal lattice, $wl$= 11 or 12 time-steps.
}}
\label{Pulsing dynamics g3 rule, sequential}
\vspace{2ex}
\end{figure}

\begin{figure}[htb]
\begin{center}
\textsf{\small
\begin{minipage}[c]{.75\linewidth} 
\begin{minipage}[c]{.45\linewidth}
\includegraphics[width=1\linewidth]{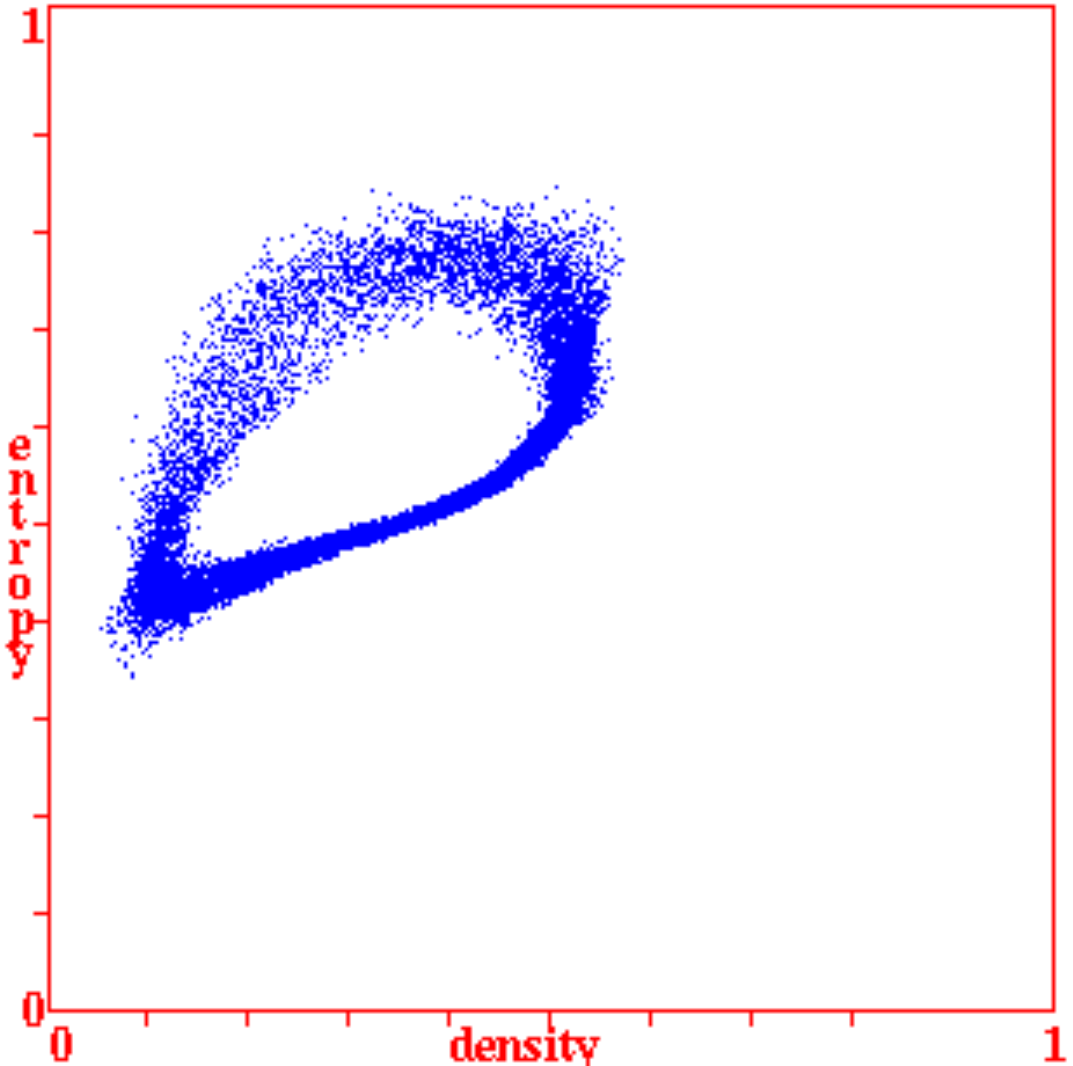}\\
(a) 200$\times$200, time-step update
\end{minipage}
\hfill
\begin{minipage}[c]{.45\linewidth}
\includegraphics[width=1\linewidth]{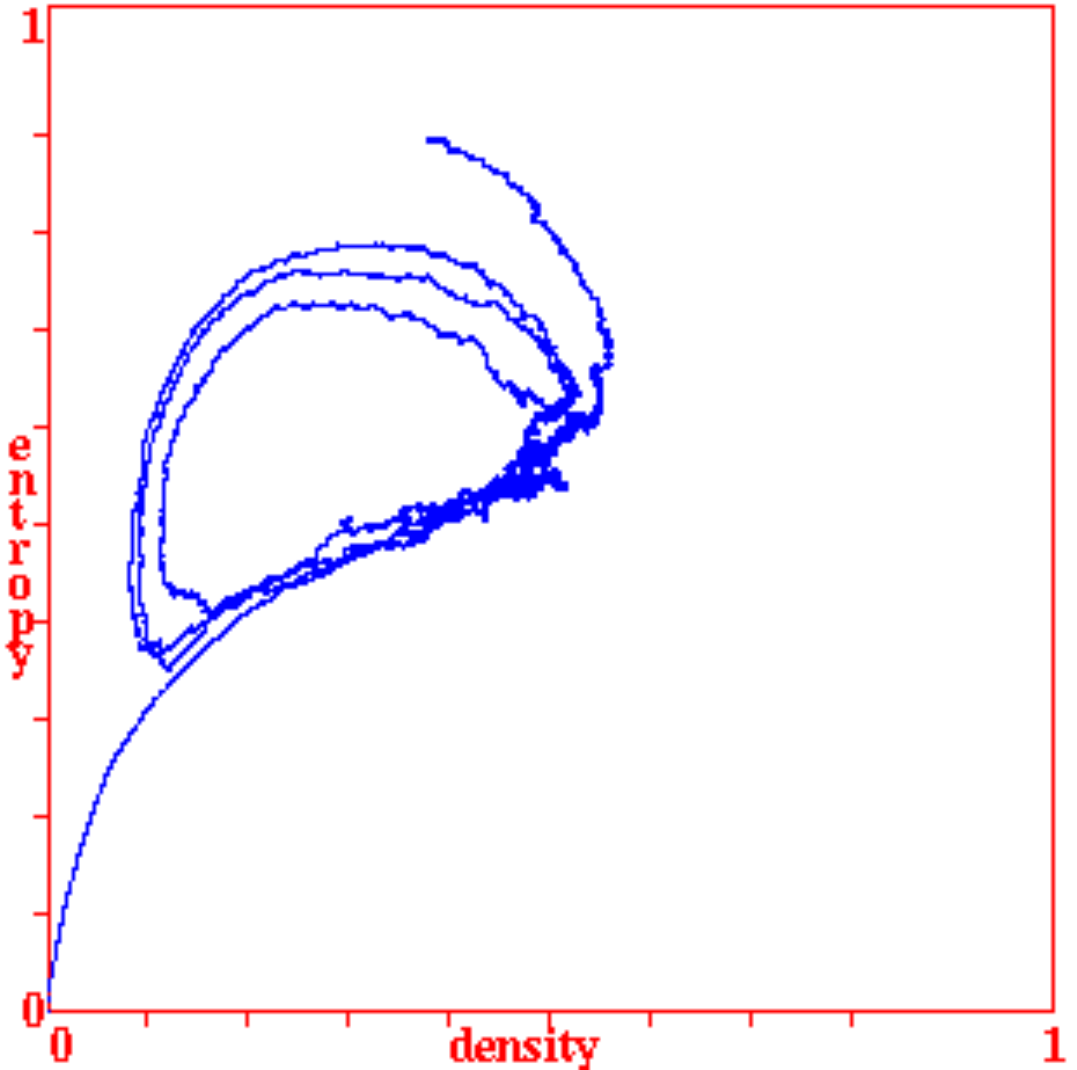}\\
(b) 50$\times$50, single-cell update
\end{minipage}
\end{minipage}
}
\end{center}
\vspace{-3ex}
\caption[\label{Pulsing dynamics g3 rule, partial order}]
{\textsf{
Sequential updating in a random order within partial order updating,
showing the entropy-density scatter plot.  $v3k7$ ``g3'' rule
with a similar waveform as in figure~\ref{Pulsing dynamics g3 rule, sequential}.
(a) Partial order limits: 1 to $n$, measures
taken at time-step intervals. (b)  Partial order limits: 1 to 1, measures
taken at each cell-update. Starting at a random initial state, pulsing 
completed only 3 cycles before measures fell to zero --- this is expected
because of the smaller network size, necessary because computation
for single-cell update is slow.
}}
\vspace{-2ex}
\label{Pulsing dynamics g3 rule, partial order}
\end{figure}

\noindent In the results so far, updating the next time-step has been
deterministic, and synchronous (in parallel) across the lattice ---
but what would be the effects of noise and asynchronicity? 
DDLab has a suite of options to introduce either or both
on-the-fly\cite{Wuensche2016}.
Two types of noise are implemented where each cell updates with a
given probability at each time-step  --- otherwise, in one
alternative the cell stays the same, and in the other its value is
assigned randomly.
For asynchronicity, the most flexible method is ``partial order''
updating where a subset of cells update (synchronously or sequentially), 
followed by the next subset --- then the ``state'' is the configuration after each
updated subset. Lower and upper limits predefine the size of each
subset between 1 and network size~$n$.  At each time-step, a random
size is set between these limits, and scattered randomly to positions
in the network --- only those are updated.
In sequential updating, each cell is updated in turn in some arbitrary
order --- then the ``state'' is the configuration when all $n$ updates
(or all cells in a partial order subset) are complete. For $n$ cells there are
$n!$ possible sequential updating orders, but the usual method is to
set a random order, re-randomised at each time-step.
Using these asynchronous and noisy updating methods, singly or in
combination, it appears that the CA pulsing model continues to
pulse whatever you throw at it.

Sequential updating may seem biologically implausible because neurons
do not wait for each one to fire in an orderly queue, but it avoids
the critique of an artificial ``synchronising mechanism'' in
synchronous models\cite{inman-harvey}. With sequential updating the
time-step becomes just a way of taking a look at the lattice at
regular intervals (and taking measures). From this point of view, any
pulsing must be a natural property of the rule, the wiring, and time seen
as a series of events.  Amazingly, it turns out that pulsing continues
when subject to sequential updating (with or without constraints on duplication),
though with a reduced waveform\footnote{Rules with a
larger RW-waveform ($wl$ and $wh$) from the pulsing case studies 
(section~\ref{Pulsing case studies}) continued to pulse with sequential updating,
but the Beehive and Spiral rules with a small RW-waveform did not.}.
From this result, it is reasonable to conjecture that ``natural pulsing'' in the
sequential case is also the driver (though stronger) when updating is
synchronous.

Figure~\ref{Pulsing dynamics g3 rule, sequential} gives an example of
sequential updating without duplication between
time-steps. 
Figure~\ref{Pulsing dynamics g3 rule, partial order} shows
two examples of sequential updating within
partial order updating (with no constraint on duplication). Note that
a partial order size of exactly one results in completely
arbitrary sequential single-cell updates, with measures taken at each cell-update.
The pulsing waveforms (for rule $v3k7$ ``g3'') are
very similar in all cases. Re-randomising wiring at each time-step or even
at each cell-update makes no significant difference.

\section{Questions on the pulsing mechanism}
\label{Questions on the pulsing mechanism}

\noindent The reaction-diffusing approach in 
section~\ref{k-totalistic rules as reaction-diffusion systems} is promising,
and explains the need for three (or more values) --- two are not enough,
but questions remain.
Unravelling the CA glider mechanism itself, or predicting glider
dynamics from a rule-table, are still unresolved questions in complex
systems --- answers would shed light on the underlying principles of
self-organisation.  The mechanism of pulsing in the CA pulsing model
is similarly unresolved --- both phenomena are emergent.
\enlargethispage{2ex}
Gliders are mobile oscillating/repeating patterns in space and time, driven
by feedback mechanisms within and between the neighborhood outputs
surrounding the glider, and within the glider itself.  Randomised
wiring disperses and synchronises these feedbacks over the whole network
--- pulsing must be a consequence, where the 3-value densities in the
disordered pattern fall into a repeating rhythm. The mobility aspect of
well formed gliders are an essential ingredient, because experiment shows
that dynamics showing up as ``complex'' in an automatic
entropy-variability search\cite{Wuensche99,Wuensche05} but lacking gliders,
do not pulse. These patterns include dynamic patches, blinkers,
as well as mobile boundaries between ordered/disorderd domains.
Further work will be required to define a well formed glider,
and the significance of mobility to pulsing. 

Unconstrained random wiring in the CA pulsing model makes pulsing
inevitable, so that an inverse search would be possible --- filtering
glider rules from the complex rules by randomising the wiring looking
for pulsing, a process which could be automated. Non-glider rules lack
glider feedback resulting in non-pulsing stable disorder when wiring
is randomised.

Starting with localised wiring, whether a regular
CA (section~\ref{Freeing one wire from CA neighboroods})
or confined random wiring (sections~\ref{Localised random wiring} -- \ref{3D systems})
there is a transition to pulsing and enhanced pulsing strength and robustness
depending on the degree and reach of the random connections.

It may be of interest to consider the 
basins of attracton\cite{Wuensche92,Wuensche2016}, their topology,
of a deterministic (noiseless) CA pulsing model. 
As the network size is reduced, pulsing will eventualy stop when the dynamics
converge on a uniform point attractor (for example, all 0s). This suggests that
pulsing does not play out on an attractor cycle, but on long
transients leading to a point attractor (even in large systems), 
and that random initial states initiate short branches to the transients.

Further questions arise regarding the diversity of waveforms
and how they relate to CA glider rules and glider dynamics; why
re-randomising at each time-step --- the annealed model\cite{Derrida}
--- makes no significant difference to the general behaviour; the
mechanism whereby confining random wiring locally results in
patchy/spiral density waves, and the robustness of pulsing to noise and
asynchronous update (section~\ref{Asynchronous and noisy updating}) --- 
especially in the case of sequential updating.

Here we have presented and documented the phenomena, listed some
unresolved questions, and provided tentative ideas on how to approach
answers rather than attempting firm explanations.

\section{Relevance to bio-oscillations}
\label{Relevance to bio-oscillations}

\noindent Pulsing --- sustained periodic oscillations --- are
ubiquitous in many dynamic bio-cellular processes based on collective
network behaviour, at a variety of scales in both time and space,
from cycles in gene expression to the rhythm of the beating heart.
Some tentative models of bio-oscillations have been suggested: reaction
diffusion\footnote{We note that pulsing from de-localising the connectivity in chemical
excitable media has been previously reported in the
Belousov-Zabotinsky Reaction (BZR) though it is not clear the significance
was recognised at the time. The BZR in a complex chemical reaction-diffusion
system (section~\ref{k-totalistic rules as reaction-diffusion systems}) 
with more than 20 chemical reactions, time delays and the
autocatalytic accumulation of HBrO$_2$. Spirals in 2D
and 3D gels are converted to whole system oscillations in solution
when stirred. Stirring presumably simulates the conversion of local to
non-local connectivity, re-randomised at each 
time-step as in Derrida's quenched model\cite{Derrida}.
In this chemical model pulsing frequency can be altered by temperature
and concentration, and maintained with a constant infusion of reagents\cite{BZ}.}, 
Hopfield networks, and attractors in discrete dynamical
networks\cite{Wuensche92,Wuensche2016}.

The CA pulsing model described in this paper, where
randomised wiring is applied to 3-value k-totalistic CA with
emergent glider dynamics, is arguably relevant to bio-oscillations,
and may serve as a model that provides pointers to the bio-mechanisms.
Oscillations can be found in all forms of life, but we focus here on
mammalian biology, and aspects of human physiology where
oscillations play a crucial role.

Clusters of excitable tissue which exhibit oscillatory behaviour
include but are not limited to:

\begin{s-itemize}
\label{osc_clumps}
\item neurohormonal systems,
\item synchronised uterine contractions,
\item the Sinoatrial Node generating the heart rate,
\item the atria and ventricular chambers of the heart,
\item Central Pattern Generators (CPGs) of the brainstem and spinal cord,
producing the following patterns of neural activity:

\begin{s-itemize}

\item the sympathetic centre in the Rostral-Ventro-Lateral Medulla
(RVLM) controlling sympathetic tone, the size of the vascular space,
venous return and hence cardiac output and its distribution.

\item the pre-B{\"o}tzinger cluster of interneurons in the ventral
respiratory centre of the medulla controlling the respiratory period.

\item the CPGs in the spinal cord underlying rhythmic motor
behaviours such as walking, swimming, and feeding.
\end{s-itemize}

\item and the basic cortical building block the
microcolumn\cite{Mountcastle} and hence perhaps the basic
physiological building block of brain function.

\end{s-itemize}

It is proposed that non-localised network connectivity combined with
biological processes similar to the glider rules described may have
been favoured by evolution for the generation of robust biological
oscillations due to the following functional advantages:

\begin{s-itemize}

\item the waveform (as defined in section~\ref{Introduction}) 
is dependant upon the rule of communication.

\item the waveform is independent of the exact wiring of
the network, i.e. random within constraints 
(sections~\ref{Freeing one wire from CA neighboroods} -- \ref{3D systems}).

\item the waveform and its phase are robust to noise, perturbation and variable transmission
(section~\ref{Asynchronous and noisy updating}).

\item there is a phase transition between disorganised (absence of waveform)
and organised (presence of a waveform) behaviour, which
occurs at a threshold of network connectivity radius relative to the
size of network (section~\ref{Localised random wiring}).

\item if connectivity radius is above threshold, the waveform is 
robust to changes in the network size,
and robustness is enhanced by increasing the radius. This built in redundancy
affords physiological reserve.

\end{s-itemize}

In theory, this results in randomly connected masses capable of
robust oscillatory behaviour in the presence of noise. The waveform
can be modified by changing the rule. Oscillatory behaviour can be
turned off and on by alterations in functional connectivity alone.
The period of oscillation can be increased and decreased
(section~\ref{The speed of a biological process}).  As a result
there is potential to store information between weakly coupled robust
controllable oscillators that is not present in non-robust oscillators
in noisy systems.

We find it significant that the above behaviour is emergent from a
simple computational model with minimal conditions. The system
requires 3 or more states, a glider rule, non-local
connectivity, a fixed number of connections, and to be thermodynamically
open. No time delay for connection distance has been
included. Periodic boundary conditions are not required.

For this model to be applicable to biological systems the following
biological equivalents are required to exist within each system:

\begin{s-itemize}

\item a biological unit with 3 (or possibly more) biological
states. Traditionally in excitable tissue these states are:
Firing (F), Refractory (R), and Ready to Fire (RF) --- (F.R.RF).

\item a biological process that has similarities to a rule with
glider behaviour for moving between these states.

\item a biological mechanism for non-local connectivity between the units.

\item and to provide variability in period and the ability to turn
oscillations off and on:

\begin{s-itemize}

\item a biological mechanism of speeding up and slowing down or even
halting the biological process underlying the rule.

\item a method of altering biological connectivity, be it functional
(short term) or structural (long term).

\end{s-itemize}
\end{s-itemize}

We entertain the following questions/possibilities which will require
further research:
Do these biological equivalents exist within biological excitable
tissues? And if they do, does the inheritance of the above properties
minimise the structural requirements of a system to fulfil its
function?  Or in other words can sophisticated behaviour be
constructed from clusters of non-locally randomly connected glider
rule system equivalents?  These could be coupled in phase by
excitatory connections, coupled out of phase by inhibitory connections
or even non-locally coupled in a random way by a glider rule system.

\section{Modeling bio-oscillations}
\label{Modeling to bio-oscillations}

CA glider rules are of interest
in modeling excitable biological media as they possess the following
similarities: by definition both gliders and action potentials are
patterns of state change that pass through a point in a medium which
after its passage is left unchanged,  they have a defined period and
form, they can be produced spontaneously and can annihilate each
other.  We propose possible additional similarities observed with the
non-localising of connectivity: the production of oscillations and the
resultant emergant properties of this system
(see section~\ref{Relevance to bio-oscillations}).

Traditionally neurones and myocytes have 3 states - (F.R.RF).  In
muscle the firing (F) state (contraction) results from increased
intracellular Calcium concentrations.  
Biological processes for moving between these states are
Membrane depolarisation/repolarisation (MD/R) 
by Voltage Gated Ion Channels (VGIC) and Calcium induced
Calcium Release (CICR), and Calcium re-uptake from intercellular stores
by the Ryannodine Receptor and SERCA respectively\cite{CaSPARKS}.
In neurones the firing (F) state is
primarily associated with MD.  MD can be spontaneous or result from
post synaptic integration of Post Synaptic Inhibitory and Excitatory
Potentials (PSIP, PSEP).  Presynaptic neurotransmitter release is a
result of increased intracellular Calcium in association with MD.

This 3 state interpretation may place restrictions on CA rules 
representing biological processes as by definition a cell cannot 
move between a Refractory and Firing state. An alternative is to equate
each state to the Nernst potential for an ion.  The Nernst Potential
is the voltage a cell membrane will move towards if membrane channels
allowing conductance of that ion are opened.
This results in 3 (or more) Voltage ``Rails'' in excitable tissue\footnote{Additional
states can be introduced for additional ion channels.
The $K^{+}$ $\approx$ -60 mV Rail could be differentiated from the RMP 
	and in neuronal systems a $Cl^{-}$ Rail at $\sim$ -80 mV could be introduced.},
so in the CA pulsing model a high density of the values 0, 1 and 2
would represent:

\begin{s-enumerate}  
\setcounter{enumi}{-1}
\item the Resting Membrane Potential (RMP) 
or most VGIC in the closed state.
\item a high density of open sodium VGIC or the $Na^{+}$ Rail $\approx$ +70 mV. 
\item a high density of open Calcium VGIC or the $Ca^{2+}$ Rail $\approx$  +120 mV.
\end{s-enumerate}

Cells in the model, dependant on scale, can represent the density of ion
channel opening in a membrane, or the membrane potential at membrane,
cellular or grid levels.

Examination of the rule-table and its input-frequency histogram
indicate the network trends through the rule-table in a series of
steps as a result of density fluctuations before returning close to
its starting configuration at period 
(section~\ref{k-totalistic rules}, figure\ref{input-histograms}).
Experimentally,  pulsing still occurs when random noise is
introduced to the deterministic system 
(section~\ref{Asynchronous and noisy updating}). 
The phase of pulsing is unaltered by randomising
wiring at each time-step. This indicates that if noise introduced to
the system is less than the density difference between time-steps,
the system will continue to pulse. The emergent pulsing behaviour is
essentially independent from, and insensitive to, initial conditions
and a degree of noise.

How this model relates to biological processes such as robust
oscillations produced by positive and negative feedback and a time
delay\cite{cao}, and other questions from 
section~\ref{Questions on the pulsing mechanism}, should be further investigated. 
Below we suggest some oscillatory physiological systems where the CA
pulsing model and its inherent properties, because of its diversity
of waveforms, might be usefully applied.

\subsection{Myometrial contractions}
\label{synchronised myometrial contractions}

Braxton Hicks contractions transition to synchronised uterine
contractions called labour. It is known that labour and pre-term
labour is associated with increased Gap Junction density.  Gap
Junctions electrically couple adjacent cells at random points in the
cell membrane.  Sophisticated computer models of the uterus report a
progression from disorganised to a single organised oscillation,
without centralised coordination as cell-cell connectivity
increases\cite{TERM}. The ability to produce global contractions is
enhanced by introducing spatial heterogeneity of
connectivity\cite{myometrial_heterogeneity}. A connectivity threshold
is observed.

This phenomenon has been generalised in the presented model.
Global synchronised oscillation appears dependant on non-localised 
connectivity for all glider rules. 
In this case, crossing the connectivity threshold likely induces
labour (section~\ref{Relevance to bio-oscillations}). 
Clinically controlling the period (section~\ref{The speed of a biological process})
can maintain labour while preserving 
uterine relaxation time and foetal oxygen delivery.

\subsection{The cardiovascular system and heartbeat}
\label{The cardiovascular system and heartbeat}

The heart could be acting as a possible instance of the CA pulsing
model operating in a real biological system. All myocardial cells in
the atria, ventricles, and conducting system, have the potential to
periodically fire by CICR, spontaneous membrane depolarisation, or
global membrane depolarisation, producing morphologically
identifiable etopic beats.  This indicates these systems over multiple
scales have similar innate and entrained periodicities and confer the
system considerable robustness.  The heart is driven primarily by an
anatomically poorly defined group of ``pacemaker'' cells called the
SinoAtrial Node (SAN). These cells have no Sodium VGIC, which could be
thought of as a modified rule.

Connectivity is effectively non-localised within constraints 
(section~\ref{Localised random wiring}) across multiple levels as follows:

\begin{s-itemize}
	
	\item the T-Tubular network non-locally connects the Cell Membrane to the
          Sarcoplasmic Reticulum (SR)\cite{TTUBE}.
	
	\item the Gap Junction network connects the cytoplasm of adjacent
          myocytes. Connectivity varies with Connexin pore
          size\cite{connexin}.
	
	\item Myocytes are arranged in a non-grid like way as they
          themselves are of differing length. There are a large number
          of other cells present\cite{I-D}.
	
	\item the ventricle requires an additional non-localised
          network to synchronise its greater mass called the Purkinje
          system. It consists of longer myocytes with less resistant,
          faster Gap Junctions (section~\ref{Localised --- but freeing one wire}).
	
\end{s-itemize}

\noindent Arrhythmias or the breakdown of the normal heart rhythm
can be categorised as disorders of the rule, the absolute number of
myocytes or connectivity, as follows:

\begin{s-itemize}
	
\item rule changes include channelopathies, acute ischaemia and severe
  electrolyte abnormality. The Vaughan-Williams classification classifies
  anti-arrhythmic medications by alterations to the rule.
	
\item loss of SAN cell mass with age typically results in alterations
  in heart rate variability (section~\ref{Pulsing case studies})
  before pacemaker failure. Larger Atrial and Ventricular chambers are
  more difficult to synchronise, increasing the risk of fibrillation
  and flutter.  Rotors can be seen in non-pulmonary vein Atrial
  Fibrillation (section~\ref{Localised random wiring}).
	
\item loss of myocyte connectivity (scarring, fatty and fibrotic
  deposits, connexin changes, inflammatory mediators, alterations in
  gene expression etc\cite{FAT}) increase the risk of arrhythmias and
  sinus node disease.  Peptides enhancing gap junction function and
  myocardial cell communication are currently under investigation as a
  new class of anti-arrhythmic drugs\cite{connexin}.

\end{s-itemize}

\subsection{The speed of a biological process}
\label{The speed of a biological process}

\noindent It is worth contrasting the steady rhythm of 
the CA pulsing model, and the period of the underlying biological oscillation it
may represent. The frequency of the bio-oscillation can change without changing the
underlying rule equivalent by altering the absolute value of the time-step and
hence period. From the CA point of view there would still be the same
number of time-steps as in a repeating waveform.  This is not modelled at
present but would become relevant when modeling two or more
biological processes with non identical rate change.

Altering the speed of a biological process is subject to
regional variance.  By way of example, release of Acetyl Choline (ACh)
from the Vagus nerve on the SAN opens GPCR Coupled Potassium leak
channels, slowing the rate of spontaneous depolarisation towards VGIC
threshold. It is unlikely the concentration of ACh and the effect of
Potassium conductance in every myocyte in the SAN is identical, yet the
heart rate slows without becoming disorganised.  Perhaps
non-localisation of connectivity, and the synchronisation it affords,
validates the extension of process change at the channel or membrane
level to the behaviour of an excitable tissue mass as a whole.

\subsection{The Central Nervous System}
\label{The Central Nervous System}


\noindent The Central Nervous System is capable of acquiring new
patterns, and reproducing them in either the long term (for example
movement/memory) or in the short term (working memory).  Neurons are
organised into functional units (CPG, brainstem nuclei, Microcolumns)
by local connectivity (dendrites, short axons, gap junctions and
synapses) with long range connectivity between units (long axons).  It
is estimated we each have $\approx2\times10^8$
microcolumms comprising $\approx100$ neurones each.

Functional connectivity is determined by neurotransmitter synaptic and
dendritic release and post synaptic integration of Inhibitory and
Excitatory Post Synaptic Potentials (IPSP and EPSP).  Regional
functional connectivity can be altered by the
neuromodulators via GPCR. Oscillatory behaviour is
thought to emerge as IPSP's are of longer duration than
EPSP's\cite{LILEY_PBM_CDR}. However non-local connectivity may be a key
component of the ``fast'' component of a fast/slow biological system
and the mean field model.  Connectivity effects on these units
are well modelled in DDLab by a 3D system with restrained random wiring with
released wires (section~\ref{3D systems}).

\subsection{Central Pattern Generators}
\label{Central Pattern Generators}

\noindent Central Pattern Generators (CPGs) are neuronal clusters that produce:

\begin{s-itemize}
	
\item Permanent Oscillations: The pre-Botzinger cluster or respiratory
  pacemaker can produce slow breathing, sniffing and gasping patterns
  dependant on input. The corresponding rhythm is projected throughout
  the cortex by noadrenergic neurones.

\item Driven Oscillations: Oscilations of the sympathetic centre in
  the RVLM are driven (at a shorter period than its own) by pulses of
  inhibition from the arterial waveform, (ie. carotid pressure sensors
  via a GABA-A interneuron) which then rebounds to its set point.
  Sympathetic nervous system abnormalities, both in set point and
  integration, are crucial in the development of cardiovascular
  disorders such as Heart Failure, Essential Hypertension and Postural
  Syncope\cite{Esler}.

\item Controlled Oscillations: Locomotion CPGs in the spinal cord are turned
  on and off by Neuromodulation.
	
\item Coupled Oscillations: Patterns of movement are stored as
  associations between GPG's and their respective muscle groups.  In a
  similar fashion it may be possible to store patterns or information
  between associated microcolumns.
	
\end{s-itemize}
The evolution of rhythmic behaviours in the invertebrate and simple
vertebrate reveals repeated building blocks such as two mutually
inhibitory half centre oscillators\cite{Katz}. However the mammalian
CPG is essentially a black box\cite{Guertin} due to the vast numbers
of neurones and their associations, and the difficulty of
obtaining simultaneous readings of their electrical activity.
Without models the situation may remain that way for some time.

The CA pulsing model demonstrates significant biological emergent
properties: sustained rhythmic oscillations, a threshold effect,
redundancy and robustness to noise. Thus the question arises whether
the pre-Botzinger cluster, the sympathetic centre in the RVLM, and
even motor CPGs need to be anything other than masses of pseudo
randomly connected neurones with associations mapped to anatomical or
physiological features, or other CPGs.

\section{Discussion}
\label{Discussion} 


\noindent To explain the pulsing mechanism is as hard as explaining the
mechanism whereby a CA rule acting on regular CA wiring is able to
generate glider dynamics.  Although its possible to discriminate
between the extremes of ordered and disordered dynamics from the
rule-table, by the $\lambda$ and $Z$ parameters\cite{Langton,Wuensche92,Wuensche99},
the link between the
CA rule and glider dynamics is still an open question, going
to the heart of the underlying principles of self-organisation. The
mechanism of the glider and glider-gun, with its many delicate
feedback loops, is also difficult to untangle.

What seems to be apparent is that for 3-value glider CA (but not
binary, \mbox{2-value)} fully random wiring makes pulsing inevitable.
Pulsing could be regarded as temporal order emerging from the
disordered patterns driven by the randomly connected CA.  Starting
with localised wiring, there is some sort of phase transition to enhanced pulsing
strength depending on the degree and reach of the random connections.

The diversity of pulsing waveforms in glider CA with random wiring may
provide models and insights into bio-oscillations in nature.
Many attributes of the CA pulsing model are reflected in oscillatory
behaviour in mammalian tissue such as the heart and central nervous system.
The model provides a classification system for oscillations in biological
systems, their formation and their breakdown according to the
(biological) rule, network size, and connectivity relative to threshold.

In this paper we have introduced and documented the pulsing
phenomena and listed the issues that require explanation. Further
systematic research and experiment is required to properly investigate
the range and scope of pulsing, its mathematical and logical
properties, the mechanisms that drive it, and its biological
significance.

\section{Acknowledgements}
\label{Acknowledgements}

Experiments and figures were made with DDLab (\url{http://www.ddlab.org/}) ---
where the rules and methods are available, so repeatable\cite{Wuensche2016}.

Thanks to \mbox{Inman Harvey} for conversations regarding asynchronous updating,
to \mbox{Terry Bossomaier} for exchanges regarding phase transitions,
and to \mbox{Paul Burt} and \mbox{Muayad Alasady} for comments regarding bio-oscillations.

\end{document}